\documentclass[a4paper,12pt]{article}
\pdfoutput=1
\usepackage{macros}

\title{A Concise Course in String Theory}
\date{}

\author{Satoshi Nawata, Runkai Tao, Daisuke Yokoyama}

\begin{document}
\maketitle
\setcounter{tocdepth}{2}
\abstract{These lecture notes offer a concise introduction to string theory, based on a
one-semester graduate course taught at Fudan University, and develop the subject from the string sigma model action to the AdS/CFT correspondence. The exposition is deliberately compact,
and is addressed to graduate students who seek a coherent overview of the foundations of
the subject rather than an exhaustive treatment.}

\vspace{2cm}

\begin{figure}[ht]\centering
  \includegraphics{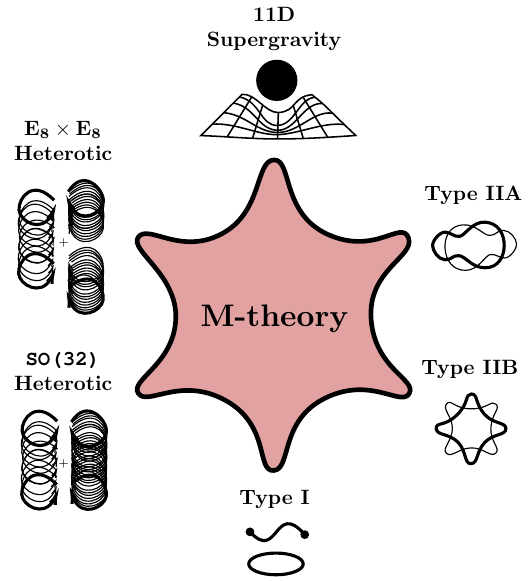}
\end{figure}

\pagebreak
\tableofcontents

\pagebreak

\section{Motivation and Overview}\label{sec:motivation}

\subsection{Why do we study string theory?}

$\bullet$ \textbf{For quantum gravity}

One of the major problems in theoretical physics is to provide a unified description of all the forces in Nature. The standard model has unified electromagnetic, weak and strong force based on quantum field theories whereas general relativity for gravity is formulated within classical physics. A quantum theory of gravity is needed to reconcile general relativity with the principles of quantum mechanics. However, it is known that the renormalization procedure does not cure ultraviolet divergence for gravity. Therefore, naive quantization of general relativity does not give a consistent theory.

String theory not only yields the first quantization of gravity but also resolves the renormalization problem of gravity by replacing point particles by vibrating strings.  To our knowledge, string theory is currently the only theory that has this property. Once we discover a candidate to unify gravity and quantum mechanics of all the forces, it is inevitable to try to understand it as well as we can although there is never any guarantee we can achieve.

Indeed, string theory has given deep insights into quantum gravity. The Bekenstein-Hawking formula \cite{Bekenstein:1973ur,Bardeen:1973gs,Hawking:1974sw} was microscopically derived by counting D-brane states with fixed mass and charge for certain (called extremal) black holes in string theory \cite{Strominger:1996sh}. The AdS/CFT correspondence \cite{Maldacena:1997re} conjectures that non-perturbative definition of string theory on AdS background is described by conformal field theory, and it partially resolves the Hawking information paradox of black holes \cite{Hawking:1976ra}. Interestingly, a generalization of the Bekenstein-Hawking formula called the Ryu-Takayanagi entanglement entropy formula \cite{Ryu:2006bv}, connects quantum gravity and quantum information theory, which is actively studied in recent years. These developments certainly pose questions to basic concepts of spacetime.

\vspace{.3cm}

\noindent$\bullet$ \textbf{Rich arena for physics theories}

String theory has had a significant impact on our understanding of established physical theories, ranging from proof of positive energy to quark confinement to quantum black holes to quantum information and more. There are several reasons for this.

First, string theory is capable of generating a vast number of quantum field theories (QFTs) in various dimensions. There are five types of string theories - Type I, IIA, IIB, Heterotic $\SO(32)$, and $E_8\times E_8$ - and M-theory is the limit of large dilaton expectation value in Type IIA string theory. These theories can contain various types of branes, such as D$p$-branes in Type II theories and M2 and M5-branes in M-theory. While there are only finitely many string theories, they can be used to "engineer" an infinite number of QFTs depending on the manifolds on which they are defined and the configurations of branes present.

Second, string theory is connected by a web of dualities, which are equivalences between different descriptions of a theory at the quantum level. The AdS/CFT correspondence is a particularly well-known example of this. These dualities can provide alternative perspectives on established physical theories and have played a crucial role in our understanding of them.

However, recent developments in the study of M5-branes have revealed that there are vast families of QFTs that do not admit a Lagrangian description and are inherently strongly coupled, making it difficult to use traditional QFT techniques to study them. Currently, we are attempting to understand these QFTs on a case-by-case basis using techniques such as dualities, dimensional reduction on certain manifolds, RG flows, and perturbation theory. However, there is no unified framework for describing these QFTs. This motivates the search for a new approach to understanding QFTs without a Lagrangian description.

%

\vspace{.3cm}
\noindent$\bullet$ \textbf{For mathematical structure}

String theory has garnered significant interest due to its ability to provide new insights and perspectives on mathematics, particularly geometry. This is because string theory has repeatedly given rise to new ways of understanding geometry as a natural consequence of seeking to understand physical theory. Some notable examples include mirror symmetry \cite{Greene:1990ud,Candelas:1990rm}, Seiberg-Witten invariants \cite{Witten:1994cg}, and the AGT relation \cite{Alday:2009aq}. In addition, string dualities have led to highly non-trivial conjectures that connect seemingly unrelated areas of mathematics.

One reason why string theory has had such a profound impact on mathematics is that it is not yet fully understood. While Einstein's theory of gravity was constructed based on Riemannian geometry, the geometric foundations of string theory are still not fully understood. However, physicists can develop new insights into geometry because physicists stumble upon a profound theory we do not understand well. There are certainly many more mathematical secrets to be uncovered within string theory.

\vspace{.3cm}
\noindent$\bullet$ \textbf{Because we do not know what it is}

As previously mentioned, one of the reasons that string theory is a fascinating topic for students is that it is based on a concept that is not yet fully understood. This can sometimes lead to criticism that string theorists do not understand the theory. However, it is also true that if we did fully understand string theory, it would likely lead to major breakthroughs in both physics and mathematics. The fact that so much is still not understood and that even relatively small pieces of progress can be significant discoveries in their own right makes working on string theory an exciting prospect. Of course, there is still a lot of work to be done in this field.

\subsection{Very very brief history}

String theory was initially developed by accident in trying to solve a different problem, subsequently developed by a long and fortunate process of tinkering. Therefore, the history of string theory itself is fascinating. For detailed history of string theory, we refer to \cite{Greene:1999kj,Schwarz,Schwarz:2011ona,cappelli2012birth,Ooguribook,Polchinski:2017vik}.

String theory was initially developed in an attempt to describe hadron physics, with the idea that mesons could be modeled as strings with charges at their ends and meson resonances as vibrational states of these strings. Although this physical picture, which first emerged from the Veneziano amplitude \cite{Veneziano:1968yb}, is now believed to be qualitatively correct at a description of strongly interacting particles, it was eventually surpassed by quantum chromodynamics (QCD) in explaining the details of strong interactions.

However, a small group of physicists continued to work on string theory in the 1970s and discovered that it contained massless spin-two particles, suggesting that it could potentially be used as a theory of gravity. Additionally, the incorporation of supersymmetry led to the removal of tachyons in bosonic string theory and the development of Type I, IIA, and IIB string theories.

In 1984, Green and Schwarz demonstrated \cite{Green:1984sg,Green:1984qs}  that the anomaly pointed out by Alvarez-Gaumé and Witten could be canceled by setting the gauge group of Type I string theory to be $\SO(32)$, marking the beginning of the first string theory revolution. This resulted in a surge of interest in the field and the development of Heterotic string theory. 

In 1995, Witten proposed  \cite{Witten:1995ex}   that the five known types of string theory were all different limits of a single theory known as M-theory, leading to the second string theory revolution. During this time, D-branes were introduced and non-perturbative aspects of string theory were extensively studied, resulting in the discovery of various dualities, a better understanding of the quantum nature of black holes, and the development of the AdS/CFT correspondence.

The second string theory revolution has led to increased research on the world-volume theories of D-branes and their connections to quantum gravity. However, the dynamics of M2-branes and M5-branes, which are fundamental degrees of freedom in M-theory, have remained a mystery. Triggered by the work of Bagger-Lambert \cite{Bagger:2006sk}, the world-volume theory for M2-branes was found to be a highly supersymmetric exquisite mixture of Chern-Simons gauge theory with some scalars and spinors \cite{Aharony:2008ug}. In contrast, the world-volume theory for M5-branes is challenging to analyze due to its lack of a Lagrangian description. However, the work of Gaiotto \cite{Gaiotto:2009we} has sparked ongoing research on the properties of M5-branes by studying them when wrapped on certain manifolds.

The following picture shows the timeline of string theory and when the standard textbooks were written. Of course, there are many other books \cite{Zwiebach:2004tj,johnson2006d,Dine:2007zp,Ibanez:2012zz,Blumenhagen:2013fgp,kiritsis2019string,tomasiello2022geometry,cecotti2023introduction} and notes \cite{Uranga,Tong:2009np,wray2011introduction,Hosomichi,Weigand}. Obviously, the textbooks do not cover developments after their publications. We will mainly follow Polchinski's textbook \cite{Polchinski}, and the reader can find more elaborate explanations in the textbooks listed above. However, I do not recommend you stick to only one book when you study a subject. Some parts are explained very well in a book, but some parts are sometimes hard to grasp for some people. Therefore, it is always a good idea to look for books, notes, and papers that suit you best.
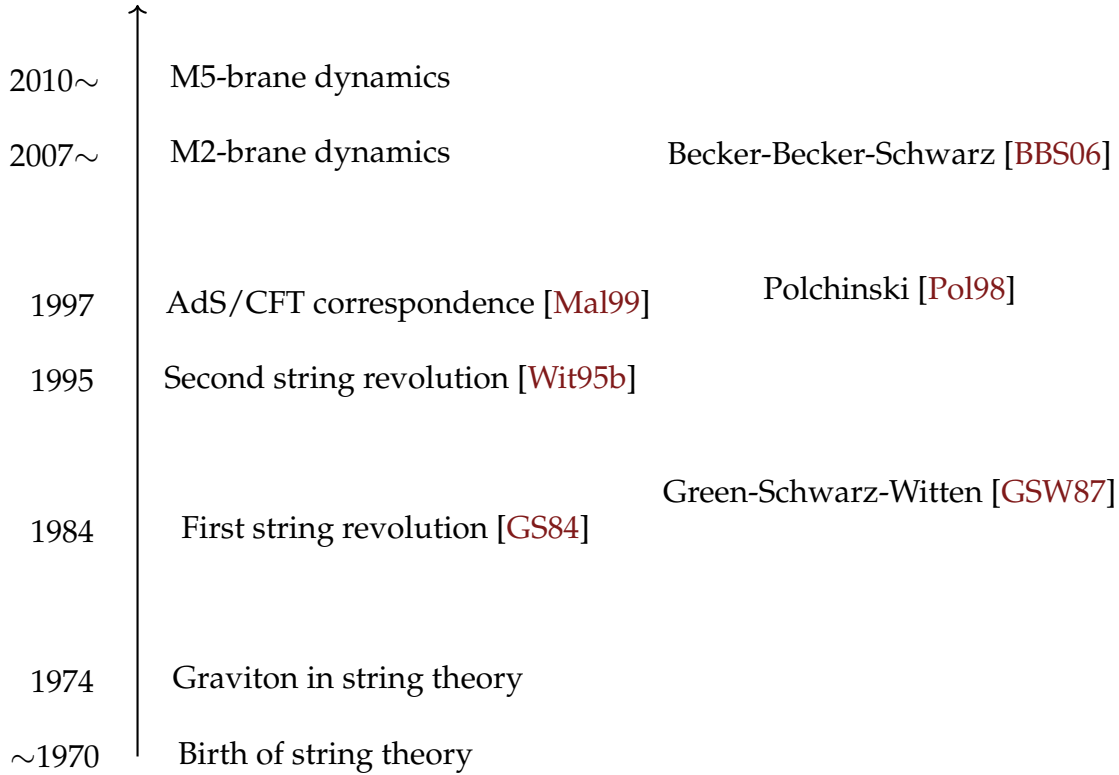
\begin{figure}[ht]\centering
\begin{tikzpicture}
\draw[->,thick]  (0,0) to (0,10);
\node at  (2.3,9) {M5-brane dynamics};
\node at  (-1.1,9) {2010$\sim$};
\node at  (10,8) {Becker-Becker-Schwarz \cite{BBS}};
\node at  (2.3,8) {M2-brane dynamics};
\node at  (-1.1,8) {2007$\sim$};
\node at  (10,6.2) {Polchinski \cite{Polchinski}};
\node at  (3.6,6) {AdS/CFT correspondence \cite{Maldacena:1997re}};
\node at  (-1,6) {1997};
\node at  (3.5,5) {Second string revolution \cite{Witten:1995ex}};
\node at  (-1,5) {1995};
\node at  (10,3.5) {Green-Schwarz-Witten \cite{GSW}};
\node at  (3.3,3) {First string revolution \cite{Green:1984sg}};
\node at  (-1,3) {1984};
\node at  (2.8,1) {Graviton in string theory};
\node at  (-1,1) {1974};
\node at  (2.5,0) {Birth of string theory};
\node at  (-1.1,0) {$\sim$1970};
\end{tikzpicture}
\caption{History of string theory and textbooks}
\end{figure}

\subsection{Very short highlights}

$\bullet$ The basic idea in string theory is that different elementary particles are all vibrational modes of a single type of string.  There are two types of strings, open and closed, and a trajectory of a string is called the \textbf{string world-sheet} which is parametrized by $(\sigma,\tau)$.
If the typical size $\ell_s$  of a string is smaller than the resolution that an accelerator can provide, we cannot see this in the experiment involving elementary particles.
$$
(\textrm{Planck scale})\  10^{-33}\;\textrm{cm}\le \ell_s\le 10^{-17}\;\textrm{cm} \  (\textrm{TeV scale})
$$
We often use the parameter
$$
\a'=\ell_s^2
$$
which is the only free parameter in string theory. As we will see, the couplings in string theory are expectation values of dynamical fields (so-called moduli) which take their value dynamically.
\begin{figure}[ht]\centering
\includegraphics[width=10cm]{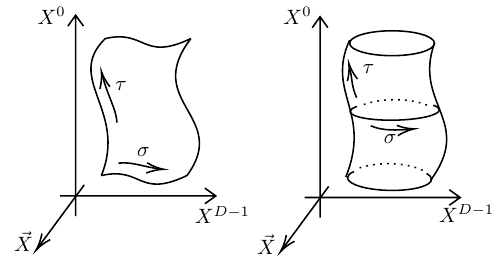}
\end{figure}

\vspace{.3cm}
\noindent $\bullet$ Application of the quantization rules provides us with the Fock space of string excitations. In bosonic string, the massless modes include among others
$$
\begin{array}{c@{\quad }c@{\quad }c@{\quad }c}
\textrm{open string}: & A_\mu & \textrm{spin 1}& \textrm{W-boson}\cr
\textrm{closed string}: & g_{\mu\nu} & \textrm{spin 2}& \textrm{graviton}
\end{array}
$$
In addition, one finds a tower of massive string excitations of mass
\begin{align}
\textrm{open bosonic}: &\quad M^2= \frac{1}{\a'} (N-1) \quad  \cr
\textrm{closed bosonic}: &\quad  M^2 = \frac{4}{\a'} (N-1)  \quad N = 0,1,2,\ldots \nonumber
\end{align}
Note that the lowest-lying state has a dimension of negative [mass]${}^2$ $(N=0)$, which is called \textbf{tachyon}. The bosonic theory is consistent only when the number of spacetime dimensions is 26. We can also refer to \cite{witten2018every}.

\vspace{.3cm}
\noindent $\bullet$ In the point-particle picture, the divergence appears when all four vertices come close to each other. On the other hand, the string world-sheet has no vertices. Thus, when we sum over all surfaces, we do not encounter configuration analogous to collapsed vertices. String theory amplitudes have no ultraviolet (short distance) divergence.  As a result, string theory provides a finite quantum theory of gravity. Moreover, this is even better than renormalizable quantum field theories since there is no divergence in the first place.
\begin{figure}[ht]\centering
\includegraphics[width=4cm]{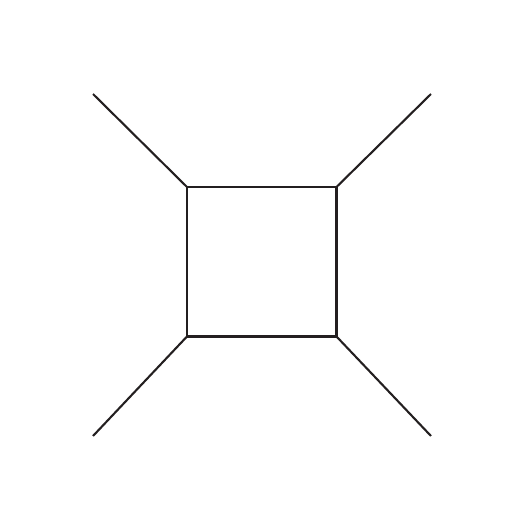} \hspace{2cm}\includegraphics[width=4cm]{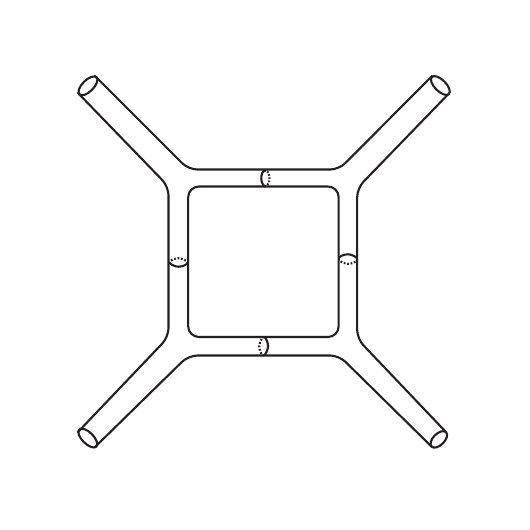}
\end{figure}

\vspace{.3cm}
\noindent $\bullet$
Strings, besides vibrating in usual spacetime, can also have some internal degrees of freedom. These internal degrees of freedom are fermionic. This can be quantized in a similar manner to that for bosonic strings. Then, a superstring can be considered as a string with symmetry between bosonic states and fermionic states. In superstring theories, the good features (e.g. emergence of gravity and UV finiteness) are preserved, and one can incorporate fermions. Moreover, as mentioned,  the tachyonic mode is absent.
\begin{enumerate}
\item the theory is consistent in 10 spacetime dimensions.
\item there are five fully consistent string theories in $d=10$.

Type IIA, Type IIB, Type I (open+closed string), $E_8\times E_8$ Heterotic, $\SO(32)$ Heterotic

\item Type II theories can have D-branes that accommodate open strings. Type IIA (resp. IIB) has D$p$-branes where $p$ is even (resp. odd), and Type I does D9, D5, D1-branes.

\item Furthermore, all the five apparently different string theories are different limits of the same underlying theory called  M-theory

\item In the low energy scale $\ll1/ \ell_s$, a theory is described by supergravity which combines supersymmetry and general relativity.

\end{enumerate}

\vspace{.3cm}
\noindent $\bullet$  How do we reduce the number of dimensions from 10 to 4? The answer is Kaluza-Klein mechanism/compactification.
If we take the 10-dimensional space of the form $\bR^{1,3}\times M$ where $M$ is a 6-dimensional compact space and its size is much smaller than the resolution of the most powerful accelerator $\sim 10^{-17}$cm, then a (9+1)-dimensional theory looks like (3+1)-dimensional.  In fact, in order for a theory to be consistent, $M$ has to be a Calabi-Yau manifold (which is endowed with a Ricci-flat metric). More importantly, the extra dimensions give ``room'' to derive the complexity of the real world from a simple setting.

\subsection{Convention}

The world-sheet coordinate indices are denoted by the alphabet letter  $a, b,\ldots$, and the target-space coordinate indices are denoted by the Greek letter $\mu,\nu,\ldots$.

The world-sheet metric is denoted by $h_{ab}$.

\subsubsection*{spacetime}
In anticipation of
string theory, we consider $D$-dimensional Minkowski space
${\bR}^{1,D-1}$. Throughout
these notes, we work with signature
$$ \eta_{\mu\nu} = {\rm diag}(-1,+1,+1,\ldots,+1)$$

\subsubsection*{World-sheet}

We use $\sigma^a$ for a local coordinate of the world-sheet and we denote the metric by $h_{ab}$ in this local coordinate so that the measure is $\sqrt{h}d^2\sigma$ for a curved world-sheet.

We often consider the case that the world-sheet is topologically a cylinder or a plane.
The Minkowski cylindrical coordinates are denoted by $(\tau,\sigma)$ and the Euclidean cylindrical coordinates $(t,\sigma)$ where the Wick rotation is performed as $\tau=-it$. In the Euclidean coordinate, we introduce the holomorphic complex coordinate $w=it+\sigma$ for the cylinder. The conformal map brings the cylindrical coordinate to the plane coordinate by $z=e^{-iw}$.
\begin{itemize}
  \item right-moving or holomorphic coordinate: $\sigma^-=\tau-\sigma$ (Minkowski cylinder), $w=it+\sigma$ (Euclidean cylinder), $z$ (Euclidean plane),
\item  left-moving or anti-holomorphic coordinate: $\sigma^+=\tau+\sigma$ (Minkowski cylinder), $\overline w=-it+\sigma$ (Euclidean cylinder), $\overline z$ (Euclidean plane),
\end{itemize}

In the plane coordinate, the holomorphic coordinates are given by
  \begin{align}\nonumber
   z = \sigma^1 +i\sigma^2 \ , &\qquad  \ol z = \sigma^1 -i\sigma^2 \ \cr
   \partial \equiv \partial_z = \frac{\partial}{\partial z} = \frac{1}{2} \left( \partial_1 -i\partial_2 \right) \ , & \qquad
   \ol\partial \equiv \partial_{\ol z} = \frac{\partial}{\partial \ol z} = \frac{1}{2} \left( \partial_1 +i\partial_2 \right) \ ,
  \end{align}
The metric is given by
  \begin{align}\nonumber
   d^2z = dzd\ol z = 2 d \sigma^1 d \sigma^2  \ ,
  \end{align}
       where the factor $2$ comes from the Jacobian.
       One can also write
  \begin{align}\nonumber
   h_{z\ol z} = h_{\ol z z} = \frac{1}{2} \ , \quad h_{z z} = h_{\ol z\ol z} = 0 \ , \quad
   h^{z\ol z} = h^{\ol z z} = 2 \ , \quad h^{z z} = h^{\ol z\ol z} = 0 \ .
  \end{align}

\subsubsection*{Parameters and constants}
\begin{itemize}
  \item String theory has only one parameter $\alpha'$, and string length and tension are written as
\be
\ell_s=\sqrt{\alpha'}~, \qquad T=\frac{1}{2\pi\alpha'}
\ee
  \item String coupling is given by an expectation value $g_s=e^\Phi$  of the Dilaton field
  \item Planck length
  \be
  \ell_P=\sqrt{\frac{\hbar G}{c^{3}}}~, \qquad \ell_P^3=g_s\ell_s^3
  \ee
\item Coupling constants of $D=11$ and $D=10$ supergravities
\be
\frac{1}{2 \kappa_{11}^{2}}=\frac{2 \pi}{\left(2 \pi \ell_{p}\right)^{9}}~,\qquad \frac{1}{2\kappa_{10}^2} = \frac{2\pi}{(2\pi \ell_s)^8}
\ee
\end{itemize}

\subsection*{Acknowledgments and disclaimer}
We are grateful to the students at Fudan University who provided valuable comments on this note during the lecture series in 2017, 2021 and 2024. Special thanks go to Jiaqi Guo and Jack Yang for their assistance with creating the figures. We also thank Y. Tachikawa for bringing the historical account in \cite{cappelli2012birth} to our attention.

Given the vast literature on string theory, it is not possible to include all relevant citations in this note. Instead, we encourage readers to delve into the original sources for a more comprehensive understanding of the subject. As mentioned above, there are already many good books on string theory, and this note covers only a small portion of the vast field of string theory. Nevertheless, we would be delighted if it serves as a concise introduction for students.

\section{Bosonic string theory}\label{sec:bosonic}

The first few sections are devoted to the study of bosonic string theory. Bosonic string theory is a framework in which the point-like particles of particle physics are replaced by one-dimensional objects called ``strings''. These strings can vibrate at different frequencies, and each frequency corresponds to a different particle.

Strings can either be loops (closed strings) or have endpoints (open strings). Closed strings give rise to gravity (graviton) while open strings can give rise to gauge fields and matter. This provides a unified description where all particles arise as different states of the same fundamental entity. One of the notable outcomes of bosonic string theory is that it requires 26 spacetime dimensions for the theory to be consistent.

One of the problematic features of bosonic string theory is the existence of tachyons, particles with imaginary mass. The presence of tachyons signals an instability in the theory. 
Another limitation is that bosonic string theory does not include fermions. Later, superstring theory will come to address these problems, which incorporates both bosons and fermions. (See \S\ref{sec:superstring}.) Nonetheless, bosonic string theory plays a role in the construction of Heterotic string theory. (See \S\ref{sec:Heterotic}.)

In this section, let us first quantize bosonic string theory by using the light-cone gauge. 
For bosonic string theory, there is a concise and wonderful lecture note \cite{Tong:2009np} by D. Tong.

\subsection{String sigma-model action}\label{sec:string-sigma}
\subsubsection*{Nambu-Goto action}

\begin{figure}[ht]\centering
\includegraphics[width=4cm]{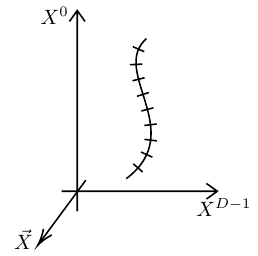}
\end{figure}

To begin with, let us recall the case of a relativistic point particle
\[
\gamma:\tau\to x^\mu(\tau)\in \bR^{1,D-1}~.
\]
The action of a relativistic point particle with mass $m$ is
\[
S=-m\int^b_ads=-m\int^b_a \Big( -\eta_{\mu\nu}\frac{dX^\mu}{d\tau}\frac{dX^\nu}{d\tau}\Big)^{\frac12}d\tau~.
\]
In a similar fashion, when a string moves through spacetime $\bR^{1,D-1}$, it traces out a two-dimensional surface known as the worldsheet, which is described by
\[
\Sigma\ni(\tau,\sigma)\to X^\mu(\tau,\sigma)\in \bR^{1,D-1}~.
\]
The analogous action, called the \textbf{Nambu-Goto action}, for the string is given by the area of a string world-sheet $\Sigma$ swept by a string
\be\label{NG-action}
S_{\textrm{NG}}=-T\int_\Sigma \textrm{Area}=-T\int_\Sigma d^2\sigma\sqrt{-\det (h_{ab})}~,\qquad h_{ab}=\eta_{\mu\nu}\partial_a X^\mu\partial_b X^\nu~
\ee
where $(\sigma^0,\sigma^1)=( \tau,\sigma)$, and $T:=1/(2\pi \a')$ is the \textbf{string tension}. The tension is usually defined as mass per unit length.
However, we do not know how to quantize the Nambu-Goto action due to the square root. Instead, for quantization of string world-sheet, we reformulate the action as follows.

\subsubsection*{String sigma-model action}
Alternatively, we remove the square root and consider the following action
\be\label{string-sigma}
 S_{\sigma}=-\frac 1{4\pi\a'}\int d^2\sigma \sqrt{-\det h}~ h^{ab}\partial_a X^\mu \partial_b X^\nu \eta_{\mu\nu}
\ee
which is called the \textbf{string sigma-model action}.\footnote{This action is called the Polyakov action in \cite{Polchinski}. However, according to \cite{BBS}, it was discovered by Brink, Di Vecchia and Howe and by Deser and Zumino several years before Polyakov skillfully used it for path-integral quantization of the string. Therefore, let us follow the notation of the progenitor, J.H. Schwarz.} Classically, the string sigma-model action is equivalent to the Nambu-Goto action.

\subsubsection*{Symmetry of $ S_{\sigma}$}
\begin{enumerate}
\item\textbf{$D$-dim spacetime Poincar\'e invariance}

The action is certainly invariant under the Poincar\'e transformation of the target space $\bR^{1,D-1}$
\[ X^\mu(\sigma)\to \L^\mu{}_\nu X^\nu(\sigma)+V^\mu \qquad \Lambda\in  \mathrm{O}(1,D-1)\]
where $\L$ is a Lorentz transformation and $V$ is the translation.

\item\textbf{World-sheet diffeomorphism invariance}

The string sigma-model is also invariant under a diffeomorphism $f:\widetilde \Sigma \to \Sigma$ of world-sheet two-dimensional manifolds. Namely, the action stays the same for $(\Sigma,h)$ and $(\widetilde \Sigma,\wt h=f^*h)$. In physics, it is most convenient to use local coordinates to express this invariance. Namely, under a local expression of the diffeomorphism
\begin{align}
\wt X^{\mu}(\wt\sigma)&=X^{\mu}(\sigma)\quad \textrm{with}\quad  \wt{\sigma}= \wt{\sigma}(\sigma)~,\cr
\wt h^{ab}&=h^{cd}\frac{\partial\wt\sigma^a}{\partial\sigma^c}\frac{\partial\wt\sigma^b}{\partial \sigma^d}~,\nonumber
\end{align}
the action is invariant $S_\sigma(X^\m,h_{ab})= S_\sigma(\wt X^\m,\wt h_{ab})$.
\item\textbf{Weyl invariance of world-sheet metrics}

A \textbf{Weyl transformation} $h\to e^{2\omega}h$ is a local scaling of the metric of a Riemannian manifold $(\Sigma,h)$ where $2\omega$ is an arbitrary function of $\Sigma$.
The action is indeed invariant under a Weyl transformation $h_{ab}\to e^{2\omega(\sigma)}h_{ab}$ of the string world-sheet $\Sigma$.

\end{enumerate}

In fact, Diff $\times$ Weyl are gauge symmetries of the theory.

\subsubsection*{Energy-momentum tensor}
The energy-momentum tensor is defined by
\begin{align}
T_{ab} \equiv&-\frac{4\pi}{\sqrt{-h}}\,\frac{\delta S}{\delta h^{ab}}\cr=&-\frac1{\a'} \Big[\partial_a X^\mu \partial_b X_\mu - \frac12h_{ab}\,\partial_c X^\mu \partial^c X_\mu\Big]
\end{align}
which is symmetric and subject to
\begin{align}\label{traceless1}
\textrm{traceless}:& \qquad T^a{}_a=0\cr
\textrm{conservation}:& \qquad \nabla_a T^{ab}=0
\end{align}
Note that the traceless condition indeed follows from the Weyl invariance (exercise), and the conservation can be shown by using the equation of motion for $X^\m$ in the following.
\begin{align}
h^{ab}\frac{\delta S}{\delta h^{ab}}=0&\ \to \ T^a{}_a=0\cr
\frac{\delta S}{\delta X^\mu}=0&\ \to \ \Delta X_\mu=-\frac{1}{\sqrt{-h}}\partial_a(\sqrt{-h}h^{ab}\partial_b X^\mu)=0~.
\end{align}
The equation of motion with respect to the metric will be discussed in \eqref{constraints}.

\subsubsection*{Gauge fixing}
The string sigma-model action $S_\sigma$ is equipped with the symmetries above, and we need to fix it for quantization. Physics does not depend on a choice of gauge fixing, but if we choose a clever gauge fixing, our life becomes much easier. As in \cite{GSW,Polchinski,BBS}, we choose the light-cone gauge in the following.

The metric $h_{ab}$ is a 2$\times$2 symmetric matrix and the world-sheet diffeomorphism invariance tells us that there is a coordinate $\sigma^a$ such that the metric is a diagonal $h_{ab}=e^\omega \eta_{ab}$. Then, the Weyl transformation brings it to the 2d Minkowski metric $\eta_{ab}$. Therefore, one can use $h_{ab}= \eta_{ab}$. If we use the light-cone coordinates on the world-sheet
\[ \sigma^\pm = \tau \pm \sigma\ ,\]
then the metric is written as
\[ds^2=-d\sigma^+d\sigma^-~.\]

However, there are residual transformations that leave the 2d Minkowski metric $\eta_{ab}$ invariant, which is called the \textbf{conformal symmetry}:
\[
\sigma^+\to f(\sigma^+) ~, \qquad \sigma^-\to g(\sigma^-)
\]
with a Weyl transformation simultaneously.
To fix the conformal symmetry, we introduce the spacetime light-cone coordinates,
\[ X^\pm = \sqrt{\frac{1}{2}}(X^0 \pm X^{D-1}) \ .\]
Then we impose
\be X^+ = x^+ + \a' p^+\,\tau   \label{lcg}~,\ee
which is called the \textbf{light-cone gauge}.

\subsubsection*{Mode Expansions}
After the gauge fixing, the equations of motion simply read
\[ \partial_{+} \partial_{-} X^{\mu}=0\quad \textrm{where} \quad \partial_{\pm}=\frac{1}{2}\left(\partial_{\tau} \pm \partial_{\sigma}\right)~.\]
The most general solution is a factorization of left- and right-mover
\be X^\mu(\sigma,\tau) = X^\mu_L(\sigma^+) + X^\mu_R(\sigma^-) \label{+-}\ee
for arbitrary functions $X^\mu_L$ and $X^\mu_R$. For a closed string, we impose a periodic condition as
\be X^\mu(\sigma,\tau)=X^\mu(\sigma+2\pi,\tau)\ .\ee
so that the left- and right-moving fields admit the Fourier expansions
\begin{align}\label{mode}
&X_{L}^{\mu}\left(\sigma^{+}\right)=\frac{1}{2} x^{\mu}+\frac{\alpha^{\prime}}{2} p^{\mu} \sigma^{+}+i \sqrt{\frac{\alpha^{\prime}}{2}} \sum_{n \neq 0} \frac{1}{n} \bar{\alpha}_{n}^{\mu} e^{-i n \sigma^{+}}, \\
&X_{R}^{\mu}\left(\sigma^{-}\right)=\frac{1}{2} x^{\mu}+\frac{\alpha^{\prime}}{2} p^{\mu} \sigma^{-}+i \sqrt{\frac{\alpha^{\prime}}{2}} \sum_{n \neq 0} \frac{1}{n} \alpha_{n}^{\mu} e^{-i n \sigma^{-}},
\end{align}
where the reality of $X^\mu$ requires that the coefficients of the Fourier modes obey
\[\alpha_{n}^{\mu}=\left(\alpha_{-n}^{\mu}\right)^{*}, \qquad \bar{\alpha}_{n}^{\mu}=\left(\bar{\alpha}_{-n}^{\mu}\right)^{*} \ .\]
This mode expansion will be very important when we come to quantum theory. If we treat the world-sheet metric dynamically, we impose  the equation of motion with respect to the worldsheet metric
\be\label{constraints}
T_{--}=-\frac{1}{\a'}\partial_-X^\mu\partial_-X_\mu =0~,\qquad T_{++}= -\frac{1}{\a'}\partial_+X^\mu\partial_+X_\mu = 0~.
\ee
We will see the implication of these constraints in quantum theory.

\subsection{Quantizations}\label{sec:quantization}
The momentum conjugate to $X^\mu$ is defined in this gauge
\[
\Pi_{\mu}=\frac{\delta S}{\delta\left(\partial_{\tau} X^{\mu}\right)}=\frac{1}{2 \pi \alpha^{\prime}} \partial_{\tau} X_{\mu}~.
\]
The canonical quantization promotes $X^\mu$ and $\Pi_\mu$ to operators that is subject to
\begin{equation}\label{XPi-cc}
\begin{aligned}
&{\left[X^{\mu}(\sigma, \tau), \Pi_{\nu}\left(\sigma^{\prime}, \tau\right)\right]=i \delta\left(\sigma-\sigma^{\prime}\right) \delta_{\nu}^{\mu}} \\
&{\left[X^{\mu}(\sigma, \tau), X^{\nu}\left(\sigma^{\prime}, \tau\right)\right]=\left[\Pi_{\mu}(\sigma, \tau), \Pi_{\nu}\left(\sigma^{\prime}, \tau\right)\right]=0}~.
\end{aligned}
\end{equation}
We translate these into commutation relations for the Fourier modes $x^\mu$, $p^\mu$, ${\alpha}_n^\mu$ and $\ol \a_n^\mu$. Using the mode expansion, we find (exercise)
\be \left[x^{\mu}, p_{\nu}\right]=i \delta_{\nu}^{\mu} \quad \textrm { and }\quad \left[\alpha_{n}^{\mu}, \alpha_{m}^{\nu}\right]=\left[\bar{\alpha}_{n}^{\mu}, \bar{\alpha}_{m}^{\nu}\right]=n \eta^{\mu \nu} \delta_{n+m, 0}~ ,\label{acom}\ee
Therefore, like harmonic oscillators, the creation operators are $\a_{-n}^{\m},\ol \a_{-n}^{\m}$ and the annihilation operators are $\a_{n}^{\m},\ol \a_{n}^{\m}$ for $n\in\bN$ so that the Hilbert space is spanned by
\be\label{bosonic-Hilb}
\a_{-n_1}^{\m_1}\cdots \a_{-n_k}^{\m_k} \ol \a_{-n_1}^{\m_1}\cdots \ol \a_{-n_k}^{\m_k}|0;k\rangle   \qquad \textrm{where} \qquad p^\m|0;k\rangle=k^\m|0;k\rangle~.
\ee

Let us consider the implication of the constraints \eqref{constraints}. Using the mode expansions \eqref{mode}, the energy-momentum tensor can be expressed as
\[
T_{--}=-\sum_n L_n^Xe^{-in\sigma^-} \qquad T_{++}=-\sum_n \ol L_n^Xe^{-in\sigma^+}~,
\]
where
\begin{align}\label{Virasoro}
L_{m}^{X}=\frac{1}{2} \sum_{n \in \mathbb{Z}} \eta_{\mu \nu}: \alpha_{m-n}^{\mu} \alpha_{n}^{\nu}: \quad \bar{L}_{m}^{X}=\frac{1}{2} \sum_{n \in \mathbb{Z}} \eta_{\mu \nu}: \bar{\alpha}_{m-n}^{\mu} \bar{\alpha}_{n}^{\nu}:
\end{align}
with $\a_0^\m=\ol \a_0^\m=\sqrt{\frac{\a'}{2}}p^\m$. We will learn that $L_m$ are generators of \textbf{Virasoro algebra}.  According to the normal-ordering $: \ :$ prescription, the lowering operators always appear to the right of the raising operators. However, it is easy to see that the normal ordering matters only in $L_0^X$ and $\ol L_0^X$.
Thus, in quantum theory, the constraint \eqref{constraints} on the energy-momentum tensor can be interpreted by
\[\left(L_{n}^{X}+A \delta_{n, 0}\right)| \textrm{phys} \rangle =0 \qquad \left(\ol L_{n}^{X}+\ol A \delta_{n, 0}\right)| \textrm{phys} \rangle =0 \qquad \textrm{for} \quad n\ge 0\]
where $A$, $\ol A$ are normal
ordering constants  so that
\[
\langle \textrm{phys}'| L_n^X | \textrm{phys} \rangle =0 \qquad  \langle \textrm{phys}'| \ol L_n^X | \textrm{phys} \rangle =0 \qquad \textrm{for} \quad n\in \bZ~.
\]
Another way to think of this constraint is as follows.
A physical amplitude will not depend on the choice of gauge
$h_{ab}(\sigma) + \delta h_{ab}$, i.e.
$$\delta\langle f \mid i\rangle=-\frac{1}{4 \pi} \int d^{2} \sigma h(\sigma)^{1 / 2} \delta h_{a b}(\sigma)\left\langle f\left|T^{a b}(\sigma)\right| i\right\rangle = 0.$$

\subsubsection*{Light-cone quantization}
To see the effect of normal ordering in $L_0$, let us consider the meaning of the light-cone gauge \eqref{lcg} in quantum theory. It is easy to see that \eqref{lcg} implies
\[
\a_n^+=0 \quad \textrm{for} \quad n\neq0~.
\]
Then, the equation of motion \eqref{constraints} with respect to the worldsheet metric
\[
 2\partial_-X^-\partial_-X^+=\sum_{i=1}^{D-2}\partial_-X^i\partial_- X^i
\]
tells us
\[
\alpha_{n}^{-}=\sqrt{\frac{1}{2 \alpha^{\prime}}} \frac{1}{p^{+}} \sum_{m=-\infty}^{\infty} \sum_{i=1}^{D-2} \alpha_{n-m}^{i} \alpha_{m}^{i}~.
\]
In particular, for $n=0$, we have
\[
M^{2}=2 p^{+} p^{-}-\sum_{i=1}^{D-2} p^{i} p^{i}=\frac{4}{\alpha^{\prime}}\left(\sum_{n>0} \sum_{i=1}^{D-2} \alpha_{-n}^{i} \alpha_{n}^{i}+\frac{D-2}{2} \sum_{n>0} n\right)~.
\]
The final term clearly diverges. Fortunately, we have nice regularization of this divergence
\be\begin{aligned}
\sum_{n=1}^{\infty} n \longrightarrow \sum_{n=1}^{\infty} n e^{-\epsilon n} &=-\frac{\partial}{\partial \epsilon} \sum_{n=1}^{\infty} e^{-\epsilon n} \\
&=-\frac{\partial}{\partial \epsilon}\left(1-e^{-\epsilon}\right)^{-1} \\
&=\frac{1}{\epsilon^{2}}-\frac{1}{12}+\mathcal{O}(\epsilon)~.
\end{aligned}\ee
The first term diverges as $\epsilon\rightarrow 0$ so that we renormalize this term away. Consequently, we obtain
\be\label{Casimir}
\sum_{n=1}^\infty n=-\frac{1}{12} \ .
\ee
Hence, we obtain
\be \label{mass2}
M^{2}=\frac{4}{\alpha^{\prime}}\left(N-\frac{D-2}{24}\right)=\frac{4}{\alpha^{\prime}}\left(\bar{N}-\frac{D-2}{24}\right)~,
\ee
where the number operators are defined by
\be
N=\sum_{i=1}^{D-2} \sum_{n>0} \alpha_{-n}^{i} \alpha_{n}^{i}, \quad \bar{N}=\sum_{i=1}^{D-2} \sum_{n>0} \bar{\alpha}_{-n}^{i} \bar{\alpha}_{n}^{i} \ .\label{numbering}
\ee
It is easy to see  from \eqref{mass2}
\be\label{lmc}
N=\bar{N}
\ee
which is called the \textbf{level-matching condition}. Moreover, if $D>2$, we have the state $|0;k\rangle$ with negative [mass]${}^2$
\be\label{closed-vaccuum} M^2 =- \frac{4}{\a'}\frac{D-2}{24}\ee
for $N=0=\ol N$ which is called the \textbf{tachyon}.

\subsubsection*{The First Excited States}

Let us look at the first excited states. The level-matching condition \eqref{lmc} requires us to act both right-  $\alpha^j_{-1}$ and left-moving  $\ol \a_{-1}^i$ creation operator on $|0;k\rangle$ simultaneously. Thus, there are $(D-2)^2$ states at $N=1=\overline N$
\be \alpha_{-1}^i\ol \a_{-1}^j\,|0;k\rangle \ ,\label{first}\ee
whose mass
\be M^2 = \frac{4}{\a'}\left(1-\frac{D-2}{24}\right)\ .\nonumber\ee
It is easy to see that the state is under a representation of the little group $\SO(D-2)\subset \SO(1,D-1)$ and therefore it should be massless. Interestingly, this is only the case if the dimension of spacetime is
\be D=26\ .\label{critical-dim}\ee
This is the critical dimension of the bosonic string; the masslessness argument is a shortcut for the statement that the light-cone Lorentz algebra closes, $[J^{i-},J^{j-}]=0$, only if $D=26$ and the intercept is $1$ \cite{GSW,Tong:2009np}. In the following sections, we will see from many different viewpoints that bosonic string theory is consistent only in this critical dimension.

Then, the states \eqref{first} transform in the ${\bf 24}\otimes {\bf 24}$ representation of $\SO(24)$, which decomposes into three irreducible representations:
\begin{align}\label{massless}
\renewcommand\arraystretch{1.2}
 \begin{array}{ll}
  \textrm{Graviton}\  G^{\mu\nu} & \alpha^\mu_{-1}\ol \alpha^\nu_{-1} |0;k \rangle  \quad (\textrm{symmetric in $\mu$ and $\nu$, and traceless} ) \ , \\[2pt]
  \textrm{$B$-field}\ B^{\mu\nu} & \alpha^\mu_{-1}\ol \alpha^\nu_{-1} |0;k \rangle  \quad (\textrm{anti-symmetric in $\mu$ and $\nu$} ) \ , \\[2pt]
  \textrm{Dilaton}\ \Phi & \alpha^\mu_{-1}\ol \alpha_{-1,\mu} |0;k \rangle \quad (\textrm{trace part} )  \ .
 \end{array}
\end{align}
Each irreducible representation gives rise to a massless field in spacetime.
As above, the traceless symmetric, anti-symmetric, and the trace part correspond to the graviton $G_{\mu\nu}$, (Kalb-Ramond) $B$-field $B_{\mu\nu}$, and the dilaton $\Phi$, respectively. Therefore, graviton arises naturally from the quantization of closed strings! These massless fields play a pivotal role throughout the lecture notes.

\subsubsection*{Open strings}
So far we have studied closed strings. Let us briefly summarize the light-cone quantization of an open string and the derivations of the results are left as an exercise to the reader.
For an open string, a world-sheet spatial coordinate spans $\sigma\in [0,\pi]$ where the boundaries of the world-sheet are at $\sigma=0,\pi$, and the action is given by
\begin{align*}
 S = \frac{1}{4\pi \alpha'} \int_{\sigma=0}^{\sigma=\pi} d\tau d\sigma
 \left( -\partial_\tau X \cdot \partial_\tau X +\partial_\sigma X \cdot \partial_\sigma X  \right) \ .
\end{align*}
Like the closed string, the mode expansion of $X^\mu = X^\mu_L(\sigma^+)
+X^\mu_R(\sigma^-)$ is given by
\begin{align}
X^\mu_L(\sigma^+) &= \frac12 x^\mu +  \a' p^\mu \,\sigma^+ + i\sqrt{\frac{\a'}{2}}\sum_{n\neq 0}
\frac{1}{n}\,\overline{\alpha}_n^\mu\, e^{-in\sigma^+} \ ,\cr
X^\mu_R(\sigma^-) &= \frac12 x^\mu +  \a' p^\mu \,\sigma^- + i\sqrt{\frac{\a'}{2}}\sum_{n\neq 0}
\frac{1}{n}\,{\alpha}_n^\mu\, e^{-in\sigma^-} \ .\label{openmode}\end{align}
where the momentum of an open string is defined as $\a_0^\mu=\sqrt{2\a'}\,p^\mu$. Note that the second term differs from the closed string \eqref{mode} by a factor of two since the spatial lengths are $l_{\text {open }}=\pi$ and $l_{\text {closed }}=2 \pi$.

The variation principle of the action leads to
\begin{align}\label{open-variation}
 0 &= \delta S
 = \frac{1}{2\pi \alpha'} \int_{\sigma=0}^{\sigma=\pi} d\tau d\sigma
 \left(- \partial_\tau X \cdot \partial_\tau \delta X +\partial_\sigma X \cdot \partial_\sigma \delta X  \right)  \nonumber\\
 &= \frac{1}{2\pi \alpha'} \int_{\sigma=0}^{\sigma=\pi} d\tau d\sigma
 \left[ -\delta X \cdot \left(- \partial_\tau \partial_\tau X +\partial_\sigma \partial_\sigma X \right)
- \partial_\tau\left(\partial_\tau X \cdot \delta X \right) + \partial_\sigma \left( \partial_\sigma X \cdot\delta X \right)  \right]  \nonumber\\
 &= \frac{1}{2\pi \alpha'} \int d\tau  \left[ \partial_\sigma X \cdot\delta X  \right]_{\sigma=0}^{\sigma=\pi} \ ,
\end{align}
where we use the equation of motion and $\delta X (t=\pm \infty)=0$ at the last equality.
This compels us to impose one of the following boundary conditions:
\begin{itemize}
\item {Neumann boundary condition}:  $\partial_\sigma X^\mu = 0 \ \ \ \ {\rm at}\ \sigma=0,\pi$
\item {Dirichlet boundary condition}:  $X^\mu = c^\mu$ (constant) {at} $\sigma=0,\pi$
\end{itemize}
which imposes the following conditions on the modes
\begin{itemize}
\item {Neumann boundary condition}:  $\alpha_n^\mu=\bar{\alpha}_n^\mu$.
\item {Dirichlet boundary condition}: $x^\mu=c^\mu, \quad p^\mu=0, \quad \alpha_n^\mu=-\bar{\alpha}_n^\mu$.
\end{itemize}
Moreover, the light-cone gauge quantization for \eqref{openmode} leads to the open string mass spectrum
\be \label{open-M2}
M^2 = 2p^+p^- - \sum_{i=1}^{D-2}p^ip^i = \frac{1}{\a'}\left(\sum_{n>0}\sum_{i=1}^{D-2} \alpha_{-n}^i\alpha_n^i + \frac{D-2}{2}\sum_{n>0} n\right)~,
\ee 
where there is the difference in the normalization $p^2$ between closed and open string: $4 p_{\text {open }}^{2}=p_{\text {closed }}^{2}$. This results in a difference by a factor of four from \eqref{mass2}. Again, in $D=26$, the open string mass spectrum becomes
\be \label{open-mass}
M^2=\frac1{\a'}(N-1)~.
\ee
so that there is the tachyon vacuum at the level $N=0$. The massless states
\be\label{open-massless}  \alpha_{-1}^i |0;k\rangle\ \ \ \ \ i=1,\ldots,D-2
\ee
at the level $N=1$ correspond to vector gauge bosons.


\subsection{Path-integral formulations}

\begin{figure}[htb]
 \centerline{\includegraphics[width=250pt]{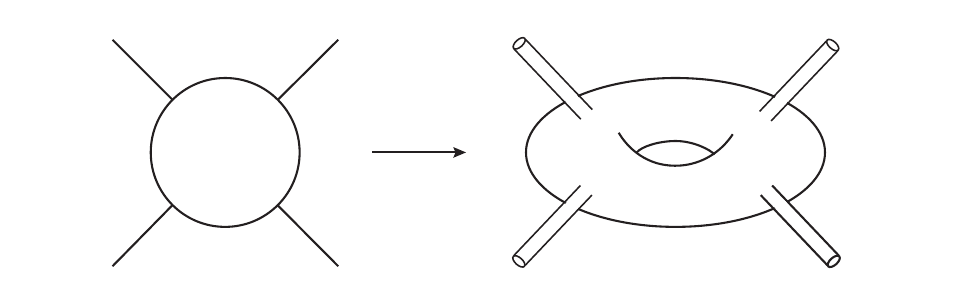}}
 \caption{A Feynman diagram of QFT (left) and a string interaction diagram (right). Each end of the cylinders emanating from the torus (or a Riemann surface in general) in the string interaction diagram corresponds to an initial/final state.}
 \label{fig1}
\end{figure}

\begin{figure}[htb]
 \centerline{\includegraphics[width=250pt]{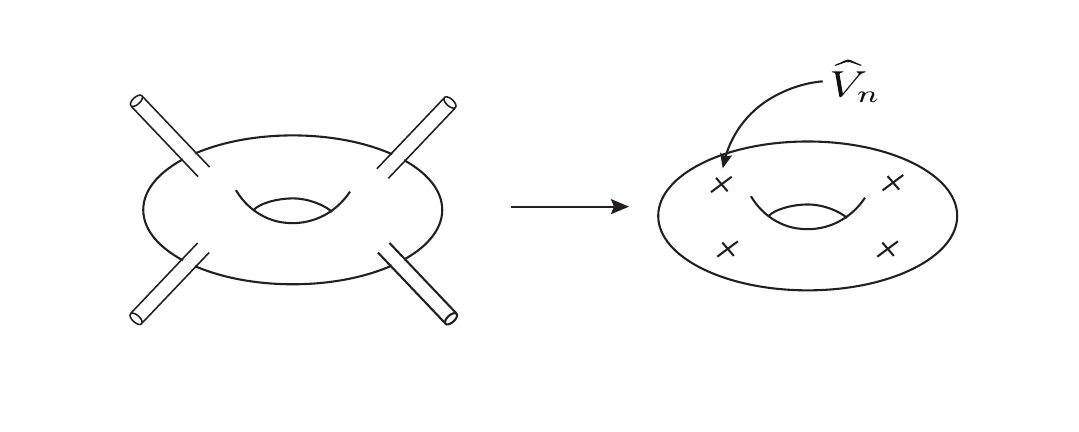}}
 \caption{String interaction in state expression (left) and operator expression (right).}
 \label{fig2}
\end{figure}

We have obtained the bosonic string spectra, and to illustrate the interaction among string states, we will introduce \textbf{path integral}.
(For more detail, see \cite[Appendix.A]{Polchinski}.)
Figure \ref{fig1} depicts an extension from quantum field theoretic (QFT) interaction to string interaction.
Each end of the cylinders in the Figure corresponds to an initial/final state of a string. The action \eqref{string-sigma} of the string sigma model is invariant under the world-sheet diffeomorphisms, and it is a conformal field theory as a result.
As we will see in the next section \S\ref{sec:2dcft}, one can make use of the \textbf{state-operator correspondence} in a 2d CFT.  Hence, instead of using the Hamiltonian formalism of string states,  we will use vertex operators $\wh V_i$ in the path integral formalism:
\begin{align}\nonumber
 A_n = \sum_g \int \left(\cD h_{ab}\right)_{g,n} \int \cD X^\mu e^{-S_\sigma [X^\mu,h_{ab}]} \wh V_1 \cdots \wh V_n \ ,
\end{align}
where $n$ is the number of in-coming/out-going strings, and $g$ is the number of holes (genus) of the Riemann surface (genus is $1$ in Figure \ref{fig2}). Here we integrate over all the metric $h_{ab}$ and field $X^\mu$ configuration up to gauge transformations, and we also sum over all the topology (genus $g$) of world-sheets.

In the path integral method, expectation values are schematically expressed in the following form
\begin{align}\nonumber
 \langle \mathcal O[X] \rangle = \int \cD X e^{iS[X]} \mathcal O [X] \ ,  \qquad
 S = \frac{1}{4\pi \alpha'} \int d\sigma d\tau \left\{ \left(\partial_\tau X\right)^2 -\left(\partial_\sigma X\right)^2\right\} \ ,
\end{align}
where $\mathcal O$ is a
gauge-invariant operator.

We usually \textbf{Wick rotate} ($\tau \to -it$) the theory so that it converges.
\begin{align}\nonumber
 \langle \mathcal O[X] \rangle = \int \cD X e^{-S_E[X]} \mathcal O [X] \ ,  \qquad
 S_E = \frac{1}{4\pi \alpha'} \int d\sigma dt \left\{ \left(\partial_t X\right)^2 +\left(\partial_\sigma X\right)^2\right\} \ .
\end{align}
The subscript $E$ will be omitted hereafter, and we will always work in the world-sheet Euclidean signature.




\section{Two-dimensional conformal field theory}\label{sec:2dcft}

The dynamics of the string's worldsheet are described by a two-dimensional conformal field theory (2D CFT). Within this framework, ``vertex operators'' represent the various vibrational states of the string, emphasizing the significance of 2D CFT in string theory research. A notable characteristic of 2D CFTs is their infinite-dimensional symmetry \cite{belavin1984infinite}, which makes the story remarkably profound.

In this section, we will learn the operator analysis in 2D CFT, covering topics such as the Operator Product Expansion (OPE) and the Ward-Takahashi identity.
Moreover, we will see Virasoro algebra, which is associated to 2d conformal symmetry.

However, it is important to acknowledge that what we are uncovering here only scratches the surface, and this subject could easily fill an entire semester. For readers intrigued by this fertile topic, we recommend consulting established references  \cite{ginsparg1988applied,francesco2012conformal,Blumenhagen:2009zz}.

\subsection{Conformal transformations}

 A CFT in any dimension is a quantum field theory invariant under conformal maps at quantum level.
A conformal map between two Riemannian manifolds $(\Sigma,h)$ and $(\wt\Sigma,h')$ is a diffeomorphism $f:\Sigma\to \wt\Sigma$ such that the pull-back metric $f^* h'$ and the original metric $h$ are related by an arbitrary function $e^{2\omega}$ of $\Sigma$ \[f^* h' =e^{2\omega} h~.\] In terms of local coordinates $(\sigma,h_{ab})$ and $(\wt\sigma,h'_{ab})$,
a \textbf{conformal transformation} $\sigma^a\rightarrow \wt{\sigma}^a(\sigma)$ relates the metrics by
\be \frac{\partial\wt\sigma^c}{\partial\sigma^a}\frac{\partial\wt\sigma^d}{\partial \sigma^b} h'_{cd}(\wt{\sigma}(\sigma)) = e^{2\omega(\sigma)} h_{ab}(\sigma)~.\label{cft}\ee
Roughly speaking, a conformal transformation is a coordinate transformation that leaves the metric invariant up to scale and thus preserves angles.  This means that the theory behaves the same at
all length scales.

A conformal transformation is a diffeomorphism under which the pull-back metric is related to the original metric by a Weyl transformation. Since the string sigma model \eqref{string-sigma} is invariant under diffeomorphisms and Weyl transformations of the world-sheet, it is conformal-invariant. Hence, we will study conformal field theory of the string world-sheet.

Note that a conformal transformation is a diffeomorphism between two Riemannian manifolds, and a Weyl transformation changes the metric
 keeping a manifold fixed. (Keep in mind the difference!) Therefore, if the metrics $(\Sigma,h)$ and $(\Sigma, e^{2\omega}h)$ are related by a Weyl transformation, then the identity map of $\Sigma$ is a conformal transformation.
\footnote{A conformal transformation is a diffeomorphism that rescales the metric, whereas a Weyl transformation rescales it at fixed points. In the gauge-fixed theory, a conformal transformation followed by the compensating Weyl transformation leaves the flat metric invariant; this is the residual symmetry of the previous section.}

\begin{figure}[ht]
	\centering
	\subfigure[$z$]{
	\includegraphics[width=0.3\textwidth]{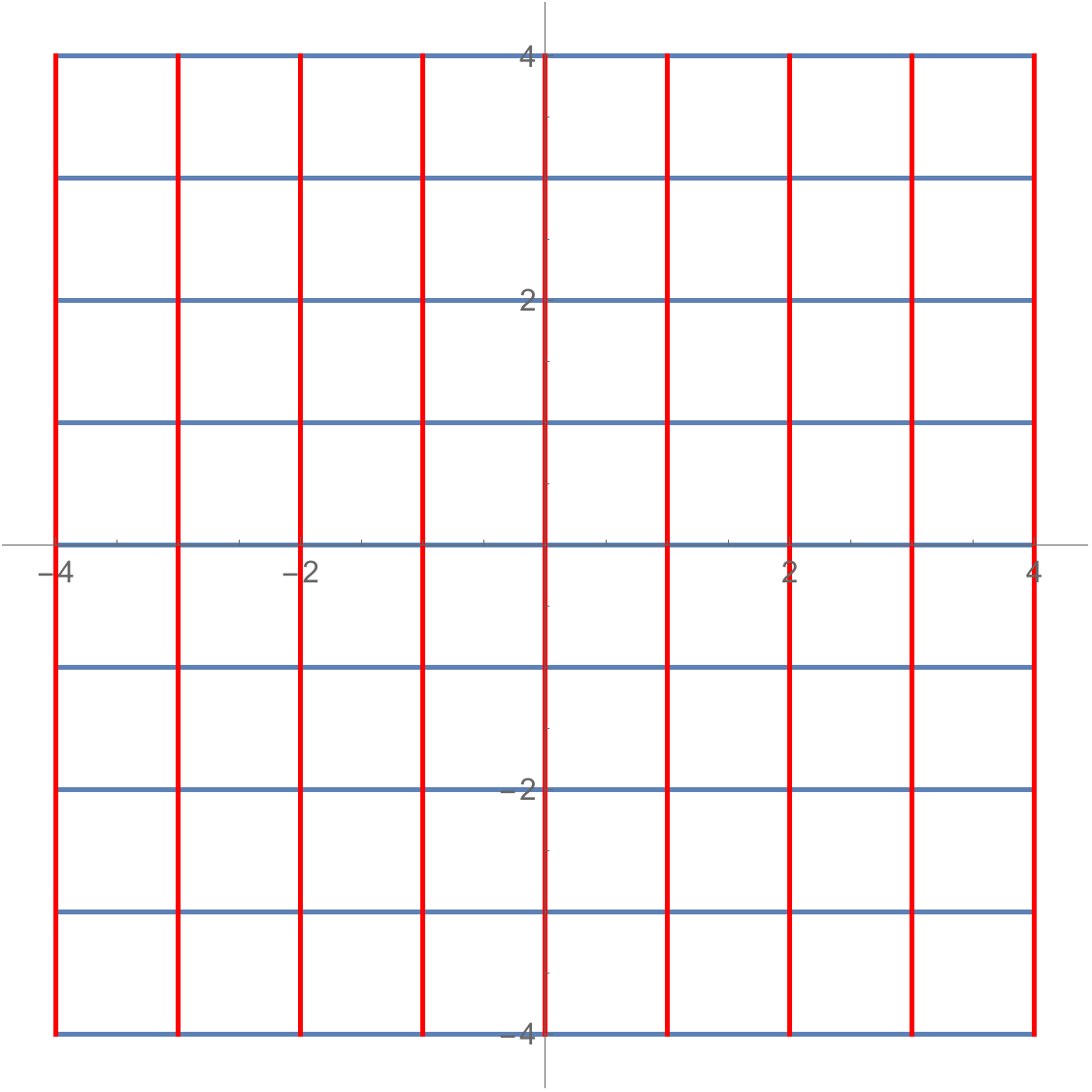}}
    \subfigure[$z'=z^2$]{
	\includegraphics[width=0.3\textwidth]{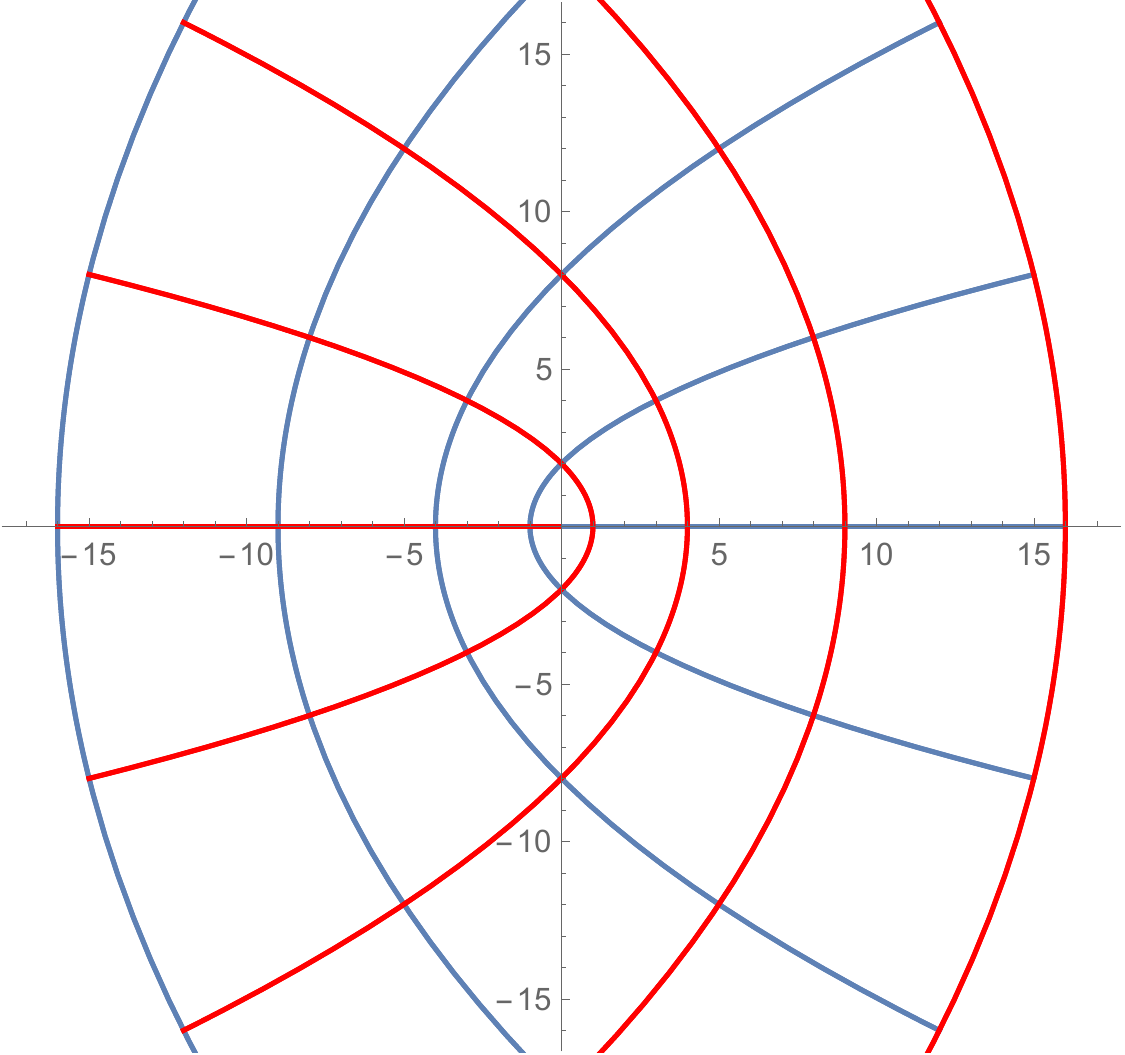}}
    \subfigure[$z'=1/z$]{
	\includegraphics[width=0.3\textwidth]{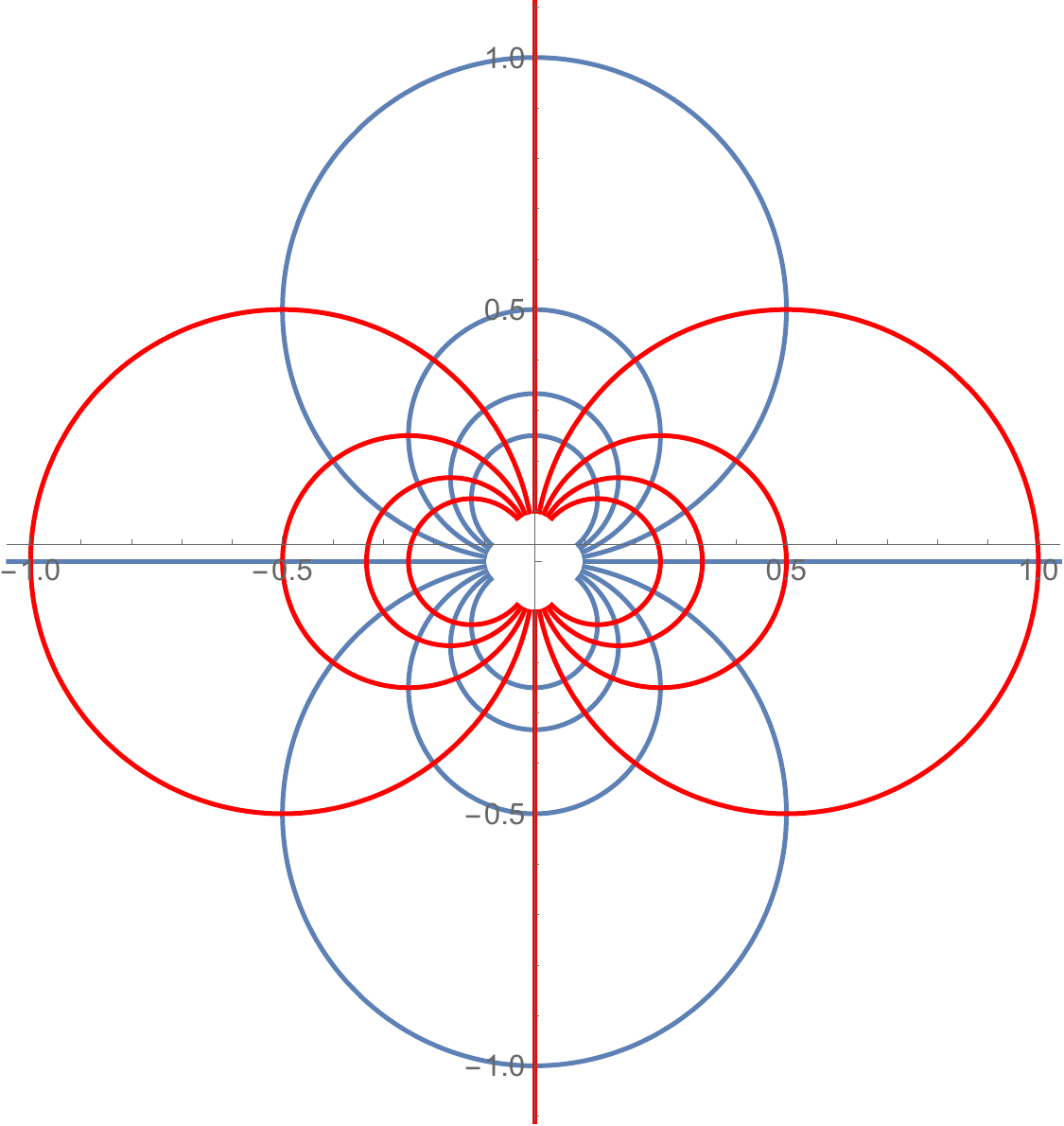}}
    \subfigure[$z'=z|z|$]{
    \includegraphics[width=0.3\textwidth]{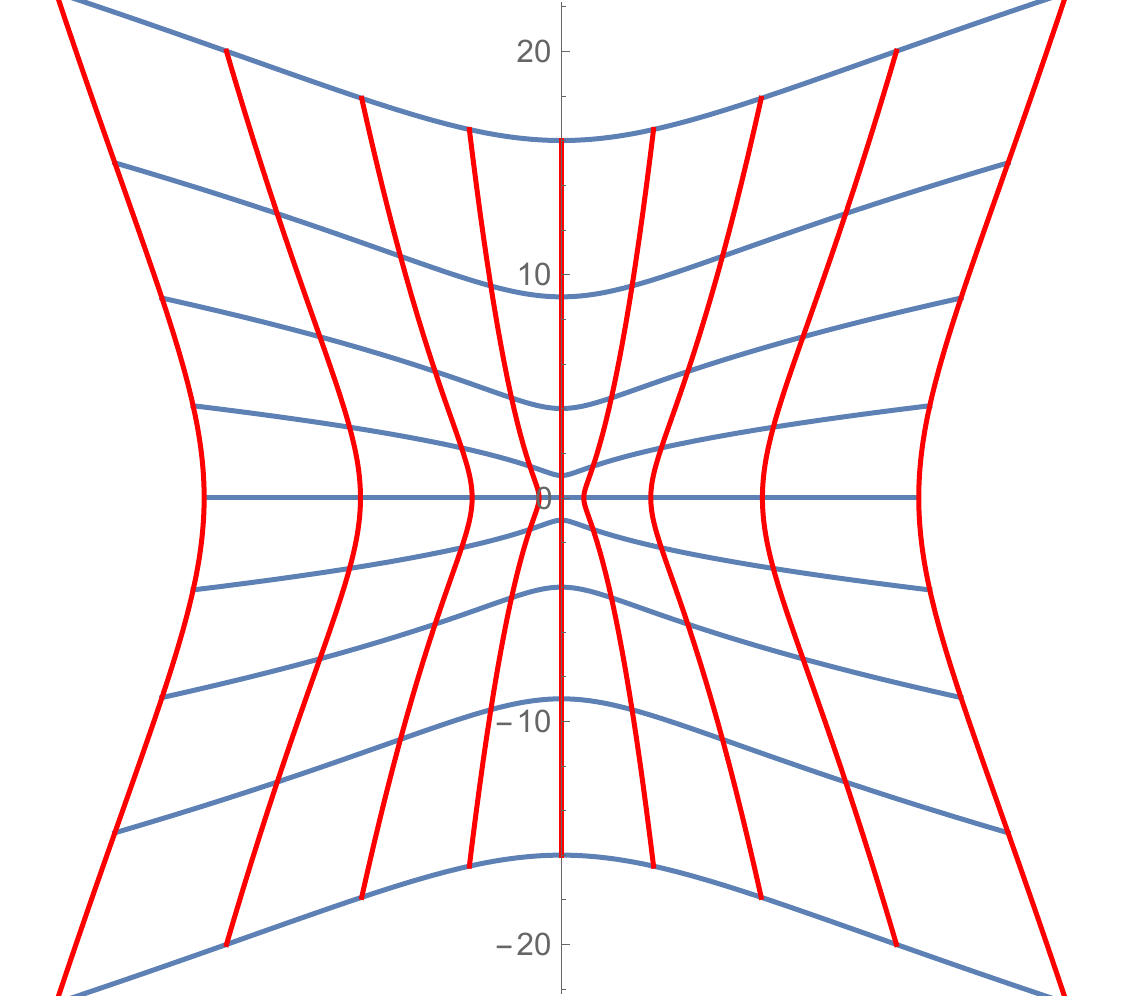}}
    	\caption{Coordinate transformation:
		The transformation from
		square lattice (a),
		onto the lattice in (b) and (c)
		is conformal,
		while the transformation
		onto the lattice in (d) is not.}
	\label{fig:2dcft-ill}
\end{figure}
\subsubsection*{2d flat space}
In complex coordinates, the 2d flat metric can be written as $ds^2 = dz d\bar{z}$. Under a holomorphic map,
\[
z\rightarrow f(z)~,
\]
the metric is transformed as
\[
ds^2 = dz d\bar{z}\quad  \rightarrow\quad  ds^2 =\frac{\partial {z}}{\partial {f}} \frac{\partial \bar{z}}{\partial \bar{f}} df d\bar{f}~.
\]

Therefore, all the holomorphic maps are conformal transformations where  $\left| \frac{\partial z}{\partial f} \right|^2$ is a conformal factor (corresponding to $e^{2\omega(\sigma)}$ in \eqref{cft}). Moreover, it is easy to show that all 2d conformal transformations are indeed holomorphic functions (exercise).  This set is infinite-dimensional, corresponding to the coefficients of the Laurent series of holomorphic functions. Due to the infinite dimensionality, the conformal symmetry becomes so powerful in two dimensions.

\subsection{State-operator correspondence}

\begin{figure}[htb]
 \centerline{\includegraphics[width=14cm]{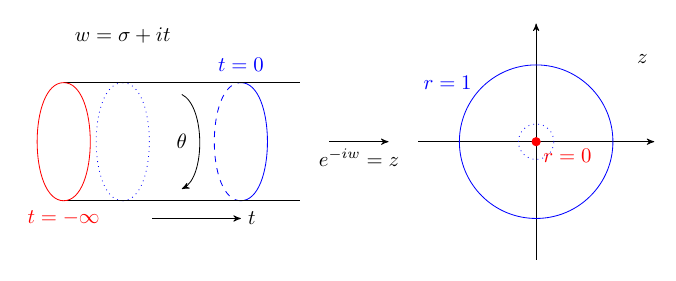}}
 \caption{State-operator correspondence. The initial state (red circle) is mapped to a local operator (red dot) at the origin.}
 \label{fig3}
\end{figure}

As we have seen, the action \eqref{string-sigma} of the string sigma model is equipped with the diffeomorphism-invariance and the Weyl-invariance. This means that it is invariant under conformal transformations.
We can use this freedom to set the world-sheet to be $\bR^2$ (recall that the topology of a propagating string is a cylinder).
\begin{align}\nonumber
 \bR^2:  \quad ds^2 &= d\sigma^1d\sigma^1+d\sigma^2d\sigma^2= dr^2 +r^2 d\theta^2 \cr
 &= r^2 \left[ d\left(\log r\right)^2 +d\theta^2 \right] \ ,\cr
 \textrm{Cylinder} : \quad ds^2 &= dt^2 +d\sigma^2 \ ,
\end{align}
where the overall factor $r^2$ in $\bR^2$ is identified with the conformal factor $e^{2\omega(\sigma)}$.
So we can identify
\begin{align}\nonumber
 t = \log r \ ,  \qquad  \sigma = -\theta \ ~,
\end{align}
and the reason for the minus sign will be clear later.
This identification tells us that there is a correspondence between
\begin{itemize} \setlength\itemsep{.1em}
 \item CFT on a cylinder and
 \item CFT on  $\bC^\times=\bC\backslash\{0\}$  ,
\end{itemize}
provided that we choose the boundary conditions to be the same.
In particular, the initial \textbf{state} of the string ($t = -\infty$) corresponds to a \textbf{local operator}
inserted at the origin, which is called vertex operator (Figure \ref{fig3}).
This is the state-operator correspondence in a CFT, which plays a crucial role.  We can express string states by the vertex operators
(Figure \ref{fig2}).

Now the reader may wonder if there is any way to express commutation relations in terms of operators
so that we can do parallel procedures of the canonical quantization in terms of operators.

\subsection{Operator product expansions in free scalar theory}\label{sec:OPE}

We study massless free scalar fields as the easiest example.
Here we adopt the Euclidean metric $\delta_{ab} = \textrm{diag}(+,+)$ though the procedure is parallel for the Lorentz metric.
We write the action in complex coordinates
\begin{align}\label{free-scalar}
 S_X = \frac{1}{2\pi\alpha'} \int d^2 z\ \partial X^\mu  \ol\partial X_\mu~.
\end{align}
Then, the equation of motion is given by $\partial\ol\partial X^\mu = 0$, which implies that $X$ is holomorphically factorized
\begin{align}\label{mode-exp}
 X^\mu(z,\ol z) &= X^\mu (z) +\ol X^\mu (\ol z)   ,   \cr
 X^\mu (z) &= \frac{1}{2}x^\mu -i \frac{\alpha'}{2} p^\mu \log z +i \sqrt{ \frac{\alpha'}{2} } \sum_{n \neq 0} \frac{1}{n} \frac{\alpha_n^\mu}{z^n} \ , \cr
 \ol X^\mu (\ol z) &=\frac{1}{2}x^\mu -i \frac{\alpha'}{2} p^\mu \log \overline z +i \sqrt{ \frac{\alpha'}{2} } \sum_{n \neq 0} \frac{1}{n} \frac{\ol \alpha_n^\mu}{\overline z^n} \ ,
\end{align}
where
\begin{align}\nonumber
 z = r e^{i\theta} = e^{\log r +i\theta} = e^{i\tau -i\sigma} = e^{i\sigma^-} \ ,  \qquad \ol z &= e^{i\sigma^+} \ .
\end{align}
This is the Euclidean version of \eqref{mode}. The (anti-)holomorphic part corresponds to the right (left) movers.

Let us consider a quantum version of the equation of motion by using a path integral formulation.
Like a normal integral, we assume that the ``total derivative'' vanishes in the path integral.
For example, in the massless free scalar case, we have
\begin{align}\nonumber
 0 = \int \mathcal D X \ \frac{\delta}{\delta X} e^{-S[X]} = \int \mathcal D X \ e^{-S[X]} \frac{1}{\pi\alpha'}\partial\ol\partial X
 = \frac{1}{\pi\alpha'} \langle \partial\ol\partial X \rangle \ ,
\end{align}
which is the Schwinger-Dyson equation (the quantum equation of motion).
The equation of motion is satisfied as an operator equation if there is no other operator nearby.

Furthermore, the same procedure can be done with operator insertion:
\begin{align}\nonumber
 0 = \int \mathcal D X \ \frac{\delta}{\delta X_\mu (z,\ol z)} \left(e^{-S[X]} X^\nu (w,\ol w) \right)
 = \Bigl\langle \frac{1}{\pi\alpha'}\partial_z\partial_{\ol z} X^\mu (z,\bar z) X^\nu(w,\bar w) + \eta^{\mu\nu}\delta^2(z-w) \Bigr\rangle \ ,
\end{align}
yielding
\begin{align}\label{operator-eom}
\Bigl\langle  \partial\ol\partial X^\mu (z,\bar z) X^\nu(w,\bar w) \Bigr\rangle= -\pi\alpha' \eta^{\mu\nu}\delta^2(z-w) \ .
\end{align}
Now we can use the
\textbf{Stokes' theorem}
\begin{align}\nonumber
 \int_D d^2z \left( \partial \ol V + \ol\partial V\right)
 = \frac{1}{i} \oint_{\partial D} \left( V dz - \ol V d\ol z \right) \ ,
\end{align}
where $D$ is an arbitrary domain (typically a disk) and $\partial D$ is its boundary.
Using Stokes' theorem, we obtain
\begin{align}
 \partial\ol\partial \log |z|^2 =2\pi \delta^2(z) \ .
 \label{eq:deltaFunc}
\end{align}
This allows us to rewrite \eqref{operator-eom} into
\begin{align}\nonumber
 X^\mu(z,\bar z) X^\nu(w,\bar w) = -\frac{\alpha'}{2} \eta^{\mu\nu} \log |z-w|^2 + \nord{X^\mu(z,\bar z)X^\nu(w,\bar w)} \ ,
\end{align}
where we introduced a normal ordering $\nord{\mathcal O}$.
This is an operator equation, and we simply omit $\langle \quad \rangle$ symbols. This expression describes the singular behavior when $ X^\mu(z,\bar z)$ approaches $ X^\nu(w,\bar w)$.

We also write it as follows:
\begin{align}\nonumber
 \nord{X^\mu(z,\bar z)X^\nu(w,\bar w)} = X^\mu(z,\bar z)X^\nu(w,\bar w) -\eta^{\mu\nu}G(z,w) \ ,
\end{align}
where $G(z,w) = -\frac{\alpha'}{2} \log|z-w|^2$.
Notice that
\begin{align}
 \partial \overline\partial \nord{X(z,\bar z)X(w,\bar w)} = 0 \ .
 \label{eq:normalEOM}
\end{align}
As we will see later, the divergent term is important and meaningful.
This is the reason why it is convenient to introduce the normal ordering so that we can separate divergent terms from the other.

The normal ordering of an arbitrary functional of operators in the free scalar theory can be expressed as follows:
\begin{align}\nonumber
 \nord{f[X]} = \exp \left[ -\frac{1}{2} \int d^2z d^2w\ G(z,w) \frac{\delta}{\delta X^\mu(z,\bar z)} \frac{\delta}{\delta X_\mu(w,\bar w)} \right]
 f[X] \ .
\end{align}
For example,
\begin{align}\nonumber
 \nord{X_1X_2X_3X_4X_5} =& X_1X_2X_3X_4X_5 -\left( G_{12}X_3X_4X_5 +\cdots \right)_\textrm{10 terms}  \cr
 &+\left( G_{12}G_{34}X_5 +\cdots \right)_\textrm{15 terms} \ ,
\end{align}
where $X_i=X(z_i)$ and $G_{ij} = G(z_i,z_j)$.\footnote{
Here the normal ordering is used to
cut off UV divergence in
a particular renormalization scheme.
It is not necessarily
equivalent to the normal
ordering
defined by the creation
and annihilation operators in the previous
section.}

Moreover, a product of normal ordered operators is given as follows:
\begin{align}\label{OPE-general}
 \nord{f[X]} \nord{g[X]} = \exp \left[ \int d^2z d^2w\ G(z,w) \left. \frac{\delta}{\delta X^\mu(z,\bar z)} \right|_{f} \left. \frac{\delta}{\delta X_\mu(w,\bar w)} \right|_{g} \right] \nord{f[X]g[X]} \ .
\end{align}
For instance, we have
\begin{align}\nonumber
 \nord{ \partial X(z,\bar z)} \nord{ X(w,\bar w)} = \nord{\partial X(z,\bar z) X (w,\bar w)} +\partial_z G(z,w)
 = \nord{\partial X (z,\bar z) X (w,\bar w)} -\frac{\alpha'}{2} \frac{1}{z-w} \ .
\end{align}
As another example, we can consider the OPE of vertex operators (See \S\ref{sec:vertex-operator})
\begin{align}\nonumber
 \nord{e^{ik\cdot X(z,\bar z)}} \nord{e^{ik'\cdot X(w,\bar w)}} &= \exp\left[ (ik) \cdot (ik') \left( -\frac{\alpha'}{2} \log|z-w|^2 \right)\right]
 \nord{e^{ik\cdot X(z,\bar z) +ik'\cdot X(w,\bar w)}} \cr &= |z-w|^{\alpha'k\cdot k'} \nord{e^{ik\cdot X(z,\bar z) +ik'\cdot X(w,\bar w)}} \ .
\end{align}

In a general field theory, a product of a pair of fields can be expanded by a single operator
\begin{align}\nonumber
 \Phi^i (z) \Phi^j (w) = \sum_{k} C^{ij}_k(z-w) \Phi^k (w) \ .
\end{align}
This is called \textbf{operator product expansion (OPE)}, and it describes the behavior when the two operators approach each other.

Typically, for a massless free scalar field theory, we have
\begin{align}\label{XXOPE}
 &X^\mu (z,\ol z) X^\nu (w,\ol w) =  -\frac{\alpha'}{2} \eta^{\mu\nu} \log |z-w|^2 +\nord{X^\mu X^\nu (w,\ol w)}  \cr
 &\qquad +\sum_{k=1}^\infty \frac{1}{k!} \left\{ (z-w)^k  \nord{\partial^k X^\mu X^\nu (w, \ol w)}
 +(\ol z-\ol w)^k  \nord{\ol\partial^k X^\mu X^\nu (w, \ol w)}\right\} \ .
\end{align}
The second and the third terms of RHS can be understood as the Taylor expansions of $\nord{X(z)X(w)}$ in terms of $z$. Note that mixing terms ($\nord{\partial^m\overline\partial^n X X}$) vanish due to the ``equation of motion'' (\ref{eq:normalEOM}).

As we will see below, divergent terms in OPE are important. Thus, we write
\begin{align}\nonumber
 X^\mu (z,\ol z) X^\nu (w,\ol w) \sim  -\frac{\alpha'}{2} \eta^{\mu\nu} \log |z-w|^2 \ .
\end{align}
The symbol $\sim$ means the regular terms are ignored.

\subsection{Conformal Ward-Takahashi identity}\label{sec:Noether}

When an action is invariant under a certain transformation $\delta$ (namely, $\delta S = 0$)
we say the theory has a (classical) symmetry.
Furthermore, if the measure of the path integral is also invariant under the
transformation, the theory has the symmetry at quantum level.
On the other hand, if the measure is not invariant, then the symmetry is anomalous.
It is non-trivial to see if the theory has an anomaly or not.

If the theory has a symmetry, Noether's theorem states that there is the corresponding conserved current $j^a$.
The space integral of its time component is the conserved charge $Q=\int_\textrm{space} j^0$, which generates the symmetry $\delta \phi = [Q, \phi]$.
Let $\delta$ be a symmetry $\delta S = 0$ and assume it acts on a field as follows: $\delta \phi (z) = \epsilon (\cdots) $, where $\epsilon$ is a small parameter.
If we promote the parameter $\epsilon$ to be world-sheet coordinate dependent (i.e. $\wh \delta \phi = \epsilon(z,\ol z) (\cdots)$), then $\wh \delta S$ is no longer zero but has to take the following form:
\begin{align}\nonumber
 \wh\delta S = \int \frac{d^2 \sigma}{2\pi} \epsilon(x,y) \partial_a j^a
 = \int \frac{d^2z}{2\pi} \epsilon(z, \ol z)
 \left( \partial \ol j +\ol\partial j \right) \ ,
\end{align}
where we introduced $j = j_{z}$ and $\ol j = j_{\ol z}$.
If the parameter is constant, $\delta S = (\textrm{total derivative}) = 0$ so that $j^a$ is \textbf{the Noether current}.
The conservation of the Noether current $\partial_a j^a = 0$ in the flat 2d implies that $j$ ($\ol j$) is a holomorphic (anti-holomorphic) function of $z$.

\subsubsection*{Ward-Takahashi identity}

Now let us study the transformation of  a point operator $\mathcal{O}(w,\ol w)$ under the symmetry $\delta$. For this purpose,
we set
\begin{align}\nonumber
 \epsilon(z,\ol z) =
 \begin{cases}
  \epsilon \quad (\textrm{const.}) \quad &\textrm{for}\ z \in D_w \ ,  \cr
  0                                      &\textrm{for}\ z \not\in D_w \ , \cr
 \end{cases}
\end{align}
where $D_w$ is a disk containing $w$. Then, the variation of the path integral
\begin{align}\nonumber
 0 = \int \cD X \ \wh \delta \left[ e^{-S} \mathcal{O}(w,\ol w) \right]
 = \int \cD X e^{-S}\  \left[ \delta \mathcal{O}(w,\ol w) -\wh \delta S \cdot \mathcal{O}(w,\ol w) \right]  .
\end{align}
Therefore, we have
\begin{align}\nonumber
 \delta \mathcal{O}(w,\ol w) &= \int \frac{d^2z}{2\pi} \epsilon(z, \ol z)
 \left( \partial \ol j(\ol z) +\ol\partial j(z) \right) \mathcal{O}(w, \ol w)  \cr
 &= \frac{\epsilon}{2\pi i} \oint_{\partial D_w} \Big( dz\ j(z) -d\ol z\ \ol j(\ol z) \Big) \mathcal{O}(w,\ol w) \ ,
\end{align}
This is called the \textbf{Ward-Takahashi identity}.

Let us see an example of the free scalar field \eqref{free-scalar}.
It is easy to see that the action of the free scalar field is invariant under the transformation $X^\mu(z,\bar z)\to X^\mu(z,\bar z)+\e^\mu$. If $\e^\mu$ is promoted to a function $\e^\mu(z,\bar z)$, then we have
\begin{align}\label{U1-current}
\wh \delta S &= \frac{-1}{2\pi \alpha'} \int d^2z\ \epsilon_\mu(z,\ol z) \Bigl[\ol\partial \left( \partial X^\mu\right) +\partial ( \ol\partial X^\mu)\Bigl] \cr
 &j^\mu = -\frac{1}{\alpha'} \partial X^\mu \ , \qquad  \ol j^\mu =-\frac{1}{\alpha'} \ol\partial X^\mu  ~.
\end{align}
It is straightforward from \eqref{XXOPE} to check that \[\delta X^\mu(w) = \epsilon_\nu \oint_{\partial D} \frac{dz}{2\pi i}\ j^\nu(z) X^\mu(w) \] is consistent.

Now, let us consider the Ward-Takahashi identity for conformal symmetry.
For an infinitesimal conformal transformation $\sigma^a\rightarrow \sigma^a + \epsilon^a(\sigma)$, the metric is transformed as
\[
\delta_{ab}\to \delta_{ab} +\partial_a\e_b+\partial_b\e_a~.
\]
Since this is a conformal transformation, it is proportional to $\delta_{ab}$ so that we have
\be \label{infinitesimal2dconformal}
\partial_a\e_b+\partial_b\e_a = (\partial_\rho \e^\rho)\delta_{ab}~.
\ee
A solution to this equation is called a \textbf{conformal Killing vector}.
The current for the conformal transformation can be written as \[j_a=T_{ab}\epsilon^b~,\]
where the straightforward calculation provides the energy-momentum tensor
\be\label{se}
T_{ab}=-2\pi\Big[\frac{\partial L}{\partial(\partial^a \phi)} \partial_b \phi -\delta_{ab} L\Big]~.
\ee
If we assume $\epsilon$ is constant, it is easy to see that the conservation of the current implies the conservation of the energy-momentum tensor:
\be\label{conservation}
\partial_a j^a=0 \quad \to\quad \partial^a T_{ab}=0~.
\ee
For general $\epsilon^a(\sigma)$, the conservation of the current gives the traceless condition of $T_{ab}$:
\begin{equation}\label{traceless}
0=\partial^a j_a=\frac12 T_{ab}(\partial^a\e^b+\partial^b\e^a )=\frac12T^a{}_a(\partial_\rho \e^\rho)\quad \to \quad T^a{}_a=0~.
\end{equation}
In the complex coordinate $z=x^1+ix^2$, the traceless condition can be written as
\[
T_{z\bar{z}} = T_{\bar{z}z} = 0
\]
and the conservation of the energy-momentum tensor can be written as
\[
\partial_{\bar{z}}T_{zz}= 0~,\qquad \partial_{z}T_{\bar{z}\bar{z}}= 0~.
\]
Thus, the non-vanishing components of the energy-momentum tensor factorize to a chiral and anti-chiral field,
\[T(z):=T_{zz} \quad \textrm{and} \quad  \bar{T}(\bar{z}):=T_{\bar{z}\bar{z}} ~.\] As a result, the Noether currents for conformal transformations $z \to z + \e(z)$ and $\bar z \to\bar  z + \bar \e(\bar z)$ are
\[
j(z)=  \epsilon(z) T(z)~,\quad \overline j(\overline z)=\overline \epsilon(\overline z)  \overline T(\overline z)  ~.
\]
The application to the Ward-Takahashi identity leads to \textbf{conformal Ward-Takahashi identity}
\begin{equation}\label{CWT}
\delta_{\epsilon,\overline \epsilon} \mathcal{O}(w,\bar{w}) = \frac{1}{2\pi i}\oint_{C_w} dz\;  \epsilon(z)T(z) \mathcal{O}(w,\bar{w})+
 \frac{1}{2\pi i}\oint_{C_{\bar w}} d\bar{z}\; \bar{\epsilon}(\bar{z}) \bar{T}(\bar{z})\mathcal{O}(w,\bar{w}) ~,
\end{equation}
where the contour integral is taken as a counter-clockwise circle both in $z$ and in $\bar z$ (thereby explaining the sign difference of the second term).

\subsection{Primary fields}

Let us first introduce some terminologies in 2d CFT.
 Fields  depending  only on $z$, i.e. $\phi(z)$, are called \textbf{chiral or holomorphic fields}  and fields $\overline \phi (\bar z)$ only depending on $\bar z$ are called \textbf{anti-chiral or anti-holomorphic  fields}.
If a field $\phi$ transforms under the scaling transformation $z\rightarrow\lambda z$ as
\begin{equation}
\phi(z,\bar{z})\rightarrow\phi'(\lambda z,\bar{\lambda}\bar{z} )=\lambda^{-h} \bar{\lambda}^{-\bar{h}} \phi( z, \bar{z})~,
\end{equation}
it is said to have \textbf{conformal dimension} $(h, \bar{h})$.
If a field transforms under a conformal transformation $z\rightarrow f(z)$ as
\begin{equation}\label{primary}
\phi(z,\bar{z})\rightarrow\phi'(f(z),\bar{f} (\bar{z}))=\left(  \frac{\partial f}{\partial z} \right)^{-h} \left( \frac{\partial \overline f}{\partial \overline z}\right)^{-\bar{h}}\phi( z,\bar{z})
\end{equation}
 it is called a \textbf{primary field} of conformal dimensions  $(h,\overline h)$. If \eqref{primary} is true only for $f\in\SL(2,\bC)/\bZ_2$, then it is called a \textbf{quasi-primary field}. Note that $\SL(2,\bC)/\bZ_2$ group acts on the holomorphic coordinate as
\[
z \mapsto \frac{a z+b}{c z+d} \qquad \begin{pmatrix}a&b\\ c&d\end{pmatrix}\in\SL(2,\bC)/\bZ_2 ~.\]

How do primary fields transform infinitesimally? Under the infinitesimal conformal transformation $z\rightarrow f(z)= z-\epsilon(z)$, we know that
\begin{align}
\left( \frac{\partial f}{\partial z} \right)^{-h}& = 1 + h \partial_z \epsilon(z) + O(\epsilon^2)~, \cr
\phi(z-\epsilon(z),\bar{z}) &= \phi(z) - \epsilon(z)\partial_z \phi(z,\bar{z}) + O(\epsilon^2)~.
\end{align}
Hence, under an infinitesimal conformal transformation, the variation of a primary field is given by
\begin{equation}
\delta_\epsilon \phi(z,\bar{z}) = \left( h\partial_z \epsilon + \epsilon \partial_z + \bar{h}\overline\partial_{\bar{z}}\bar{\epsilon} +\bar{\epsilon} \overline\partial_{\bar{z}} \right)  \phi(z,\bar{z}) \label{eq:threetwothree}.
\end{equation}

Consequently, using simple complex analysis
\begin{align}
(\partial_w \e(w))\phi(w,\bar w)&=\frac1{2\pi i}\oint_{C_w} dz~ \frac{\e(z) \phi(w,\bar w)}{(z-w)^2}\cr
\e(w)(\partial_w\phi(w,\bar w))&=\frac1{2\pi i}\oint_{C_w} dz ~\frac{\e(z)\partial_w \phi(w,\bar w)}{z-w}~,
\end{align}
one can read off the OPE of a primary operator $\phi$ of conformal dimension $(h,\tilde{h})$ with the energy-momentum tensor $T$ (anti-chiral part $\overline  T$ can be obtained by complex conjugate)
\be\label{Tprimary-OPE} T(z)\,\phi(w,\overline w) =h\frac{\phi(w,\overline w)}{(z-w)^2} + \frac{\partial_w \phi(w,\overline w)}{z-w} + \textrm{regular terms}
\cdots \ee
In general, the OPE of an operator $\cO$ of conformal dimension $(h,\tilde{h})$ with the energy-momentum tensor $T$ and $\overline  T$ takes the form
\[ T(z)\,{\cal O}(w,\overline w) =\quad \cdots + h\frac{{\cal O}(w,\overline w)}{(z-w)^2} + \frac{\partial{\cal O}(w,\overline w)}{z-w} +
\cdots \]

One of the main interests in a CFT is to calculate correlation functions of primary fields. Indeed, the conformal Ward-Takahashi identity can be applied to a correlation function of primary fields
\[
\langle T(z) \phi_1(w_1,\bar{w}_1)\cdots \phi_n(w_n,\bar{w}_n)\rangle=\sum_{i=1}^n \Big(\frac{h_i}{(z-w_i)^2} + \frac{\partial_{w_{i}}}{z-w_i} \Big)\langle\phi_1(w_1,\bar{w}_1)\cdots \phi_n(w_n,\bar{w}_n)\rangle~.
\]
Moreover, conformal symmetry is so powerful that it determines the forms of two-point and three-point functions of primary fields (exercise).

\vspace{.4cm}

\noindent $\bullet$ \textbf{2-point function}

For chiral primary operators $\phi_i$ with conformal dimension $h_i$ ($i=1,2$), their 2-point function  is of form
\be\label{2-pt}
\langle \phi_1(z_1)\phi_2(z_2)\rangle =\delta_{h_1h_2}\frac{d_{12}}{(z_1-z_2)^{2h_1}}
\ee
If $d_{12}$ is non-degenerate, the fields can be normalized such that $d_{12}=\delta_{12}$.

\vspace{.4cm}
\noindent $\bullet$ \textbf{3-point function}

A 3-point function is also completely fixed by conformal invariance up to the appearance of a \textbf{structure constant} $C_{123}$ (exercise)
\be\label{3-pt}
\langle \phi_1(z_1)\phi_2(z_2)\phi_3(z_3)\rangle =\frac{C_{123}}{(z_1-z_2)^{h_1 +h_2 -h_3}(z_2-z_3)^{h_2 +h_3 -h_1}(z_3-z_1)^{h_3 +h_1 -h_2}}
\ee
The structure constant depends on a CFT and, in general, it is not easy to determine it.

\vspace{.4cm}
\noindent $\bullet$ \textbf{Multi-point function}

The computation of multi-point functions involves \textbf{conformal blocks} with the 3-point function. The details are explained in \cite{francesco2012conformal,Blumenhagen:2009zz}.

\subsection*{Free scalar field}

Now let us study conformal Ward-Takahashi identity in the simplest example, the free scalar field \eqref{free-scalar}.
Let us recall that the energy-momentum tensor in 2d free scalar  theory  is
\be T_{a b}^{X}=-\frac{1}{\alpha^{\prime}}\left(\partial_{a} X^{\mu} \partial_{b} X_{\mu}-\frac{1}{2} \delta_{a b}\left(\partial_{c} X^{\mu} \partial^{c} X_{\mu}\right)\right)\ ,\label{classicalt}\ee
As in \eqref{+-}, since the equation of motion for $X^\mu$ is $\partial_z \overline \partial_{\bar z} X^\mu=0$, the classical solution holomorphically factorizes as 
\[X^{\mu}(z, \bar{z})=X^{\mu}(z)+\bar{X}^{\mu}(\bar{z})~.\] 
In \eqref{U1-current}, we find the conserved holomorphic and anti-holomorphic $\U(1)$ current 
\be j^\mu(z) := -\partial X^\mu(z)/\alpha' ~,\quad \textrm{and} \quad \bar{j}^\mu(\bar{z}) := -\overline \partial \overline X^\mu(\overline z)/\alpha' ~.\ee
Moreover, the energy-momentum tensor becomes much simpler in complex coordinates. It is simple to check that $T^X_{z\bar z}=0$ while
\be \label{TX} T^X(z) = -\frac{1}{\a'}\,\partial X^\mu(z)\partial X_\mu(z)~, \quad   \overline T^X(\overline z) = -\frac{1}{\a'}\,\overline \partial \overline X^\mu(\overline z)\overline \partial \overline X_\mu(\overline z)~.\ee

From the definition \eqref{primary}, one can see that $X(z,\overline z)$ is a primary field of conformal dimension $(0,0)$. However, since the conformal dimension is of $(0,0)$, the two-point function does not exactly take the form \eqref{2-pt}. Indeed, the OPE $XX$ tells us that the propagator takes the form
\[
\langle X^\mu(z,\overline z)X^\nu(w,\overline w) \rangle=-\frac{\a'}{2}\eta^{\mu\nu}\log|z-w|^2~.
\]
Also, for each $\mu$, the currents $j^\mu(z)$, $\overline j^\mu(\overline z)$ are primary fields of conformal dimension (1,0) and (0,1),
respectively. Focusing only on the holomorphic part, an immediate check is their correlation function
\[
\langle \partial X^\mu(z) \partial X^\nu(w) \rangle =-\frac{\a'}{2}\frac{\eta^{\mu\nu}}{(z-w)^2}~,
\]
which takes the form \eqref{2-pt}. To convince ourselves completely, we need to compute the OPE with the energy-momentum tensor by Wick's theorem %
\begin{align}
T^X(z)\,\partial X^\mu(w) &= -\frac{1}{\a'}: \partial X^\nu(z)\partial X_\nu(z): \partial X^\mu(w)\cr
&= \frac{\partial X^\mu(w)}{(z-w)^2}
+\frac{\partial^2 X^\mu(w)}{z-w} + \textrm{regular terms} \cdots
\end{align}
This is indeed the OPE for a primary operator of conformal dimension $h=1$.

Finally, let us check to see the $TT$ OPE of the energy-momentum tensors. This can be done by applying the Wick contractions, and the result is
\begin{align}\label{scalartt} T^X(z)\,T^X(w) &= \frac{1}{\alpha^{\prime\,2}}\ :\partial X^\mu(z)\,\partial X_\mu(z):\ :\partial X^\nu(w)\,\partial X_\nu(w): \cr
&= \frac{D/2}{(z-w)^4} + \frac{2T^X(w)}{(z-w)^2} +
\frac{\partial T^X(w)}{z-w} + \ldots \end{align}
 Therefore, the energy-momentum tensor is an operator of conformal dimension $(h,\overline{h})=(2,0)$. Because there is a higher singular term proportional to $(z-w)^{-4}$, the energy-momentum tensor is not a primary field. In fact, this is a general property of the $TT$ OPE in all 2d CFTs.

\subsection{Virasoro algebra}

For the free scalar field, we have already seen that $T$ has
conformal dimension $(h,\tilde{h})=(2,0)$. This remains true in any CFT. The reason for this is simple:
 $T_{ab}$ has dimension  $\Delta =2$ because we obtain the energy by integrating over space.
It has spin $s=2$ because it is a symmetric two-tensor. However, these two pieces of information
are equivalent to the statement that $T$ is an
operator of conformal dimension $(2,0)$. This means that the
$TT$ OPE takes the form,
\[ T(z)\,T(w) = \ldots + \frac{2T(w)}{(z-w)^2} + \frac{\partial T(w)}{z-w} + \ldots \]
and a similar one for $\bar{T}\bar{T}$. What other terms could we have in this expansion? Since
each term has dimension $\Delta=4$, dimensional analysis indeed tells us that the singular part of the OPE takes the form
\[T(z)\,T(w) = \frac{c/2}{(z-w)^4}+ \frac{2T(w)}{(z-w)^2} + \frac{\partial T (w)}{z-w}
+\ldots \]

From the OPE of the energy-momentum tensor, one can see its variation under an infinitesimal conformal transformation $z\to z -\e(z)$
\begin{align}\label{inf-T-variation}
\delta_\e T(w)&= \frac{1}{2\pi i}\oint_{C_w} dz~ \e(z)T(z)T(w)\cr
&=\e(w)\partial T(w)+2\e'(w)T(w)+\frac{c}{12}\e'''(w)
\end{align}
One can verify by a straightforward computation that this is the infinitesimal version of
 the following transformation under finite transformation $z\to w(z)$:
\begin{equation}
T'(w) = \left( \frac{\partial w}{\partial z} \right)^{-2} \left[T(z) - \frac{c}{12} S\left( w,z \right) \right], \label{eq:finitettransform}
\end{equation}
where the \textbf{Schwarzian derivative} $S$ is defined as
\begin{equation}
S(w,z) := \frac{1}{(\partial_z w)^2} \left((\partial_z w)(\partial_z^3 w)-\frac32(\partial_z^2 w)^2  \right).
\end{equation}
Using the mapping from the plane to the cylinder $z=e^{-iw}$ ($w=\sigma+it$) as in Figure \ref{fig3}, the energy-momentum tensor is transformed as
\begin{equation}
T_{cyl}(w) =- z^2 T(z) + \frac{c}{24}.
\end{equation}
The Laurent mode expansion of the energy-momentum tensor on the cylinder is therefore
\begin{equation}\label{Casimir2}
T_{cyl}(w) =- \sum_{n\in \mathbb{Z}} \left( L_n - \frac{c}{24}\delta_{n,0} \right) e^{inw}~.
\end{equation}
Including the contribution from the anti-holomorphic sector, the Hamiltonian is defined by
\[ H\equiv \int d\sigma\ T_{\tau\tau} = -\int d\sigma\, (T_{ww} + \bar{T}_{\bar w\bar  w})=L_0+\overline L_0-\frac{c}{12}~.\]
This tells us that the ground state energy on the cylinder is
\[ E = -\frac{c}{12}~.\]
This is indeed the (negative) Casimir energy on a cylinder. In the string sigma model, each coordinate $X^\mu$ of the target space gives rise to a free scalar theory, and it is easy to see from the computation  \eqref{scalartt} that each coordinate $X^\mu$ contributes to the central charge by
$c=1$. Thus, each target dimension yields the energy density $E= -1/12$ as we have seen in the quantization of bosonic string theory \eqref{Casimir}. Moreover, we have the central charge \be\label{cX} c^X=D\ee for the string sigma model with the target space $\bR^{1,D-1}$.

\begin{figure}\centering
\includegraphics[width=10cm]{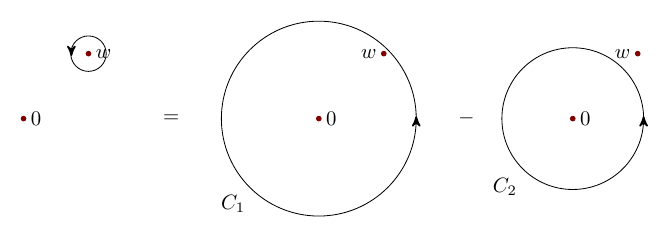}
\caption{Contours are chosen according to the radial ordering.}\label{OPE-com-relation}
\end{figure}

In the Euclidean flat space, the mode expansion of the energy-momentum tensor is expressed as
\[
T(z)= \sum_{n\in \mathbb{Z}}\frac{L_n}{{z}^{n+2}}~,
\]
where the shift by two in the exponent of $z$ is due to the conformal dimension of the energy-momentum tensor.
It is natural to find the commutation relation of the generators $L_m$.  The commutator can be computed by
\[ [L_m,L_n] = \left(\oint\frac{dz}{2\pi i}\oint\frac{dw}{2\pi i}\ -\ \oint\frac{dw}{2\pi i}\oint\frac{dz}{2\pi i}\right)\,z^{m+1}w^{n+1}\,R(T(z)T(w))\]
Here we impose the radial ordering
\[R(A(z) B(w)):=\begin{cases}A(z) B(w) &\text { for }|z|>|w| \\ B(w) A(z) &\text { for }|w|>|z|\end{cases}\]
according to which the contour is chosen as in Figure \ref{OPE-com-relation}.
Clever manipulation of the contour makes life easier
\begin{align}
[L_m,L_n] &= \oint \frac{dw}{2\pi i}\oint_w\frac{dz}{2\pi i} \ z^{m+1}w^{n+1}\,T(z)\,T(w) \\
&= \oint\frac{dw}{2\pi i}\,\textrm{Res}\left[z^{m+1}w^{n+1}\left(\frac{c/2}{(z-w)^4}+\frac{2T(w)}{(z-w)^2}
+\frac{\partial T(w)}{z-w} + \ldots\right)\right]    \label{LL}
\end{align}
A simple computation (exercise) leads to the \textbf{Virasoro algebra}
\[ [L_m,L_n] = (m-n)L_{m+n} + \frac{c}{12}m(m^2-1)\delta_{m+n,0}\]
For instance, the Virasoro generators of the free scalar theory are expressed in terms of the modes as in  \eqref{Virasoro}.

\subsection{Vertex operators}\label{sec:vertex-operator}

We shall connect 2d conformal field theory to string theory, considering vertex operators.
The state-operator correspondence tells us that there are corresponding local operators for the tachyon and the massless states
\eqref{massless} in the bosonic closed string theory.
The tachyon state is just a vacuum state with a certain momentum $k^\mu$. Therefore, the corresponding operator is
\be\label{tachyon-vertex}
|0;k \rangle \quad \leftrightarrow \quad : e^{ik X(0,0)}:~.
\ee
This can be easily verified from the operator analogue of $p^{\mu} |0;k \rangle = k^{\mu} |0;k \rangle$, namely,
\begin{align}\nonumber
\partial X^{\mu}(z): e^{i k \cdot X(0)}: \ \sim \   \frac{-i\alpha'k^{\mu}}{2z}: e^{i k \cdot X(0)}:~.
\end{align}
The first excited states are obtained by acting $ \alpha^\mu_{-1}\tilde\alpha^\nu_{-1}$ on the vacuum $|0;k \rangle$. Each mode in \eqref{mode-exp} can be extracted by the Fourier transformation of $\partial X^\mu(z)$ and the corresponding operator is
\begin{align}\nonumber
 \alpha^\mu_{-m} = i\sqrt{\frac{2}{\alpha'}} \oint \frac{dz}{2\pi i} z^{-m} \partial X^\mu(z)
 \quad \to \quad
 i\sqrt{\frac{2}{\alpha'}} \frac{1}{(m-1)!} \partial^m X^\mu(0)  ~.
\end{align}
Thus, we have operators corresponding to the massless states
\[
\zeta_{\mu\nu}\alpha^\mu_{-1}\ol \alpha^\nu_{-1} |0;k\rangle \quad \leftrightarrow \quad  \zeta_{\mu\nu} \partial X^\mu \ol\partial X^\nu :e^{ikX}:
\]
where $\zeta_{\mu\nu}$ are the polarization tensors subject to $k^\mu \zeta_{\mu\nu} = 0$.


The string amplitude is
\begin{align}\label{amplitude}
 A_n = \sum_g \int \left(\cD h_{ab}\right)_{g,n} \int \cD X^\mu e^{-S_\sigma [X^\mu,h_{ab}]} \prod_{i=1}^n \int d^2z \sqrt h V_i  \ .
\end{align}
Here $\hat V_i =\int d^2z \sqrt h V_i$ is an operator corresponding to a string state. It is integrated out over the world-sheet to be Diff$\times$Weyl invariant.

\subsubsection*{Mass from vertex operator}
Now, let us consider a constant scaling $z \to \lambda z$ and $\ol z \to \ol\lambda \ol z$.
Under the scaling, a field transforms as $\phi(z,\ol z) \to \lambda^{-h}\ol\lambda^{-\ol h} \phi(z,\ol z)$,
which should compensate for the scaling of the measure $dzd\ol z \to \lambda\ol\lambda dzd\ol z$.
Namely, $h=\ol h=1$.
Conformal dimensions of the vertex operators are
\begin{align}\nonumber
\renewcommand\arraystretch{1.2}
 \begin{array}{lll}
  \textrm{Name} & \mathcal O & (h, \ol h) \\\hline
  \textrm{Tachyon} & e^{ik\cdot X} & (\frac{\alpha'k^2}{4},\frac{\alpha'k^2}{4}) \\
  \textrm{1st excited states} & \zeta_{\mu\nu} \partial X^\mu \ol\partial X^\nu e^{ikX} & (1+\frac{\alpha'k^2}{4},1+\frac{\alpha'k^2}{4}) \\
 \end{array}
\end{align}
The consistency condition for the tachyon leads to
\begin{align}\nonumber
 M^2 = -k^2_\textrm{Tachyon} = -\frac{4}{\alpha'} \ .
\end{align}
Similarly,  the first excited states lead to the massless condition
\begin{align}\nonumber
 M^2 = -k^2_\textrm{1st} = 0 \ .
\end{align}
Both results are consistent with the analysis in \eqref{closed-vaccuum} and \eqref{massless}.
Here one may consider that one can obtain the string spectrum
in arbitrary spacetime dimensions.
However, as we will show in the next section, the on-shell condition only holds in $D=26$ where the Weyl transformation is a quantum symmetry.

\section{Weyl anomaly}\label{sec:Weyl}

The classical action \eqref{string-sigma} of the string sigma model is invariant under the Weyl symmetry, and the Weyl invariance implies the traceless condition \eqref{traceless1} of the energy-momentum tensor. However, we will see that there can be an anomaly in the Weyl symmetry if the string world-sheet is curved. As we will see, it is characterized by the central charge $T^a_{~a} = -\frac{c}{12} R^{(2)}$ where $R^{(2)}$ is the world-sheet Ricci scalar curvature.

Moreover, in string theory, a target space can also be a non-trivial curved spacetime where the action becomes
\begin{align}\nonumber
 S = \frac{1}{4\pi \alpha'}\int  d^2z \sqrt h\ h^{ab} \partial_a X^\mu \partial_{b} X^\nu G_{\mu\nu}(X) +\cdots \ .
\end{align}
This is no longer a free theory, and it describes an interacting non-linear theory, called the \textbf{(string) non-linear sigma model}. In an interacting quantum field theory,
the breakdown of scale invariance is described in terms of a $\beta$ function. In the non-linear sigma model, the anomaly of the Weyl invariance due to the curved target spacetime is also characterized by $\beta$-functions.

\subsection{Weyl anomaly from curved world-sheet}
Although the string sigma model is classically Weyl-invariant, in quantum theory, the trace of the energy-momentum tensor can be nonzero due to anomaly.
Nonetheless, it is
\begin{itemize}\setlength\itemsep{.1pt}
 \item world-sheet diff invariant,
 \item zero on a flat world-sheet $\bR^2$,
 \item world-sheet mass dimension two.
\end{itemize}
Hence, they constrain the form of the trace of the energy-momentum tensor as
\begin{align}\nonumber
 T^a_{~a} = k R^{(2)} \ ,
\end{align}
where $R^{(2)}$ is the world-sheet Ricci scalar curvature. In this subsection, we shall determine the coefficient $k$.

Given a curved world-sheet Riemann surface $(\Sigma,h_{ab})$ with non-trivial metric, we can always take a local coordinate (called \textbf{conformal gauge}) of the world-sheet Riemann surface in which the metric is a conformally flat
\begin{align}\label{conformal-gauge}
 ds^2 = e^{2\omega(\sigma^1,\sigma^2)} (d\sigma^1d\sigma^1+d\sigma^2d\sigma^2) = e^{2\omega(z,\ol z)} dzd\ol z \ .
\end{align}
This local coordinate is called an \textbf{isothermal coordinate}.
In the isothermal coordinate, the scalar curvature and non-trivial Christoffel symbols are expressed as
\begin{align}\label{conformal-gauge2}
 R^{(2)} = -8 e^{-2 \omega} \partial \bar{\partial} \omega \ , \qquad \Gamma_{z z}^{z}=2 \partial \omega, \quad \Gamma_{\bar{z}\bar{z}}^{\bar{z}}=2 \bar{\partial} \omega
\end{align}
and the other Christoffel symbols are zero $\Gamma^a_{bc}=0$.
Therefore, the diagonal element of the energy-momentum tensor becomes
\begin{align}\nonumber
 T_{z\ol z} = \frac{1}{4} e^{2\omega}\ T^a_{~a} = -2k \partial\ol\partial \omega \ .
\end{align}
Also, the conservation $\nabla_a T^a_{~ b} = 0$ of the energy-momentum tensor now becomes
\begin{align}\nonumber
 0 =& \nabla^z T_{zz} +\nabla^{\ol z} T_{\ol zz} = g^{\ol z z} [\nabla_{\ol z} T_{zz} +\nabla_{z} T_{\ol zz}]\cr
 =&g^{\ol z z} [\ol\partial  T_{z z}+(\partial -2 \partial \omega)(-2 k \partial \bar{\partial} \omega)]\cr
 =& g^{\ol z z} [\ol\partial \left( T_{zz} -2k\left( \partial\partial\omega -\partial\omega\partial\omega \right) \right)]~.
\end{align}
This implies that $T_{zz}$ deviates from a holomorphic operator $T(z)$
\be\label{T-curved}
T_{zz} = 2k\left( \partial\partial\omega -\partial\omega\partial\omega \right)+T(z) ~.
\ee

On a curved world-sheet, the conformal Ward-Takahashi identity \eqref{CWT} becomes
\begin{align}\nonumber
 \delta_{\epsilon,\ol\epsilon} \mathcal O (w,\ol w) &= \int_{\Sigma} \frac{d^2z}{2\pi i}\sqrt{h}
 \left\{\left(\nabla^{{z}}\epsilon^z T_{z z}+\nabla^{\bar z} \epsilon^z T_{\bar{z} z}\right)+\left(\nabla^{\bar z} \ol\epsilon^{\ol z}T_{\bar{z}\bar{z}}+\nabla^{z} \ol\epsilon^{\ol z}T_{z\bar{z}}\right) \right\} \mathcal O (w,\ol w) \nonumber\cr
 &= \int_{\Sigma} \frac{d^2z}{2\pi  i}
 \left\{\left(\nabla_{\bar{z}}\epsilon(z) T_{z z}+\nabla_{z} \epsilon(z)T_{\bar{z} z}\right)+\left(\nabla_{z} \ol\epsilon(\ol z)T_{\bar{z}\bar{z}}+\nabla_{\bar{z}} \ol\epsilon(\ol z)T_{z\bar{z}}\right) \right\} \mathcal O (w,\ol w) \nonumber\cr
 &= \int_{\Sigma}\frac{d^2z}{2\pi i}
 \left\{ \ol\partial\left(\epsilon(z)T(z)\right) +\partial\left(\ol\epsilon(\ol z) \ol T(\ol z)\right) \right\} \mathcal O (w,\ol w) \ ,
\end{align}
where we use the current conservation \[T_{z\bar z}(\nabla^z \e^{\bar z}+\nabla^{\bar z} \e^{z})=0\] from the second and third line. This means that $T(z)$ obeys the flat version of the conformal Ward-Takahashi identity \eqref{CWT}.
This gives the transform property of $T(z)$ as in the flat space \eqref{inf-T-variation}
\begin{align}\nonumber
 \delta_\epsilon T(z) = \epsilon(z)\partial T(z) +2\partial\epsilon(z) T(z) +\frac{c}{12} \partial^3 \epsilon(z) \ .
\end{align}
On the other hand, \eqref{T-curved} yields
\begin{align}
 \delta_\epsilon T(z) = \delta_\epsilon T_{zz} -2k\left( \partial\partial\delta_\epsilon\omega -2\partial\omega\partial\delta_\epsilon\omega \right) \ .
 \label{eq:curvedTransfT}
\end{align}
Using (finite) transformations,
\begin{align}\nonumber
 &z \to \wt z = z -\epsilon(z) \ , \cr
 &T_{zz}  \to \wt T_{\wt z\wt z} = \left( \partial_{z} \wt z \right)^{-2} T_{zz} \ , \cr
 &\omega(z) \to \wt \omega (\wt z) = \omega(z) -\frac{1}{2} \log \left| \partial_z \wt z \right|^2 \ ,
\end{align}
(\ref{eq:curvedTransfT}) can be expressed as
\begin{align}
 \delta_\epsilon T(z) = \epsilon(z)\partial T(z) +2\partial\epsilon(z) T(z) -k \partial^3 \epsilon(z) \ .
\end{align}
Finally, we see that the Weyl anomaly is proportional to the central charge
\begin{align}\label{Weyl-anomaly}
 T^a_{~a} = kR^{(2)} = -\frac{c}{12} R^{(2)}  ~.
\end{align}
We have seen in \eqref{cX} that the central charge of the string sigma model is equal to the dimension $D$ of the target space. Hence, naively looking, the bosonic string theory would suffer from the Weyl anomaly. However, as we will see in \S\ref{sec:ghost}, we introduce ``ghost CFT'' to fix gauge in the world-sheet path integral, and the ghost CFT has $c^{\rm gh}=-26$.
As a result, the bosonic string theory is Weyl invariant at quantum level only if $D=26$.

\subsection{Non-linear sigma model}

In \S\ref{sec:bosonic}, we start with a flat $D$-dimensional target space, and the quantization of bosonic string naturally leads to the graviton \eqref{massless}.
Therefore, in general, we can consider that the target spacetime is curved with a non-trivial metric so that the action is
\begin{align}\nonumber
 S[X^\mu, h_{ab}] = \frac{1}{4\pi \alpha'} \int d^2\sigma \left( \sqrt h h^{ab} \partial_a X^\mu \partial_b X^\nu G_{\mu\nu}(X)
 \right) \ .
\end{align}
This is called the non-linear sigma model. In this subsection, we provide a concise overview of the non-linear sigma model where the massless fields \eqref{massless} receive quantum corrections. Although we present key results, many of their derivations are not included here. For more details, we direct the reader to \cite[\S3.4]{GSW}.

If we consider a perturbation from the flat-metric in this action
\begin{align}\nonumber
 G_{\mu\nu}(X) = \eta_{\mu\nu} + f_{\mu\nu} (X) \ ,
\end{align}
then the partition function becomes
\begin{align}\nonumber
 Z &= \int \cD h_{ab} \cD X^\mu e^{-S} \nonumber\cr
 &= \int \cD h_{ab} \cD X^\mu e^{-S_0} \left(
 1+\frac{1}{4\pi \alpha'} \int d^2\sigma \left( \sqrt h h^{ab} \partial_a X^\mu \partial_b X^\nu f_{\mu\nu}(X) \right) +\cdots
 \right) \ .
\end{align}
Notice that the perturbative part is the graviton operator
with wave function $f_{\mu\nu}(X) = \zeta_{\mu\nu} e^{ik\cdot X}$.

In a curved target, the theory is no longer free, so that we need to take into account various quantum effects. We will see that $\beta$-functions encode the anomaly of scaling invariance at quantum level.
Note that the discussion here is brief, and the reader is referred to \cite{Callan:1989nz} for the details.
As a 2d field theory, we can consider the vacuum expectation value (VEV) for $X$,
which we set to $X_0$:
\begin{align}\nonumber
 \wh X(\sigma,\tau) = X_0 +X(\sigma,\tau) \ .
\end{align}
On the other hand, $X_0$ is a certain point in spacetime and
we will expand the metric around this point:
\begin{align}\nonumber
 G_{\mu\nu} (X) = G_{\mu\nu} -\frac{1}{3} R_{\mu\lambda\nu\rho} X^\lambda X^\rho +\mathcal O(X^3) \ ,
\end{align}
where $G_{\mu\nu}$ and $R_{\mu\lambda\nu\rho}$ are a metric and a Riemann curvature tensor of the target spacetime at $X_0$, respectively.
In a field-theoretic sense, these can be understood as coupling constants
\begin{align}\nonumber
 S &= \frac{1}{4\pi \alpha'} \int d^2\sigma \ \partial^a X^\mu \partial_a X^\nu
 \left( G_{\mu\nu} -\frac{1}{3} R_{\mu\lambda\nu\rho} X^\lambda X^\rho +\cdots
 \right)  \nonumber\cr
 &\to \frac{1}{2} \int d^2\sigma \ \partial^a X^\mu \partial_a X^\nu
 \left( G_{\mu\nu} -\frac{2\pi \alpha'}{3} R_{\mu\lambda\nu\rho} X^\lambda X^\rho +\mathcal O(\alpha'^2)
 \right) \ .
\end{align}
Here we rescale the field $X \to \sqrt{2\pi \alpha'} X$ so that
the expansion looks like a ``stringy expansion''.

\subsubsection*{Perturbation theory for non-linear sigma model}

We want to check if the theory (non-linear sigma model) has Weyl anomaly.
As we briefly saw, it is an interacting theory, and
interacting theories have generally non-trivial $\beta$-functions:
\begin{align}\nonumber
 \beta[g] \equiv E \frac{\partial}{\partial E} g(E) = \frac{\partial}{\partial (\log E)} g(E) \ ,
\end{align}
where $g$ is a coupling constant and $E$ is a characteristic energy scale.
When we consider a global scaling of coordinate:$z \to \wt z = \lambda z = (1-\epsilon) z = e^{-\epsilon} z$,
energy scales oppositely: $E \to \wt E = \frac{1}{ \lambda} E = e^{\epsilon} E$.
So the $\beta$-function can be written as
\begin{align}\nonumber
 \beta[g] = \frac{\partial}{\partial \epsilon} g(\epsilon) \ .
\end{align}
The variation of the action is expressed in two ways:
\begin{align}\nonumber
 &\delta_\epsilon S =
 \begin{cases}
  \int \frac{d^2\sigma}{2\pi} \sqrt{h} \delta_\epsilon h^{ab}  T_{ba} = -\epsilon \int \frac{d^2\sigma}{2\pi} T^a_{~a} \ , \\[2pt]
  \frac{1}{4\pi \alpha'} \int d^2\sigma \ \partial^a X^\mu \partial_a X^\nu
  \left( \epsilon \frac{\partial}{\partial \epsilon} G_{\mu\nu}(\epsilon) +\cdots \right) \ ,
 \end{cases}
\end{align}
where the first variation is a formal transformation of the theory, and in quantum regime,
it should be proportional to the trace part of the energy-momentum tensor.
On the other hand, the second variation is the actual theory with an assumption that
$\epsilon$ dependence of the theory is only in the coupling constants.
Identifying them, we have
\begin{align}\nonumber
 T^a_{~a} = -\frac{1}{2\alpha'} \beta[G_{\mu\nu}] \partial^b X^\mu \partial_b X^\nu +\cdots \ .
\end{align}
This shows that the anomaly is parametrized by $\beta$-functions.

Let us consider perturbation theory, namely
loop corrections to two-point function etc, so that we can see if the theory is anomalous.
\begin{align}\nonumber
 &\langle X^\mu (\sigma_1) X^\nu (\sigma_2) \rangle \nonumber\cr & \qquad =
 \int \frac{d^2k}{(2\pi)^2} \frac{2\pi \alpha'}{k^2} e^{ik\cdot (\sigma_1-\sigma_2)}
 \left\{ G^{\mu\nu} +\frac{2\pi \alpha'}{3} R^{\mu\nu} \left(
 \int \frac{d^2p}{(2\pi)^2} \frac{1}{p^2} + \frac{1}{k^2}\int \frac{d^2p}{(2\pi)^2}
 \right)+\cdots \right\} \ .
\end{align}
Among terms proportional to $R^{\mu\nu}$ in the integrand, the first integral provides a logarithmic and the second does quadratic divergence.
Since the quadratic divergence is discarded by dimensional regularization, we focus on the logarithmic divergence and introduce regularization parameters:
\begin{align}\nonumber
 \int_E^\Lambda \frac{d^2p}{(2\pi)^2} \frac{1}{p^2} = \frac{1}{2\pi} \log
 \left( \frac{\Lambda}{E} \right) \ ,
\end{align}
where $\Lambda$ is an ultra-violet (UV) energy scale supposed to be $\infty$,
and $E$ is an infra-red (IR) energy scale supposed to be our life energy scale, which is very low ($\sim 0$).

The divergence can be subtracted by counter terms as follows.
The bare action $\wh S = S +S_\mathrm{ct}$ describes UV physics at the energy scale $\Lambda$ whereas the physical action $S$ describes IR physics at the energy scale $E$. They are related by renormalization
\begin{align}\nonumber
 \wh S = \frac{1}{4\pi \alpha'} \int d^2\sigma \ \partial^a \wh X^\mu \partial_a \wh X^\nu
 \left( \wh G_{\mu\nu} -\frac{1}{3} \wh R_{\mu\lambda\nu\rho} \wh X^\lambda \wh X^\rho +\cdots
 \right)
\end{align}
with
\begin{align}\nonumber
 &\wh X^\mu = Z^\mu_\nu X^\nu \ , \quad Z^\mu_\nu = \delta^\mu_\nu +
 \sum_{n=1}^\infty \alpha'^n Z_{\nu,(n)}^\mu (\Lambda/E) \ , \cr
 &\wh G^{\mu\nu} = G^{\mu\nu} +\sum_{n=1}^\infty \alpha'^n G_{(n)}^{\mu\nu} (\Lambda/E) \ ,
 \textrm{etc.}
\end{align}
Note that the bare action only depends on $\Lambda$ (not on $E$), and hence,
the bare coupling constants ($\wh G_{\mu\nu}$ etc) only depend on high energy $\Lambda$.
The counter terms lead to other contributions
\begin{align}
 \langle X^\mu (\sigma_1) X^\nu (\sigma_2) \rangle  \sim
 \left\{ G^{\mu\nu} +\frac{\alpha'}{3} R^{\mu\nu} \log \left( \frac{\Lambda}{E} \right)
 -\alpha' \left(G_{(1)}^{\mu\nu} +Z_{(1)}^{\mu\nu} +Z_{(1)}^{\nu\mu}
 \right)+\cdots \right\} \ .
 \label{eq:renom2pt}
\end{align}
Unfortunately, the equation above cannot fix the ratio between $G_{(1)}$ and $Z_{(1)}$.
We need further information like a 4-point function to determine the ratio.
We simply list the result:
\begin{align}\nonumber
 &G^{\mu\nu}_{(1)} = R^{\mu\nu} \log \left( \frac{\Lambda}{E} \right) \ , \quad
 Z^{\mu\nu}_{(1)} =  -\frac{1}{3} R^{\mu\nu} \log \left( \frac{\Lambda}{E} \right) \ ,
\end{align}
which does cancel the divergent term in (\ref{eq:renom2pt}).
From the result we can derive the $\beta$-function:
\begin{align}\nonumber
 &G_{\mu\nu} (E,\Lambda) = \wh G_{\mu\nu} (\Lambda) - \alpha' R_{\mu\nu} \log \left( \frac{\Lambda}{E} \right) \ , \cr
 &\beta[G_{\mu\nu}] = \frac{\partial}{\partial (\log E)} G_{\mu\nu} (E,\Lambda) = \alpha' R_{\mu\nu} \ .
\end{align}
Therefore, for the theory to be anomaly-free, the spacetime is required to be Ricci-flat ($R_{\mu\nu}=0$).

\subsubsection*{Non-linear sigma model (general)}

If we incorporate all the massless states \eqref{massless}, we can write the most general form of the non-linear sigma model for a closed string as
\begin{align}\label{NLSM-general}
 &S_{\textrm{closed}} = \frac{1}{4\pi \alpha'} \int d^2\sigma \left( \sqrt h h^{ab} \partial_a X^\mu \partial_b X^\nu G_{\mu\nu}(X)
 +i \varepsilon^{ab} \partial_a X^\mu \partial_b X^\nu B_{\mu\nu}(X)
 +\alpha' \sqrt h R^{(2)} \Phi(X)
 \right) \ .
\end{align}
Note that the $B$-field is a higher dimensional analogue of gauge fields (mathematically called ``gerbe'').
       Its gauge transformation is given by
  \begin{align}\label{B-gauge}
   \delta B_{\mu\nu} = \partial_\mu \Lambda_\nu -\partial_\nu \Lambda_\mu \ ,
  \end{align}
       and the field strength $H_{\mu\nu\lambda}$ is defined as
   \begin{align}\nonumber
    H_{\mu\nu\lambda} = \partial_\mu B_{\nu\lambda}
    +\partial_\nu B_{\lambda\mu} +\partial_\lambda B_{\mu\nu} \ .
   \end{align}
The $B$-field often is the source of ``stringy'' effects, and it plays an important role on many occasions.

Again, the Weyl anomaly is characterized by $\beta$-functions for the massless fields
\begin{align}\nonumber
 T^a_{~a} = -\frac{1}{2\alpha'} \beta[G_{\mu\nu}] \partial_a X^\mu \partial^a X^\nu
 -\frac{i}{2\alpha'} \beta[B_{\mu\nu}] \varepsilon^{ab} \partial_a X^\mu \partial_b X^\nu
 -\frac{1}{2} \beta [\Phi] R^{(2)}
\end{align}
where $\beta$-functions are expressed as
\begin{align}\nonumber
 &\beta[G_{\mu\nu}] = \alpha' R_{\mu\nu} +2\alpha' \nabla_\mu \nabla_\nu \Phi
 -\frac{\alpha'}{4}H_{\mu\lambda\rho}H_\nu^{\ \lambda\rho} +\mathcal O(\alpha'^2) \ , \cr
 &\beta[B_{\mu\nu}] = -\frac{\alpha'}{2} \nabla^\lambda H_{\lambda\mu\nu}
 +\alpha' \nabla^\lambda \Phi H_{\lambda\mu\nu} +\mathcal O(\alpha'^2) \ , \cr
 &\beta[\Phi] = \frac{D-26}{6} -\frac{\alpha'}{2} \nabla^2\Phi
 +\alpha' \nabla^\lambda\Phi \nabla_\lambda\Phi
 -\frac{\alpha'}{24}H_{\mu\lambda\rho}H^{\mu\lambda\rho} +\mathcal O(\alpha'^2) \ .
\end{align}
In order for the theory to be non-anomalous, all the $\beta$-functions must be zero, and for a flat target with constant dilaton and $H=0$, the vanishing of $\beta[\Phi]$ again requires $D=26$. Nonconstant dilaton backgrounds can modify this conclusion, as explored in the problems.
The perturbative quantum field theory on the world-sheet can describe the classical gravity from the vanishing of $\beta$ function.
\begin{enumerate}
\item   The equation
$\beta[G_{\mu\nu}]=0$
is analogous to Einstein's equation
    with source terms from the
    $B$-field
    and the dilaton.
\item
    The equation $\beta[B_{\mu\nu}]=0$ is the generalization of Maxwell's equation.
\end{enumerate}

The condition that $\beta$-functions vanish ($\beta=0$) is equivalent to the equation of motion for the following spacetime effective action
\footnote{
 We are in the regime where the perturbation theory works well. Namely, the length scale of the target space is long compared to the string scale. We thus can ignore the internal structure of the string and write the following
low energy effective theory for quantum gravity.}
  \begin{align}\label{bosonic-NS}
   S_\mathrm{eff} = \frac{1}{2\kappa_D^2} \int d^D X \sqrt{-G} e^{-2\Phi} \left[
   \frac{2(D-26)}{3\alpha'} +R -\frac{1}{12} H_{\mu\lambda\rho}H^{\mu\lambda\rho}
   +4 \nabla^\lambda\Phi \nabla_\lambda\Phi +\mathcal O(\alpha'^2)
   \right] \ .
  \end{align}

\begin{figure}[htb]
\centerline{\includegraphics[width=400pt]{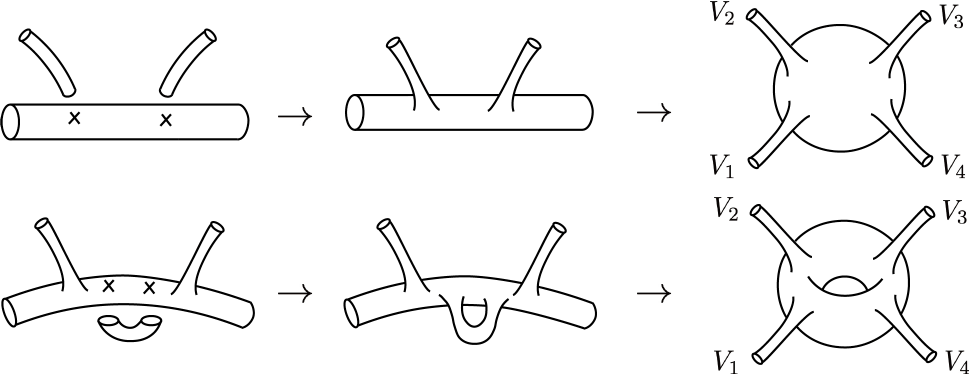}}
 \caption{4-point amplitude example. The upper one is a construction of 4pt tree amplitude from cylinders. The lower one is a construction of 4-point 1-loop amplitude from the tree amplitude.}
\label{stringcoupling2.eps}
\end{figure}

The action \eqref{NLSM-general} tells us another important fact in string theory. The dilaton field is coupled to the world-sheet Ricci scalar.
If the dilaton takes an expectation value $\langle \Phi(X)\rangle=\Phi$, the dilaton action gives the Euler characteristic of the world-sheet Riemann surface
  \begin{align}\nonumber
   S_\mathrm{dilaton} &= \frac{1}{4\pi \alpha'} \int d^2\sigma \left( \alpha' \sqrt{h} R^{(2)} \Phi(X) \right)  \nonumber\cr
   &\to \frac{1}{4\pi} \int d^2\sigma \left( \sqrt{h} R^{(2)} \Phi \right)
   = \Phi (2-2g)
  \end{align}
  In other words, the 2d gravity is topological with no physical degrees of freedom.
       Defining $g_s = e^\Phi$, and the dilaton part of the action becomes
  \begin{align}\nonumber
   e^{-S_\mathrm{dilaton}} = g_s^{2g-2} \ .
  \end{align}
  This implies that $g_s$ can be understood as a string coupling.
An $n$-point string tree amplitude can be understood as $n-2$ cylinders attaching to a cylinder.
       Hence, the amplitude should be proportional to $g_s^{n-2}$.
A higher loop (higher genus) amplitude can be derived by attaching $g$ cylinders to the tree amplitude,
       and the amplitude should be $\wh A_{n,g} \propto g_s^{n-2+2g}$. (See Figure \ref{stringcoupling2.eps}.)
       Usually, vertex operators are re-normalized so that $g_s^n$ is included in the definition of $V_1 \cdots V_n$.
       Therefore, the re-normalized amplitude should be
  \begin{align}\nonumber
   A_{n,g} \propto g_s^{2g-2} \ ,
  \end{align}
       which coincides with the dilaton action. In conclusion, the string coupling $g_s$ is just the expectation value of the dilaton field.

\section{BRST quantization}\label{sec:BRST}

We have been studying bosonic string theory, but there are several caveats.

\vspace{.3cm}
\noindent$\bullet$ In the light-cone quantization \eqref{lcg}, Lorentz invariance is not manifest.  Can we quantize strings in a way that is manifestly Lorentz invariant?

\vspace{.3cm}

\noindent$\bullet$ We have seen that the Weyl symmetry of the string sigma model is anomalous \eqref{Weyl-anomaly} on a curved world-sheet. How can the bosonic string be anomaly-free?

\vspace{.3cm}

\noindent$\bullet$ Although we learned that string amplitude is expressed via Feynman path integral
\eqref{amplitude},
we do not know how to perform this path integral. In particular, the path integral is endowed with huge gauge symmetries, world-sheet diffeomorphism and Weyl symmetry. How can we treat integration measure and fix gauge in the path integral?

\vspace{.3cm}

To answer these questions, we will study the quantization procedure via path integral, which is often called \textbf{modern covariant quantization}. This method uses  the analog of the
Faddeev-Popov method of gauge theories. Furthermore, the physical state condition is imposed via the BRST symmetry.

After gauge-fixing the reparametrization and Weyl symmetry, the integral over $h_{ab}$ turns into a path-integral over ghost CFT, which has $c^{g}=-26$. Therefore, in order for the theory to be Weyl anomaly-free, the matter part of the theory has to be a CFT with $c^X=26$.

\subsection{Quantization via path integral}\label{sec:ghost}

\begin{figure}[ht]\centering
\includegraphics[width=9cm]{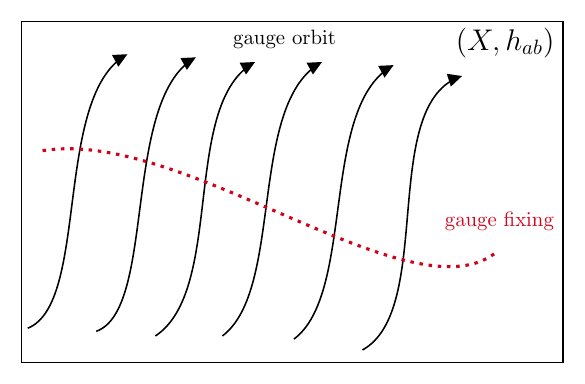}
\caption{We need to fix a gauge in the field configuration space. Physically inequivalent configurations are depicted by the dotted line.}\label{fig:gauge}
\end{figure}

\subsubsection*{Faddeev-Popov gauge fixing}
As expressed in \eqref{amplitude}, the integration measure can be written as $[\cD X_\mu][\cD h_{ab}]_{g,n}$ where the scalar fields and the metric on 2d surfaces with genus $g$ and $n$ marked points are integrated out. However, if there is no anomaly, this integral has the world-sheet diffeomorphisms and Weyl symmetry under which the field configurations are transformed as
\bea\label{global}
X_\mu^\zeta(\wt\s)=&X_\mu(\s) \cr
h_{ab}^\zeta(\wt\s)=&e^{2\omega(\s)}  \frac{\partial \s^c}{\partial \wt\s^a}\frac{\partial \s^d}{\partial \wt\s^b} h_{cd}(\s)~,
\eea
where $\zeta$ indicates Diff$\times$Weyl.
These symmetries are redundant and integrating along these directions just gives rise to the volume of the symmetry group. (In Figure \ref{fig:gauge}, the solid arrows schematically draw gauge redundancy and  the dotted line shows physically distinct configurations.) Thus, we need to carry out gauge-fixing. Thankfully, there is a
standard method to fix gauge, introduced by Faddeev and Popov. A basic idea is to insert the identity of the following form in the path integral:
\be1= \int {\cal D}\zeta\ \delta(h-\hat{h}{}^{\,\zeta}) \det \left(\frac{\delta \hat{h}{}^{\,\zeta} }{\delta \zeta}\right)\label{dfp}\ee
where the Jacobian
factor  $\det \left(\frac{\delta \hat{h}{}^{\,\zeta} }{\delta \zeta}\right)$ is called \textbf{Faddeev-Popov determinant} and we denote it by $\Delta_{FP}[\hat{h}]$. Since ${\cal D}\zeta$ is a gauge-invariant measure, $\Delta_{FP}[\hat{h}]$ is independent of $\zeta$ so that it can be factored out of the integral above. The insertion of this identity into the path integral fixes the metric as $\hat{h}$ because of the delta functional. Because $\int {\cal D}\zeta$ integral only contributes an infinite multiplicative factor, we can discard this integral. Therefore, after Faddeev-Popov gauge fixing, the
form of the path integral can be schematically written as
\be Z[\hat{h}] = \int {\cal D}X\ \Delta_{FP}[\hat{h}]\,e^{-S_{\s}[X,\hat{h}]}\label{fullpf}\ee
We should make several remarks about this procedure.

\vspace{.3cm}
\noindent$\bullet$ First, the conformal transformations are residual gauge symmetries not fixed above. We have to throw away these residual gauge symmetries in the path integral in order to avoid over-counting. Indeed we will be careful to fix this extra residual gauge freedom when computing string amplitudes.

\vspace{.3cm}
\noindent$\bullet$ Second, there are caveats related to global properties of the world-sheet Riemann surface $\Sigma_{g,n}$. In fact, metrics on a Riemann surface encode the information of ``shape'' of the Riemann surface $\Sigma_{g,n}$ called \textbf{complex moduli}, which is not accounted for by local gauge transformations $\zeta$. The space which parametrizes ``shape'' of the Riemann surface is called \textbf{moduli space of Riemann surface} $\Sigma_{g,n}$ which is $(6g-6+2n)$-dim$_\bR$:
\[
\cM_{g,n}:=\frac{\left[\cD h_{ab}\right]_{g,n}}{\textrm{Diff}\times\textrm{Weyl}}
\]
Therefore, the path integral involves integrals over $\cM_{g,n}$ as well. At this moment, we postpone both the issues and will come back to them in \S\ref{sec:amplitude} on string amplitudes.

\vspace{.3cm}
Now let us take an infinitesimal version of \eqref{global} where
a Weyl transformation is parameterized by $\omega(\sigma)$ and an
infinitesimal diffeomorphism by $\delta\sigma^\alpha = \e^\alpha(\sigma)$. Subsequently, the change of the metric under a gauge transformation is read off
\be\label{gauge-orbit}\delta \hat{h}_{ab}  = 2\omega\hat{h}_{ab} + \nabla_a \e_b + \nabla_b \e_a:=2\wt \omega \hat{h}_{ab} +(P\cdot \e)_{ab}~,\ee
where we decompose it into
\begin{align}
(P\cdot \e)_{ab}&=\nabla_a \e_b + \nabla_b \e_a-h_{ab}(\nabla\cdot \e) \cr
\wt \omega&=\omega+\frac12 (\nabla\cdot \e) ~.
\end{align}

Indeed, the operator $P$ maps vectors $\e_a$ to symmetric traceless two-tensors $(P\cdot \e)_{ab}$. Thus, the Faddeev-Popov determinant can be written as
\[
\Delta_{FP}[\hat{h}]=\det \frac{\delta(P\cdot \e,\wt\omega)}{\delta(\e,\omega)}=\det \left| \begin{matrix}P&0\\\ast &1
\end{matrix}
\right|=\det P~.
\]

To compute $\det P$, we use \textbf{Faddeev-Popov ghosts}, which can be understood as an infinite-dimensional version of the following integral.
Given a matrix $M_{ij}$, its determinant can be expressed as a Grassmann integral
\[
\int \prod_{i=1}^nd\psi_i d\theta_i  \exp(\theta_i M_{ij}\psi_j) = \det M~.
\]
where $\theta, \psi$ are Grassmann variables. Accordingly, we introduce anti-commuting fermionic fields, $c^a$ (ghosts) and $b_{ab}$ (anti-ghost) where $b_{ab}$ transforms as a symmetric traceless tensor and $c^a$ as a vector. Then, we can express
\[
\Delta_{FP}[\hat{h}] =\int{\cal D} c{\cal D} b\, \exp\left(\frac{i}{2\pi}\int d^2\sigma \sqrt{-\hat{h}}\, b^{ab} (P\cdot c)_{ab}\right):= \int{\cal D}c{\cal D}b\ \exp[i S_{g}]~,
\]
where the ghost action can be written as
\be\label{ghost1} S_{g}=\frac{1}{2 \pi} \int d^{2} \sigma \sqrt{-\hat{h}} b^{a b} \nabla_{a} c_{b} ~.\ee
Even though the $bc$ ghost fields were introduced to fix a gauge, they look like dynamical fields with the action above. Consequently, the Faddeev-Popov gauge fixing results in a fermionic 2d CFT, usually called \textbf{$bc$ ghost CFT}.

Let us make some remarks about the equations of motion of $S_{g}$:

\vspace{.3cm}
\noindent$\bullet$ The equation of motion for $c_a$ is given by
\be\label{CKV}
P\cdot c = \nabla_a c_b +\nabla_b c_a
-h_{ab} (\nabla \cdot c) = 0~.
\ee
Therefore the solutions for $c$ are in one-to-one correspondence with the \textbf{conformal Killing vectors}, which are the generators of the residual symmetry.

\vspace{.3cm}
\noindent$\bullet$ The equation of motion for $b_{ab}$ is
\be\label{metric-moduli} \nabla_a b^{ab}=0 ~.\ee
We will understand the geometric meaning of these equations when discussing the moduli
space of Riemann surfaces.

\vspace{.3cm}
To understand the properties of the $bc$ ghost CFT, it is convenient to use the Euclidean signature so that we will perform Wick rotation in what follows. Then, the factor of $i$ in the action disappears.
The expression for the full partition function \eqref{fullpf} is
\be Z[\hat{h}]=\int {\cal D}X{\cal D}c{\cal D}b\ \exp\left(-S_{\s}[X,\hat{h}] -
S_{g}[b,c,\hat{h}]\right)~.\label{total}\ee

\subsubsection*{$bc$ ghost CFT}

Now let us study the $bc$ ghost CFT more in detail. For this purpose, we pick the conformal gauge \eqref{conformal-gauge} for a world-sheet metric. Using \eqref{conformal-gauge2}, the ghost action can be written as
\begin{align}
S_{g}& = \frac{1}{2\pi}\int d^2z\ \left( b_{zz}\nabla_{\bz}c^z + b_{\bz\bz}\nabla_z c^{\bz}\right)\cr
&=\frac{1}{2\pi} \int d^2z\ b_{zz}\,\partial_{\bz}c^z +
b_{\bz\bz}\,\partial_zc^{\bz}\end{align}
For the sake of simplicity, let us define
\begin{align} b= b_{zz}~,\qquad \bar{b}=b_{\bar{z}\bar{z}}~,\qquad
c=c^z~,\qquad \bar{c}=c^{\bar{z}}~.\nonumber\end{align}
Then, the action simplifies to
\be\label{ghost-action}
S_{g} = \frac{1}{2\pi}\int d^2z\ \left(b\,\overline \partial c + \bar{b}\,\partial\bar{c}
\right)~,\ee
which gives the equations of motion
\be \label{ghost-holo}\overline \partial b = \partial\bar{b} = \overline \partial c = \partial\bar{c}=0\ee
Thus, we see that $b$ and $c$ are holomorphic fields, while $\bar{b}$ and $\bar{c}$ are
anti-holomorphic.

To compute the OPEs of the $bc$ ghost fields, we use the path integral techniques  in \S\ref{sec:OPE}:
\be
0=\int \mathcal{D} c \mathcal{D} b \frac{\delta}{\delta b(z)}\left[e^{-S_{g}} b(w)\right]=\int \mathcal{D} c \mathcal{D} b \, e^{-S_{g}}\left[-\frac{1}{2 \pi} \bar{\partial} c(z) b(w)+\delta(z-w)\right]~,\nonumber
\ee
which tells us that
\be \overline \partial c(z)\,b(w) = 2\pi \,\delta(z-w)~.\nonumber\ee
We can perform a similar computation for $c(z)$, which yields
\be \overline \partial b(z)\,c(w) = 2\pi\,\delta(z-w)~.\nonumber\ee
We can integrate both of these equations using \eqref{eq:deltaFunc}.  Then, we obtain the $bc$ OPE
\begin{align} b(z)\,c(w) &= \frac{1}{z-w} + \ldots\nonumber\\ c(w)\,b(z) &= \frac{1}{w-z} + \ldots\nonumber\end{align}
In fact, the second equation follows from the first equation and Fermi statistics. Hence, the
OPEs of $b(z)\,b(w)$ and $c(z)\,c(w)$ are trivial for the obvious reason.

In any CFT, it is of the utmost importance to find the form of the energy-momentum tensor.
The energy-momentum tensor is obtained via Noether's theorem with respect to world-sheet transformations $\delta z = \e(z)$, under which
\[\delta b = (\e\partial+2(\partial\e))b~,\qquad \d c = (\e\partial-(\partial\e))c~.\]
Indeed both $b$ and $c$ are primary fields with conformal dimensions $h=2$ and $h=-1$, respectively, which can be easily seen from their index structure $b_{zz}$ and $c^z$.
From these rules, one can deduce the form of the energy-momentum tensor
\be\label{Tgh} T^{g}(z) =- 2:b(z)\partial c(z): + :c(z)\partial b(z):~.\ee
In fact, this form can be obtained from the first principle, namely the variation of the action under the metric (exercise).

The OPEs of $b$ and $c$ with the stress tensor are
\begin{align}
&T^{g}(z) c(w)=-\frac{c(w)}{(z-w)^{2}}+\frac{\partial c(w)}{z-w}+\ldots \cr
&T^{g}(z) b(w)=\frac{2 b(w)}{(z-w)^{2}}+\frac{\partial b(w)}{z-w}+\ldots
\nonumber\end{align}
so that $b$ and $c$ are primary fields of conformal dimension $2$ and $-1$, respectively. Consequently, they admit the mode expansions
\[
b(z)=\sum_{m\in\bZ}\frac{b_m}{z^{m+2}}\qquad c(z)=\sum_{m\in\bZ}\frac{c_m}{z^{m-1}}~.
\]
The ghost OPEs give the commutation relations
\be\label{ghost-comm}\left\{b_{m},c_{n}\right\}=\delta_{m+n,0}~, \quad\left\{c_{m}, c_{n}\right\}=\left\{b_{m}, b_{n}\right\}=0~.\ee
Also, the Virasoro generators of the $bc$ ghost are expressed as
\be \label{L-bc}
L^{g}_m=\sum_{n\in\bZ}(2m-n):b_nc_{m-n}:+a^g\delta_{m,0}~.
\ee
The constant $a^g$ in the zero mode $L_0^{g}$ is determined by the Casimir energy $\frac{c^g}{24}$ in \eqref{Casimir2} and the commutation relation of the $bc$ modes $\frac1{12}=-\sum_{n>0}n$
\[
a^g=\frac{-26}{24}+\frac1{12}=-1~.
\]
This is consistent with the commutation relation $L_0=[L_1,L_{-1}]$ (\cite[(2.7.20)]{Polchinski}).

Finally, we can compute the $TT$ OPE  (exercise)
\be T^{g}(z)\,T^{g}(w) = \frac{-13}{(z-w)^4}+\frac{2T^{g}(w)}{(z-w)^2} + \frac{\partial T^{g}(w)}{z-w} +\ldots\nonumber\ee
Now one can read off the  central charge of the $bc$ ghost system, which is
\be c^{g}=2(-13)=-26 ~.\nonumber\ee

We learned that the Weyl symmetry is anomalous unless $c=0$.
Since the Weyl symmetry is a gauge symmetry, the theory must be Weyl anomaly-free. Since the total central charge of the string sigma model and ghost theory \eqref{total} is given by $c=c^X +c^{g}$, the dimension of the target space must be $D=26$.
Again, we obtain the critical dimension of the bosonic string theory!

\subsection{BRST quantization}\label{sec:BRSTsub}

In 4d Yang-Mills theory, the Lagrangian with Faddeev-Popov ghosts has the continuous symmetry, called \textbf{BRST symmetry} (Becchi-Rouet-Stora-Tyutin). The BRST symmetry is generated by a nilpotent charge $Q_B$ $(Q_B^2=0)$ that commutes with the Hamiltonian. The nilpotency of the BRST charge has substantial
consequences. Since the BRST transformations come from the gauge symmetry, all physical states must be BRST-invariant. Hence, we require a physical state to be annihilated by $Q_B$
\[
Q_{B} |\textrm{phys}\rangle=0~.
\]
However, one can always add a state of the form $Q_{B} | \chi \rangle$ since this state will be annihilated by $Q_B$ because of the nilpotent BRST charge. However, this state is orthogonal to all physical states, and therefore it is a \textbf{null state}.
Thus, two states related by
\[ | \psi'\rangle = |\psi\rangle + Q_B |\chi\rangle \]
have the same inner products with all the physical states and they are thus indistinguishable.
This is the remnant in the gauge-fixed version of the original
gauge symmetry. As a result, the Hilbert
space of physical states is isomorphic to the $Q_B$-cohomology, i.e.
\be \label{BRST-quantization}
\cH^{\textrm{phys}}\cong\frac{\cH^{Q_B\textrm{-closed}}}{\cH^{Q_B\textrm{-exact}}}~.
\ee
This covariant way of determining the physical Hilbert space with ghosts is known as \textbf{BRST quantization}.

We are now ready to apply this formalism to the bosonic string \cite[\S4]{Polchinski}. Combining \eqref{free-scalar} and \eqref{ghost-action}, the action we consider is
\[
S_X+S_{g}= \frac{1}{2\pi}\int d^2z\ \left( \frac1{\alpha'}\partial X^\mu  \ol\partial X_\mu+b\,\overline \partial c + \bar{b}\,\partial\bar{c}\right)~.
\]
This is invariant under the following BRST transformations:
\begin{align}\nonumber
&\delta_B X^\mu = i \epsilon ( c \partial + \bar{c} \bar{\partial} ) X^\mu \,,
\cr
&\delta_B c=  i\epsilon  c \partial c  \qquad \delta_B \bar c=  i\epsilon \bar{c} \bar{\partial} \bar c \,,
\\
&\delta_B b =  i \epsilon ( T^X + T^{g} ) \qquad \delta_B \bar b =  i \epsilon ( \overline T^X + \overline T^{g} )\,, \label{BRST-trans}
\end{align}
where the explicit forms of the energy-momentum tensors can be found in \eqref{TX} and \eqref{Tgh}. Note that we impose \eqref{ghost-holo} here.  This exhibits typical features of the BRST transformation: a bosonic field is transformed into a fermionic field and vice versa, and the ghost field $b$ is transformed into the energy-momentum tensor.
Noether's theorem tells us that there is a classical current associated to the BRST symmetry, and the holomorphic part of the BRST current takes the form (exercise)
\begin{align}\label{jB}
j_B &= c(z)T^X(z) + \frac12 : c(z)T^{g}(z): +\frac32:\partial^2c(z):\\
&= c(z)T^X(z) + : b(z)c(z)\partial c(z) : +\frac32:\partial^2c(z):~.\nonumber
\end{align}
and the BRST charge is defined by
\[ Q_B = \oint \frac{dz}{2\pi i} ~ j_B \,.\]
It follows from \eqref{BRST-trans} that
\be\label{Q-b}
\{Q_B,b(z) \}=( T^X + T^{g} ) \to \{Q_B,b_m\}=L_m^X+L_m^{g}~.
\ee
Computing $j_Bj_B$ OPE, one can convince oneself that the BRST charge is
nilpotent  $Q_B^2=0$ if and only if  $D=26$ (exercise).
Therefore, $j_B$ is a primary field with $h=1$, and $Q_B$ is a conserved charge at quantum level only when $D=26$.
Furthermore,
we can express the BRST charge in terms of the $X^{\mu}$ Virasoro
operators and the ghost oscillators as
\be
Q_B = \sum_n c_n (L^X_{-n}-\d_{n,0}) + \sum_{m,n} \frac{m-n}{2} : c_m c_n
b_{-m-n} :  \,,
\label{27}\ee
where the normal ordering constant comes from $\{Q_B,b_0\}=L_0^X+L_0^{g}$.
In the case of closed strings, there is the anti-holomorphic part $\bar Q_B$, and the total
BRST charge is $Q_B+\bar Q_B$.

We will find the physical open string states in the BRST context.
According to our previous discussion, the physical states will have
to be annihilated by the BRST charge, and not be of the form
$Q_B|\phantom{\chi}\rangle$.

First, we have to describe our extended Hilbert space which includes
the ghosts.
As far as the $X^{\mu}$ oscillators are concerned, the situation is
the same as in \eqref{bosonic-Hilb}, so we need to consider only the ghost Hilbert space.
Indeed the ghost commutation relations \eqref{ghost-comm} generate a two-state spin system $|\uparrow\rangle, |\downarrow\rangle$ where
\bea\label{ghost-Hilb}
&b_0 | \downarrow \rangle = 0\,, \qquad b_0 |\uparrow\rangle =
|
\downarrow
\rangle\,, \cr
&c_0 | \uparrow \rangle
= 0\,, \qquad c_0 | \downarrow\rangle = |
\uparrow
\rangle \,,\cr
&b_{n>0}|\uparrow\rangle=b_{n>0}|\downarrow\rangle=c_{n>0}|\uparrow\rangle=c_{n>0}|\downarrow\rangle=0~.
\eea

The full Hilbert space will be a tensor product of the two  $|k,\uparrow\rangle, |k,\downarrow\rangle$ by acting creation operators where $|k\rangle=|0;k\rangle$ denotes the vacuum of the matter theory.
From the light-cone quantization, we know that there is only one vacuum called Tachyon. Therefore, we have to pick the ghost vacuum among the two spin states.
For this purpose, we further impose one more condition, namely
\be
b_0|{\rm phys}\rangle =0\,.
\label{b0-annihilate}\ee
This is sometimes called the \textbf{Siegel gauge} \cite[\S3.2]{GSW}.
Under this condition, \eqref{ghost-Hilb} tells us that the correct ghost vacuum is
$|\downarrow\rangle$.
We can now create states from this vacuum by acting with the negative
modes
of the ghosts $b_m,c_n$. Note that $c_0 | \downarrow\rangle = |
\uparrow
\rangle$ does not satisfy the
Siegel condition \eqref{b0-annihilate}.
Also, \eqref{b0-annihilate} yields the condition
\be\label{L0-annihilate} 0=\{Q_B,b_0\}|{\rm phys}\rangle=(L_0^X+L_0^{g})|{\rm phys}\rangle ~.\ee
Therefore, if a physical state  is at level $N$ with momentum $k$, $|{\rm phys}\rangle=|N;k\rangle$, then we have
\be \label{k_N}
k^2 = \frac{1-N}{\alpha'}
\ee
which is consistent with the light-cone gauge. As in \eqref{numbering}, we can define the number operator with the ghost sector as
\begin{equation}
N=\sum_{n>0} \eta_{\mu\nu} :\alpha_{-n}^\mu \alpha_{ n}^\nu : +n\left(: b_{-n} c_n:+: c_{-n} b_n:\right)
\end{equation}
so that the Virasoro zero mode is expressed as
\be 
L_0= \a' k^2+ N-1~.
\ee 

Now, let us impose the physical condition \eqref{BRST-quantization} for open-string states level by level. To this end, let us explicitly write the open-string mode expansions of the zero modes of the Virasoro generators
\begin{align}\label{open-Virasoro}
L^X_0&=\alpha^{\prime} p^{2}+\sum_{n=1}^{\infty} \alpha_{-n} \cdot \alpha_{n}=\a'p^2+\a_{-1}\cdot\a_1+\cdots~,\cr
L^{g}_0&=-1+\sum_{n \in \bZ} n: b_{-n} c_{n}:=-1+b_{-1}c_{1}+c_{-1}b_{1}+\cdots~,
\end{align}
(Compare with \eqref{Virasoro} and \eqref{L-bc}.)
At level zero, there is only the vacuum state and the BRST quantization leads to
$|k,\downarrow\rangle$
\[
 Q_B | k,\downarrow \rangle = ( L^X_0 - 1 ) c_0 | k,\downarrow\rangle= ( L^X_0 - 1 ) | k,\uparrow\rangle=0\,.
\]
and it is not $Q_B$-exact.

At the first level, the possible operators that can act on the vacuum $| k,\downarrow\rangle$ are $\alpha^\mu_{-1}$,
$b_{-1}$ and $c_{-1}$. The most general state of this form is then
\be
|\psi\rangle = ( \zeta \cdot \alpha_{-1} + \beta b_{-1} + \gamma c_{-1} )
| k,\downarrow\rangle \,, \label{n=1}
\ee
which has 28 parameters: a 26-dimensional vector $\zeta_\mu$ and two more constants $\beta$, $\gamma$. First we note that \eqref{k_N} yields the massless
condition  $k^2 =0$. In addition, the BRST condition demands
\[
 0 = Q_B |\psi\rangle = 2 ( (k\cdot \zeta) c_{-1} + \beta
k\cdot \alpha_{-1} ) |k,\downarrow \rangle \,.
\]
This only holds if $k\cdot\zeta=0$ and
$\beta=0$. Therefore the BRST condition removes the unphysical anti-ghost excitations as well as all polarizations that are not orthogonal to the momentum, thereby eliminating two out of the $26+2$ original states. Hence, there are only 26 parameters left.

Next, we have to make sure that this state is not $Q_B$-exact: a general state $|\chi\rangle$
is of the same form as \eqref{n=1}, but with parameters
$\zeta'^{\mu}$, $\b'$ and $\g'$. Thus, the most general $Q_B$-exact state at
this level
with $k^2=0$ will be
\[ Q_B |\chi \rangle = 2 ( k\cdot \zeta' c_{-1} + \beta'
k\cdot\alpha_{-1} ) | k,\downarrow\rangle \,. \]
This means that the $c_{-1}$ part in \eqref{n=1} is BRST-exact and that
the polarization has the equivalence relation $\zeta_\mu \sim
\zeta_\mu + 2 \beta' k_\mu $. This leaves us with the 24 physical
degrees of freedom we expect for a massless vector particle in 26 dimensions. In sum, the physical state at level one is
 \[
 \{|\zeta;k\rangle ~; \quad k\cdot\zeta=0\}/~ \zeta_\mu \sim
\zeta_\mu + 2 \beta' k_\mu ~.
 \]

The same procedure can be followed for the higher levels of open string states. It can be proved that $\cH_{\textrm{light-cone}}$ is isomorphic to the $Q_B$-cohomology, and the inner product is positive-definite. In the
case
of the closed string, we have to use $Q_B+\bar Q_B$ for the BRST quantization.

\section{Bosonic string amplitudes}\label{sec:amplitude}

Finally, we will discuss the string amplitude, which provides information about various features of string theory. The string amplitude can be expressed as follows:
\begin{align*}
 A_n = \sum_g \int \frac{[\cD h_{ab}]_{g,0}}{(\textrm{Diff $\times$ Weyl})} \int \cD X^\mu e^{-S_\sigma [X^\mu,h_{ab}]} \prod_{i=1}^n \int d^2\sigma \sqrt h V_i  \ .
\end{align*}
In this section, we will examine the degrees of freedom in the metric integration of the string amplitude after (Diff $\times$ Weyl) gauge fixing. Specifically, for a closed oriented Riemann surface $\Sigma_{g,n}$, the dimension of the integration is
\begin{align*}
\dim \frac{[\cD h_{ab}]_{g,n}}{(\textrm{Diff $\times$ Weyl})} = 6g -6 +2n \ .
\end{align*}

\begin{figure}[htb]
\centerline{\includegraphics[width=300pt]{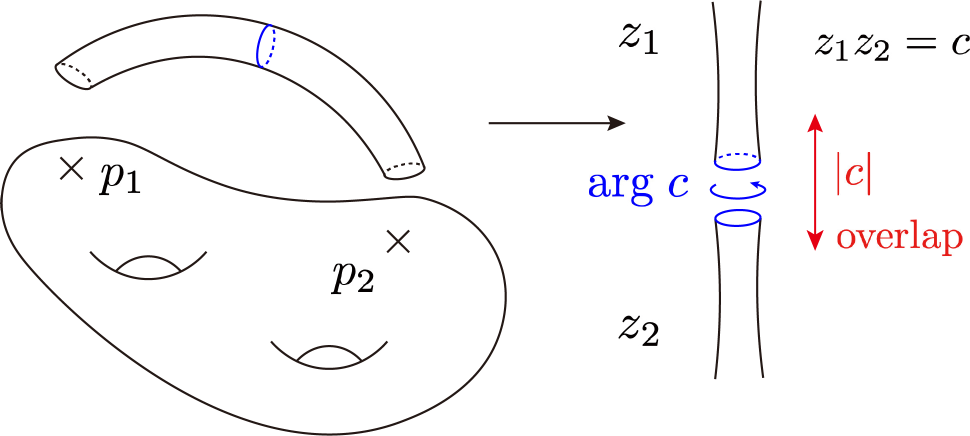}}
\caption{Attaching a pair of cylinders to $\Sigma_g$, and patching the pair of cylinders by $z_1z_2=c$.}
\label{g-increase}
\end{figure}

To understand the dimension of integration in the string amplitude, it is helpful to consider the term $6g$. Even after gauge fixing (e.g., fixing the metric locally), there is still freedom to change the "shape" of the world-sheet, known as metric moduli. If we increase the number of genus $g$, we can do so by attaching a cylinder to a Riemann surface $\Sigma_{g}$, as described in \S\ref{sec:Weyl}. To do this, we must specify the locations of the cylinder's endpoints and its shape. The endpoints are denoted by two complex points $p_1$ and $p_2$, and around these points, we introduce complex coordinates $z_1$ and $z_2$. We also impose the condition $z_1 z_2 = c$, where $c$ is a constant whose phase specifies the twist and magnitude specifies the length (see Figure \ref{g-increase}). In total, there are an additional six parameters that describe the shape of the world-sheet.  The term $2n$ comes from the integration of the vertex operators, which is necessary for the vertex operator to be Diff $\times$ Weyl-invariant.

\subsection{Teichm\"uller space and moduli space of Riemann surfaces}

\begin{figure}[htb]\centering
\includegraphics[width=5cm]{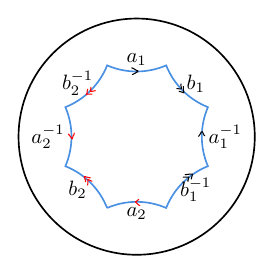}\qquad\raisebox{.5cm}{\includegraphics[width=8cm]{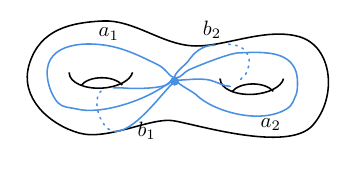}}
\caption{A $4g$-polygon with geodesic arcs and each angle $2\pi/4g$ in the Poincar\'e disc $\bD$ ($g=2$)}
\label{genus-two}
\end{figure}

Let us briefly introduce Teichm\"uller space $\cT_g$ and moduli space $\cM_g$ of Riemann surfaces of genus $g$. For the sake of brevity, we focus on an oriented closed Riemann surface without punctures. It is known that every Riemann surface admits a Riemannian metric of constant curvature:
\begin{description}
\item[positive] A two-sphere $S^2$ ($g=0$) with a fixed radius in $\bR^3$ has positive constant curvature with respect to the induced metric from the standard metric of $\bR^3$
  \item[zero] A two-torus $T^2$ ($g=1$) constructed as a quotient space $T^2=\bC/\Gamma$ has zero curvature where $\Gamma$ is a two-dimensional lattice.
  \item[negative] A Riemann surface of genus $g>1$ can be constructed from a $4g$-polygon with geodesic arcs and each angle $2\pi/4g$ in the Poincar\'e disc $\bD$ where the edges are identified via relation \[a_{1} b_{1} a_{1}^{-1} b_{1}^{-1} a_{2} b_{2} a_{2}^{-1} b_{2}^{-1} \cdots a_{g} b_{g} a_{g}^{-1} b_{g}^{-1}=1~.\]
  (See Figure \ref{genus-two}.) Hence, it admits a negative constant curvature.
\end{description} 
A Riemann surface is a complex one-dimensional manifold. In short, the moduli space of Riemann surfaces classifies Riemann surfaces up to isometries while Teichm\"uller space classifies Riemann surfaces up to isometries isotopic to the identity map. We will see its meaning below.

Let us first consider the case of genus one. As mentioned, a torus is constructed from a quotient space $\bC_w/\Gamma_\tau$. As far as ``shape'' is concerned, we can set one edge to $1\in \bC$ without loss of generality. Then, an identification is given by
\be \label{torus-identification}w \cong w+2 \pi(m+n \tau)~,\qquad (m,n)\in\bZ\times \bZ~,\ee
where $\tau = \tau_1 +i\tau_2$ with $\tau_2 > 0$. In this coordinate, the metric is flat $ds^2=dwd\overline w$ so that $w$ is a conformally-flat local coordinate.
Another way to see ``shape'' of a torus is
\[(\sigma^{1}, \sigma^{2}) \cong (\sigma^{1}, \sigma^{2})+2 \pi(m, n)\]
with a Riemannian metric
\begin{equation}\label{torus-metric}
d s^{2}=\left|d \sigma^{1}+\tau d \sigma^{2}\right|^{2}~.
\end{equation}
In fact, the holomorphic coordinate $w$ and $(\sigma^1,\sigma^2)$ are related by
\be\label{c-str-torus}
w=\sigma^1+\tau \sigma^2~,
\ee
so that $\tau$ encodes the information about holomorphicity. Thus, the parameter $\tau$ is called a complex structure that describes the ``shape'' of the torus, roughly speaking. Since \eqref{torus-metric} is invariant under $\tau_2\to -\tau_2$, the space of complex structures $\tau$ is therefore the upper half plane $\bH$, which can be identified with the Teichm\"uller space $\cT_{g=1}$ of a torus.

\begin{figure}[htb]\centering
\raisebox{0cm}{\includegraphics[width=5cm]{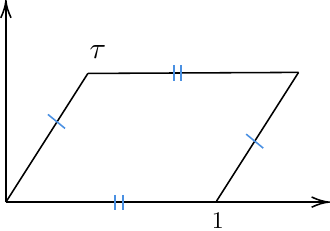}}
\qquad\qquad
\includegraphics[width=10cm]{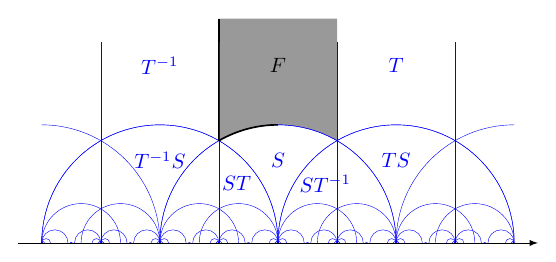} \caption{The Teichm\"uller space of a torus is the upper half-plane, and the mapping class group $\PSL(2,\bZ)$ acts on it. The moduli space is the fundamental region $F$ (the shaded region). }\label{modular}
\end{figure}

There are global transformations that leave ``shape'' as it is. The lattice $\Gamma_\tau$ is invariant under transformations
\begin{align}
 T: \tau \to \tau+1 \ , \quad S: \tau \to -\frac{1}{\tau} \ .
 \label{eq:ModularTra}
\end{align}
These transformations $T,\ S$ generate the \textbf{modular transformation}
\begin{align} \label{eq:ModularTra2}
 \tau \to \frac{a\tau+b}{c\tau+d} \ , \quad
 \begin{pmatrix}
  a & b \cr c & d
 \end{pmatrix} \in \PSL(2,\bZ) \ ,
\end{align}
where $\PSL(2,\bZ)=\SL(2,\bZ)/\{\pm\}$.
Hence, the modular transformations do not change the ``shape'' of a torus, but they act on the Teichm\"uller space $\bH$ as in Figure \ref{modular}. As a result, the moduli space of complex structures of a torus is
\be
\cM_{g=1}=\cT_{g=1}/\PSL(2,\bZ)~,
\ee
which is the shaded region in Figure \ref{modular}, called the \textbf{fundamental region}.

Now let us consider a Riemann surface of genus $g>1$. The Poincar\'e disk with the hyperbolic metric
\[\Bigl(\bD,ds^2=\frac{4(dx^2+dy^2)}{(1-x^2-y^2)^2}\Bigr)\]
has negative constant curvature, and geodesics are portions of circles that intersect the disk boundary at right angles. As briefly mentioned above, a Riemann surface of genus $g>1$ can be constructed by a $4g$-polygon with geodesic arcs and each angle $2\pi/4g$ in the Poincar\'e disk. However, not all Riemann surfaces of genus $g>1$ with negative constant curvature (hyperbolic Riemann surfaces) are constructed in this way. The \textbf{moduli space} of Riemann surfaces indeed parametrizes hyperbolic Riemann surfaces that are not related by an isometry.

A hyperbolic Riemann surface of genus $g>1$ admits a \textbf{pants decomposition} by cutting along $3g-3$ simple closed curves into $2g-2$ pants. In fact, one can take geodesic arcs for $3g-3$ simple closed curves and measure their lengths with respect to the hyperbolic metric. The ``shape'' of each pants is uniquely determined by the lengths of three edges. Thus, when we construct a hyperbolic Riemann surface by gluing pants, the ``shape'' will be determined by twist angles at gluing along the simple closed curves. These data will determine how to construct a hyperbolic Riemann surface from $2g-2$ pants, and the collection of the length coordinates $l_i$ and the twist coordinates $\theta_i$
\begin{equation}
\left(l_{1}, \cdots, l_{3 g-3} ; \theta_{1}, \cdots, \theta_{3 g-3}\right)
\end{equation}
are referred to as  \textbf{Fenchel-Nielsen coordinates}. The Fenchel-Nielsen coordinates indeed parametrize the space of ``complex structure'' called \textbf{Teichm\"uller space} $\cT_g$, and we have the bijection
\[\cT_g\cong \bR_+^{3g-3}\times  \bR^{3g-3}\]
In other words, $\cT_g$ can be identified with an open disk in $\bR^{6g-6}$. Moreover, the Teichm\"uller space is endowed with a complex structure and K\"ahler form
\bea
\omega=\sum_{i=1}^{3g-3} d l_i\wedge d\theta_i~,
\eea
referred to as the \textbf{Weil-Petersson form} so that it can be considered as a K\"ahler manifold.

\begin{figure}[htb]
\centerline{\includegraphics[width=15cm]{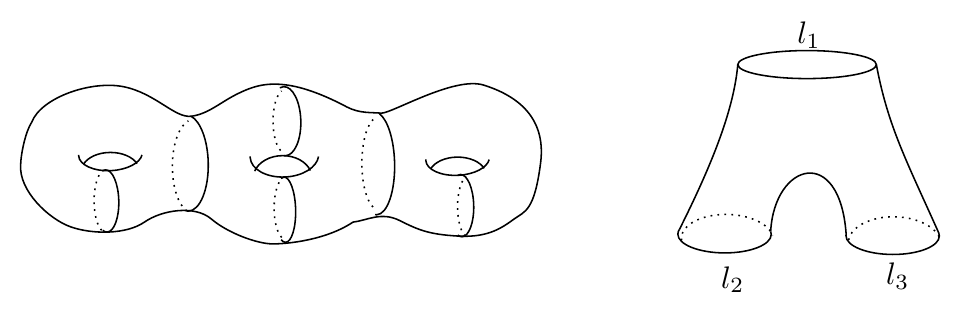}}
\caption{A pants decomposition of Riemann surface}
\label{Pants-decomp}
\end{figure}

\begin{figure}[htb]
\centerline{\includegraphics[width=13cm]{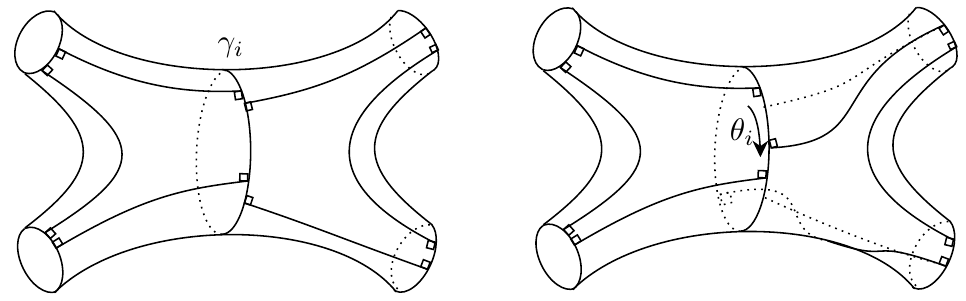}}
\caption{twist coordinates}
\label{angle}
\end{figure}

\begin{figure}[htb]
\centerline{\includegraphics[width=12cm]{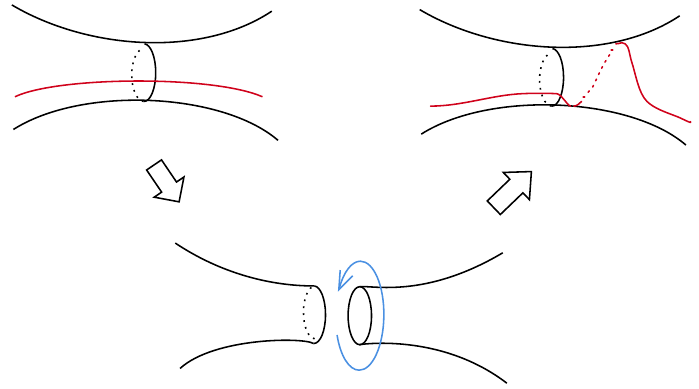}}
\caption{The Dehn twist along a simple closed curve}
\label{Dehn}
\end{figure}

\begin{figure}[htb]
\centerline{\includegraphics[width=12cm]{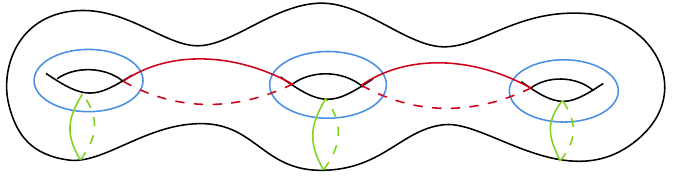}}
\caption{Choice of $3g-1$ cycles for Dehn twists. ($g=3$ here)}
\label{Dehn-cycles}
\end{figure}

The relation between the Teichm\"uller space $\cT_g$ and the moduli space $\cM_g$ is given by the mapping class group $\textrm{MCG}_g$. The Teichm\"uller space classifies hyperbolic Riemann surfaces up to isometries isotopic to the identity map. However, there are isometries that are not isotopic to the identity map. (In physics, they are sometimes called \textbf{large gauge transformations}.) The \textbf{mapping class group} appears as the quotient of orientation-preserving diffeomorphisms by diffeomorphisms isotopic to the identity map
\[\textrm{MCG}_g =\frac{\textrm{Diff}_+(\Sigma_g)}{\textrm{Diff}_0(\Sigma_g)}~.\]
Although this definition looks horrendous, the theorem of Lickorish states that the mapping class group $\textrm{MCG}_g$ of Riemann surfaces of genus $g$ is a discrete group generated by Dehn twists along $3g-1$ simple closed curves shown in Figure \ref{Dehn-cycles}. A Dehn twist is an isometry that generates a $2\pi$-twist along a simple closed curve $C$ of a hyperbolic Riemann surface $\Sigma_g$ (Figure \ref{Dehn}). Consequently,
the moduli space of Riemann surfaces of genus $g$ can be identified as the quotient space
\[\cM_g=\cT_g/\textrm{MCG}_g~. \]
In the case of a torus, $\cT_{g=1}$ is the upper half-plane and $\textrm{MCG}_g$ is $\PSL(2,\bZ)$, so that $\cM_{g=1}$ is the fundamental region.

We can also discuss $\cT_g$ and $\cM_g$ from the viewpoint of complex analytics very briefly. More elaborate explanations can be found in \cite{Nelson:1986ab} for string theory. A Riemann surface $\Sigma$ is defined by a local coordinate system $\left\{\left(U_{j}, z_{j}\right)\right\}_{j \in J}$ that are patched by biholomorphic mappings
\[
z_{k} \circ z_{j}^{-1}: z_{j}\left(U_{j} \cap U_{k}\right) \rightarrow z_{k}\left(U_{j} \cap U_{k}\right)~.
\]
A local coordinate system defines a complex structure on $\Sigma$. Given Riemann surfaces $R$ and $S$, a biholomorphic map $f: R \rightarrow S$ is a holomorphic map $f:R\to S$ which has the holomorphic inverse mapping $f^{-1}: S \rightarrow R$. The Teichm\"uller space is the space of complex structures modulo biholomorphic equivalence. One can reconstruct a complex structure from a real two-dimensional oriented closed manifold with a Riemannian metric, which can be expressed on a local coordinate $(U,\sigma^a)$ as
\[
ds^2=h_{ab}d\sigma^ad\sigma^b~.
\]
Writing $z=\sigma^1+i\sigma^2$, the metric takes the form
\be \label{metric} d s^{2}=\lambda|d z+\mu d \bar{z}|^{2} ~.\ee
where
\[
\lambda =\frac{1}{4}\left(h_{11}+h_{22}+2 \sqrt{h_{11} h_{22}-h_{12}^{2}}\right)~, \qquad
\mu =\frac{h_{11}-h_{22}+2 i h_{12}}{h_{11}+h_{22}+2 \sqrt{h_{11} h_{22}-h_{12}^{2}}}~.
\]
Moreover, we can choose a conformally-flat coordinate $w$ such that
\[\rho d w d\overline w=\rho\left|\partial  w\right|^{2}\left|d z+\frac{\overline{\partial}w}{\partial  w} d \bar{z}\right|^{2}~.\]
Comparing \eqref{metric}, the existence of a conformally-flat coordinate is indeed equivalent to the existence of a solution to
\be\label{Beltrami}\mu=\frac{\overline \partial w}{\partial w}~.\ee
This is referred to as the \textbf{Beltrami equation}, and there is always a solution $w$. Moreover, $\left\{\left(U_{j}, w_{j}\right)\right\}_{j \in J}$ defines a complex structure, and we write the resulting Riemann surface $\Sigma$. For example, in the case of a torus, we define a complex structure \eqref{c-str-torus} from a Riemannian metric \eqref{torus-metric}. Even if we perform a Weyl transformation on the metric, the resulting Riemann surface $\Sigma'$ is biholomorphically equivalent to the Riemann surface $\Sigma$ constructed from the original metric. Therefore, the Teichm\"uller space can be identified with
\be \label{Teichmuller}
\cT_g=\frac{\textrm{Met}}{\Weyl\times\Diff_0  }~,
\ee
where $\textrm{Met}$ is the space of metrics on real two-dimensional Riemannian manifolds of genus $g$.

In fact, $\mu=\mu_{\bar z}{}^z$ in \eqref{Beltrami} can be considered as a $(-1,1)$-form, called \textbf{Beltrami differential}, which encodes the space of metrics up to Weyl transformations. However, there are gauge equivalent configurations due to diffeomorphisms. To distinguish non-trivial variation in the space $L_{(-1,1)}^{\infty}(\Sigma_g)$ of Beltrami differential, Teichm\"uller has introduced a pairing $(\cdot,\cdot)$ with a \textbf{holomorphic quadratic differential} (or $(2,0)$-form) $ \varphi_{z z} (z)dz\otimes dz$:
\be\label{pairing}(\mu, \varphi) \equiv \int_{\Sigma_g} \mu_{\bar{z}}{}^{z} \varphi_{z z} \mathrm{~d} z \wedge \mathrm{d} \bar{z}~.\ee
Then, Teichmüller's Lemma states that the subspace of Beltrami differentials for a trivial variation is given by
\[
\mathcal{N}=\left\{\mu \mid(\mu, \varphi)=0\quad  \text{for}\ \ \forall\varphi\in H^0(\Sigma_g,K^{\otimes2}) \right\}~.
\]
Hence, the holomorphic tangent space of the Teichm\"uller space at $\Sigma$ can be identified with
\be\label{tangent} T_{\Sigma_g}^{(1,0)} \mathcal{T}_{g}\cong L_{(-1,1)}^{\infty}(\Sigma_g) / \mathcal{N}~,\ee
and the holomorphic cotangent space with the space of holomorphic quadratic differentials
\be\label{cotangent} T_{\Sigma_g}^{*(1,0)} \mathcal{T}_{g}\cong H^0(\Sigma_g,K^{\otimes2}) \ee
Note that $K$ is the cotangent bundle on $\Sigma_g$, and it has degree $2 g-2$ so that the Riemann-Roch formula tells us \[\dim H^0(\Sigma_g,K^{\otimes2})=\operatorname{deg}(K^{\otimes 2})-g+1=3 g-3~.\]
The space of anti-holomorphic quadratic differentials has the same dimension.

\subsection{Gauge fixing and string amplitudes}\label{sec:CKV-metric}

\subsubsection*{Conformal Killing vectors \& Moduli space of metrics}

The gauge transformation of a world-sheet metric under diffeomorphism and Weyl transformation is given in \eqref{gauge-orbit}. Moreover, conformal Killing vectors $\e$ satisfying
\begin{equation}
  P\cdot \e=0~.
\end{equation}
do not change the metric infinitesimally. The existence of conformal killing vectors means that there are zero modes of the ghost $c$-field as pointed out in \eqref{CKV}.  Thus, we need extra care even after introducing the $bc$ ghost.
In addition, the physically inequivalent change of a metric is
perpendicular to \eqref{gauge-orbit} as in Figure \ref{fig:gauge}. Writing the change $\delta^\perp h_{ab}$, we have
\begin{align*}
 0 &= \int d^2\sigma \sqrt h \ \delta^\perp h_{ab} \left[
 \left(P\cdot \epsilon\right)^{ab} +2\wt\omega h^{ab} \right] \cr
 &= \int d^2\sigma \sqrt h\ \left[
 \left(P^T \cdot \delta^\perp h \right)_a \epsilon^a +2\wt\omega h^{ab}\delta^\perp h_{ab} \right] \ .
\end{align*}
To satisfy the orthogonality for arbitrary $\epsilon$ and $\omega$,
it is required that
\begin{align*}
 h^{ab} \delta^\perp h_{ab} = 0 \ , \qquad \left(P^T \cdot \delta^\perp h \right)_a =: -\nabla^b \delta^\perp h_{ba} = 0 \ .
\end{align*}
Therefore, solutions to this equation are equivalent to zero modes of the ghost $b$-field \eqref{metric-moduli}. Since these zero modes are absent in the action,
we need to insert appropriate zero modes to derive a non-trivial amplitude
(because $\int db \cdot 1 = 0=\int dc \cdot 1 $).
In a conformal gauge, equations for conformal Killing vectors and metric moduli become
\begin{align*}
 &\partial \ol\epsilon = \ol\partial \epsilon = 0 \ , \cr
 &\partial \delta^\perp h_{\ol z\ol z} = \ol\partial \delta^\perp h_{zz} = 0 \ .
\end{align*}
Hence, variations of the metric moduli correspond to holomorphic quadratic differentials, which can be identified with the holomorphic cotangent space \eqref{cotangent} of the Teichm\"uller space at $\Sigma$.

Let us look at some examples.
For a two-sphere, we have the stereographic projections, and the two local coordinates are related by $w=1/z$, yielding
\[
\begin{aligned}
\delta w &=\frac{\partial w}{\partial z} \delta z=-z^{-2} \delta z \cr
\delta h_{ww} &=\left(\frac{\partial w}{\partial z}\right)^{-2} \delta h_{z z}=z^{4} \delta h_{z z}
\end{aligned}
\]
CKV $\delta z$ is holomorphic at $w=0$ if it grows no more rapidly than $z^{2}$ as $z \rightarrow \infty$. On the other hand, there are no moduli for the metric
\begin{align*}
 &\delta^\perp h_{zz} = \delta^\perp h_{\ol z\ol z} = 0 \ , \cr
 &\epsilon = a_0 +a_1 z + a_2 z^2 \ , \cr
 &\ol\epsilon = a_0^* +a_1^* \bar z + a_2^* \bar  z^2 \ .
\end{align*}
Therefore, there are 6 CKVs and no modulus.

On the torus, the only holomorphic doubly periodic functions are the constants, so there are two real moduli and two real CKVs.
\begin{align*}
\delta^\perp h_{zz} = a \ , \qquad \epsilon = b \ 
\end{align*}
where $b$ represents translations along two directions.

As we have seen, the dimension of the metric moduli is $6g-6$ for a Riemann surface of genus $g>1$. However, there is no CKV. Thus, if we write the dimension of the conformal Killing group by $\mu$ and the dimension of the metric moduli by $\nu$, we summarize the results in Table \ref{table:CKV-metric}. Thus, the dimension of the Teichm\"uller space is $\dim \cT_g =\nu$, while the difference $\nu-\mu=6g-6$ is fixed by the Riemann-Roch theorem for every genus.

\begin{table}[htbp]\centering
\begin{tabular}{c|c|c|c}
 & $g=0$ & $g=1$ & $g \ge 2$ \cr\hline
 $\mu$ & $6$ & $2$ & $0$ \cr
 $\nu$ & $0$ & $2$ & $6g-6$
\end{tabular}
\caption{The number of zero modes of $b$ and $c$.}
\label{table:CKV-metric}
\end{table}

\subsubsection*{Ghost number anomaly}
We can derive the conclusion that we need to insert the ghost fields in correlation functions to have non-trivial results by considering the ghost number anomaly.
The ghost action in \eqref{ghost-action} is invariant under the transformation
\[\delta_g c = \epsilon c~, \qquad \delta_g b = -\epsilon b~.\]
The corresponding Noether current is a ghost number current $j = cb$ under which we can define
the ghost number $[c]=1$ and $[b]=-1$. At a flat world-sheet, the current satisfies the conservation law $\ol\partial j = 0$.
However, it has an anomaly in a curved world-sheet proportional to the scalar curvature like in \eqref{Weyl-anomaly}
\begin{align}\label{Ghost-anomaly}
 \nabla^z j_z = \kappa \cdot R^{(2)} \ .
\end{align}

Let us determine the coefficient $\kappa$.
The OPE of the energy-momentum tensor and the ghost number current leads to
\begin{align*}
 T(z)j(w) &= -\nord{(2b\partial c + \partial b c) (z)} \nord{cb(w)} \nonumber\cr
 &\sim \frac{-3}{(z-w)^3} +\frac{ j(w)}{(z-w)^2} +\frac{\partial j(w)}{z-w} \ .
\end{align*}
Its infinitesimal version is
\begin{align}\label{conformal-j}
 &\delta j = \epsilon \partial j +\partial\epsilon j -\frac{3}{2}\partial^2 \epsilon \ . 
\end{align}

On the other hand, in the conformal gauge \eqref{conformal-gauge}, \eqref{Ghost-anomaly} can be expressed as
\begin{align*}
 j_z = -4\kappa \partial\omega +j(z) \ ,
\end{align*}
where $j(z)$ is the holomorphic current. Using this expression, we find a transformation of $j(z)$ as
\[
j(z)=\frac{\partial \tilde{z}}{\partial z} \tilde{j}(\tilde{z})+2 \kappa \frac{\partial}{\partial z} \ln \frac{\partial \tilde{z}}{\partial z}~,
\]
which yields
\begin{align*}
 &\delta j(z) = \epsilon \partial j(z) +\partial\epsilon j(z) +2\kappa\partial^2 \epsilon \ .
\end{align*}
Comparing this with \eqref{conformal-j}, we have $\kappa = -\frac{3}{4}$.

The zero modes of the \( b \)-field correspond to the dimension of the moduli space of Riemann surfaces, \(\Sigma_g\), while the zero modes of the \( c \)-field match the number of conformal Killing vectors. Given that the zero modes of the \( bc \)-ghost field are fermionic, we must include an equivalent number of \( bc \)-ghost fields in the Grassmannian path integral to prevent it from vanishing. 

The ghost number anomaly puts a constraint on the non-vanishing ghost correlation functions. Let us consider the variation of the following correlation function under the ghost number symmetry $\delta_g$
\begin{align*}
\delta_g \Bigl\langle  \prod_{j=1}^{\nu/2} b(z_j) \prod_{i=1}^{\mu/2} c(z_i)\Bigr\rangle
 = \frac{\m-\n}{2} \Bigl\langle \prod_{j=1}^{\nu/2} b(z_j) \prod_{i=1}^{\mu/2} c(z_i) \Bigr\rangle \ .
\end{align*}
On the other hand, the Ward-Takahashi identity becomes
\begin{align*}
\delta_g \Bigl\langle \prod_{j=1}^{\nu/2} b(z_j)\prod_{i=1}^{\mu/2} c(z_i)  \Bigr\rangle
 =  &- \Bigl\langle \bigl(\int \frac{d^2z}{2\pi}\sqrt h \nabla^z j_z \bigr)\prod_{j=1}^{\nu/2} b(z_j)\prod_{i=1}^{\mu/2} c(z_i) \Bigr\rangle  \nonumber\cr
 =  &\Bigl\langle \bigl(\int \frac{3d^2z}{8\pi}\sqrt h R^{(2)} \bigr)\prod_{j=1}^{\nu/2} b(z_j)\prod_{i=1}^{\mu/2} c(z_i) \Bigr\rangle\cr
 =  & (3-3g) \Bigl\langle\prod_{j=1}^{\nu/2} b(z_j)\prod_{i=1}^{\mu/2} c(z_i) \Bigr\rangle \ .
\end{align*}
Therefore, we can conclude that $\# c - \# b = 3-3g$, and similarly we have $\#\ol c - \#\ol b = 3-3g$. This is consistent with $\mu-\nu=6-6g$ as in Table \ref{table:CKV-metric}.

\subsubsection*{General closed-string amplitude}

Now let us heuristically derive the closed string amplitude. If the conformal Killing group is present, we insert the ghost zero modes $\# c +\# \ol c \equiv \mu$ at specified positions to fix the gauge. If enough numbers of vertex operators are present, the gauge fixing can be done by pairing $\mu/2$ vertex operators with the ghost zero modes at specific positions
\[
c\bar cV(\wh\sigma_j) \qquad (j=1,\ldots, \mu/2)~.
\]
Moreover, we need to integrate over the physically inequivalent world-sheet metric, which is the moduli space of the world-sheet Riemann surface $\cM_{g}$. This can be done by using the non-degenerate pairing with the cotangent and tangent bundle of the moduli space introduced in \eqref{pairing}. Since the $b$ and $\bar b$ zero modes are sections of (anti-)holomorphic cotangent bundle of the moduli space, they are paired with  (anti-)holomorphic tangent vectors as
\begin{align*}
 (b, \partial_k \wh h) = \int \frac{d^2\sigma}{4\pi} \sqrt{\wh h}\ b_{ab} \frac{\partial}{\partial t_k} \wh h^{ab}(t)~,\qquad  (\bar b, \overline\partial_k \wh h) = \int \frac{d^2\sigma}{4\pi} \sqrt{\wh h}\ b_{ab} \frac{\partial}{\partial \overline t_k} \wh h^{ab}(t)~ .
\end{align*}
where $t_k$ are coordinates of $\cM_g$.

\begin{figure}[ht]\centering
  \includegraphics{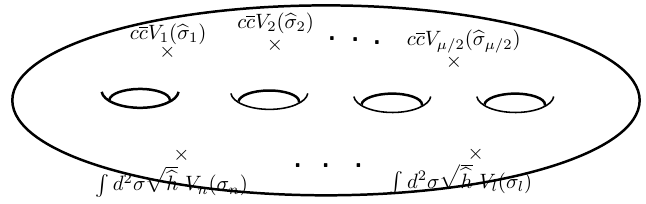}
  \caption{$\mu/2$ vertex operators are paired with $c\overline c$ at fixed positions and the others are integrated over the Riemann surface.}
  \label{fig:amplitude}
\end{figure}

With these insertions, the general closed-string amplitude in the bosonic string theory is written as
\begin{align}\label{closed-amplitude}
 A_{g,n}=&\int_{\cM_g} d^{\nu}t \int \cD[c, \bar c , b, \bar b , X]  \ e^{-S_\sigma [X,\wh h] -S_\textrm{gh}[b,c] -\Phi\chi(\Sigma_g)} \cr
 &\qquad \prod_{k=1}^{\nu/2} (b, \partial_k \wh h)   (\bar b, \overline\partial_k \wh h)\prod_{i=1}^{\mu/2}g_s  c\ol cV_i(\wh\sigma_i)
 \prod_{l=\mu/2+1}^{n}\int d^2\sigma_l \sqrt{\wh h}  g_sV_l(\sigma_l)  \ ,
\end{align}
where $\chi(\Sigma_g)$ is the Euler characteristic ($2-2g$), and $\Phi$ is a vacuum expectation value of a dilaton that gives the string coupling $g_s =e^\Phi$. Let us now consider explicit examples and carry out computations.

\subsection{Tree amplitude}

The first example is the tree-level amplitude in string theory: a sphere with $n\ge 3$ punctures.
Since we have $\mu =6$ and $\nu = 0$ for a sphere, we must insert at least three vertex operators to fix conformal Killing vectors. Considering a sphere $S^2\cong \bC\cup \{\infty\}$, we give the flat metric locally. Then, the amplitude formula becomes
\begin{align*}
 A_{g=0,n} &= g_s^{-2} \int \cD[c, \bar c , b, \bar b , X]  \ e^{-S_\sigma [X,\wh h] -S_\textrm{gh}[b,c]}
 \prod_{i=1}^{3}c\ol c(z_i,\ol z_i)g_s  V_i(z_i,\ol z_i)  \prod_{l=4}^{n}\int d^2z_l  g_s  V_l(z_l,\ol z_l) \cr
 &= g_s^{n-2} \prod_{l=4}^{n}\int d^2z_l  \Bigl\langle \prod_{j=1}^n V_j(z_j,\ol z_j) \Bigr\rangle_X
 \Bigl\langle c\ol c(z_1,\ol z_1) c\ol c(z_2,\ol z_2) c\ol c(z_3,\ol z_3) \Bigr\rangle_{bc} \ ,
\end{align*}
where we fix the positions of three operators $c\ol c(z_i,\ol z_i)g_s  V(z_i,\ol z_i)$ for $i=1,2,3$.
As a result, the amplitude factorizes into the matter sector and the ghost sector.

Let us compute when all the vertex operators are Tachyons \eqref{tachyon-vertex}:
\begin{align*}
 V_j (z_j,\ol z_j) = \nord{e^{ik_j\cdot X(z_j,\ol z_j)}} \ .
\end{align*}
Then, the OPE of the Tachyon vertex operators can be computed from \eqref{OPE-general}, which yields
\begin{align*}
 \Bigl\langle \prod_{j=1}^nV_j (z_j,\ol z_j)\Bigr\rangle_X = C_X (2\pi)^D \delta^D(\sum k_i) \prod_{i < j}^n |z_{ij}|^{\alpha'k_i\cdot k_j} \ .
\end{align*}

To compute the ghost sector,
we perform an integral over the zero modes
\begin{align*}
 c(z) = c_{0} + c_{1} z +c_{2} z^2 \ , \cr
\ol c(\ol z) = \ol c_{0} + \ol c_{1}\ol z +\ol c_{2} \ol z^2 \ .
\end{align*}
Then, the integral is indeed very simple
\begin{align*}
 \Bigl\langle c\ol c(z_1,\ol z_1) c\ol c(z_2,\ol z_2) c\ol c(z_3,\ol z_3) \Bigr\rangle_{bc}
 =& C_{bc} \int \prod_{i=0}^2 d\ol c_i dc_i \ c(z_1)\ol c(\ol z_1) c(z_2)\ol c(\ol z_2) c(z_3)\ol c(\ol z_3)\cr
 =&C_{bc}
\det \left|\begin{array}{ccc}
1 & 1 & 1 \cr
z_{1} & z_{2} & z_{3} \cr
z_{1}^{2} & z_{2}^{2} & z_{3}^{2}
\end{array}\right| \det \left|\begin{array}{ccc}
1 & 1 & 1 \cr
\bar{z}_{1} & \bar{z}_{2} & \bar{z}_{3} \cr
\bar{z}_{1}^{2} & \bar{z}_{2}^{2} & \bar{z}_{3}^{2}
\end{array}\right|\cr
 =& C_{bc} |z_{12}|^2 |z_{23}|^2 |z_{31}|^2 \ ,
\end{align*}
where $C_{bc}$ is the normalization constant independent of positions.

\subsubsection*{Shapiro-Virasoro amplitude}

Let us explicitly perform the four-point amplitude
\begin{align*}
 A_{0,4} = g_s^{2} C_\textrm{4pt} (2\pi)^D \delta^D(\sum k_i) \int \prod_i d^2z_i
 \prod_{i < j}^4 |z_{ij}|^{\alpha'k_i\cdot k_j} \prod_{i < j}^3 |z_{ij}|^2 \ .
\end{align*}
To fix the conformal Killing vectors, we can set $(z_1,z_2,z_3)$ to $(0,1,\infty)$.
Then the expression is reduced to
\begin{align}\label{Virasoro-Shapiro}
 A_{0,4} = g_s^{2} C_\textrm{4pt} (2\pi)^D \delta^D(\sum k_i)
 B\left( -\frac{\alpha' s}{4}-1,  -\frac{\alpha' t}{4}-1,  -\frac{\alpha' u}{4}-1 \right) \ ,
\end{align}
where 
\be 
B(a,b,c) = \int d^2z |z|^{2a-2} |1-z|^{2b-2} = \pi \frac{\Gamma(a)\Gamma(b)\Gamma(c)}{\Gamma(a+b)\Gamma(b+c)\Gamma(c+a)} \quad (a+b+c=1)~.
\ee
Here we use the Mandelstam variables 
\begin{align*}
 &s = -k_{1+2}^2 = -k_{3+4}^2 = -2 k_1\cdot k_2 -\frac{8}{\alpha'} \ , \cr
 &t = -k_{1+3}^2 = -k_{2+4}^2 = -2 k_1\cdot k_3 -\frac{8}{\alpha'} \ , \cr
 &u = -k_{1+4}^2 = -k_{2+3}^2 = -2 k_1\cdot k_4 -\frac{8}{\alpha'} \ , 
\end{align*}
with $k_{i+j}^2=(k_{i}+k_{j})^2$, satisfying the constraint
\be 
s+t+u = -\frac{16}{\alpha'}~.
\ee
In \eqref{Virasoro-Shapiro}, the Mandelstam variables $s,t,u$ are symmetric under permutations.
This is called the \textbf{Shapiro-Virasoro amplitude}. It is easy to see that Shapiro-Virasoro amplitude has a permutation symmetry among $s,t, u$.
For the derivation of the $B$ function, the reader can refer to \cite[\S7(Vol.1 p.386 and p.373)]{GSW}.

\begin{figure}[htb]
\centerline{\includegraphics[width=10cm]{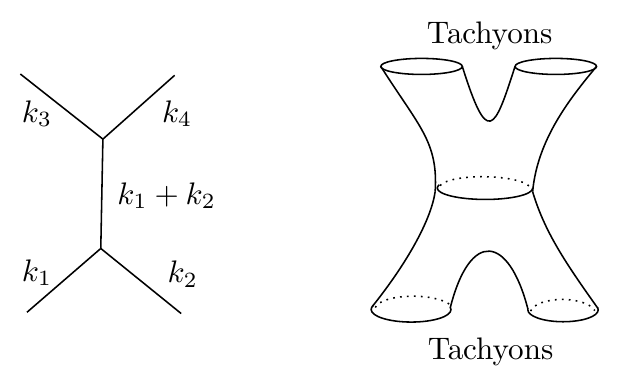}}
\caption{The s-channel of the Shapiro-Virasoro amplitude.}
\label{s-channel.eps}
\end{figure}

The $\Gamma$-function has a pole at non-positive integers. This shows that the amplitude in \eqref{Virasoro-Shapiro} has poles in the $s$-channel:
\begin{align*}
 -\frac{\alpha's}{4}-1 \in \bZ_{\le 0} , \qquad s = -k_{1+2}^2 = M^2 = \frac{4}{\alpha'} (n-1) \quad (n \in \bZ_{\ge 0}) \ ,
\end{align*}
where $M$ is the mass of a propagating state as in Figure \ref{s-channel.eps}. Since the mass spectra are the same as in \eqref{mass2}, this implies that the propagating states are tachyon, graviton, $B$-field, etc.
The four-point amplitude can be divided into two three-point amplitudes as in Figure \ref{s-channel.eps}.
At the $s$-channel tachyon pole, it becomes
\begin{align*}
 A_{0,3} &= a_3 (2\pi)^D \delta^D (\sum k_i) \ , \qquad a_3 \simeq C g_s \ , \cr
 A_{0,4} &= a_4 (2\pi)^D \delta^D (\sum k_i) \ , \qquad a_4 \simeq -\frac{4\pi}{\alpha'} \frac{1}{s+\frac{4}{\alpha'}} C g_s^2
 \ .
\end{align*}
We can fix the overall constant from unitarity
\begin{align*}
 &a_4 = \frac{(a_3)^2}{s+\frac{4}{\alpha'}} \ , \qquad
 \therefore \quad C = -\frac{4\pi}{\alpha'} \ .
\end{align*}

Although we have considered a closed-string tree-level amplitude here, the result for the open-string amplitude, the so-called \textbf{Veneziano amplitude} \cite{Veneziano:1968yb}, historically came first. It is obtained by the same steps on the disk, which we map to the upper half-plane with the open-string vertex operators $:e^{ik\cdot X}:$ on the real axis. The matter correlator is $\prod_{i<j}|y_{ij}|^{2\alpha'k_i\cdot k_j}$, the ghost zero modes are again absorbed by three $c$ insertions, and $\mathrm{PSL}(2,\bR)$ fixes $(y_1,y_2,y_3)=(0,1,\infty)$. The remaining integral over $y_4$ is divided into the ranges of the cyclic orderings of the four insertions; with the tachyon mass $M^2=-1/\alpha'$ the ordering $(1,2,3,4)$ gives
\[
I(s,t)=\int_0^1dy\,y^{-\alpha's-2}(1-y)^{-\alpha't-2}=B(-\alpha's-1,-\alpha't-1)~,
\]
and the full amplitude is $I(s,t)+I(t,u)+I(u,s)$. Its poles at $\alpha's=N-1$ reproduce the open-string spectrum \eqref{open-mass}, and the residue of the massless pole at $t=0$ fixes the gauge coupling in terms of the open-string coupling $g_o$. The discovery of the Veneziano amplitude is widely acknowledged as the starting point for the developments leading to string theory. The very first examples of string amplitudes by Veneziano \cite{Veneziano:1968yb}, Virasoro \cite{Virasoro:1969me}, Shapiro \cite{Shapiro:1970gy}, as well as Koba and Nielsen \cite{Koba:1969rw,Koba:1969kh} appeared long before the formulation of string theory. The early history of string theory before the superstring revolution in 1984 is recounted in \cite{cappelli2012birth}.

\subsection{One-loop amplitude}\label{sec:1-loop}

Let us consider a torus amplitude which is analogous to one-loop amplitude so that it encodes quantum corrections. First, we will use the cylindrical holomorphic coordinate $w=\sigma+it$ in Figure \ref{fig3} with periodicity \eqref{torus-identification}.

As seen in \S\ref{sec:CKV-metric}, there are two constant conformal Killing vectors and metric moduli. Note that an integral (with an appropriate normalization) over a torus automatically picks up zero modes
\[
c(0) = \int \frac{d^2w}{4\pi^2\tau_2}c(w)
\]
(The other zero modes $\ol c(0)$, $b(0), \ol b(0)$ are similarly obtained.)

%
Let us consider an infinitesimal deformation of the flat metric
\begin{align*}
 ds^2 = dwd\ol w \to d(w+\epsilon\ol w) d(\ol w+\ol\epsilon w) \ ,
\end{align*}
where $\delta h_{ww} = \ol\epsilon$ and $\delta h_{\ol w\ol w} = \epsilon$. One can set a new conformally-flat coordinate
 $\wt w=w+\epsilon \ol w$.
This changes the period of a torus to
\[
\wt w=\wt w+(2\pi (1+\epsilon),2\pi(\tau+\epsilon\ol\tau))~.
\]
Consequently, the complex modulus becomes
\[\wt \tau = \frac{\tau+\epsilon\ol\tau}{1+\epsilon} \simeq \tau -2i\epsilon\tau_2\]
so that $\delta \tau = -2i\epsilon\tau_2$. Thus, a tangent vector of the metric moduli  is
\begin{align*}
 \partial_\tau h_{\ol w\ol w} =  \frac{\delta h_{\ol w\ol w}}{\delta \tau} = \frac{i}{2\tau_2} \ ,
\end{align*}
so that we have
\begin{align*}
 (b, \partial_\tau  h) = \int \frac{d^2w}{2\pi} \sqrt{h}\ b_{ww}(w) h^{ w\ol w}h^{ w\ol w}  \frac{\partial}{\partial \tau}  h_{\ol w\ol w} (\tau) =2\pi ib_{ww}(0)\ .
\end{align*}

The torus amplitude is expressed as an integral over the fundamental region $F$ in Figure \ref{modular} with $bc$ zero modes insertion
\begin{align*}
 A_{1,n}
 &= g_s^{n} \frac{1}{2} \int_F d^2\tau   \Bigl\langle (b, \partial_\tau h) (\ol b, \partial_{\ol \tau} h)
 c(0)\ol c(0)  V_1(0) \prod_{j=2}^{n}\int d^2w_j \sqrt h V_j  (w_j,\ol w_j)\Bigr\rangle \ .
\end{align*}
where $\frac{1}{2}$ is due to the symmetry $w\to -w$. This $\bZ_2$ symmetry is a discrete subgroup of CKG. Since the zero mode on a torus can be obtained by integrating an operator out, the expression can be manipulated into
\begin{align*}
 A_{1,n}
 &= g_s^{n} \frac{1}{2}  \int_F \frac{d^2\tau}{\tau_2}
 \bigl\langle b(0)\ol b(0)c(0)\ol c(0) \bigr\rangle_{bc}
 \Bigl\langle \prod_{j=1}^{n}\int d^2w_j \sqrt h V_j (w_j,\ol w_j)\Bigr\rangle_X \ .
\end{align*}

\subsubsection*{Torus partition function for bosonic string}
Let us explicitly compute the amplitude without vertex operators ($n=0$):
\begin{align*}
 A_{1,0}
 &= \frac{1}{2} \int_F \frac{d^2\tau}{\tau_2}
 \bigl\langle b(0)\ol b(0)c(0)\ol c(0) \bigr\rangle_{bc}
 \bigl\langle 1 \bigr\rangle_X \ .
\end{align*}

The Feynman path integral over a torus with complex structure $\tau$ can be written in terms of the Hamiltonian formalism
\begin{align*}
 \bigl\langle 1 \bigr\rangle_X = \Tr \exp \left[ 2\pi i \left( \tau_1 P +i\tau_2 H \right)\right]
 = \Tr \left[ q^{L_0^X -\frac{c^X}{24}} \ol q^{\ol L_0^X -\frac{c^X}{24}} \right] \ ,
\end{align*}
where $q = e^{2\pi i\tau}$. Note that the space and time translations are generated by the zero modes of the Virasoro generator \eqref{Virasoro} \[P = L_0^X -\ol L_0^X~, \qquad H = L_0^X +\ol L_0^X -\frac{c^X}{12}~.\]

The trace is taken over the Hilbert space \eqref{bosonic-Hilb}. Writing the holomorphic Fock space with a fixed target coordinate as
\begin{equation}
|n_1,n_2,n_3,\cdots;k\rangle=\a_{-1}^{n_1}\a_{-2}^{n_2}\a_{-3}^{n_3}\cdots|0;k\rangle, \quad \mbox{with}\quad  n_i\in \mathbb{Z}_{\ge0}~,
\end{equation}
its generating function can be written as
\bea
\label{free-boson-process}
\mbox{Tr}_{\cH^X}(q^{L_0^X-\frac{c^X}{24}})&=q^{-\frac{1}{24}} \sum_{n_1=0}^\infty \sum_{n_2=0}^\infty \cdots \left\langle k;n_1,n_2,\cdots\left| q^{L_0^X} \right| n_1,n_2,\cdots ;k\right\rangle  \nonumber\\
&=q^{\frac{\a'}{4}k^2-\frac{1}{24}} \prod_{n=1}^\infty \frac1{1-q^n}=q^{\frac{\a'}{4}k^2}\frac{1}{\eta(\tau)}~,
\eea
where $\eta(\tau)$ is the \textbf{Dedekind eta-function}.
Hence, we have the bosonic torus partition function
\begin{align*}
 \bigl\langle 1 \bigr\rangle_X &=
 \int \frac{d^Dxd^Dk}{(2\pi)^D} \exp \left( -4\pi\tau_2 \frac{\alpha'}{4} k^2 \right)
 \left| \frac{q^{-\frac{1}{24}}}{\prod_{n \ge 1} (1-q^n)} \right|^{2D}  \nonumber\cr
 &= i \frac{V_{D}}{(2\pi \ell_s)^{D}} \left(\tau_2\right)^{-D/2} \left|\eta(\tau)\right|^{-2D} \ ,
\end{align*}
where $i$ in the second line is from Wick-rotation of spacetime momentum $k^0 \to ik^0_E$.

Similarly, we can write the ghost partition function over a torus with complex structure $\tau$ as
\begin{align*}
 \bigl\langle b(0)\ol b(0)c(0)\ol c(0) \bigr\rangle_{bc}
 = \Tr [(-1)^F b_0 \ol b_0 c_0 \ol c_0 q^{L_0^{g} -\frac{c^g}{24}} \ol q^{\ol L_0^{g} -\frac{c^g}{24}}  ]
\end{align*}
with $c^g=-26$, and  the zero mode of the Virasoro generator \eqref{L-bc} is
\be
L^{g}_0=-\sum_{n\in\bZ}n:b_nc_{-n}:-1~.
\ee
Here $F$ is the fermion number operator; $F=1$ for fermion and $F=0$ for boson. The insertion of $(-1)^F$ means that the periodic boundary condition is imposed on the ghost fields along the time circle.
The trace is taken over the Hilbert space spanned by physical states that are constructed by acting with creation operators $b_{-n},c_{-m}$ for $n,m\in \bZ_{>0}$ (as well as anti-holomorphic operators) on the $bc$ ghost vacuum $|\downarrow\downarrow\rangle$
\[
 \Tr_{\cH^g} [(-1)^F b_0 c_0  q^{L_0^{g} -\frac{c^g}{24}} ] = q^{-1-\frac{c^g}{24}}\prod_{n \ge 1} (1-q^n)^2
 \]
Thus, we have
\begin{align*}
 \bigl\langle b(0)\ol b(0)c(0)\ol c(0) \bigr\rangle_{bc}
 = q^{-1-\frac{c^g}{24}} \ol q^{-1-\frac{c^g}{24}} \prod_{n \ge 1} (1-q^n)^2 (1-\ol q^n)^2 = |\eta(\tau)|^4 \ .
\end{align*}
In conclusion, the torus amplitude is
\begin{align}\label{Boson-torus}
 A_{1,0}
 &= \frac{iV_D}{(2\pi \ell_s)^{D}} \int_F \frac{d^2\tau}{2\tau_2}  \left(\tau_2\right)^{-D/2} \left|\eta(\tau)\right|^{-2(D-2)} \ .
\end{align}
At the critical dimension $D=26$, the expansion of the Dedekind eta-function
\[\left|\eta(\tau)\right|^{-48} \simeq \left|q^{-1} +24 +\mathcal O(q)\right|^2~,\]
gives the spectrum of the bosonic string theory. Furthermore, it is invariant under $\PSL(2,\bZ)$. In fact, the measure $\frac{d^2\tau}{\tau_2^2}$ is invariant under $\PSL(2,\bZ)$, and $\tau_2^{\frac12}|\eta(q)|^2$ is also invariant under $\PSL(2,\bZ)$ (see \S\ref{sec:modular}).

Note that the limit $\tau_{2} \rightarrow 0$ describes the ultraviolet (UV) regime since the Euclidean time becomes small. The integral over the fundamental domain $F$ avoids this UV region. Consequently, there is no UV divergence in the one-loop amplitude of the bosonic string theory thanks to the modular invariance of a torus.

\section{Superstring theories}\label{sec:superstring}

We have seen so far that bosonic strings suffer from two major problems:

\vspace{.3cm}
\noindent $\bullet$  Their spectrum always contains a tachyon. In that respect, their vacuum is unstable.

\vspace{.3cm}
\noindent $\bullet$    They do not contain spacetime fermions. This lack of fermionic states is in
contrast to observations and makes the bosonic string unrealistic.
\vspace{.3cm}

Both of these challenges are remedied in superstring theory.
Supersymmetry is a symmetry that exchanges bosons and fermions. The world-sheet superstring theory consists of a bosonic and a fermionic sector. The bosonic sector is identical to the world-sheet theory of the bosonic string. Therefore, we can view our efforts up to now as a preliminary study of one-half of the superstring theory.
In fact, we will see that the presence of fermions resolves the problem of Tachyon. Moreover, we will learn that the critical dimension of superstring theory is $D=10$.

There are five superstring theories as follows, and we will study them in this order.

\begin{description}
\item{\textbf{Type IIA \& IIB}}

Oriented string theories that can incorporate open strings if there are D-branes. IIA: Ramond ground states with opposite chirality, and D$p$-branes ($p$ even). IIB: Ramond ground states with the same chirality, and D$p$-branes ($p$ odd).

\item{\textbf{Type I}}

An open and closed unoriented string theory, including Yang-Mills degrees of freedom with $\SO(32)$ gauge group, that can incorporate D1, D5, D9-branes.

\item{\textbf{Heterotic $\SO(32)$ \& $E_8\times E_8$}}

Type II right-movers \& bosonic left-movers, including Yang-Mills degrees of freedom with either $\SO(32)$ or
$E_8\times E_8$ gauge group.
\end{description}

There exist two major formulations of superstring theory. Both formulations enjoy supersymmetry on the world-sheet and in spacetime, but they differ in the following respect:

\vspace{.3cm}
\noindent $\bullet$  In the \textbf{Ramond-Neveu-Schwarz (RNS) formulation}, supersymmetry is manifest on the world-sheet, but not in spacetime.

\vspace{.3cm}
\noindent $\bullet$  In the \textbf{Green-Schwarz (GS) formulation} \cite[\S5]{GSW} \cite[\S5]{BBS}, supersymmetry is manifest in spacetime, but not on the world-sheet .
\vspace{.3cm}

More recently, the pure-spinor formalism \cite{Berkovits:2004px} has been developed as yet another approach to superstring.
In this section, let us discuss the RNS formalism of superstring theory.

\subsection{RNS formulation}

With the complex coordinate convention, the action becomes
\begin{align}\label{action}
 S^{\textrm{m}} = \frac{1}{4\pi} \int d^2 z\ \Big( \frac{2}{\alpha'} \partial X^\mu  \overline\partial X_\mu+\psi^\mu\overline\partial\psi_\mu+\ol\psi^\mu\partial\ol\psi_\mu\Big)
 \end{align}
where the equations of motion tell us $\psi^\mu(z)$ (resp. $\ol\psi^\mu(\bar z)$) is chiral (resp. anti-chiral). The action is invariant under \textbf{supersymmetric transformation} (exercise)
\be\label{susy-trans}
\delta X^\mu=-\sqrt{\frac{\a'}{2}}\,(\e \psi^\mu+\overline \e \ol\psi^\mu)~,\qquad
\delta \psi^\mu =\sqrt{\frac{2}{\a'}}\, \e \partial X^\mu~, \qquad \delta \ol\psi^\mu =\sqrt{\frac{2}{\a'}}\,\overline\e \overline \partial X^\mu~.
\ee
The Noether theorem implies that there are currents for the supersymmetry
\be
T_F(z)=i\sqrt{\frac{2}{\a'}}\psi^{\mu}\partial X_{\mu}\;,\qquad \ol T_F(z)=i\sqrt{\frac{2}{\a'}}\ol
\psi^{\mu}
\bar \partial X_{\mu}\,.\label{274}\ee
which are called \textbf{supercurrents}. Indeed, $\psi^\mu$ and $\ol\psi^\mu$  are primary fields of conformal dimension, $(\frac12,0)$ $(0,\frac12)$, respectively, and therefore  their OPEs are
\be\label{OPE}
X^\mu(z,\bar z)X^\nu(0,0)\sim -\frac{\a'}{2} \eta^{\mu\nu}\ln |z|^2~,\quad \psi^\mu(z) \psi^\nu(0)\sim \frac{\eta^{\mu\nu}}{z}~,\quad \ol\psi^\mu(\bar z) \ol\psi^\nu(0)\sim \frac{\eta^{\mu\nu}}{\bar z}~.
\ee
Using the OPEs, one can show the supersymmetric transformation \eqref{susy-trans} (exercise).

The energy-momentum tensor of the action \eqref{action} is
\begin{align}
T_B(z) =-\frac1{\a'}\,\partial X^\mu \partial X_\mu-\frac12
\psi^\mu\partial\psi_\mu
\end{align}
along with their complex conjugates $\ol T_B$, $\ol T_F$.
Their OPEs can be computed by using  \eqref{OPE}
\begin{align}
T_B(z)T_B(w) &\sim \frac{3D}{ 4(z-w)^4}+\frac{2T_B(w)}{ (z-w)^2}
  +\frac{\partial_w T_B(w)}{z-w}\cr
T_B(z) T_F(w) &\sim \frac{3T_F(w)}{ 2(z-w)^2}+\frac{\partial_w T_F(w)}{
  z-w}\cr
T_F(z)T_F(w) &\sim \frac{D}{ (z-w)^3}+\frac{2T_B(w)}{z-w}\,\,.
\end{align}
and similarly for the anti-chiral part. The central charge of the theory is
\be \label{c}c^{\textrm{m}}=\frac32D\ee
where each scalar and fermion contributes 1 and 1/2, respectively.

\subsubsection*{Ramond vs Neveu-Schwarz}

In superstring theory, the fermionic fields on the closed string may be either periodic or anti-periodic on the circle around the string, corresponding to two different spinor bundles.
It is conventional to denote these spin structures by \textbf{Ramond (R)} and \textbf{Neveu-Schwarz (NS)}, defined as follows.
\begin{align}
\psi^\mu(t,\sigma+2\pi)=+\psi^\mu
  (t,\sigma) &\qquad\qquad\textrm{ R: periodic
on cylinder}\cr
\psi^\mu(t,\sigma+2\pi)=-\psi^\mu
  (t,\sigma) &\qquad\qquad\textrm{ NS:
anti-periodic on cylinder}
\end{align}
As in Figure \ref{fig3}, the mapping $z=e^{-iw}$ from the cylinder $w=\sigma+it$ to the 2-plane $z\in \bC$ is a conformal map. Under the conformal map, the primary field $\psi^\mu$ with conformal dimension $(\frac12,0)$ is transformed as
\[
\psi^\mu(z) =\Big(\frac{dz}{dw}\Big)^{-\frac12}\psi^\mu(w)=\textrm{const}\times e^{-i\frac{w}{2}}\psi^\mu(w)~.
\]
Hence the (anti-)periodicity assignments are reversed
between the cylinders and the plane:
\begin{align}
\psi^\mu(e^{2\pi i}z)=-\psi^\mu(z)
 &\null\qquad\qquad\hbox{\rm R: anti-periodic on plane}\cr
\psi^\mu(e^{2\pi i}z)=+\psi^\mu(z)
  &\null\qquad\qquad \hbox{\rm NS: periodic on plane}
  \end{align}
The boundary conditions for anti-chiral fields $\ol\psi^\mu$ are defined in a similar fashion.

As in the bosonic string, one can  decompose $\psi^\mu$ and $\ol\psi^\mu$ in
modes
\[
\psi^\mu(z) =\sum\limits_{n\in\bZ+\nu}\frac{\psi_n^\mu}{  z^{n+1/2}}~,\qquad
\ol\psi^\mu(\bar{z}) =\sum\limits_{n\in\bZ+\nu}\frac{\ol{\psi}_n^\mu}{ \bar{z}^{n+1/2}}
\]
where $\nu$ takes the values 0 (R) and $\frac12$ (NS). The canonical quantization leads to the algebra
\be \label{canonical-fermion}
\{\psi_m^\mu,\psi_n^\nu\}
  =\eta^{\mu\nu}\delta_{m+n,0}\qquad
\{\ol{\psi}_m^\mu,\ol{\psi}_n^\nu\}
 =\eta^{\mu\nu}\delta_{m+n,0}\,\,.
\ee

The mode expansion must be carried out with care
here since we must distinguish between Ramond and
Neveu-Schwarz sectors.
\begin{align}
T_B(z)=\sum\limits_{n\in\bZ}\frac{L_n^{\textrm{m}}}{z^{n+2}}~,\qquad T_F(z)=\sum_{r\in \bZ+\nu}\frac{G_r^{\textrm{m}}}{ z^{r+3/2}}~,
\end{align}
where the generators can be written in terms of the modes (exercise)
\begin{align}
L_m^{\textrm{m}}&=\frac12\sum_{n\in \bZ}\eta_{\m\n}:\a_{m-n}^\mu \a_n^\nu :+\frac14\sum_{r\in \bZ+\nu} (2r-m)\eta_{\m\n}:\psi_{m-r}^\m\psi_r^\n:+a^{\textrm{m}}\delta_{m,0} \cr
G_r^{\textrm{m}}&=\sum_{n\in \bZ}\eta_{\m\n}\, \a_n^\m\psi_{r-n}^\n~.
\end{align}
The normal ordering constant $a^{\textrm{m}}$ can be determined like in the bosonic string theory. Each periodic
boson contributes $-\frac1{24} $. The fermionic contributions are
\begin{align}\label{RNS-normalordering}
-\tfrac12\sum_{r=0}^\infty r=\tfrac1{24} &\qquad \textrm{R-sector}~,\cr
-\tfrac12\sum_{r=0}^\infty (r+\tfrac12)=-\tfrac1{48} &\qquad \textrm{NS-sector}~.
\end{align}
Including the shift $\tfrac1{24}c^{\textrm{m}}=\tfrac{1}{16}D$ gives
\begin{align}\label{zero-m}
&a^{\textrm{m}}=\tfrac1{24}c^{\textrm{m}}+\Big(-\tfrac1{24}+\tfrac1{24}\Big)D=\tfrac{1}{16}D&\textrm{R-sector}~,\cr
&a^{\textrm{m}}=\tfrac1{24}c^{\textrm{m}}+\Big(-\tfrac1{24}-\tfrac1{48}\Big)D=0 &\textrm{NS-sector}~.
\end{align}
In fact,  the generators $L_m^{\textrm{m}}$ and $G_r^{\textrm{m}}$ form the algebra called the
\textbf{ $\cN=1$ superconformal
algebra} with central charge \eqref{c} (exercise):
\begin{equation}\label{N1SCA}
\begin{aligned}
 \left[L_m, L_n\right]=&(m-n) L_{m+n}+\frac{c}{12}\left(m^3-m\right) \delta_{m,-n}~, \\
 \left\{G_r, G_s\right\}=&2 L_{r+s}+\frac{c}{12}\left(4 r^2-1\right) \delta_{r,-s}~, \\
 \left[L_m, G_r\right]=&\frac{m-2 r}{2} G_{m+r} ~.
\end{aligned}
\end{equation}
where we omit the upper script ${}^{\textrm{m}}$.

\subsubsection*{Ghost CFT}

In the bosonic string theory, we studied the BRST quantization with the Faddeev-Popov $bc$ ghost. In superstring theory, ghost fields also appear with their supersymmetric partners:
\[
S^{\textrm{gh}}=\frac{1}{2\pi}\int d^2z (b\overline \partial c+\beta\overline \partial \g)
\]
where $b,c$ are fermionic and $\beta,\gamma$ are bosonic fields.  Hence, the standard method tells us the $\beta\g$ OPEs
\[
  \g(z)\b(w)=-\b(z)\g(w)\sim\frac{1}{z - w}
\]
We have seen that the conformal dimensions of $X$ and $\psi$ differ by $\frac12$. This is the same for the ghost fields. Since the $b$ and $c$ ghosts have conformal dimensions $2$ and $-1$ respectively, $\beta$ and $\g$ are primary fields of conformal dimensions $(\frac32,0)$ and $(-\frac12,0)$ respectively. Hence, the form of the energy-momentum tensor and the supercurrent are
\begin{align}
T_B^{\textrm{gh}}(z) &=: (\partial b) c : -2\, \partial: bc :+ : (\partial \b) \g : -\frac32\, \partial: \b\g :\cr
T_G^{\textrm{gh}}(z) &= (\partial\b)c+\frac32\b\partial c-2b \g~.
\end{align}
Then, the $TT$ OPE determines the central charge of the ghost SCFT. The $bc$ system contributes $-26$ to the central charge as we know, while  the
$\beta\gamma$
system contributes $+11$. Hence, the total central charge
\[
c^{\textrm{tot}}=c^{\textrm{m}}+c^{\textrm{gh}}=\tfrac{3}{2} D-26+11~.
\]
Then, we happily obtain the critical dimension $D=10$ of superstring theory if we impose the Weyl-anomaly-free condition $c^{\textrm{tot}}=0$. In the following, we assume $D=10$.

Now, let us write the Virasoro generators of the ghost SCFT. The $\b\g$ ghosts have the same boundary condition as the fermionic fields $\psi^\mu\ol\psi^\mu$ so that we have the mode expansions
\[
\beta(z)=\sum_{r\in \bZ+\nu}\frac{\beta_r}{z^{r+\frac32}}~,\quad \g(z)=\sum_{r\in \bZ+\nu}\frac{\g_r}{z^{r-\frac12}}~,
\]
which satisfy the commutation relation
\[
[\g_s,\b_r]=\delta_{r,-s}~.
\]
Using these modes, we express the Virasoro generator of the ghost SCFT as
\begin{align}
L_m^{\textrm{gh}}&=\sum_{n\in \bZ}(2m-n):b_nc_{m-n}:+\frac12\sum_{r\in\bZ+\nu}(m+2r):\b_{m-r}\g_r:+a^{\textrm{gh}}\d_{m,0}\cr
G_r^{\textrm{gh}}&=-\sum_{n\in \bZ}\Big[\frac12(n+2r)\b_{r-n}c_n +2b_{n}\g_{r-n}\Big]
\end{align}
They also satisfy the $\cN=1$ superconformal algebra \eqref{N1SCA}.
Again, using the commutation relations of the ghost modes, one can determine the normal ordering constant
\begin{align}\nonumber
&a^{\textrm{gh}}=\tfrac{-15}{24}+\Big(\tfrac1{12}-\tfrac1{12}\Big)=-\tfrac58& \textrm{R-sector}~,\cr
&a^{\textrm{gh}}=\tfrac{-15}{24}+\Big(\tfrac1{12}+\tfrac1{24}\Big)=-\tfrac12 & \textrm{NS-sector}~.
\end{align}
where two fermions and bosons contribute $\frac1{12}$ and $-\frac1{12}$, respectively, in the R sector while their contributions are $\frac1{12}$ and $\frac1{24}$, respectively, in the NS sector.
Combining them with \eqref{zero-m} at $D=10$, we have the total vacuum energy
\begin{align}\nonumber
&a^{\textrm{tot}}=0 \qquad\quad \textrm{R-sector}~,\cr
&a^{\textrm{tot}}=-\tfrac12 \qquad  \textrm{NS-sector}~.
\end{align}
In the R sector, the vacuum energy is zero so that the Tachyon is absent. On the other hand, Tachyon is still present in the NS sector. This will be projected out by the GSO projection, as we will see below.

\subsection{Physical spectrum and the GSO Projection}

Before discussing the GSO projection, let us study the fermionic spectrum generated by fermionic modes $\psi^\mu_r$.  We first consider the NS spectrum since it is simpler. Since $r$ takes half integers, we can define the ground state of the NS sector as
\[
\psi^\mu_r|0;k\rangle_{\textrm{NS}}=0 \quad \textrm{for} \ r>0~.
\]
When we include the ghost part of the vertex operator, it contributes to the total fermion number $F$, so that the matter plus ghost ground state has the odd fermion number
\be\label{NS-vac}
(-1)^{F}|0;k\rangle_{\textrm{NS}}=-|0;k\rangle_{\textrm{NS}}~.
\ee
Because there exist the zero modes $\psi^\mu_0$, the R sector is more subtle. In fact, the canonical commutation relation \eqref{canonical-fermion} of the zero modes satisfies the \textbf{Clifford algebra}
\[
\{\sqrt{2}\psi_0^\mu,\sqrt{2} \psi_0^\nu\}=2\eta^{\mu\nu}~.
\]
Therefore, we can identify them with Gamma matrices $\Gamma^\mu=\sqrt{2}\psi_0^\mu$, and the ground state of the R sector becomes the spin representation of $\SO(1,D-1)$. For mathematics of Clifford algebra, spin group, and spin representations, we refer to \cite[Appendix B]{Polchinski}. However, we can heuristically understand the spin representation as follows. The following basis for this representation is often convenient:
\begin{align}\label{raising-lowering}
\Gamma^{\pm}_i &= \frac{1}{\sqrt 2}\left ( \psi^{2i}_0\pm i \psi^{2i+1}_0\right
) \qquad i=1,\ldots,4 \cr
 \Gamma^{\pm}_0 &= \frac{1}{\sqrt 2}\left ( \psi^{1}_0 \pm \psi^{0}_0\right )
\end{align}
In this basis, the Clifford algebra takes the form
\be
\{ \Gamma^{+}_i, \Gamma^{-}_j \}=\delta_{ij}~,\qquad  \{ \Gamma^{+}_i, \Gamma^{+}_j \}=0=\{ \Gamma^{-}_i, \Gamma^{-}_j \}~.
\ee
The $\Gamma^{\pm}_i$, $i = 0,\ldots, 4$ act as raising and lowering
operators, generating the $2^5= 32$ Ramond ground states:
\be\label{Ramond-32}
|s_0,s_1,s_2,s_3,s_4 ;k\rangle = |\mathbf{s};k\rangle
\ee
where each of the $s_i$ is $\pm\frac12$, and where
\be
\Gamma^{-}_{i} | -\tfrac12 , -\tfrac12 , -\tfrac12 , -\tfrac12 , -\tfrac12 ;k\rangle = 0
\ee
while $\Gamma^{+}_i$ raises $s_i$ from $-\frac12$ to $\frac12$. One can further define the chirality operator
\be\label{Gamma11}
\Gamma_{11}=(2)^{5} \psi_0^0\psi_0^1\psi_0^2\cdots\psi_0^9~,
\ee
which acts on $|\mathbf{s}\rangle$ as
\[
\Gamma_{11}|\mathbf{s};k\rangle=(-1)^{ F}|\mathbf{s};k\rangle =\left\{\begin{array}{ll}+|\mathbf{s};k\rangle &\qquad\textrm{even \# of}\ -\frac12 \cr -|\mathbf{s};k\rangle &\qquad \textrm{odd \# of} \ -\frac12 \end{array} \right.~.
\]
Hence, the Dirac representation $\bf 32$ decomposes into a
$\bf 16_s$ with an even number of $-\frac12$'s and $\bf 16_c$ with an odd number.
\be\label{32}
\bf 32=16_s\oplus16_c~.
\ee

Now, let us study the physical spectrum of superstring theories, using the GSO projection. We start with the open string. As in \S\ref{sec:BRSTsub}, we use the BRST quantization for the constraints on physical states. We define the BRST current and charge as
$$
Q_{\mathrm{B}}=\frac{1}{2 \pi i} \oint d z j_{\mathrm{B}}~.
$$
where
$$
j_{\mathrm{B}}=c T_B^{\mathrm{m}}+\gamma T_F^{\mathrm{m}}+\frac{1}{2}\left(c T_B^{\mathrm{gh}}+\gamma T_F^{\mathrm{gh}}\right)~.
$$
Using the OPE relation, it is straightforward, though tedious, to show that the $L_n$ and $G_r$ can be obtained by
$$
\left\{Q_{\mathrm{B}}, b_n\right\}=L_n, \quad\left[Q_{\mathrm{B}}, \beta_r\right]=G_r .
$$
In the BRST quantization, the space of physical states is isomorphic to the $Q_B$-cohomology \eqref{BRST-quantization}. Therefore, a physical state $|\psi\rangle$ is annihilated by the positive modes
\be\label{physical-cond}
L_n|\psi\rangle=0 \quad (n>0)~,\qquad G_r|\psi\rangle=0 \quad (r> 0)~.
\ee
The physical states are defined modulo
\be\label{QB-exact}
L_n|\chi\rangle\cong 0~,\qquad G_r|\chi\rangle\cong 0 ~,\qquad \textrm{for}\ \ n,r<0~.
\ee
As in \eqref{b0-annihilate} and \eqref{L0-annihilate}, we also impose the condition that the zero modes annihilate the physical state, which gives
the mass-shell condition
\begin{equation}
\label{physical 2}
    L_0 \ket{\psi} 
    = (\alpha' k^2 + N + a^\text{tot}) 
    \ket{\psi} =0.
\end{equation}
In addition, in the $\mathrm{R}$ sector we impose
$$
\beta_0|\psi\rangle=G_0|\psi\rangle=0~.
$$
As at the end of \S\ref{sec:BRSTsub}, by taking $Q_B$-cohomology, the action of all the ghost modes $b,c,\b,\g$ is null, namely not included in the space of physical states.

Furthermore, in the RNS theory, we need to impose the \textbf{GSO (Gliozzi-Scherk-Olive)  projection} in order to have an equal number of bosonic and fermionic states at each mass level.

In the NS sector, the GSO projection is just to remove states with odd fermion numbers so that the GSO projection operator on the NS sector is expressed as
\be\label{GSO-NS}
 P_{\textrm{GSO}}^{\textrm{NS}}=\frac{1+(-1)^{F_\text{NS}}}{2} \qquad \textrm{NS sector}~.
\ee
 At  level 0, from
 the $L_0$ condition, the lowest state $|0;k\rangle_{\textrm{NS}}$ is tachyonic 
 \begin{equation}
     m^2=-k^2 = - 
     \frac{1}{2 \alpha'}
 \end{equation}
Fortunately, the GSO projection removes this state because it has an odd fermion number as in \eqref{NS-vac}. 
The NS ground state
is at level $\frac12$, where we have a massless state with vector polarization
\[
|\zeta;k\rangle_{\textrm{NS}}=\zeta\cdot \psi_{-\frac12}|0;k\rangle_{\textrm{NS}}~,
\]
with even fermion number so we need to keep it in the spectrum. The physical state conditions \eqref{physical-cond}, \eqref{physical 2} are
\begin{align}
0&=L_0|\zeta;k\rangle_{\textrm{NS}}=\a'k^2 |\zeta;k\rangle_{\textrm{NS}},\cr
0&=G_{\frac12}|\zeta;k\rangle_{\textrm{NS}}=\sqrt{2\a'}~ k\cdot \zeta|0;k\rangle_{\textrm{NS}}~,
\end{align}
while the null state at this level leads to
\[
G_{-\frac12}|0;k\rangle_{\textrm{NS}}=\sqrt{2\a'}~ k\cdot \psi_{-\frac12}|0;k\rangle_{\textrm{NS}}~.
\]
Therefore, we have
\[
m^2= -k^2=0~,\quad  k\cdot \zeta=0~,\quad \zeta^\mu\cong \zeta^\mu+k^\mu~.
\]
Thus, the NS ground state is massless with degrees of freedom for 8 spacelike polarizations, forming the vector representation ${\bf 8_v}$ of $\SO(8)$.

A Ramond ground state can be expressed with spinor polarization
\[
|u;k\rangle_{\textrm{R}}= u_{\mathbf{s}}|\mathbf{s};k\rangle_{\textrm{R}}~.
\]
The mass-shell condition \eqref{physical 2} implies that the Ramond ground state is massless
\[
0 = L_0\ket{u;k}_R
= \alpha' k^2 
\ket{u;k}_R,
\]
while the physical condition 
\eqref{physical-cond} 
\[
0=G_{0}|u;k\rangle_{\textrm{R}}=\sqrt{\a'}~  u_{\mathbf{s}} k\cdot\G_{\mathbf{s}\mathbf{s}'} |\mathbf{s}';k\rangle_{\textrm{R}}
\]
leads to the Dirac equation
\[
u~ k\cdot\G=0~.
\]
By choosing the momentum vector $k^\mu=(k,k,0,\ldots,0)$, this amounts to
\be \label{fermon-light-cone}
\G_0^+u=0 \quad  \longrightarrow \quad s_0 =+ \tfrac12~,
\ee
giving 16 degeneracies $|+,s_1,s_2,s_3,s_4\rangle$ for the physical
Ramond vacuum.  This is a representation $\bf 16$ of $\SO(8)$ which again
decomposes into ${\bf 8_s}$ with an even number of $-\frac12$'s and ${\bf
8_c}$ with an odd number:
\begin{equation}\label{8spin}
  \bf 16=8_s\oplus 8_c~.
\end{equation}
In the R sector, the GSO projections will pick one of these two irreducible representations, and therefore the GSO projection operators can be written as
\be\label{GSO-R}
P^{\textrm{R}\pm}_{\textrm{GSO}}=\frac{1\pm(-1)^{F_R}}{2}\qquad \textrm{R sector}~.
\ee
where $F_R=\sum_{i=0}^4(\frac12+s_i)$.
Indeed, the two choices
$\bf 8_s$ and $\bf 8_c$ differ by the spacetime parity redefinition.

\subsection{Torus partition functions for superstring theory}

Let us see the role of the GSO projections in the
torus (one-loop) partition function of superstring theory. Although the computation is very similar to that in \S\ref{sec:1-loop}, fermionic torus partition functions are more interesting due to boundary conditions along circles.

Since the critical dimension of superstring theory is $D=10$, the bosonic contribution can be read off from \eqref{Boson-torus}. Consequently, the torus partition function for superstring theory can be written as follows
\begin{align*}
 Z = \frac{iV_{10}}{(2\pi \ell_s)^{10}} \int \frac{d^2\tau}{2\tau_2^2} \
 \frac{1}{\tau_2^4 |\eta(\tau)|^{16}} Z_\mathrm{F}(\tau) \ol Z_\mathrm{F}(\ol\tau) \ ,
\end{align*}
where $Z_F$ is $\psi$ contribution and $\ol Z_F$ is $\ol\psi$ contribution.

As usual, we will focus on the holomorphic sector $Z_F$, and we will sum up all the possible boundary conditions for fermions.
Since the NS \& R boundary conditions give rise to disconnected Fock spaces, we can introduce a relative phase $e^{i\theta}$ between them. Hence, the partition function in the Hamiltonian formalism is written as follows.
\begin{align*}
 &Z_\mathrm{F}(\tau) = Z_\mathrm{NS}^{+} +Z_\mathrm{NS}^{-} +e^{i\theta} Z_\mathrm{R}^{+} +e^{i\theta} Z_\mathrm{R}^{-} \ , \cr
 &Z_\mathrm{S}^{\pm} = \Tr \left[ (\pm)^F e^{2\pi i \tau H_\mathrm{S}}\right] \ ,
\end{align*}
where S is either NS or R,
 and the superscript is an option for
periodicity of world-sheet time $t$ direction ($-$ is periodic).
In the light-cone gauge, the ``Hamiltonian''s for fermions  are
\begin{align}\label{Hamiltonian-fermion}
 H_\mathrm{NS} &= L_0  = \sum_{r=\frac{1}{2}}^\infty r \psi_{-r} \cdot \psi_r -\frac{D-2}{48} \ , \cr
 H_\mathrm{R} &= L_0  = \sum_{r=1}^\infty r \psi_{-r} \cdot \psi_r +\frac{D-2}{24} \ ,
\end{align}
where $D$ should be $10$ for superstring, and the light-cone gauge means that the spacetime indices $i$ run from $2$ to $9$,
namely \[\psi \cdot \psi = \sum_{i=2}^9 \psi^i \psi^i.\]

Although the ground state of the NS sector is unique, the R sector has vacuum degeneracies as in \eqref{Ramond-32} before imposing any physical condition.
However, the physical state condition requires $s_0=+\frac{1}{2}$ \eqref{fermon-light-cone} so that
there are $2^4 = 16$ degeneracies on which the operator $(-1)^{F_R}$ acts as
\[
(-1)^{F_R} |\bfs ; k \rangle =(-1)^{\sum_{i=0}^4 \left(\frac{1}{2} +s_i\right)}|\bfs ; k \rangle
\]
Therefore, we have
\begin{align*}
\sum_{\bfs} \langle \bfs; k| (\pm 1)^{F_R} |\bfs ; k \rangle =
 \begin{cases}
  16 \qquad &\textrm{for $+$ sign}   \cr
  0 \qquad &\textrm{for $-$ sign,}
 \end{cases} ,
\end{align*}
where sum over $s_i\ (i=1,2,3,4)$ is understood.
In the presence of the fermion zero modes, the index $\Tr (-1)^{F_R}$ vanishes because $s_i=\pm\frac12$ gives the opposite sign. (This is the generic feature of the index.)

\begin{figure}[ht]
	\centering
	\includegraphics[width=0.8\linewidth]{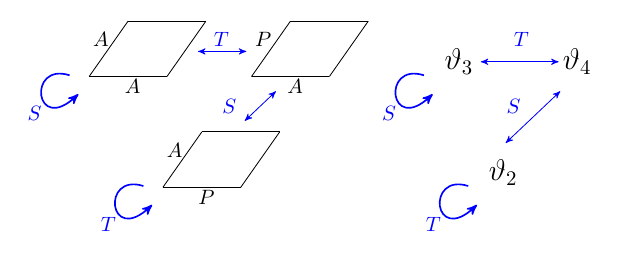}
	\caption{The $T$ and $S$ transformations of $\vartheta$-functions. $A$ and $P$ represent anti-periodic and periodic boundary conditions of fermions.}
	\label{fig:modular-theta}
\end{figure}

Counting the eigenvalues of \eqref{Hamiltonian-fermion} in the fermionic Fock spaces with a given boundary condition, the torus partition functions are therefore written as
\begin{align*}
 &Z_\mathrm{NS}^{+} = \Tr \left[ e^{2\pi i \tau H_\mathrm{NS}}\right] = q^{-\frac{1}{6}} \prod_{n=1}^\infty (1+q^{n-\frac{1}{2}})^8
 = \left(\frac{\vartheta_3(\tau)}{\eta(\tau)}\right)^4 \ , \cr
 &Z_\mathrm{NS}^{-} = \Tr \left[ (-1)^F e^{2\pi i \tau H_\mathrm{NS}}\right] = - q^{-\frac{1}{6}} \prod_{n=1}^\infty (1-q^{n-\frac{1}{2}})^8
 = - \left(\frac{\vartheta_4(\tau)}{\eta(\tau)}\right)^4 \ , \cr
 &Z_\mathrm{R}^{+} = \Tr \left[ e^{2\pi i \tau H_\mathrm{R}}\right] = 16 q^{\frac{1}{3}} \prod_{n=1}^\infty (1+q^{n})^8
 = \left(\frac{\vartheta_2(\tau)}{\eta(\tau)}\right)^4 \ , \cr
 &Z_\mathrm{R}^{-} = \Tr \left[ (-1)^F e^{2\pi i \tau H_\mathrm{R}}\right] =0 q^{\frac{1}{3}} \prod_{n=1}^\infty (1-q^{n})^8
 = 0 \left(\frac{\vartheta_1(\tau)}{\eta(\tau)}\right)^4 = 0 \ ,
\end{align*}
where $q = e^{2\pi i \tau}$, and modular functions are summarized in \S\ref{sec:modular}. Remarkably, we obtain the Jacobi theta functions $\vartheta_i(\tau)$ ($i=1,\ldots,4$) depending on boundary conditions. Moreover, the $\SL(2,\bZ)$ action on the boundary condition over a torus easily tells us their modular properties as in Figure \ref{fig:modular-theta}.
Note that the minus sign in $Z_\mathrm{NS}^{-}$ comes from the fact  \eqref{NS-vac} that the NS vacuum is fermionic.

Therefore, the torus partition function for fermions is given by
\begin{align*}
 Z_\mathrm{F}(\tau) = \frac{1}{\eta^4(\tau)} \left\{  (\vartheta_3(\tau))^4 - (\vartheta_4(\tau))^4 +e^{i\theta} (\vartheta_2(\tau))^4 \right\}
\end{align*}
In order for $ Z_\mathrm{F}(\tau)$ to be modular invariant, the relative phase factor is uniquely determined as $e^{i\theta}=-1$ so that
\begin{align*}
 Z_\mathrm{F}(\tau) = \frac{1}{\eta^4(\tau)} \left\{  (\vartheta_3(\tau))^4 - (\vartheta_4(\tau))^4 -(\vartheta_2(\tau))^4 \right\} = 0 \ .
\end{align*}
In fact, the partition function is zero due to supersymmetry and this is called the Jacobi-Riemann identity \eqref{Jacobi-Riemann}. The partition function is zero so that it is trivially modular invariant.

The combination derived above is indeed expressed by using the GSO projection operators \eqref{GSO-NS} \eqref{GSO-R}:
\begin{align*}
 Z_\mathrm{F}(\tau) =  &2 \Tr \left[ \frac{1+(-1)^F}{2} e^{2\pi i \tau H_\mathrm{NS}}\right]
 -2 \Tr \left[ \frac{1\pm(-1)^F}{2} e^{2\pi i \tau H_\mathrm{R}}\right]  \nonumber\cr
 =  &2 \Tr \left[ P^\mathrm{NS}_\mathrm{GSO} e^{2\pi i \tau H_\mathrm{NS}}\right]
 -2 \Tr \left[ P^{\mathrm{R},\pm}_\mathrm{GSO} e^{2\pi i \tau H_\mathrm{R}}\right] \ .
\end{align*}
and the minus sign in the R sector indeed comes from spacetime spin-statistics.
Note that a choice of the GSO projections $\pm$ in the R sector does not affect the result due to the fermion zero modes. In other words, the GSO projection is compatible with the modular-invariance of the torus partition function.

\subsection{Modular functions}\label{sec:modular}
It is quite amusing to see that the modular functions such as the Dedekind eta function and Jacobi theta functions defined in the 19th century naturally arise in string theory.
Here we summarize the definition and the basic properties of the modular functions describing torus partition functions.
The reader may also refer to \cite[\S 7.2]{Polchinski} and \cite[\S 4.2]{Blumenhagen:2009zz}.

The infinite product forms of them are
\begin{align}\label{eta-theta}
 &\eta(\tau) = q^{\frac{1}{24}} \prod_{n=1}^\infty (1-q^n) \ , \cr
 &\vartheta_2(\tau) = 2 q^{\frac{1}{8}} \prod_{n=1}^\infty (1-q^n) (1+q^n)^2 \ , \cr
 &\vartheta_3(\tau) = \prod_{n=1}^\infty (1-q^n) (1+q^{n-\frac{1}{2}})^2 \ , \cr
 &\vartheta_4(\tau) = \prod_{n=1}^\infty (1-q^n) (1-q^{n-\frac{1}{2}})^2 \ ,
\end{align}
where $q=e^{2\pi i \tau}$.
The first one is called the \textbf{Dedekind eta function}, and
the others are the \textbf{Jacobi theta functions}.

Their $T$- and $S$-transformations are read off
\begin{align}\label{S-T}
 &\eta(\tau+1) = e^{i\pi/12} \eta(\tau) \ ,   &\eta(-1/\tau) = \sqrt{-i\tau} \eta(\tau) \ ,\cr
 &\vartheta_2(\tau+1) = e^{i\pi/4} \vartheta_2(\tau) \ ,  &\vartheta_2(-1/\tau) = \sqrt{-i\tau} \vartheta_4(\tau) \cr
 &\vartheta_3(\tau+1) = \vartheta_4(\tau) \ , &\vartheta_3(-1/\tau) = \sqrt{-i\tau} \vartheta_3(\tau)  \cr
 &\vartheta_4(\tau+1) = \vartheta_3(\tau) \  , &  \vartheta_4(-1/\tau) = \sqrt{-i\tau} \vartheta_2(\tau) \ .
\end{align}

There are two important identities. One is
called the \textbf{Jacobi-Riemann identity}:
\begin{equation}\label{Jacobi-Riemann}
 (\vartheta_3(\tau))^4 = (\vartheta_2(\tau))^4 +(\vartheta_4 (\tau))^4 \ .
\end{equation}
The other is \textbf{Jacobi identity}:
\begin{equation}
 \vartheta_2(\tau) \vartheta_3(\tau) \vartheta_4(\tau) = 2 \eta^3(\tau) \ ,
\end{equation}
from which the modular property of the Dedekind eta function follows.

\section{Type II superstring theories}\label{sec:TypeII}
We have seen that the triality $\bf 8_v$, $\bf 8_s$, $\bf 8_c$ of the eight-dimensional irreducible representations of $\SO(8)$ appear as the massless spectrum of the RNS superstring theory after the GSO projections. One way to explain the critical dimension $D=10$ of superstring theory is that the little group $\SO(8)$ of the Lorentz group $\SO(1,9)$ of the spacetime has these special eight-dimensional irreducible representations. As explained, the GSO projections \eqref{GSO-R} in the R sector pick one of the irreducible spinor representations $\bf 8_s$, $\bf 8_c$ of $\SO(8)$. In this section, we will obtain superstring theories of two different types, depending on the sign in the GSO projection operators. The analysis of massless fields in Type II theories predicts extended objects, called \textbf{D-branes} \cite{Polchinski:1995mt}.

\subsection{Type II superstrings}\label{sec:Type2}
Like bosonic closed string theory, massless spectrum even in closed superstring can be determined by taking tensor products of massless states in the left- and right-moving sector. However, we now have a choice between  $\bf 8_s$ and $\bf 8_c$ in the R sector. Since the GSO-projected NS and R sectors both have integer levels, level matching is the same condition $N=\bar N$ as in the bosonic string, and independent projections on the two chiralities reproduce the torus partition function $|Z_{\rm F}|^2$ of \S\ref{sec:superstring}. Hence, there are two inequivalent ways to construct superstring theory: the same (IIB) or opposite (IIA) choices on the right- and left-moving spectrum.
These lead to the massless sectors
\begin{align}\label{IIA-IIB}
{\rm Type ~ IIA\colon} & ({\bf 8_v}\oplus{\bf 8_s}) \otimes
   ({\bf 8_v}\oplus{\bf 8_{c}}) \nonumber\\
{\rm Type ~ IIB\colon} & ({\bf 8_v}\oplus{\bf 8_s}) \otimes
   ({\bf 8_v}\oplus{\bf 8_{s}})
\end{align}
of $\SO(8)$. Although one can also choose
\begin{align}
{\rm Type ~ IIA'\colon} & ({\bf 8_v}\oplus{\bf 8_c}) \otimes
   ({\bf 8_v}\oplus{\bf 8_{s}}) \nonumber\\
{\rm Type ~ IIB'\colon} & ({\bf 8_v}\oplus{\bf 8_c}) \otimes
   ({\bf 8_v}\oplus{\bf 8_{c}})
\end{align}
they are equivalent after the spacetime parity redefinition. By the construction, Type IIB theory is chiral whereas IIA theory is not chiral. Also, the tensor products of the NS-NS and R-R sectors provide massless bosonic fields whereas those of the NS-R and R-NS sectors give massless fermionic fields.

\begin{table}[ht] \centering
\begin{tabular}{ |c|c|c| }
 \hline
 IIA & ${\bf 8_v}$ & ${\bf 8_c}$ \\ \hline
 ${\bf 8_v}$ & ${\bf 1} \oplus {\bf 28}  \oplus {\bf 35}$ & ${\bf 8_s}\oplus{\bf 56_c}$ \\
  & $\phi \ \ B_{\mu\nu} \ \ G_{\mu\nu}$ & $\lambda^- \ \ \psi^+_m$ \\ \hline
  ${\bf 8_s}$ & ${\bf 8_c}\oplus{\bf 56_s}$ & ${\bf 8_v} \oplus
{\bf 56_t}$ \\
    & $\lambda^+ \ \ \psi^-_m$ & $C_n \ \ C_{nmp}$ \\
 \hline
\end{tabular}
\hspace{2cm}
\begin{tabular}{ |c|c|c| }
 \hline
 IIB & ${\bf 8_v}$ & ${\bf 8_s}$ \\ \hline
 ${\bf 8_v}$ & ${\bf 1} \oplus {\bf 28}  \oplus {\bf 35}$ & ${\bf 8_c}\oplus{\bf 56_s} $ \\
  & $\phi \ \ B_{\mu\nu} \ \ G_{\mu\nu}$ & $\lambda^+ \ \ \psi^-_m$  \\ \hline
  ${\bf 8_s}$ & ${\bf 8_c}\oplus{\bf 56_s} $ & ${\bf 1} \oplus {\bf 28}  \oplus {\bf 35}_+$ \\
    & $\lambda^+ \ \ \psi^-_m$  & $C \ \ C_{mn}\ \ C_{mnpq}$ \\
 \hline
\end{tabular}
\caption{Massless fields in Type IIA and IIB theory.}\label{tab:masslessII}
\end{table}

Let us first look at the massless bosonic sector.    In the NS-NS sector, this is the same as bosonic string theory
\be
{\bf 8_v} \otimes {\bf 8_v} = \phi \oplus B_{\mu\nu} \oplus G_{\mu\nu}
={\bf 1} \oplus {\bf 28}  \oplus {\bf 35} .
\ee
Thus, the new ingredients come from the R-R sector, and the IIA and IIB spectra are respectively
\begin{align}
{\bf 8_s} \otimes {\bf 8_c} &= [1] \oplus [3] = {\bf 8_v} \oplus
{\bf 56_t} \nonumber\\
{\bf 8_s} \otimes {\bf 8_s} &= [0] \oplus [2] \oplus [4]_+
= {\bf 1} \oplus {\bf 28}  \oplus {\bf 35}_+ ~.
\end{align}
Here $[n]$ denotes the $n$-th antisymmetric representation of
$\SO(8)$, and we associate R-R $n$-form $C_{(n)}$ to it. 
Also, here  $[4]_+$ means its R-R field strength $G_{(5)}=dC_{(4)}$ is self-dual
\be\label{SD-5form} \ast G_{(5)}=G_{(5)}~.\ee
 Note that the representations
$[n]$ and $[8-n]$ are related by the Hodge dual so that they are related by contraction with the
8-dimensional $\epsilon$-tensor. As we will see next, these R-R fields are associated to D-branes in Type II theories, and $[n]$ and $[8-n]$ are related by the electro-magnetic duality.

Let us next look at the massless fermionic sector.
The tensor products of the NS-R and R-NS sectors are given by
\begin{align}
{\bf 8_v} \otimes {\bf 8_c} &=
{\bf 8_s}\oplus{\bf 56_c} \nonumber\\
{\bf 8_v} \otimes {\bf 8_s} &=
{\bf 8_c}\oplus{\bf 56_s}~.
\end{align}
The $\bf 56_{s,c}$ correspond to gravitinos $\psi^\pm_{m,\a}$ that are superpartners of gravitons, and they have vector and one spinor indices by construction. As we will see in \S\ref{sec:supergravity}, they will give spacetime supersymmetry.  The $\bf 8_{s,c}$ are dilatino $\lambda^\pm_{\a}$ which are superpartners of the dilaton field.

\subsection{Introduction to D-branes}\label{sec:intro-Dbrane}

\begin{figure}[htb]
\centerline{\includegraphics[width=350pt]{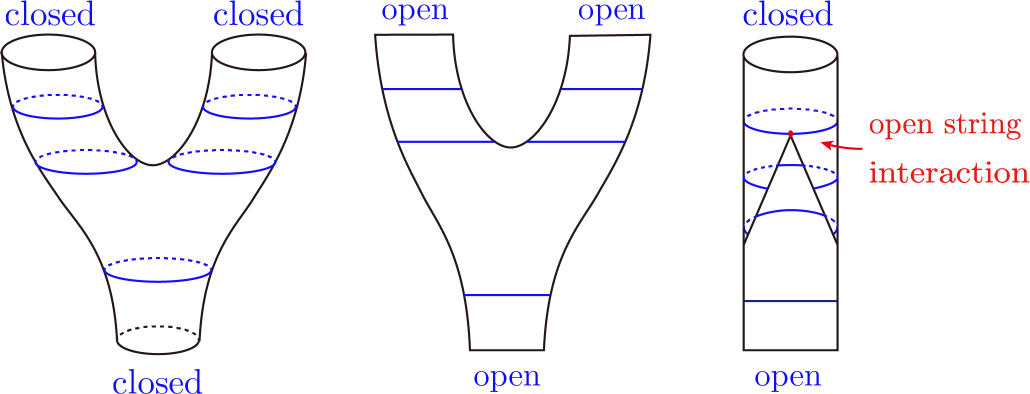}}
\caption{Open string interaction induces closed string}
\label{OpenClosed}
\end{figure}

The massless spectrum from closed strings analyzed above does not incorporate gauge fields because gauge fields arise from an open string.
To incorporate open strings in Type II theories, we need to introduce D-branes.
Note that closed superstring theories are consistent in themselves as we see in Heterotic string theory \S\ref{sec:Heterotic}.
However, only open string theory is inconsistent,
and it requires closed strings as well (see Figure \ref{OpenClosed}).

\subsubsection*{Boundary conditions \& D-brane}

\begin{figure}[ht]\centering
\raisebox{1cm}{\includegraphics[width=6cm]{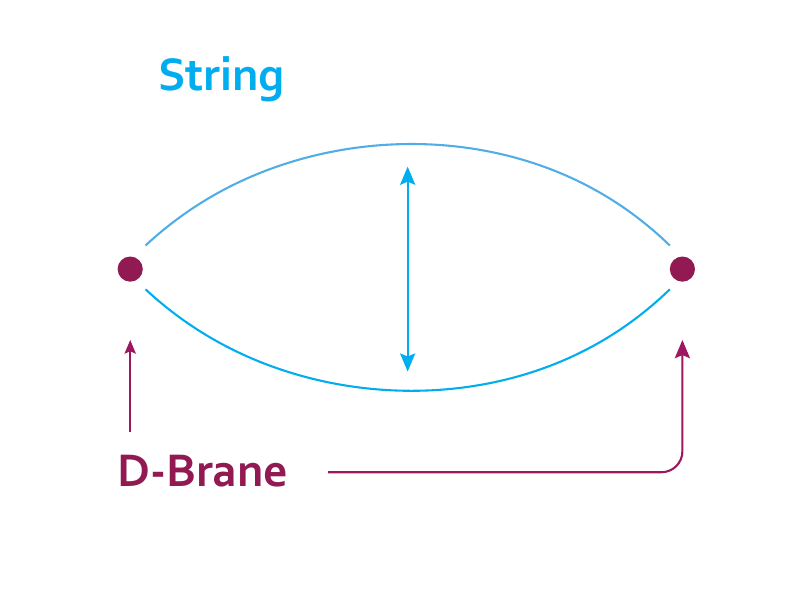}}\hspace{1cm} \includegraphics[width=8cm]{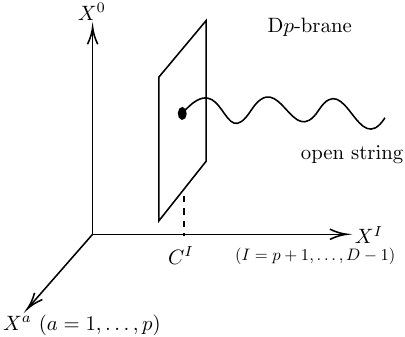}
\caption{D-brane must exist at the ends of the open string so that momentum can escape from the string.}
\label{BoundaryObj}
\end{figure}

As seen in \eqref{open-variation}, we have imposed the two types of boundary conditions for open strings.
In fact, the Dirichlet boundary condition (Figure \ref{BoundaryObj}) does not conserve the momenta of open strings (exercise).
It implies that there must be an object into which the momentum goes.
We call this object a D-brane where the D stands for Dirichlet and
brane comes from the membrane. Indeed, we can impose the Neumann boundary condition to some coordinates and the Dirichlet to the other coordinates:
\begin{align*}
 &\textrm{Neumann condition on } X^a \quad (a=0,1,\ldots,p)  \\
 &\textrm{Dirichlet condition on } X^I \quad (I=p+1,\ldots,D-1)
\end{align*}
The corresponding D-brane is called a \textbf{D$p$-brane}, which extends to a $p$-dimensional subspace or a $(p+1)$-dimensional spacetime.
Now we can visualize a configuration of D-brane and string as in Figure \ref{BoundaryObj}.
A D-brane is a dynamical object as it should receive the momentum so that it has action and interactions with string. (See \S\ref{sec:D-brane}.)
On the other hand, in order to give the Dirichlet boundary condition $X^I = c^I$,
a D-brane must be infinitely heavier than a string.

\subsubsection*{Chan-Paton factor}

We can consider not only a single D-brane but also a stack of D-branes,
and a string now has a choice of D-branes on which it ends.
Let us label this option $i$ ($i=1,\ldots,n$), which is called \textbf{Chan-Paton factor}.
As an open string has two endpoints, Chan-Paton degree of freedom is specified by
\begin{align}\label{CP-factors}
 |N;k \rangle  \quad  \to \quad |N;k;ij \rangle \ .
\end{align}
Now we have $n^2$ massless vector states in both bosonic string and superstring theory.
As usual, we use $n \times n$ Hermitian matrices $T^a$ normalized to
\begin{align*}
 \Tr (T^a T^b) = \delta^{ab} \ ,
\end{align*}
which constitute a complete set for the data of the open string endpoints:
\begin{align}\label{CP-factor2}
 |N=1;k;a \rangle = T^a_{ij} |N=1;k;ij \rangle \ .
\end{align}
These states correspond to a $\U(n)$ gauge field. (You can check it by a three-point amplitude.)

\begin{figure}[htb]
\centerline{\includegraphics[width=13cm]{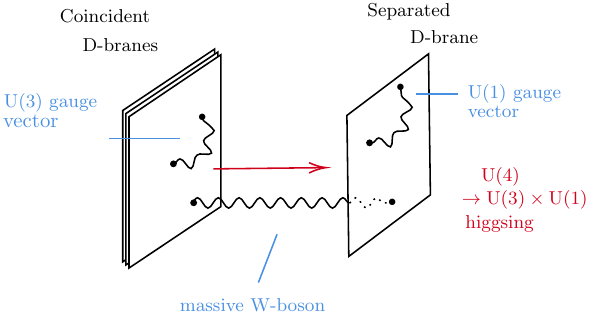}}
\caption{Many D-branes and Chan-Paton factors. $\U(n)$ gauge group and Higgs mechanism.}
\label{ManyD}
\end{figure}

Note that in order to realize $\U(n)$ gauge group, the D-branes must coincide at a point.
Otherwise, the gauge group is broken by Higgs mechanism (see Figure \ref{ManyD}).

We will see in \S\ref{sec:orientifold} that other types of gauge groups ($\SO$ or $\Sp$) can be realized in string theory with orientifold planes.

\subsubsection*{D-branes in IIA/IIB superstring theory}

D-branes arise as boundary conditions of open strings, and they carry gauge fields. Since D-branes are intrinsic to string theory, it reveals many interesting facets so that we will further investigate their properties.
In \S\ref{sec:Type2}, we saw that Type II superstring theories are endowed with R-R fields,
which are anti-symmetric tensors analogous to gauge fields.
Like an electron/monopole is coupled to the gauge fields,
D-branes are coupled to the R-R fields in Type II theories.

To use the analogy of electromagnetism, let us quickly review an electron/monopole interacting with the electromagnetic fields in the $(3+1)$-dim spacetime.
An electron is expressed by a source
\[J^\mu_e = (\rho, \bfj) = (q_e \delta^3(\bfr-\bfr(t)), \partial_t \bfr(t) \rho )~,\]
and its coupling to the gauge field $A$ is given by
\begin{align*}
 S_J = \int A_\mu J^\mu d^4x = q_e \int_\gamma A_\mu dx^\mu = q_e \int_\gamma A \ ,
\end{align*}
where $q_e$ is an electric charge, and $\gamma$ is the world-line of the electron. Since the Maxwell equation is \be\label{Maxwell}\ast d \ast F_{(2)}= J_e~,\ee we can obtain the electric charge by
\[
q_e=\int_{S^{2}} \ast F_{(2)}~.
\]
On the other hand, the electromagnetic duality
\be\label{Maxwell2}
 d  F_{(2)}= J_m\ee tells us that
a magnetic monopole is a source $J_m= q_m \delta^3(r)$ of a magnetic flux.
Therefore, the magnetic charge can be measured as
\begin{align*}
 q_m= \int_{S^2} F_{(2)} = \int_{B} d F_{(2)} \ ,
\end{align*}
where $S^2$ is a surface around the monopole.
It is well-known that the charges of the electron and monopole obey the Dirac quantization condition (for instance, see \cite[\S13.3]{Polchinski}):
\be
q_e q_m\in 2\pi \bZ~.
\ee

\begin{figure}[htb]
\centerline{\includegraphics[width=15cm]{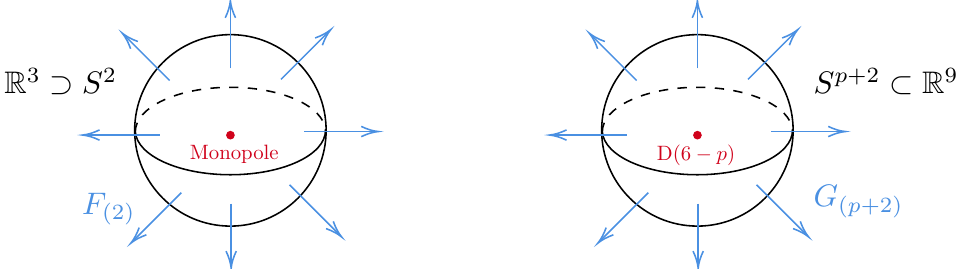}}
\caption{Higher dimensional analog of monopole: magnetic flux of D$(6-p)$-brane.}
\label{mono}
\end{figure}

In a similar fashion, a D$p$-brane ($p\le 3$) supported on a world-membrane $M_{p+1}$  is electrically coupled to the R-R $(p+1)$-form $C_{(p+1)}$ as
\begin{equation}\label{RR-coupling}
 S_{p} = \mu_{p}\int_{M_{p+1}} C_{(p+1)} \ .
\end{equation}
where $\mu_{p}$ is the charge of the D$p$-brane.
An exterior derivative of the R-R potential
gives its R-R field strength $G_{(p+2)} = d C_{(p+1)}$, and its electromagnetic dual is given by its Hodge dual
\[
\widetilde{G}_{(8-p)}=\ast G_{(p+2)}
\]
From Gauss's law, the charge is given by the flux of $\widetilde{G}_{(8-p)}$ over a sphere $S^{8-p}$ around the D$p$-brane
\[
\mu_{p}=2\kappa_{10}^2\int_{S^{8-p}} \widetilde{G}_{(8-p)}~,
\]
where $2\kappa_{10}^{2}=(2 \pi)^{7} \alpha^{\prime 4}$.
The magnetic dual of the D$p$-brane is a D$(6-p)$-brane, and its magnetic charge is accordingly given by
\[
\mu_{6-p}=2\kappa_{10}^2\int_{S^{p+2}} G_{(p+2)}~.
\]
They must obey the Dirac quantization condition
\be\label{Dirac-quantization}
2\kappa_{10}^2\mu_{p}\mu_{6-p}\in 2\pi \bZ~.
\ee
Writing the flux $\wt G_{(8-p)}=d\wt C_{(7-p)}$, the D$(6-p)$-brane is magnetically coupled to  $\wt C_{(7-p)}$.
Now, reading off R-R fields from Table \ref{tab:masslessII}, we see that there are D$p$-branes for even $p$ in Type IIA and for odd $p$ in Type IIB theory.

\begin{figure}[htb]
\centerline{\includegraphics[width=10cm]{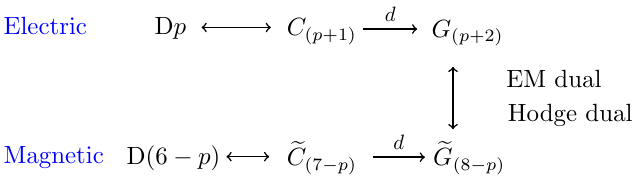}}
\caption{Electro-magnetic duality and D-brane.}
\label{EleMagD}
\end{figure}


Even from the NS-NS sector, we can predict the presence of an extended object in string theory.
A string or a fundamental string, denoted as F$1$, is electrically coupled to the $B$-field.
On the other hand, there is an object magnetically coupled to the $B$-field,
which is called \textbf{NS$5$-brane}. 

In conclusion, we summarize extended objects in Type II superstring theory in Table~\ref{table:branetypeII}.
\begin{table}[htbp]
 \centering{
\begin{tabular}{l|ccc}
  IIA & $B_{(2)}$ & $C_{(1)}$ & $C_{(3)}$ \\
\hline
  Electric & F$1$ & D$0$ & D$2$ \\
  Magnetic & NS$5$ & D$6$ & D$4$ \\
\end{tabular} \hspace{12pt}
\begin{tabular}{l|cccc}
  IIB & $B_{(2)}$ & $C_{(0)}$ & $C_{(2)}$ & $C_{(4)}^+$ \\
\hline
  Electric & F$1$ & D$(-1)$ & D$1$ & D$3$ \\
  Magnetic & NS$5$ & D$7$ & D$5$ & D$3$ \\
\end{tabular}
}  \caption{Extended objects in Type II superstring theory}\label{table:branetypeII}
\end{table}
\vspace{2pt}

Note that a D$(-1)$-brane is an object localized in time (called an instanton). There are D$8$-branes in Type IIA theory, which are non-dynamical
so that there is no corresponding R-R anti-symmetric tensor. Moreover, these branes can be understood as a decay of the space-filling D9-brane and anti-D9-brane pair D9-$\overline{\textrm D9}$ \cite{Sen:1998tt}, which admits a beautiful mathematical interpretation \cite{Witten:1998cd} by K-theory.

\section{T-duality}\label{sec:Tdual}

We have seen that the critical dimensions $D$ of bosonic and supersymmetric string theory are $D=26$ and $D=10$, respectively. To obtain an effective theory in lower dimensions, we can make use of \textbf{Kaluza-Klein compactifications} where  the
true spacetime takes the form of a direct product $M_{d} \times K_{D-d}$, where
$M_d$ is the $d$-dimensional Minkowski spacetime, and $K_{D-d}$ is a very tiny compact manifold. As we will see, an effective theory in $M_d$ still sees interesting ``stringy'' effects in this Kaluza-Klein scheme.

First, let us focus on the simplest form of compactification, known as \textit{toroidal compactifications}, denoted by $K=T^{D-d}$. A torus is essentially a product of circles ($S^1$) and is characterized by its flat geometry. Consequently, the nonlinear sigma model for a torus can be described by a free 2d CFT. Intriguingly, such simple compactifications introduce the concept of \textbf{T-duality}. Furthermore, \textbf{Heterotic string theories}, which we will explore in \S\ref{sec:Heterotic}, have been developed using these toroidal compactifications. To gain an understanding of the fundamental properties, we will first examine the toroidal compactifications in bosonic string theory. The toroidal compactifications of superstring theories are equally fascinating and merit detailed exploration in relation to string dualities later on.

The concept of D-branes was introduced as boundary conditions of open strings. T-duality gets particularly rich when we include D-branes so that we will study their properties more in detail.
The explanation of T-duality in Polchinski's lecture notes \cite{Polchinski:1996na} is so elegant that we just simply follow it in this section.

\subsection{\texorpdfstring{$S^1$}{S1} compactification in closed bosonic string}

To begin with, let us first study the simplest case of the spacetime $\bR^{1,24}\times S^1$ where we compactify 25-th direction on a circle $S^1$ of radius $R$.  For closed strings, we have the
familiar mode expansion \eqref{mode-exp}.
Now let us take a close look at the zero modes which can be written as
\[
X^{\mu}(z, \bar{z})=x^{\mu}-i \sqrt{\frac{\alpha^{\prime}}{2}}\left(\alpha_{0}^{\mu}+\bar{\alpha}_{0}^{\mu}\right) t+\sqrt{\frac{\alpha^{\prime}}{2}}\left(\bar{\alpha}_{0}^{\mu}-\alpha_{0}^{\mu}\right) \sigma+\text { oscillators. }
\]
where the spacetime momentum of the string is
\be\nonumber
p^\mu =
\frac1{\sqrt{2\a'}}(\alpha^\mu_0 + {\ol\alpha}^\mu_0)~.
\ee

Under $\sigma \to \sigma+2\pi$,
the oscillator term is periodic and $X^\mu(z,\bar{z})$
changes by $2\pi\sqrt{(\a'/2)}(\ol\alpha^\mu_0-{\alpha}^\mu_0).$
For a non-compact spatial direction $\bR^{1,24}$, $X^\mu$ is single-valued $X^\mu(t,\s)= X^\mu(t,\s+2\pi)$,
which requires
\be\nonumber
\alpha^\mu_0={\ol\alpha}^\mu_0=\sqrt{\frac{\a'}{2}}p^\mu~,\qquad \mu=0,1,\ldots,24~.
\ee

\begin{figure}[ht]\centering
\includegraphics{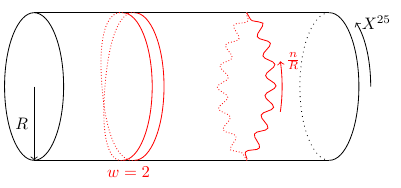}
\end{figure}

On the other hand, since the 25-th direction is put on the circle $S^1$ of radius $R$, it has a period
 $X^{25}\sim X^{25}+2\pi R$. Hence, the momentum $p^{25}$ can take the values $n/R$ for $n\in\bZ$ where $n$ is called \textbf{Kaluza-Klein momentum}.
Also, under $\sigma\sim\sigma+2\pi$, $X^{25}(z,\bar{z})$
can change by $2\pi wR$ where $w$  is called
 the \textbf{winding number}.  Thus, we have
\[
\alpha_{0}^{25}+\bar{\alpha}_{0}^{25}=\frac{2 n}{R} \sqrt{\frac{\alpha^{\prime}}{2}}, \qquad \bar{\alpha}_{0}^{25}-\alpha_{0}^{25}=\sqrt{\frac{2}{\alpha^{\prime}}} w R
\]
implying
\begin{align}\label{25a}
\alpha^{25}_0 = \left(\frac{n}{R}-\frac{wR}{\a'}\right)
\sqrt{\frac{\a'}{2}} ~, \qquad
{\ol\alpha}^{25}_0 =
\left(\frac{n}{R}+\frac{wR}{\a'}\right)\sqrt{\frac{\a'}{2}}~.
\end{align}

Now, let us study their mass spectrum. The mass formula for the string with one dimension compactified on a circle can be interpreted from a 25-dimensional viewpoint in which one regards each of the Kaluza-Klein momenta, which are given by $n$, as distinct particles. Thus, the mass formula is given by
\bea 
M^2 = -\sum_{\mu=0}^{24}p^\mu p_\mu  =&\frac{2}{\alpha^{\prime}}(\alpha_0^{25})^2+\frac{4}{\alpha^{\prime}}(N-1)\cr 
=&\frac{2}{\alpha^{\prime}}(\bar{\alpha}_0^{25})^2+\frac{4}{\alpha^{\prime}}(\bar{N}-1) .
\eea
where $\mu$ runs only over the non-compact dimensions. Here $N$ and $\ol N$ are the right and left number operators \eqref{numbering}. Using \eqref{25a}, the difference between the two expressions above yields
\be\label{level-matching}
N-\ol N=nw~,
\ee
so that the level-matching condition is modified due to the $S^1$ compactification. In a similar fashion, the mass can be expressed in terms of the Kaluza-Klein momentum and the winding number
\be\label{mass}
M^2=\frac{n^2}{R^2}+\frac{w^2R^2}{\a'^2}+\frac2{\a'}(N+\ol N-2)
\ee

The mass spectra \eqref{mass} of the theories at radius $R$ and $\a'/ R$ are identical
when the winding and Kaluza-Klein modes are interchanged
$n \leftrightarrow w$. This symmetry of the bosonic string theory is called \textbf{T-duality}.
From the viewpoint of strings, a circle of radius $R$ is equivalent to that of radius $\a'/ R$, which suggests a minimal length $\sqrt{\a'}$, one of the ways in which strings soften short-distance physics. This shows that strings see geometry in an unprecedented and intriguing way, and this is one of the remarkable phenomena called ``stringy geometry''.

From Equation \eqref{25a}, it is evident that this interchange results in:
\be
\alpha_0^{25} \rightarrow -\alpha_0^{25}, \quad \bar{\alpha}_0^{25} \rightarrow \bar{\alpha}_0^{25}
\label{tzemo}.
\ee
In fact, under the T-duality transformation, not only the zero mode but also the entire right-moving segment of the compact coordinate reverses its sign:
\be
X^{\prime25}(z,\bar{z}) = -X^{25}(z) + \overline X^{25}(\bar{z})\ . \label{onesidep}
\ee
Remarkably, the energy-momentum tensor, the operator product expansions (OPEs), and all correlation functions remain invariant under this transformation. In other words, T-duality, which connects the theories at radii $R$ and $\a'/R$, represents an exact symmetry in perturbative closed string theory.

Due to T-duality, a theory with a compactification radius of $R$ is equivalent to one with a radius of $\a'/R$. This equivalence suggests the existence of a "minimal radius," \(R = \sqrt{\a'}\), known as the self-dual radius in string theory. At the self-dual radius, the duality \(R \to \a'/R\) maps \(R\) back to itself, potentially leading to intriguing phenomena. We will explore the physics at the self-dual radius in the following.

\subsubsection*{Self-dual radius: $R = \protect \sqrt {\a'}$}

As we know, the massless spectra of bosonic string theory include gravitons. Hence, let us see the effect of the $S^1$ compactification on the gravitons. In the Kaluza-Klein mechanism  $M_{25}\times S^1$, the metric is decomposed into compact and non-compact spacetime direction
\bea\label{KK-metric}
ds^2 = G_{MN} dx^M dx^N =&\wt G_{\mu\nu} dx^\mu dx^\nu + G_{25,25}(dx^{25}+ A_\mu dx^\mu)^2 ~.\cr 
\wt G_{\mu\nu}=&G_{\mu\nu}-G_{25,25}A_\mu A_\nu
\eea
where the fields $G_{\mu\nu}$, $G_{25,25}$, and $A_\mu$ are allowed to depend only on the non-compact coordinates $x^\mu$ ($\mu=0,1,\ldots,24$). Under a coordinate transformation
\[x'^{25} = x^{25} + \lambda (x^\mu)  \]
the part $G_{\mu,25}=G_{25,\mu}$ of the metric transforms as
\[A'_\mu = A_\mu - \partial_\mu \lambda~ . \]
Thus, it behaves as $\U(1)$ gauge field, and gauge transformations arise as part of the higher-dimensional coordinate transformation. On the other hand, the part $G_{25,25}$ of the metric behaves as a scalar field. Indeed, writing $G_{25,25}=e^{2\s}$, the Ricci scalar for the metric  \eqref{KK-metric} can be written as
\[
R_{26} = R_{25} - 2e^{-\s}\nabla^2 e^\s - \frac14e^{2\s}F_{\m\n}F^{\m\n}~ .
\]
Actually, it is straightforward to see the corresponding vertex operators at generic radius $R$:
\begin{align}
\partial X^\mu \bar \partial X^\nu e^{i k\cdot X}	&\quad \longleftrightarrow \quad G_{\mu\nu}, B_{\mu\nu}, \phi \cr
\partial X^\mu \bar{\partial} X^{25} e^{i k\cdot X}, \ \partial X^{25} \bar{\partial} X^\mu e^{i k\cdot X}& \quad \longleftrightarrow \quad A^\mu, B_{\mu, 25}\cr
\partial X^{25} \bar{\partial} X^{25} e^{i k\cdot X}&\quad \longleftrightarrow \quad e^\sigma
\end{align}
where $\nu = 0, \ldots, 24$  runs the coordinate indices for $M_{25}$.  In fact, the middle line indicates that the theory has $\U(1)_\ell \times \U(1)_r$ gauge symmetry at generic radius $R$.

However, at the self-dual radius $R=\sqrt{\a'}$, the mass formula \eqref{mass} becomes
\[
M^2=\frac{1}{\a'}(n^2+w^2+2(N+\ol N-2))~,
\]
so that the massless spectra actually get enlarged. In addition to the generic solution $n=w=0$, $N=\ol N =1$, there are also massless modes as in the following table:
$$
\begin{tabular}{ccccc}
&$n$&$w$&$\ol  N$&$N$\\\hline
A&$\pm1$&$\pm1$&$0$&$1$\\
B&$\pm1$&$\mp1$&$1$&$0$\\
C&$\pm2$&$0$&$0$&$0$\\
D&$0$&$\pm2$&$0$&$0$
\end{tabular}$$

\noindent Hence, the states corresponding to A and B contain four new gauge bosons with vertex operators
\[
\bar{\partial} X^\mu e^{\pm 2iX^{25}(z)/\sqrt{\a'}} e^{i k\cdot X} ~,  \qquad\qquad  {\partial} X^\mu e^{\pm 2i\overline X^{25}(\bar z)/\sqrt{\a'}} e^{i k\cdot X}~.
\]
Thus, $\U(1)_\ell\times \U(1)_r$ gets enhanced to $\SU(2)_\ell\times \SU(2)_r$. Consequently, the T-duality can be understood as the Weyl group $\bZ_2$ of $\SU(2)_r$.
Indeed, C and D also give rise to new gauge bosons (exercise).
It is expected that the new gauge bosons must combine with the
old into a non-Abelian theory. In fact, if one can define the current
\[
j^\pm(z)=j^1(z)\pm i \, j^2(z):=e^{\pm 2iX^{25}(z)/\sqrt{\a'}} \qquad j^3(z):=i\, \partial X^{25}(z)/\sqrt{\a'}~,
\]
they satisfy the OPEs (exercise)
\[
j^a(z) j^b (0) \sim \frac { k\delta^{ab} } {2z^2}
		+ \frac {i {\epsilon^{abc}} j^c(0)} {z}~.
\]
with $k=1$. Here $\epsilon^{abc}$ is the structure constant
of $\SU(2)$.  This is precisely the definition of $\SU(2)$ affine Lie
algebra with level $k=1$.  The same story is repeated for the left movers.  Hence we see that we have
an enhancement of gauge symmetry from $\U(1)_\ell \times \U(1)_r$
to $\SU(2)_\ell \times \SU(2)_r$ at $R = \sqrt{\a'}$.

In fact, when the theory moves away from the self-dual radius $R=\sqrt{\a'}$, the $\SU(2)_\ell \times \SU(2)_r$  gauge symmetry is Higgsed.
The world-sheet action is deformed by turning on the marginal operator
\[
V_{a\ol a}:=j_a \bar j_{\ol a} e^{ik\cdot X}~,
\]
which is equivalent to giving the VEV of the $(3,3)$-component of the Higgs field. As a result, when the theory is away from the self-dual
radius, the $\SU(2)_\ell \times \SU(2)_r$ gauge symmetry is spontaneously broken down to a $\U(1)_\ell \times \U(1)_r$.

\subsection{T-duality of open strings}\label{sec:Tdual-open}

Now let us consider T-duality on the open string spectrum in the $S^1$ compactification. At the end of \S\ref{sec:quantization}, we briefly studied open string spectra. There, we learned that Neumann boundary condition on the $X^\mu$-direction is achieved by imposing $ \alpha_n^\mu = \overline{\alpha}_n^\mu$ while Dirichlet boundary condition is by $\alpha^\mu_n = -\overline{\alpha}_n^\mu$. Since T-duality along the 25-th direction transforms the modes as in \eqref{tzemo}, it therefore exchanges Neumann to Dirichlet condition \cite{Dai:1989ua,Leigh:1989jq}:
\[
\left.\partial_{\sigma} X^{25}\right|_{\sigma=0, \pi}=\left.0 \quad \rightarrow \quad \partial_{t} X^{\prime 25}\right|_{\sigma=0, \pi}=0~.
\]
Suppose that an open string with Neumann boundary condition has KK momentum $n$ along the 25-th direction, and we perform T-duality on that circle. Then, a simple calculation
\begin{align}
X^{\prime 25}(\pi)-X^{\prime 25}(0) &=\int_{0}^{\pi} d \sigma \partial_{\sigma} X^{\prime 25}=i \int_{0}^{\pi} d \sigma \partial_{t} X^{25} \\
&=2 \pi \alpha^{\prime} p^{25}=\frac{2 \pi \alpha^{\prime} n}{R}=2 \pi n R^{\prime}
. \label{deltax}
\end{align}
tells us that the $X^{25}$ coordinate of the open string endpoints is fixed after T-duality. T-duality transforms the KK momentum $n$ to winding number $n$  as in the figure below.
This can be interpreted as follows. Open strings can freely move on a space-filling D25-brane. After T-duality, the space-filling D25-brane becomes a D24-brane with its $X^{25}$ coordinate fixed. As a result, all the endpoints of open strings are constrained to the fixed $X^{25}$-direction whereas they are free to move in the other 24 spatial dimensions.

More generally, the boundary condition of open strings can be imposed at any dimension. Since T-duality interchanges Neumann and Dirichlet boundary conditions, T-duality tangent to a D$p$-brane brings it to a D$(p-1)$-brane. On the other hand, T-duality orthogonal to a D$p$-brane turns it into a D$(p+1)$-brane.

\begin{figure}[ht]\centering
\includegraphics[width=13cm]{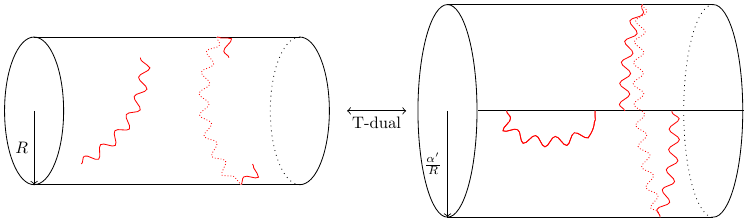}
\end{figure}

We can ask a question which position of the $X^{25}$ coordinate a D24-brane is located in the figure above. This question is related to Chan-Paton factors we encountered in \S\ref{sec:intro-Dbrane}. Suppose that there are $K$ space-filling D25-branes, which give rise to $\U(K)$ spacetime gauge fields. In the current setting, 
$X^{25}$ direction is compactified on the circle $S^1$ so that we can include a Wilson loop on $S^1$ as
\be 
A_{25}=\textrm{diag}\{\theta_1,\theta_2,\ldots,\theta_K\}/2\pi R
\ee 
The insertion of the Wilson loop breaks the gauge group as $\U(K) \to \U(1)^K$, and the broken gauge group is abelian so that we can write it as
\be\nonumber
A_{25}=-i \Lambda^{-1} \partial_{25} \Lambda, \quad \Lambda=\operatorname{diag}\{e^{i X^{25} \theta_{1} / 2 \pi R}, e^{i X^{25} \theta_{2} / 2 \pi R}, \ldots, e^{i X^{25} \theta_{K} / 2 \pi R}\}\ .
\ee
Then, under the translation $X^{25}\to X^{25}+2\pi R$, an open string state $|N=1;n;ij \rangle$ is shifted by a phase
\be
e^{i(\theta_j-\theta_i)}~,
\ee
which means its momentum is shifted by $(\theta_j-\theta_i)/2\pi$ from an integer $n$. Under T-duality, the momentum is transformed into the winding number. Therefore, $\theta_i$ can be understood as the $X^{25}$ coordinate of the $i$-th D24-brane after T-duality.
Namely, the open string state $|N=1;n;ij \rangle$ is mapped to an open string of length 
\be\nonumber
X^{\prime 25}(\pi)-X^{\prime 25}(0)=\left(2 \pi n+\theta_{j}-\theta_{i}\right) R^{\prime}~.
\ee
There are in general $K$ D24-branes at different positions as schematically depicted in the following figure.
\begin{figure}[ht]\centering
\includegraphics[width=7cm]{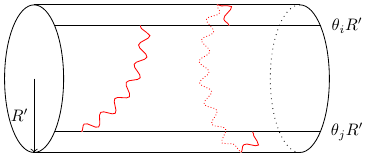}
\end{figure}

Then, the $(D-1)=24$-dimensional mass of this open string is
\begin{align}
M^{2} &=\left(p^{25}\right)^{2}+\frac{1}{\alpha^{\prime}}(N-1) \\
&=\left(\frac{\left[2 \pi n+\left(\theta_{j}-\theta_{i}\right)\right] R^{\prime}}{2\pi \alpha^{\prime}}\right)^{2}+\frac{1}{\alpha^{\prime}}(N-1)~ .\nonumber
\end{align}
Hence, massless gauge bosons arise only from open strings $|N=1;n=0;ii \rangle$ whose endpoints are on the same D-brane. 

Due to the string tension, open strings that stretch between different D-branes become massive (gauge field). Therefore, this can be understood as the Higgs mechanism in which $X^{25}$ expectation values (coordinates) of $K$ D24-branes take different values, and the gauge group is broken as $\U(K) \to \U(1)^K$. This situation is called \textbf{Coulomb phase}, where the Goldstone bosons are ``eaten'' by gauge fields $|N=1;n=0;ij \rangle$ which become massive.

Let us make an important remark. So far, we have treated D-branes just as rigid boundary conditions. However,  D-branes are dynamical so that they can fluctuate in shape and position. As we will see in \S\ref{sec:D-brane}, \S\ref{sec:BH}, \S\ref{sec:AdSCFT}, the gravitational and gauge dynamics on a stack of D-branes makes string theory very intriguing.

\subsection{T-duality of Type II superstrings}

Let us examine how T-duality operates on Type II superstrings compactified on \(M_{9} \times S^1\). We observe that T-duality performs a parity transformation on the right-moving sector of the worldsheet, transforming
$$
X^9(z) \quad \longleftrightarrow \quad-X^{\prime 9}(z)
$$
The superconformal invariance requires
$$
\psi^9 \leftrightarrow-\psi^{\prime 9} .
$$
This alteration reverses the chirality of the right-moving R-sector ground state by swapping the raising and lowering operators \(\psi^8 \pm i\psi^9\) (refer to \eqref{raising-lowering}). Essentially, T-duality acts as a spacetime parity operation on one side of the worldsheet, reversing the chiralities of the right- and left-moving ground states. Consequently, the Type IIA theory with a compactification radius \(R\) is T-dual to the Type IIB theory with radius \(\alpha'/R\).

Chiral spinors are defined through the relation:
\[
\Gamma_{11} \psi_{\pm} = \pm \psi_{\pm}
\]
with \(\bar{\psi}_{\pm} = \psi_{\pm}^\dagger \Gamma^0\). The vertex operators for the R-R field strengths \(G_{\mu_1 \cdots \mu_{p+2}}\) are represented as spinor bilinears:
\[
\mathrm{IIA}: \bar{\psi}_{-}^L \Gamma^{\mu_1 \cdots \mu_{p+2}} \psi_{+}^R, \quad \mathrm{IIB}: \bar{\psi}_{+}^L \Gamma^{\mu_1 \cdots \mu_{p+2}} \psi_{+}^R,
\]
where \(\psi^R\) and \(\psi^L\) are derived from the right and left movers, respectively. The antisymmetric product of \((p+2)\) gamma matrices is denoted by \(\Gamma^{\mu_1 \cdots \mu_{p+2}}\).

\begin{figure}[ht]\centering
  \includegraphics[width=12cm]{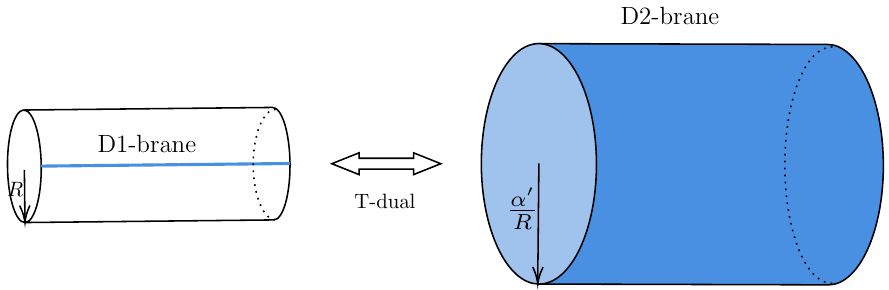}
  \caption{D$1$-brane and its T-dual D$2$-brane on $\bR \times S^1$.}
  \label{fig:D1-D2-T-dual}
\end{figure}

Given that IIA and IIB theories have distinct R-R fields, T-duality in the 9th direction transforms one into the other, impacting the spin fields as follows:
\[
\psi^R_{\alpha}(z) \to \beta_9\,\psi^R_{\alpha}(z), \quad \overline{\psi}^L_{\alpha}(\bar{z}) \to \overline{\psi}^L_{\alpha}(\bar{z}) ,
\]
where \(\beta_9 = \Gamma^{11}\Gamma^9\) represents the 9-reflection on spinors. The $\Gamma_{11}$ just gives sign $\pm 1$ when it acts on spinor fields because the $\mathrm{R}$ ground states have definite chirality. The $\Gamma^9$ adds a ``9'' to the set $\mu_1 \cdots \mu_p$ if none is present, or removes one if it is present via $\left(\Gamma^9\right)^2=1$. This is how T-duality acts on the RR field strengths, adding or subtracting the index for the dualized dimensions. Thus, if we start from the IIA string we get the IIB RR fields and vice versa. Consequently, the R-R fields are transformed as:
\begin{align}\label{RR-Tduality}
C_9 & \to C \cr
C_{\mu}, C_{\mu \nu 9} & \to C_{\mu 9}, C_{\mu \nu} \cr
C_{\mu \nu \lambda} & \to C_{\mu \nu \lambda 9}.
\end{align}
Figure \ref{fig:D1-D2-T-dual} illustrates the behavior of D-branes under T-duality, and Table \ref{tab:Tdual} lists the corresponding D-brane configurations in Type II theories that are dual to each other.

\begin{table}[ht]\centering
  IIB \qquad
\begin{tabular}{c|cccccccccc}
& {0} & {1} & {2} & {3} & {4} & {5} & {6} & {7} & {8} & {9} \\
\hline  NS5  & $\times$ & $\times$ & $\times$ & $\times$ & $\times$ & $\times$ & & & & \\
\hline D1 & $\times$ & $\times$ &  & & & & & & &  \\
\hline D5 & $\times$ & $\times$ &  & & & & $\times$ & $\times$ & $\times$ &$\times$
\end{tabular}\\
\tikzset{every picture/.style={line width=0.75pt}}


\begin{tikzpicture}[x=0.75pt,y=0.75pt,yscale=-1,xscale=1]

\draw    (203,138) -- (203,175.5) ;
\draw [shift={(203,177.5)}, rotate = 270] [color={rgb, 255:red, 0; green, 0; blue, 0 }  ][line width=0.75]    (10.93,-3.29) .. controls (6.95,-1.4) and (3.31,-0.3) .. (0,0) .. controls (3.31,0.3) and (6.95,1.4) .. (10.93,3.29)   ;
\draw [shift={(203,136)}, rotate = 90] [color={rgb, 255:red, 0; green, 0; blue, 0 }  ][line width=0.75]    (10.93,-3.29) .. controls (6.95,-1.4) and (3.31,-0.3) .. (0,0) .. controls (3.31,0.3) and (6.95,1.4) .. (10.93,3.29)   ;

\draw (210,144) node [anchor=north west][inner sep=0.75pt]  [align=left] {T-dual on $X^{6}$};

\end{tikzpicture}\\
 IIA \qquad
\begin{tabular}{c|cccccccccc}
& {0} & {1} & {2} & {3} & {4} & {5} & {6} & {7} & {8} & {9} \\
\hline  NS5  & $\times$ & $\times$ & $\times$ & $\times$ & $\times$ & $\times$ & & & & \\
\hline D2 & $\times$ & $\times$ &  & & & &$\times$ & & &  \\
\hline D4 & $\times$ & $\times$ &  & & & & & $\times$ & $\times$ &$\times$
\end{tabular}\\
\tikzset{every picture/.style={line width=0.75pt}} 

\begin{tikzpicture}[x=0.75pt,y=0.75pt,yscale=-1,xscale=1]

\draw    (203,138) -- (203,175.5) ;
\draw [shift={(203,177.5)}, rotate = 270] [color={rgb, 255:red, 0; green, 0; blue, 0 }  ][line width=0.75]    (10.93,-3.29) .. controls (6.95,-1.4) and (3.31,-0.3) .. (0,0) .. controls (3.31,0.3) and (6.95,1.4) .. (10.93,3.29)   ;
\draw [shift={(203,136)}, rotate = 90] [color={rgb, 255:red, 0; green, 0; blue, 0 }  ][line width=0.75]    (10.93,-3.29) .. controls (6.95,-1.4) and (3.31,-0.3) .. (0,0) .. controls (3.31,0.3) and (6.95,1.4) .. (10.93,3.29)   ;

\draw (210,144) node [anchor=north west][inner sep=0.75pt]  [align=left] {T-dual on $X^{2}$};

\end{tikzpicture}\\
  IIB \qquad
\begin{tabular}{c|cccccccccc}
& {0} & {1} & {2} & {3} & {4} & {5} & {6} & {7} & {8} & {9} \\
\hline  NS5  & $\times$ & $\times$ & $\times$ & $\times$ & $\times$ & $\times$ & & & & \\
\hline D3 & $\times$ & $\times$ & $\times$ & & & &$\times$ & & & \\
\hline D5 & $\times$ & $\times$ & $\times$ & & & & & $\times$ & $\times$ &$\times$
\end{tabular}\\
\tikzset{every picture/.style={line width=0.75pt}}  

\begin{tikzpicture}[x=0.75pt,y=0.75pt,yscale=-1,xscale=1]

\draw    (203,138) -- (203,175.5) ;
\draw [shift={(203,177.5)}, rotate = 270] [color={rgb, 255:red, 0; green, 0; blue, 0 }  ][line width=0.75]    (10.93,-3.29) .. controls (6.95,-1.4) and (3.31,-0.3) .. (0,0) .. controls (3.31,0.3) and (6.95,1.4) .. (10.93,3.29)   ;
\draw [shift={(203,136)}, rotate = 90] [color={rgb, 255:red, 0; green, 0; blue, 0 }  ][line width=0.75]    (10.93,-3.29) .. controls (6.95,-1.4) and (3.31,-0.3) .. (0,0) .. controls (3.31,0.3) and (6.95,1.4) .. (10.93,3.29)   ;

\draw (210,144) node [anchor=north west][inner sep=0.75pt]  [align=left] {S-dual};

\end{tikzpicture}\\
 IIB \qquad
\begin{tabular}{c|cccccccccc}
& {0} & {1} & {2} & {3} & {4} & {5} & {6} & {7} & {8} & {9} \\
\hline  D5  & $\times$ & $\times$ & $\times$ & $\times$ & $\times$ & $\times$ & & & & \\
\hline D3 & $\times$ & $\times$ &  $\times$ & & & &$\times$ & & &  \\
\hline NS5 & $\times$ & $\times$ &  & & & &  & $\times$ & $\times$ &$\times$
\end{tabular}
\caption{D-branes and dualities in Type II theories. The last two configurations will appear in \S\ref{sec:HW}.}
\label{tab:Tdual}
\end{table}

\section{Unoriented strings and Type I theory}\label{sec:TypeI}

\begin{figure}[htb]
 \centerline{\includegraphics[width=15cm]{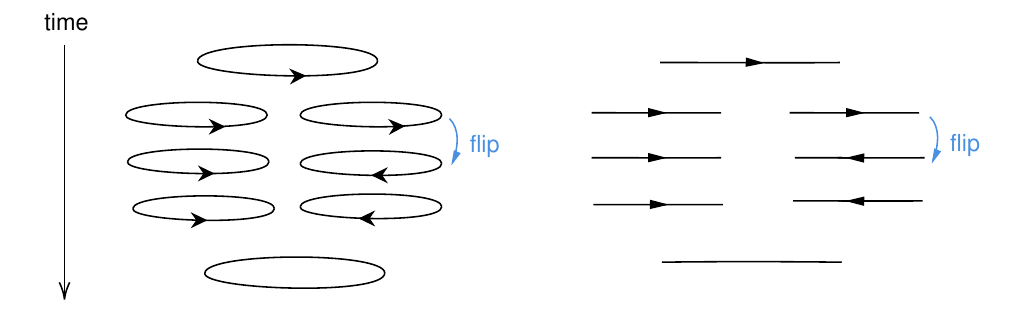}}
\caption{Unoriented processes. The left is closed string one, and the right is open string one.}
\label{unoriented-WS}
\end{figure}

So far we have considered oriented world-sheet Riemann surfaces.
However, it is quite natural to consider processes in Figure \ref{unoriented-WS}, which yields an unoriented world-sheet.
Therefore, we will consider the so-called \textbf{unoriented string},
which leads to Type I superstring theory. As we will see below, an unoriented string theory needs to incorporate (unoriented) open strings.

Open strings can end on D-branes in Type II theories. However, without D-branes, Type II theories cannot incorporate open strings
due to supersymmetry as follows. In fact, $D=10$
$\cN=2$ supersymmetry requires $\textbf{8}_v$ to have two superpartners
as for a graviton to have two gravitinos in Table \ref{tab:masslessII}.
On the other hand, after the GSO projection, open strings can have one of the following massless states:
\begin{align*}
 &P_\mathrm{GSO}: \quad \mathrm{NS}, \ \mathrm{R+} = \textbf{8}_v +\textbf{8}_s \ ,  \\
 &\wt P_\mathrm{GSO}: \quad \mathrm{NS}, \ \mathrm{R-} = \textbf{8}_v +\textbf{8}_c \ ,
\end{align*}
where $\textbf{8}_v$ represents a gauge field and $\textbf{8}_{s,c}$ are gauginos.
Hence, the gauge field has only one superpartner so that it cannot exist in Type II theories without D-branes.

In fact, the existence of (parallel) D-branes breaks a half of supersymmetries so that open strings can end on D-branes. To consider unoriented string theory, we will introduce an \textbf{orientifold plane} or \textbf{O-plane} that breaks a half of supersymmetry.
Although the argument below mainly focuses on the bosonic string for simplicity, it is straightforward to apply for superstrings.

\subsection{Unoriented string theory}\label{sec:orientifold}

As in Figure \ref{unoriented-WS}, a world-sheet parity symmetry exchanges
left- and right-movers for closed strings and a reversal of the spatial direction for open strings.

Let us first take a look at a world-sheet parity flip operator $\Omega$ on a bosonic closed string. It exchanges the left and right-moving modes which ends up with a reversal of  the spatial direction up to the overall sign:
\begin{align}\label{O-closed}
 \Omega :
 \begin{cases}
  X(t,\sigma) &\leftrightarrow \quad X(t,-\sigma) \\
  b(t,\sigma) &\leftrightarrow \quad \ol b(t,-\sigma)  \\
  c(t,\sigma) &\leftrightarrow \quad -\ol c(t,-\sigma)  \\
 \end{cases} \ , \qquad\qquad  \Omega :
  \begin{cases}
   \alpha^\mu_n &\leftrightarrow \quad \ol \alpha^\mu_n \\
   b_n &\leftrightarrow \quad \ol b_n  \\
   c_n &\leftrightarrow \quad \ol c_n  \\
  \end{cases} \ ,
\end{align}
where $t$ is the Euclidean world-sheet time, not $\tau$. 
 Only $\Omega$ invariant states such as the tachyon vacuum, gravitons and dilaton survive under this projection, while the antisymmetric $B$-field is projected out. 
 Since a closed string is electrically coupled to the $B$-field, this implies that an unoriented closed string is not stable and it should decay.
 We define the space-time after the projection as a space-time-filling $\textrm{O}25$-plane.
 This is a special case of orientifold that will be introduced soon. 

Now, let us introduce a space-filling D$25$-brane and consider an open string with the Neumann boundary condition. As in \eqref{open-variation}, the Neumann boundary condition identifies the left and right-moving modes. Then, the world-sheet parity operator acts as
\begin{align}\label{O-open}
 \Omega :
 \begin{cases}
  X(t,\sigma) &\leftrightarrow \quad X(t,\pi-\sigma) \\
  b(t,\sigma) &\leftrightarrow \quad \ol b(t,\pi-\sigma)  \\
  c(t,\sigma) &\leftrightarrow \quad -\ol c(t,\pi-\sigma)  \\
 \end{cases} \ ,\qquad\qquad
 \Omega :
 \begin{cases}
  \alpha^\mu_n &\leftrightarrow \quad (-1)^n \alpha^\mu_n \\
  b_n &\leftrightarrow \quad (-1)^n b_n  \\
  c_n &\leftrightarrow \quad (-1)^n c_n  \\
 \end{cases} \ ,
\end{align}
Now recall that we can incorporate Chan-Paton degrees of freedom \eqref{CP-factors} for open string states. The orientation flip $\Omega$ exchanges the Chan-Paton indices $i\leftrightarrow j$. Hence,
writing an open string state as in \eqref{CP-factor2}
\begin{align*}
 |N,k;T\rangle \equiv |N,k;ij\rangle T_{ij} \ ,
\end{align*}
$\Omega$ acts on the state as
\begin{align}\label{Omega-sign}
 \Omega |N,k;T\rangle = (-1)^N|N,k;T^t\rangle \ .
\end{align}
Therefore, it imposes the condition
\bea
 T^t = T & \quad \textrm{for } N \textrm{ even }\cr
 T^t =- T & \quad \textrm{for } N \textrm{ odd }~.
\eea
In particular, a gauge boson arises at level one \eqref{open-massless} so that we need to impose the anti-symmetric condition $ T^t =- T$ on the Chan-Paton factor. Consequently, it gives rise to a $\SO(n)$ gauge field.

In general, when the world-sheet parity is flipped, we can also transform the Chan-Paton
index by a matrix $P$ at the same time.
\[\Omega|N ; k ; i j\rangle=(-1)^NP_{jj^{\prime}}\left|N ; k ; j^{\prime}i^{\prime} \right\rangle P_{i^{\prime}i}^{-1}\]
The projection operator is subject to $\Omega^2=1$, which gives the constraint
$$\Omega^{2}|N ; k ; i j\rangle=[(P^{T})^{-1} P]_{i i^{\prime}}\left|N ; k ; i^{\prime} j^{\prime}\right\rangle(P^{-1} P^{T})_{j^{\prime} j}~,\quad\Rightarrow \quad P^tP^{-1}=\pm1$$
Furthermore, there is a $\U(n)$ gauge equivalence relation $T \sim \wt T = UT U^{-1}$ $(U\in \U(n))$, which amounts to  $\wt P \sim U P U^t$.
Taking the gauge equivalence into account, $P^tP^{-1}=1$ leads to $P=1$, in which the gauge group becomes $\SO(n)$ as above. On the other hand,
for $P^tP^{-1}=-1$, the rank needs to be even $n=2k$, and we can take a basis such that
\[P = i
       \begin{pmatrix}
        0 & -\textbf{1}_{k\times k}  \\
        \textbf{1}_{k\times k} & 0
       \end{pmatrix}\]
In this situation, the gauge Lie algebra becomes $\Sp(k)$ so that $T^t=-PTP$. The choice of the gauge group is related to the sign of charge and tension for the orientifold plane. We then denote orientifold plane $\mathrm{O}^+$ or $\mathrm{O}^-$ 
for gauge group $\Sp(k)$ and $\SO(n)$ respectively.
We will discuss it in detail at the end of this section.

\subsubsection*{Unoriented superstring spectrum}

Now, let us consider the unoriented closed string in superstring theories.
Since $\Omega$ exchanges the left- and right-moving modes, both must have the same spectra. Hence, it can only act on Type IIB theory but not on Type IIA theory. (See \eqref{IIA-IIB}.) 
In Table \ref{tab:masslessII}, the flip $\Omega$ projects out the $B$-field $[2]$ in the NS-NS sector,
as well as a half of the NS-R R-NS sector $\textbf{8}_c + \textbf{56}_s$ (only the diagonal part survives).
Supersymmetry requires that the number of bosons and fermions be the same,
which implies that $[0]$ and $[4]_+$ are eliminated
and only the second rank anti-symmetric field $[2]$ survives in the R-R sector. This implies that $\Omega$ projects out D$(-1)$-, D$3$-, D$7$-branes.   D$1$- and D$5$-branes survive, which are electrically and magnetically coupled to the R-R two-form $C_{(2)}$, respectively.
The remaining states are
\begin{align}\label{massless-type1}
 [0] \oplus [2] \oplus(2) \oplus\textbf{8}_c \oplus\textbf{56} =& \textbf{1} \oplus\textbf{28} \oplus\textbf{35} \oplus\textbf{8}_c \oplus\textbf{56}_s  \cr
 & \Phi \quad C_{\mu\nu} \ \ G_{\mu\nu}  \ \  \lambda^+  \ \ \ \psi_\mu^- \
\end{align}
This gives us 
$10$d $\mathcal{N}=1$ supergravity multiplet.
However, this theory is anomalous. 
We thus need to incorporate unoriented open strings to cancel this anomaly \cite{Green:1984sg,Green:1984qs}.
In conclusion, Type I is an unoriented, open plus closed superstring theory in which the massless spectrum is
\begin{align*}
  [0] \oplus [2] \oplus (2) \oplus\textbf{8}_c \oplus\textbf{56} \oplus(\textbf{8}_v \oplus\textbf{8}_s)_{\SO(n) \textrm{ or } \Sp(n)} \ .
\end{align*}
It can be
shown that $\mathcal{N}=1$ supersymmetric Yang-Mills theory with 
gauge group
$\SO(32)$ or $E_8 \times E_8$ can be 
the additional sector to cancel this
anomaly. However, 
$E_8\times E_8$ 
gauge group can not 
be constructed from 
unoriented open string by the Chan-Paton factors.
The gauge group $\SO(32)$ is the only option for 
type I string theory.
 From the world-sheet perspective, the anomaly cancellation corresponds to the \textbf{tadpole cancellation}.
In what follows, we will investigate the tadpole cancellation in bosonic string theory, and will see that the gauge group needs to be $\SO(2^{13})$. A similar analysis tells us that the gauge group must be  $\SO(32)$ in Type I superstring theory.

\subsection{Amplitude in unoriented string theory}

As above, unoriented open strings give rise to either $\SO(n)$ or $\Sp(n)$ massless gauge fields.
However, unless the tadpole cancellation is satisfied, the theory is still anomalous. We will see it in bosonic string theory.

\subsubsection*{Cylinder}
\begin{figure}[htb]
\centerline{\includegraphics[width=\textwidth]{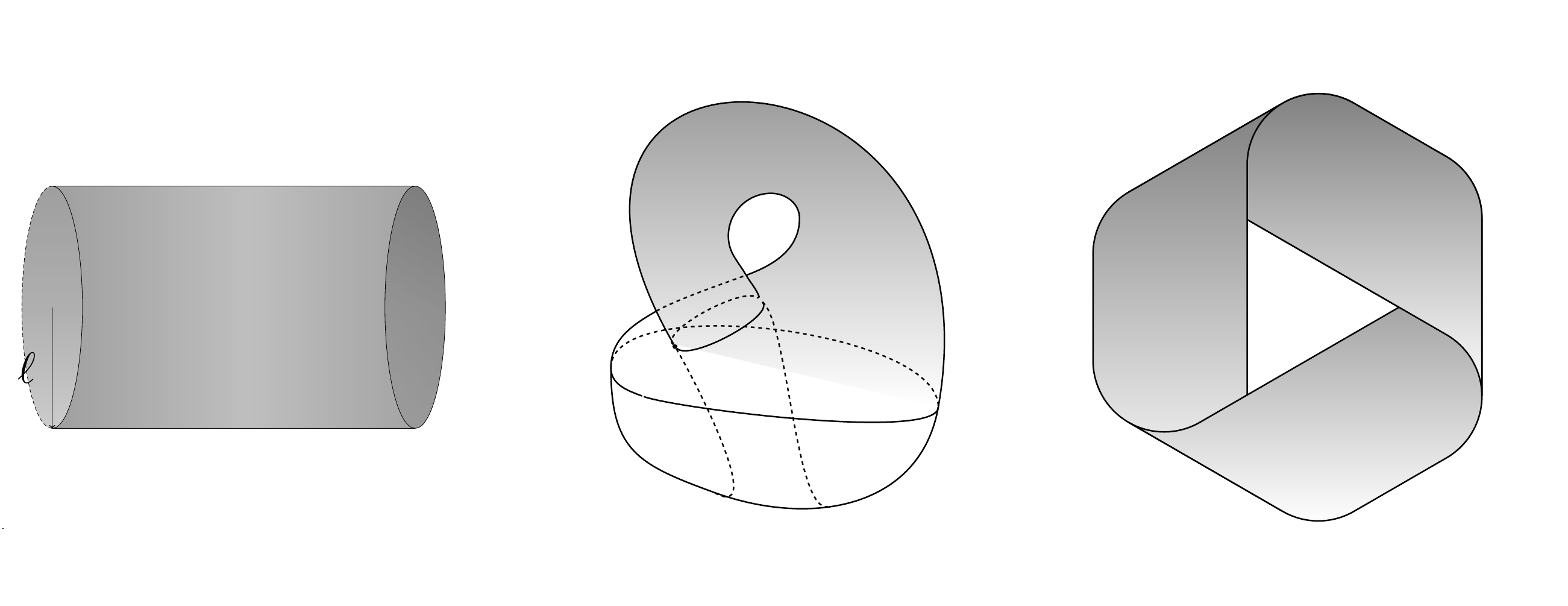}}
\caption{Cylinder, Klein bottle and M\"obius strip.}
\label{oriented-unoriented}
\end{figure}

First, let us evaluate an ``oriented'' open string one-loop amplitude where a world-sheet is a cylinder (or annulus) as in the left of Figure \ref{oriented-unoriented}. We consider that there are $n$ D$25$-branes (or D$9$-branes for superstring) so that all of the boundary conditions are of Neumann type. The evaluation is parallel to \S\ref{sec:amplitude}.
It is clear from Figure \ref{oriented-unoriented} that we notice the following facts:
\vspace{-4pt}
\begin{itemize}
 \setlength{\itemsep}{0pt}
 \item The range is $0 \le \Re~w \le \pi$, the period is $w \sim w +2\pi il$.
 \item There is a real modulus $l$ so that the amplitude needs one $b$ zero mode insertion.
 \item There is a time translational isometry, a shift of $\Im~w$ so that  the amplitude needs one $c$ zero mode insertion.
\end{itemize}
Consequently, the cylinder partition function is written as
\begin{align*}
 A_{0,C} = \int \frac{dl}{2l} \langle b_0 c_0 \rangle_\mathrm{gh} \langle 1 \rangle_\mathrm{mat} \ ,
\end{align*}
where the ghost zero modes in open strings are read off
\begin{align*}
 &b_0 = \frac{1}{2\pi} \int_0^{\pi} \left[ dw\ b(w) +d\ol w\ \ol b(\ol w)\right] \ , \\
 &c_0 = \frac{i}{2\pi} \int_0^{\pi} \left[ dw\ c(w) -d\ol w\ \ol c(\ol w)\right] \ .
\end{align*}

Using operator formalism, we can derive each contribution as
\begin{align*}
 \langle 1 \rangle_\mathrm{mat} &= n^2\ \Tr \left[ q^{L_0^X -\frac{c}{24}} \right]= n^2 \cdot \frac{iV_{26}}{(2\pi)^{26}} (2l\alpha')^{-13} \cdot \eta(il)^{-26} \ ,  \\
 \langle b_0 c_0 \rangle_\mathrm{gh} &= \Tr \left[ (-1)^F b_0 c_0 q^{L_0^g-\frac{c}{24}} \right] = \eta(il)^2\ ,
\end{align*}
where $q=e^{-2\pi l}$ and the Virasoro generators are in \eqref{open-Virasoro} for open strings. Note that $n^2$ comes from the trace over the Chan-Paton degrees of freedom.
Therefore, the cylinder partition function  is
\begin{align}\label{cylinder}
 A_{0,C} = n^2 \cdot \frac{iV_{26}}{(2\pi \ell_s)^{26}}\int_0^\infty \frac{dl}{2l} \frac{1}{(2l)^{13} \eta(il)^{24}} \ ,
\end{align}
where $\ell_s = \sqrt{\alpha'}$ is the string length.

\begin{figure}[htb]
 \centerline{\includegraphics[width=13cm]{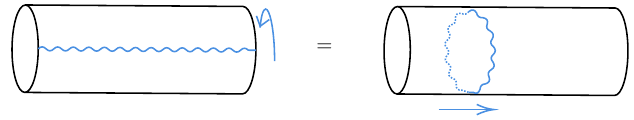}}
 \caption{Pictorial description of the equivalence between open string one-loop and closed string propagation.}
 \label{fig:openClosedAmp}
\end{figure}

\begin{figure}[htb]
 \centerline{\includegraphics[width=5cm]{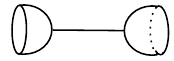}}
 \caption{Intermediate propagation is replaced by particles (lines).}
 \label{masslessProp}
\end{figure}

Let us look into its physical implication. When $l\to 0$, the cylinder becomes a thin tube, and it can be regarded as a propagation of a closed string as in Figure \ref{fig:openClosedAmp}. This is called the \textbf{open-closed duality}.
       This is justified by rewriting the amplitude. Using the modular property \eqref{S-T} of the Dedekind $\eta$-function, we can rewrite it in terms of $l=s^{-1}$ as
\be\label{cylinder2}
  A_{0,C} = \frac{n^2}{2^{13}} \cdot \frac{iV_{26}}{(2\pi \ell_s)^{26}} \int_0^\infty \frac{ds}2\ \eta(is)^{-24} \ ,
\ee
where
\[\eta(is)^{-24} = q^{-1} +24 +\cdots \equiv \sum_{N=0}^\infty \mathcal N_N q^{N-1} \quad (q = e^{-2\pi s}) \ .\]
The leading term is divergent due to the closed string tachyon,
while the second term gives a divergence when
$s\to \infty$.
The UV divergence $l\to 0$ in an open string is now replaced by the IR divergence $s \to \infty$ of a closed string propagation, which can be understood as particle propagations (sum of lines) as follows (see also Figure \ref{masslessProp}).
 \begin{align*}
  \int_0^\infty ds \sum_{N=0}^\infty \mathcal N_N e^{-2\pi s(N-1)}
  \sim  \sum_{i} \int_0^\infty ds\ e^{-s(k^2+m^2_i)} = \left. \sum_i \frac{1}{k^2+m^2_i} \right|_{k=0} \ .
 \end{align*}
       We can see that the IR divergence is from massless particle propagation (graviton etc.),
       which is absorbed or emitted from D$25$-branes. This point will be revisited at the end of \S\ref{sec:DBI}.
 In conclusion, the divergence is due to the existence of the D$25$-branes,
       which have definite tension (this is why they emit graviton/dilaton).
To cure this divergence, we will consider unoriented string amplitudes when world-sheets are Klein bottle and M\"obius strip as in  Figure \ref{oriented-unoriented}. The reason will become clear later.

\subsubsection*{Klein bottle amplitude}
\begin{figure}[htb]
\centerline{\includegraphics[width=8cm]{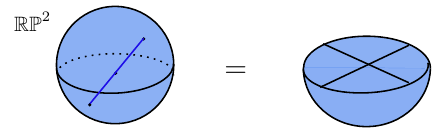}}
\caption{$\mathbb{RP}^2$ is formed by identifying antipodal points on $S^2$, akin to connecting the southern hemisphere to the equator.}
\label{crosscap}
\end{figure}

\begin{figure}[htb]
\centering
\includegraphics[width=13cm]{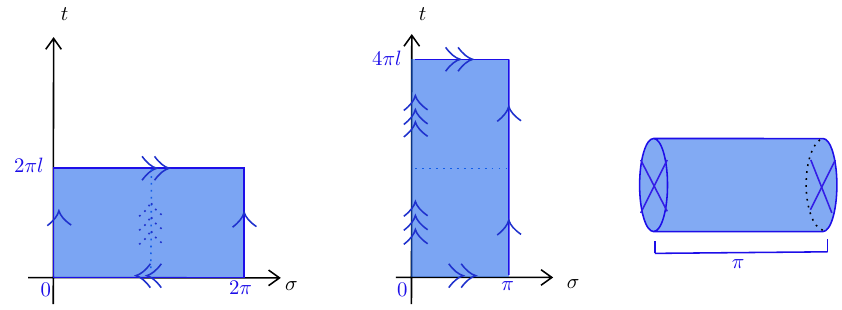}
\caption{Representation of the Klein bottle as a cylinder with cross-cap boundaries at both ends.}
\label{fig:KleinBottle}
\end{figure}

We proceed to evaluate the one-loop amplitude of an unoriented closed string propagating on a Klein bottle world-sheet, as depicted in Figure \ref{fig:KleinBottle}. The Klein bottle arises from identifying the top and bottom edges of the string world-sheet via the world-sheet parity operator.

The amplitude takes the form:
\begin{align*}
A_{0,K} &= \int \frac{dl}{2l}
\operatorname{Tr} \biggl[ \Omega (-1)^F \frac{1}{2}(b_0+\overline{b}_0) \frac{1}{2}(c_0+\overline{c}_0) q^{L_0 +\overline{L}_0 -\frac{c}{12}} \biggr] \quad (q=e^{-2\pi l}) \ .
\end{align*}
Using \eqref{O-closed}, we can express it as:
\begin{align*}
A_{0,K} &= \int \frac{dl}{2l}
\operatorname{Tr} \biggl[ (-1)^F b_0 c_0 q^{2L_0^{\text{tot}} -\frac{c^{\text{tot}}}{12}} \biggr] \ ,
\end{align*}
where $L_0$ and $c$ denote the Virasoro generators and the central charge for closed strings, respectively. Hence, the resulting expression becomes:
\begin{align}\label{klein}
A_{0,K} =& \frac{iV_{26}}{(2\pi \ell_s)^{26}} \int \frac{dl}{2l} \frac{1}{l^{13} \eta(2il)^{24}} \cr
=& 2^{13} \frac{iV_{26}}{(2\pi \ell_s)^{26}} \int \frac{ds}{2} \eta(is)^{-24} \ .
\end{align}
where we apply the modular $S$-transformation ($2l=s^{-1}$) from the first to second line. 
Similarly, we encounter a tadpole divergence originating from the Klein bottle.

\subsubsection*{M\"obius strip amplitude}
\begin{figure}[htb]
\centerline{\includegraphics[width=13cm]{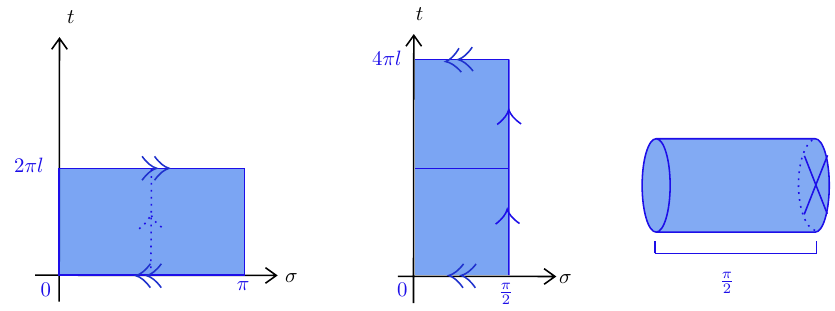}}
\caption{M\"obius strip.}
\label{fig:MobiusStrip}
\end{figure}

Finally, let us evaluate the one-loop amplitude of an unoriented open string on a M\"obius strip world-sheet, as depicted in Figure \ref{fig:MobiusStrip}. The M\"obius strip exhibits several key characteristics:
\vspace{-4pt}
\begin{itemize}
 \setlength{\itemsep}{0pt}
 \item The spatial range is $0 \le \sigma \le \pi$, with the time period $2\pi l$ together with the orientation flip: $(t,\sigma) \sim (t+2\pi l,\pi -\sigma)$.
 \item There is a real modulus $l$; the amplitude needs one $b$ zero mode insertion.
 \item There is a real isometry, which is the time translation; the amplitude needs one $c$ zero mode insertion.
\end{itemize}

Thus, the M\"obius strip amplitude is given by
\begin{align*}
 A_{0,M} =  \int \frac{dl}{2l}
 \Tr \left[ \Omega (-1)^F b_0 c_0 q^{L_0 -\frac{c}{12}} \right] \qquad (q=e^{-2\pi l}) \ .
\end{align*}
The trace is over Hilbert space of the matter and ghost sectors on the strip
as well as the Chan-Paton factors, which give in total $n^2$ degeneracy for each state.
We can divide the effect of $\Omega$ into two parts as follows.
\begin{align*}
 \Omega |\Phi;T\rangle = |\Omega\Phi;PT^tP \rangle
 \equiv \Omega_\Phi \cdot \Omega_T |\Phi;T\rangle \ .
\end{align*}
Let us see the $\Omega_T$, which is defined as
\begin{align*}
 \Omega_T = \frac{PT^tP}{T} \ .
\end{align*}
For $\mathrm{SO}(n)$, where $P=1$, we obtain:
\begin{align*}
    \Omega_{T,\mathrm{SO}} = \frac{T^t}{T} =
    \begin{cases}
        +1 & (\textrm{for symmetric } T)  \\
        -1 & (\textrm{for anti-symmetric } T)
    \end{cases},
\end{align*}
resulting in:
\begin{align*}
    \operatorname{Tr}_{T,\mathrm{SO}} \left[ \Omega \right] = \frac{n(n+1)}{2} -\frac{n(n-1)}{2} = n.
\end{align*}
In the case of $\Sp(n)$ (exercise), we have
\begin{align*}
 \Tr_{T,\Sp} \left[ \Omega \right] = -n \ .
\end{align*}
Since the action of $\Omega$ to a state of level $N$ introduces the sign $(-1)^N$ \eqref{Omega-sign}, the contribution from the matter and ghost part can be evaluated as
\begin{align*}
 \Tr_\Phi \left[ \Omega (-1)^F b_0 c_0 q^{L_0 -\frac{c}{12}} \right]
 &= \frac{iV_{26}}{(2\pi \ell_s)^{26}} \frac{1}{(2l)^{13}}
 \cdot q^{-1} \prod_{N=1}^\infty \frac{(1-(-q)^N)^2}{(1-(-q)^N)^{26}}  \cr
 &= \frac{iV_{26}}{(2\pi \ell_s)^{26}} \frac{1}{(2l)^{13}} \cdot \frac{-1}{\eta(il+\frac{1}{2})^{24}} \ .
\end{align*}
Therefore, the M\"obius strip amplitude is
\begin{align*}
 A_{0,M} &= \pm n \cdot \frac{iV_{26}}{(2\pi \ell_s)^{26}} \int \frac{dl}{2l}
 \frac{1}{(2l)^{13}} \cdot \frac{-1}{\eta(il+\frac{1}{2})^{24}} \ ,  \\
 &= \mp 2n \cdot \frac{iV_{26}}{(2\pi \ell_s)^{26}} \int \frac{ds}{2}
 \eta(is+\frac{1}{2})^{-24} \ ,
\end{align*}
where, by setting $2l=s^{-1}$, we apply the modular property \eqref{S-T} to the following expression:
\bea 
 q^{-1} \prod_{N=1}^\infty (1-(-q)^N)^{-24}  = -\eta(il+\frac{1}{2})^{-24}= \vartheta_3(2il)^{-12}\eta (2il)^{-12}~.
 \eea

\begin{figure}[htb]
\centerline{\includegraphics[width=\textwidth]{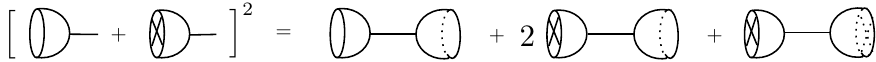}}
\caption{Pictorial expression for the unoriented open string amplitude.}
\label{tadpoleCancel}
\end{figure}

To sum up, three amplitudes are
(introduced additional $\frac{1}{2}$ factor for Cylinder as an unoriented amplitude)
\begin{align*}
 A_{0,C} &= \frac{n^2}{2^{13}} \cdot \frac{iV_{26}}{(2\pi \ell_s)^{26}}
 \int_0^\infty \frac{ds}{2} \eta(is)^{-24} \ ,  \\
 A_{0,K}
 &= 2^{13} \cdot \frac{iV_{26}}{(2\pi \ell_s)^{26}} \int \frac{ds}{2} \eta(is)^{-24} \ ,  \\
 A_{0,M}
 &= \mp 2n \cdot \frac{iV_{26}}{(2\pi \ell_s)^{26}} \int \frac{ds}{2}
 \eta(is+\frac{1}{2})^{-24} \ ,
\end{align*}
where
\begin{align*}
 &\eta(is)^{-24} = q^{-1} +24 +\mathcal O(q) \qquad (q = e^{-2\pi s}) \ ,  \\
 &\eta(is+\frac{1}{2})^{-24} = -q^{-1} +24 +\mathcal O(q) \ .
\end{align*}
As we saw in the oriented string case, the massless states lead to IR singularity.
On the other hand, in the unoriented open string case, we have
\begin{align*}
 \frac{1}{2^{13}} \left[ n\mp 2^{13} \right]^2 \cdot \frac{iV_{26}}{(2\pi \ell_s)^{26}}
 \int_0^\infty \frac{ds}{2} \cdot ( 24+\cdots) \ ,
\end{align*}
where the upper (minus) sign corresponds to SO($n$) and the lower (plus) sign to Sp($n$). Therefore, the unoriented IR divergence vanishes only for $\SO(2^{13})=\SO(8192)$. This cancellation can be illustrated in Figure \ref{tadpoleCancel}.
The cross cap shows another object (other than D-brane) that absorbs and emits gravitons etc., which is called O-plane.
In this situation, it should be space-filling. Hence, it is O$25^-$-plane, and a single O$25^-$-plane cancels the tension of $2^{13}$ D$25$-branes.

Although our discussion was in the bosonic string,
parallel argument perfectly works for superstring. Here we give the final results for the cylinder, Klein bottle, and M\"obius strip amplitude in R-R sectors (exercise):
\bea 
\begin{aligned}
& A_{0, C}=n^2 \cdot \frac{i V_{10}}{\left(2 \pi \ell_s\right)^{10}} \int_0^{\infty} \frac{d l}{8 l} \frac{\vartheta_2(il)^4}{(2l)^{5} \eta(i l)^{12}} \ , \\
& A_{0, K}=2^{5} \cdot \frac{i V_{10}}{\left(2 \pi \ell_s\right)^{10}} \int_0^{\infty} \frac{d l}{8 l} \frac{\vartheta_2(2il)^4}{(2l)^{5} \eta(2i l)^{12}} \ , \\
& A_{0, M}=\mp  n \cdot \frac{i V_{10}}{\left(2 \pi \ell_s\right)^{10}} \int_0^{\infty} \frac{d l}{8 l} \frac{\vartheta_2(i l+\frac12)^4}{(2l)^{5} \eta(i l+\frac12)^{12}}\ ,
\end{aligned}
\eea 
respectively. Consequently, we have the unoriented superstring amplitude 
\be 
\frac{1}{2^{5}}\left[n \mp 2^{5}\right]^2 \cdot \frac{i V_{10}}{\left(2 \pi \ell_s\right)^{10}} \int_0^{\infty} \frac{d s}{8} \cdot ( 16+\cdots)
\ee 
where \textbf{the IR divergence vanishes for $\SO(2^5)=\SO(32)$} \cite{Polchinski:1987tu}. This is another way to show the Green-Schwarz anomaly cancellation \cite{Green:1984sg,Green:1984qs}.
This means that
\begin{align*}
 \textrm{Type I} = \textrm{Type IIB} +32 \textrm{ D$9$-branes} +\textrm{O$9^-$-plane} \ .
\end{align*}
Note that in the superstring case, D-branes and O-plane have RR-charge
in addition to tension, which has relations
\begin{align*}
 T_{\mathrm{O}9^\pm} = \pm 32 \cdot T_{\mathrm{D}9} \quad (\textrm{tension}) \ , \quad
 \mu_{\mathrm{O}9^\pm} = \pm 32 \cdot \mu_{\mathrm{D}9} \quad (\textrm{RR-charge}) \ .
\end{align*}

\subsection{T-duality of Type I theory}\label{sec:TypeI'}

\begin{figure}[htb]
\centerline{\includegraphics[width=8cm]{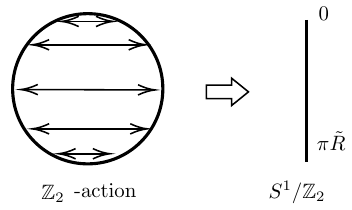}}
\caption{$\ZZ_2$ orbifold of $S^1$.}
\label{s1orbifold}
\end{figure}

Let us recall T-duality.
Consider $X^{i}$ is $S^1$ compactified and T-duality acts as follows:
\begin{align*}
 T_{i}:  \quad X^{i}(z,\ol z) = X^{i}(z) +\ol X^{i}(\ol z) \quad\to\quad
 X^{\prime i}(z,\ol z) = -X^{i}(z) +\ol X^{i}(\ol z) \ .
\end{align*}
On the other hand, the orientation flip acts as follows:
\begin{align*}
 \Omega: \quad X^{i}(z,\ol z) = X^{i}(z) +\ol X^{i}(\ol z) \quad\to\quad
 X^{i}(\ol z,z) = \ol X^{i}(z) +X^{i}(\ol z) \ .
\end{align*}
Therefore, in the T-dual coordinate $X'$, the orientation flip acts as
\begin{align*}
 \Omega: \quad X^{\prime i}(z,\ol z) \quad\to\quad
 -X^{\prime i}(\ol z,z) \ .
\end{align*}
This is understood as spacetime \textbf{orbifold}
as well as world-sheet orientation flip (see Figure \ref{s1orbifold}),
which is called \textbf{orientifold}.
Therefore, the dual space is not $S^1$ but $S^1/\ZZ_2$
with radius $\wt R = \frac{\alpha'}{R}$.
Note that there are two fixed points where O-planes sit and induce the spacetime reversal and the orientation flip.

\begin{figure}[htb]
\centerline{\includegraphics[width=11cm]{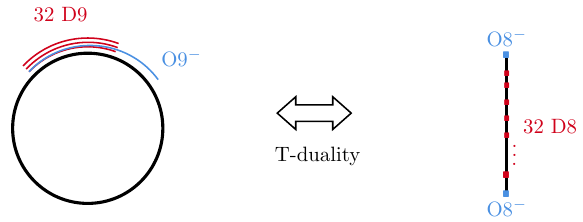}}
\caption{T-dual of Type I superstring theory.}
\label{type1dual}
\end{figure}

Let us consider Type I superstring theory with $X^9$ compactified on $S^1$
and take T-duality along the $S^1$.
With a proper Wilson line
\begin{align*}
 A_9 = i
 \begin{pmatrix}
  & -a_1 & & & \\
  a_1 & & & & \\
  & & & -a_2 & \\
  & & a_2 & & \\
  & & & & \ddots
 \end{pmatrix}
\end{align*}
D$8$-branes sit at different points in $\ZZ_2$ symmetric way (see Figure \ref{type1dual}).
Note that an O$9^-$-plane splits into two O$8^-$-planes.
Accordingly, tension and RR-charge reduce by $2$.
In the end, the T-dual of Type I on $S^1$ is
\begin{align}
 \textrm{Type I' } = \textrm{ Type IIA on } S^1/\ZZ_2 \quad \textrm{with} \quad  2 \textrm{ O$8^-$-plane }
 + (16+16) \textrm{ D$8$-branes }.
\end{align}

\subsection{\texorpdfstring{O$p$}{Op}-plane}

Of course, one can consider further T-dualities along other directions.  In particular, an
orientifold $p$-plane (O$p$) in Type II string theory is defined \cite{Dabholkar:1996pc,Dabholkar:1997zd} as the $\bZ_2$ involution
\footnote{Note that we have already seen the special case $p=9$, where the space-time is fixed under the involution. }
\be
\mathrm{O} p: \quad \mathbb{R}^{1, p} \times \mathbb{R}^{9-p} / {I}_{9-p} \Omega \cdot\begin{cases} 1 & p=0,1 \bmod 4 \\ (-1)^{F_{L}} & p=2,3 \bmod 4,\end{cases}
\ee
where ${I}_{9-p}$ is the involution of all coordinates in the transverse space $\mathbb{R}^{9-p}$ and $F_L$ is the left-moving spacetime fermion number operator. The presence of $(-1)^{F_{L}}$ for $p=2,3 \bmod 4$ can be understood from the requirement that the generator squares to one on fermions. Due to the spacetime involution ${I}_{9-p}$ transverse to the O$p$-plane, the number of D$p$-branes parallel to the O$p$-plane is effectively doubled. Therefore, a stack of $n$ D$p$-branes along the O$p^-$-plane leads to an $\SO(2n)$ gauge theory (Dynkin type $D_{n}$) in Type II theory. To realize an $\SO(2n+1)$ gauge group (Dynkin type $B_{n}$), we need a combination of an O$p^-$-plane and $\frac12$ D$p$-brane, which is denoted by $\wt{\mathrm{O}p}^-$. Consequently, $n$ D$p$-branes with  $\wt{\mathrm{O}p}^-$-plane give rise to an $\SO(2n+1)$ gauge theory as a world-volume theory. As we will see in \S\ref{sec:Sdual}, it is natural to incorporate a $\wt{\mathrm{O}p}^+$-plane in Type II theory, which also gives rise to an $\Sp(k)$ gauge theory like an ${\mathrm{O}p}^+$-plane. They behave differently under the S-duality of Type IIB theory.

Each T-duality doubles the number of O-planes, and hence,
reduces the tension and the R-R charges.
Namely, we have the following relations:
\begin{align*}
 T_{\mathrm{O}p^\pm} = \pm 2^{p-5} \cdot T_{\mathrm{D}p} \quad (\textrm{tension}) \ , \quad
 \mu_{\mathrm{O}p^\pm} = \pm 2^{p-5} \cdot \mu_{\mathrm{D}p} \quad (\textrm{R-R charge}) \ .
\end{align*}
Here a D$p$-brane is counted together with its image, whereas the O$9$ relation above counts the 32 D$9$-branes of the covering space; the two conventions differ by a factor of two. Since an $\wt{\mathrm{O}p}^-$-plane is  a combination of an O$p^-$-plane and $\frac12$ D$p$-brane, there is $\frac12$-shift in R-R charge as summarized in Table \ref{tab:Op}.
O$p$-planes bring richness to string theory such as quiver gauge theories,
the topology of $\RP^p$, real K-theory, Dirac quantization conditions, etc, which are beyond the scope of this lecture. We refer to \cite{Witten:1998xy,Hanany:2000fq,Hanany:1999sj,Tachikawa:2018njr} and references therein.

\begin{table}[ht]\centering
\begin{tabular}{|l|l|l|}
\hline O-plane & gauge group & RR-charge \\
\hline$\mathrm{O} p^{-}$ & $\SO(2 n)$ & $-2^{p-5}$ \\
 $\mathrm{O} p^{+}$ & $\Sp(2 n)$ & $+2^{p-5}$ \\
 $\widetilde{\mathrm{O}p}^{-}$ & $\SO(2 n+1)$ & $-2^{p-5}+\frac{1}{2}$ \\
 $\widetilde{\mathrm{O} p}^{+}$ & $\Sp(2 n)$ & $+2^{p-5}$ \\
\hline
\end{tabular}
\caption{Gauge groups and R-R charges of O$p$-planes.}\label{tab:Op}
\end{table}

\section{Heterotic string theories}\label{sec:Heterotic}

We now embark on the study of \textbf{Heterotic string theories} \cite{Gross:1984dd,Gross:1985fr,Gross:1985rr}. Heterotic string is a hybrid construction of the left-moving sector of the 26-dimensional bosonic string and the right-moving sector of 10-dimensional superstring. In the Heterotic formulation, the 16 extra bosonic degrees of freedom from the left-moving sector are compactified on specific 16-dimensional tori, giving rise to two possible gauge groups: $\SO(32)$ or $E_8\times E_8$. The choice of 16-dimensional tori is of crucial importance, as these tori possess exceptional properties. Alternatively, we can describe the left-moving sector in terms of 32 free fermions, whose current algebra corresponds to either $\SO(32)$ or $E_8\times E_8$ at level $k=1$. This combination of two distinct types of modes has been aptly named \textbf{heterosis}, highlighting the unique hybridization inherent in the Heterotic string theory.

\subsection{Bosonic construction}

\subsubsection*{Toroidal compactifications}

\begin{figure}[ht]\centering
\includegraphics[width=12cm]{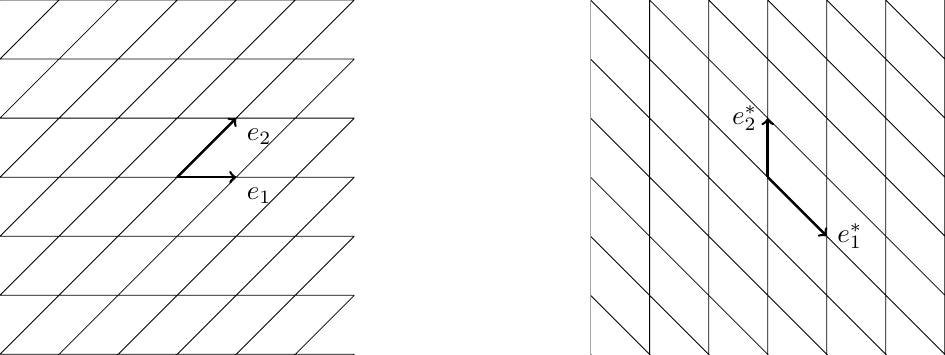}
\caption{Lattice and its dual lattice}\label{fig:dual-lattice}
\end{figure}

In \S\ref{sec:Tdual}, we explored the compactification of bosonic string theory on a circle $S^1$. Now, we will extend our analysis to the compactification of bosonic string theory on a $d$-dimensional torus $T^d$, resulting in an effective theory with $(26-d)$ dimensions. The torus is defined by identifying points in the $d$-dimensional internal space according to
\be \label {torusequiv}
	X^I \sim X^I + 2 \pi e^I_i w^i =X^I+2\pi W^I~, \qquad \textrm{for} \quad w^i \in \bZ~.
	\ee
where capital letters denote the compact dimensions. The vectors $\mathbf{e}_i= \{e_i^I\}$  ($i= 1\cdots d$) are $d$ linearly independent vectors known as \textbf{vielbein}, which generate a $d$-dimensional lattice $\Lambda$. Moreover, the vielbein brings the metric into the standard Euclidean form
\begin{align}
	G_{ij} =\bfe_i\cdot\bfe_j= e^I_i e^J_j \delta_{IJ}, \qquad 	X^I \equiv e^I_i X^i\end{align}
The torus $T^d$ on which we perform the compactification is obtained by dividing $\mathbb{R}^d$ by $\Lambda$
\[T^d = \frac{{\bR}^d }{2\pi \Lambda}~.\]
The momentum $p^I$ conjugate to the coordinates $X^I$ on the torus is quantized as $\mathbf{p} \cdot \mathbf{W} \in\mathbb{Z}$. Hence, the momentum $p$ takes values on the dual lattice $\Lambda^*$ (Figure \ref{fig:dual-lattice})
\[	\Lambda^* \equiv \{ e^{*Ii} n_i; \quad n_i \in \bZ \}, \qquad
		 G^{i j}=\bfe^{*i}\cdot\bfe^{*j}= e_I^{*i} e_J^{*j}\delta^{IJ}. \]

The closed string in the compact directions $X^I(z,\bar{z}) = X_R^I(z) + \bar{X}_L^I(\bar{z})$ obeys the condition
\[
X ^I( \s+2\pi,\t)=X ^I( \s,\t)+2\pi W^I
\]
where $W^I$ serves as an analogue of winding numbers. The mode expansion for the compact directions is given by
\begin{align}
X_R^I(z)&=x^I-i\sqrt{\frac{\ap}{2}}p_R^I
\ln z+i\sqrt{\frac{\ap}{2}}\sum_{m\neq0}\frac{\alpha^I_m}{mz^m}, \nonumber\\
\overline X_L^I(\bz)&={\overline x}^I-i\sqrt{\frac{\ap}{2}}{p^I_L}
\ln \bz+i\sqrt{\frac{\ap}{2}}\sum_{m\neq0}\frac{{\overline\alpha}^I_m}{m\bz^m}.\nonumber
\end{align}
where the zero modes are
\begin{align}
\bfp_L := p^I_L &=\frac{1}{\sqrt{2}} \left[\sqrt{\a'}p^I + \frac {W^I}{\sqrt{\a'}}\right] =\frac{1}{\sqrt{2}}\left[ \sqrt{\ap}e^{* I i} n_i + \frac {e^I_i} {\sqrt{\ap}}  w^i\right]~,\cr
\bfp_R := p^I_R &=\frac{1}{\sqrt{2}} \left[\sqrt{\a'}p^I - \frac {W^I}{\sqrt{\a'}}\right] =\frac{1}{\sqrt{2}}\left[ \sqrt{\ap}e^{* I i} n_i - \frac{e^I_i} {\sqrt{\ap}}  w^i\right]~.
\end{align}
The mass formula and the level-matching condition are now
\begin{align}\label{mass-toroidal}
\ap M^2&=2(N+\overline N-2) + (\ap p_Ip^I+\frac1{\ap} W_IW^I)\cr
&=2(N+\overline N-2) +(\ap n_in_jG^{ij}+\frac1{\ap} w^iw^jG_{ij})\cr
N-\overline N&=p_I W^I=n_iw^i
\end{align}
As previously observed, the expressions for $p_L$ and $p_R$ suggest \textbf{T-duality} between the winding number $W^I$ and the momentum $p^I$. In fact, \textbf{T-duality} corresponds to an equivalence between a pair of compactification lattices $\bfe_i$ and
$\bfe'_i$ related as  $\sqrt{\ap}\bfe'_i = \frac{\bfe^{*i}}{\sqrt{\ap}}$. These two compactifications yield the same spectrum since their allowed values of momenta are related as
\be \label {tdualp}
		\bfp_L \leftrightarrow \bfp_L';\qquad \bfp_R \leftrightarrow -\bfp_R'\ee
by interchanging the labels $n_i$ and $w^i$.

Now let us combine the zero modes into the $(d+d)$-dimensional vectors $\bfP=(\bfp_L,\bfp_R)$.
This construction treats $\Lambda$ and $\Lambda^*$ on equal footing
as
\[
\bfP=\bfE^{* i} n_i + \bfE_j w^j,
\]
where
\[
	\bfE_j = \frac1{\sqrt{\ap}}( \bfe_j,  - \bfe_j  )~, \quad
	\bfE^{* i} = \sqrt{\ap}(\bfe^{*i}, \bfe^{*i}  ).
\]
Note that the length scale of the lattice is normalized by the string length, given as $\sqrt{\alpha'} = \ell_s$. Therefore, $\mathbf{P}$ takes values in a $(d+d)$-dimensional lattice $\Gamma_{d,d}$ spanned by $\{\mathbf{E}^{*i}\}$ and $\{\mathbf{E}_j\}$. This lattice satisfies the following properties:
\begin{itemize}\setlength{\parskip}{-0.1cm}
\item \textbf{Lorentzian} if the signature of the metric $G$ is $((+1)^d,(-1)^d)$,
\item \textbf{integral} if $v\cdot w \in \bZ$ for all $v,w\in \G_{d,d}$,
\item \textbf{even} if $\G_{d,d}$ is integral and $v^2$ is even for all $v\in\G_{d,d}$,
\item \textbf{self-dual} if $\G_{d,d} =(\G_{d,d})^*$,
\item \textbf{unimodular} if $\textrm{Vol}(\G_{d,d}) =  | \det G| = 1$.
\end{itemize}
In fact, the metric of this lattice is defined by
\[
\bfP\cdot\bfP'=(\bfp_L\cdot\bfp'_L-\bfp_R\cdot\bfp'_R)=n_iw'^{i}+n'_iw^{i}
\]
which confirms that it is Lorentzian. Since $\mathbf{P}\cdot\mathbf{P} \in 2\mathbb{Z}$, it is also even. The derivation of the self-duality property will be left as an exercise. The unimodular property $\textrm{Vol}(\G_{d,d})=\textrm{Vol}(\G_{d,d})=1$ immediately follows from  the self-dual property.
The lattice $\G_{d,d}$  in the torus compactification of the string is called \textbf{Narain lattice}.

The partition function of the bosonic string compactified on a torus $T^d$ can be expressed as
\[	Z^{\textrm{bos}}_{\G_{d,d}} = \frac 1 {\t_2^{(24-d)/2}| {\eta (q)}|^{48}}
		\sum_{(\bfp_R, \bfp_L) \in \G_{d,d}} q^{\frac{1}{2} \bfp_R^2}
								\bar q^{\frac{1}{2} \bfp_L^2}
\]
where $|\eta(q)|^{48}$ represents the contribution from the bosonic oscillators and $\tau_2^{(24-d)/2}$ arises from the integration over the non-compact momenta. This expression can be readily extended to the Type II string compactified on $T^d$
\[
Z^{\textrm{Type II}}_{\G_{d,d}} = \frac 1 {\t_2^{(8-d)/2}| {\eta (q)}|^{24}} \frac14\Big| -\vartheta_2^4(\tau) + \vartheta_3^4(\tau) - \vartheta_4^4(\tau)\Big|^2
		\sum_{(\bfp_R, \bfp_L) \in \G_{d,d}} q^{\frac{1}{2} \bfp_R^2}
								\bar q^{\frac{1}{2} \bfp_L^2}
\]
which vanishes by virtue of the Jacobi-Riemann identity.

\subsubsection*{Heterotic strings}\label{sec:Heterotic-bosonic}

After studying toroidal compactifications, we are now ready to introduce the $D=10$ Heterotic string.
As mentioned at the beginning, the Heterotic string combines the left-moving sector of the 26-dimensional bosonic string with the right-moving sector of the 10-dimensional superstring. The left-moving bosonic string is compactified on a 16-dimensional torus, where the momenta of the additional chiral bosons $X^I(\bar{z})$ take values on a 16-dimensional lattice $\Gamma_{16}$. Thus, the partition function of the Heterotic string can be expressed as
\begin{equation}\label{het-pf}
	Z^{\textrm{het}}(\tau )  = \frac 1 {\t_2^{4}\eta (q)^{12}\eta (\bar q)^{24}} \Big( -\vartheta_2^4(\tau) + \vartheta_3^4(\tau) - \vartheta_4^4(\tau)\Big)
		\sum_{\bfp_L \in \G_{16}} 	\bar q^{\frac{1}{2} \bfp_L^2}
\end{equation}
Here, $\eta(q)^{8}\eta(\bar{q})^{24}$ represents the contribution from the bosonic oscillators, the $\tau_2^{4}$ factor arises from the zero modes of the noncompact transverse coordinates, and $\vartheta^4_i/\eta(q)^{4}$ comes from the world-sheet fermions. The most interesting part of this partition function is the lattice sum
\[
P(\bar\tau):=\sum_{\bfp_L \in \G_{16}} 	\bar q^{\frac{1}{2} \bfp_L^2}
\]
In order for the partition function \eqref{het-pf} to be invariant under the modular transformation $\textrm{SL}(2,\mathbb{Z})$, the modular transformations of $\eta$ and $\vartheta_i$ reveal the following properties:
\[
T:P(\bar\t+1)=P(\bar\t)~,\qquad S:P(-1/\bar\t)=\bar\t^8P(\bar\t)~.
\]
The invariance under the $T$-transformation implies that $\mathbf{p}_L^2 \in 2\mathbb{Z}$, requiring the lattice $\Gamma_{16}$ to be \textbf{even}. For the $S$-transformation, we employ the Poisson resummation formula
\[	\sum_{\bfp\in\Lambda}e^{-\pi\a(\bfp+\bfx)^2+2\pi i \bfy\cdot (\bfp+\bfx)}=\frac{1}{\textrm{Vol}(\Lambda) \a^{\dim \Lambda /2}}\sum_{\bfq\in\Lambda^*}e^{-2\pi i \bfq\cdot \bfx-\frac{\pi}{\a}(\bfy+\bfq)^2}	\]
which amounts to
\[
P(-1/\bar\t)=\frac{\bar\t^8}{\textrm{Vol}(\G_{16})}\sum_{ \bfp_L\in(\G_{16})^*} 	\bar q^{\frac{1}{2} \bfp_L^2}~.
\]
This requires that the lattice $\G_{16}$ is \textbf{self-dual}, \textit{i.e.} $(\G_{16})^*=\G_{16}$ so that  $\textrm{Vol}(\Lambda)=1$.

It turns out that there are only two even self-dual Euclidean lattices in 16 dimensions
\begin{itemize}\setlength{\parskip}{-0.1cm}
\item the root lattice of $E_8\times E_8$
\item the weight lattice of $\Spin(32)/\bZ_2$
\end{itemize}
The metric $G_{ij}$ of the root lattice of $E_8$ is the Cartan matrix of $E_8$ \footnote{ Due to the limited scope of our discussion, we are unable to delve into exceptional Lie algebras or the classification of semi-simple Lie algebras in detail \cite{kirillov2008introduction}. However, if you would like to develop some intuition about weight and root lattices, I recommend referring to Figure 7.3, Figure 8.1, and Figure 8.2 in \cite{kirillov2008introduction} for the example of $A_2$. For a deeper understanding of the structure of $E_8$ in the context of string theory, I suggest exploring Section 6 of \cite{GSW}.}:
\begin{equation}\nonumber
\left (
\begin{smallmatrix}
 2 & -1 &  0 &  0 &  0 &  0 &  0 & 0 \\
-1 &  2 & -1&  0 &  0 &  0 &  0 & 0 \\
 0 & -1 &  2 & -1 &  0 &  0 &  0 & 0 \\
 0 &  0 & -1 &  2 & -1 &  0 &  0 & 0 \\
 0 &  0 &  0 & -1 &  2 & -1 &  0 & -1 \\
 0 &  0 &  0 &  0 & -1 &  2 & -1 & 0 \\
 0 &  0 &  0 &  0 &  0 & -1 &  2 & 0 \\
 0 &  0 & 0 &  0 &  -1 &  0 &  0 & 2
\end{smallmatrix}\right ),	\qquad\qquad	 \raisebox{-.5cm}{\includegraphics[width=5cm]{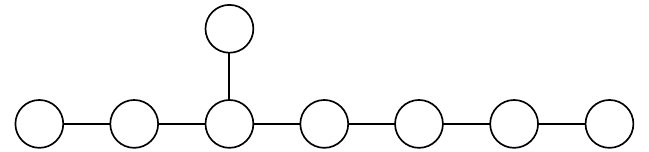}}
\end{equation}

Let us examine the massless fields in Heterotic string theory in more detail. 
As usual, there is the tachyonic vacuum of the bosonic string. At the massless level, we have oscillator excitations $\overline \a_{-1}^\mu|0\rangle$, $\overline \a_{-1}^I|0\rangle$ in the left-moving sector. The former transform like spacetime vectors while the internal oscillator excitations correspond to the left-moving part of the Abelian $\U(1)^{16}$ gauge boson. They form the \textbf{Cartan subalgebra} of $E_8 \times E_8$ or $\SO(32)$. Since the mass formula is given by 
\be 
\a' M^2 = 2(N+\overline N -2) + \mathbf{p}_L^2~,
\ee 
additional massless modes come from $\mathbf{p}_L^2=2$ for $N=1$, $\overline N=0$. Both the root lattice of $E_8\times E_8$ and the weight lattice of $\Spin(32)/\bZ_2$  contain 480 vectors of (length)$^2=2$ and generate the 496-dimensional non-Abelian gauge bosons of these groups. It is remarkable that, despite Heterotic strings being closed strings, gauge fields emerge due to the presence of the extra 16-dimensional torus. This phenomenon can be understood as a unique \textbf{stringy effect}, and the gauge groups are restricted to either $E_8 \times E_8$ or $\mathrm{SO}(32)$ for the consistency of the theory.

 Moreover, we have observed that the $\mathrm{SO}(32)$ gauge group arises in Type I string theory. This connection is not coincidental, as we will explore in \S\ref{sec:dualities}. It turns out that Type I and Heterotic $\mathrm{SO}(32)$ strings are related through \textbf{S-duality}, further highlighting the profound interplay between different string theories.

As a result, the massless spectra of Heterotic string are as follows

\vspace{.3cm}
\noindent $\bullet$ Gravitons, $B$-fields, dilaton in $D=10$
\[
\psi^\mu_{-\frac12}|0\rangle_{\textrm{NS}}\otimes\overline\alpha_{-1}^\nu|0\rangle
\]
$\bullet$  their supersymmetric partners, gravitino and dilatino
\[
|\mathbf{s}\rangle_{\textrm{R}}\otimes\overline\alpha_{-1}^\nu|0\rangle
\]
$\bullet$ 496 gauge bosons of  $E_8\times E_8$ or $\SO(32)$
\[
\psi^\mu_{-\frac12}|0\rangle_{\textrm{NS}}\otimes\overline\alpha_{-1}^I|0\rangle~,\qquad \psi^\mu_{-\frac12}|0\rangle_{\textrm{NS}}\otimes|\bfp_L^2=2\rangle
\]
$\bullet$  496 supersymmetric partners, gaugini
\[
|\mathbf{s}\rangle_{\textrm{R}}\otimes\overline\alpha_{-1}^I|0\rangle~,\qquad |\mathbf{s}\rangle_{\textrm{R}}\otimes|\bfp_L^2=2\rangle
\]
Indeed, Heterotic string theory is $D=10$ $\cN=1$ supergravity coupled to $D=10$ $\cN=1$  $E_8\times E_8$ or $\SO(32)$  super-Yang-Mills theory so that it has 16 real supersymmetric charges.

\subsection{Fermionic construction}

The 16 bosonic fields compactified on the self-dual lattice can be described by fermionic fields, which is called \textbf{fermionization}. Therefore, let us delve into the fermionic construction of Heterotic string theory.

The world-sheet action of Heterotic string theory is given by the sum of the matter and ghost actions:
\begin{align}
 S^{\textrm{m}} &= \frac{1}{4\pi} \int d^2 z\ \Big( \frac{2}{\alpha'} \partial X^\mu  \overline\partial X_\mu+\psi^\mu\overline\partial\psi_\mu+\overline \lambda^A\partial\overline\lambda_A\Big)\cr
S^{\textrm{gh}}&=\frac{1}{2\pi}\int d^2z \ (b\overline \partial c+\bar b \partial \bar c+\beta\overline \partial \g)
\end{align}
where $\mu$ represents the 10-dimensional indices and the right-moving sector is supersymmetric. For the theory to be Weyl-anomaly free, the total central charge $c^{\textrm{tot}}$ must vanish:
\[
c_R=10+5-26+11=0~,\qquad c_L=10-26+c^{\lambda}=0\ \Rightarrow\ c^\lambda=16~.
\]
Since each Majorana-Weyl anti-chiral fermion $\overline\lambda^A$ contributes $\frac12$ to the central charge, we need 32 left-moving fermions $\overline\lambda^A$ in the action.

Interestingly, there are two possible boundary conditions for the left-moving fermions $\overline\lambda_A$ that yield consistent string theories. If we impose the same boundary condition on all fermions, we obtain the Heterotic $\mathrm{SO}(32)$ gauge group. On the other hand, if we impose one boundary condition on half of the fermions and the opposite boundary condition on the other half, we obtain the Heterotic $E_8 \times E_8$ gauge group.

\subsubsection*{Heterotic $\SO(32)$ (HO)}
In the case of Heterotic $\mathrm{SO}(32)$ (HO), the same boundary condition is imposed on all left-moving fermions
\begin{align}
\overline \lambda^A(t,\sigma+2\pi)=+\overline \lambda^A
  (t,\sigma) &\qquad\qquad\textrm{ R: periodic
on cylinder}\cr
\overline \lambda^A(t,\sigma+2\pi)=-\overline \lambda^A
  (t,\sigma) &\qquad\qquad\textrm{ NS:
anti-periodic on cylinder}
\end{align}
This leads to a global $\mathrm{SO}(32)$ symmetry that rotates the $\overline\lambda^A$ fermions ($A=1,\ldots,32$). In order for the theory to be consistent, we need to impose the GSO projection on the left-moving sector. In Heterotic string theory, we select states with odd fermionic numbers in the NS sector and states with even fermionic numbers in the R sector.
\bea 
P_{\textrm{NS}}^{\textrm{H}}:=&\frac{1-(-1)^F}{2}~,\cr
P_{\textrm{R}}^{\textrm{H}}:=&\frac{1+(-1)^F}{2}~.
\eea
Additionally, we must impose the level matching condition:
\be\label{level-matching2}
N-a=\overline  N-\overline a
\ee
where the normal ordering constants in the left-moving sector are
\[
{\overline a}_{\textrm{NS}}=\frac{8}{24}+\frac{32}{48}=1~,\qquad {\overline a}_{\textrm{R}}=\frac{8}{24}-\frac{32}{24}=-1~.
\]
Here the first term comes from the left-moving bosonic field $\overline X^i$ whereas the second term depends on the boundary condition \eqref{RNS-normalordering} of $\overline\lambda^A$. As a result, the R sector contains only massive states. 

Under the GSO projection, the Tachyon state $ |0\rangle_{\textrm{NS}}$ in the NS sector is preserved, but there is no corresponding state in the right-moving sector. Therefore, it does not satisfy the level matching condition \eqref{level-matching2}, and the left-moving Tachyon is excluded from the spectrum. Then, the first excited states after the GSO projection in the NS sector are
\begin{align}
\overline \a^i_{-1} |0\rangle_{\textrm{NS}} \,&,\qquad (\mathbf{8_v},\mathbf{1}) \cr
\overline \lambda^A_{-1/2} \overline \lambda^B_{-1/2} |0\rangle_{\textrm{NS}}\,&, \qquad  (\mathbf{1},\textbf{adj}) \nonumber
\end{align}
where the bold letters denote the representations of $\mathrm{SO}(8)\times \mathrm{SO}(32)$. The adjoint representation $\textbf{adj}$ of $\mathrm{SO}(32)$ corresponds to the antisymmetric representation with dimension $32 \times 31/2 = \textbf{496}$. 

The following table summarizes the massless spectrum of Heterotic $\mathrm{SO}(32)$ (HO) theory. The first row represents the $D=10$ $\mathcal{N}=1$ supergravity multiplet, while the second row shows the $\mathcal{N}=1$ gauge multiplet in the adjoint representation of $\mathrm{SO}(32)$, as we have seen in the bosonic construction.
\begin{table}[ht] \centering
\begin{tabular}{ |c|c|c| }
 \hline
Left$\backslash $Right & ${\bf 8_v}$ & ${\bf 8_s}$ \\ \hline
 $ (\mathbf{8_v},\mathbf{1})$ & ${\bf 1} \oplus {\bf 28}  \oplus {\bf 35}$ & ${\bf 8_c}\oplus{\bf 56_s}$ \\
  & $\phi \ \ B_{\mu\nu} \ \ G_{\mu\nu}$ & $\lambda^+ \ \ \psi^-_m$ \\ \hline
  $(\mathbf{1},\textbf{496})$ & $\SO(32)$ gauge boson &  $\SO(32)$ gaugini \\
    & $A^\mu_{[A,B]}$ & $\eta_{[A,B]}$ \\
 \hline
\end{tabular}
\caption{Massless field contents of Heterotic string theory}
\label{tab:heterotic}
\end{table}

\subsubsection*{Heterotic $E_8\times E_8$ (HE)}
The second Heterotic string theory is obtained by dividing the $\overline \lambda^A$ into two sets of 16 with independent boundary conditions,
\[
\overline\lambda^A(t,\sigma+2\pi)=\left\{ \begin{matrix}\e_1 \overline\lambda^A(t,\sigma) & \quad~ A=1,\ldots,16\\ \e_2 \overline\lambda^A(t,\sigma) & \qquad A=17,\ldots,32  \end{matrix}\right.
\]
where $\e_i=\pm1$. Therefore, in the left-moving sector,  we need to take the following boundary conditions into account
\[
(\textrm{NS}_1,\textrm{NS}_2)~,\quad (\textrm{R}_1,\textrm{NS}_2)~,\quad (\textrm{NS}_1,\textrm{R}_2)~,\quad (\textrm{R}_1,\textrm{R}_2)~.
\]
Consequently, the global symmetry is broken to $\mathrm{SO}(16)_1\times\mathrm{SO}(16)_2$. The GSO projection is applied independently to the two sets of left-movers. 
We also apply for the level-matching condition \eqref{level-matching2}. The normal ordering constant in each boundary condition is
\[
{\overline a}_{\textrm{NS}_1,\textrm{NS}_2}=1~,\qquad {\overline a}_{\textrm{R}_1,\textrm{NS}_2}={\overline a}_{\textrm{NS}_1,\textrm{R}_2}=\frac{8}{24}+\frac{16}{48}-\frac{16}{24}=0~,\qquad {\overline a}_{\textrm{R}_1,\textrm{R}_2}=-1~.
\]
Similarly to Heterotic $\mathrm{SO}(32)$, the $(\textrm{R}_1,\textrm{R}_2)$ boundary condition contains only massive states. Although the Tachyon state $ |0\rangle_{\textrm{NS}_1,\textrm{NS}_2}$ in the NS sector is preserved under the GSO projection, it does not satisfy the level-matching condition \eqref{level-matching2}, so it is not present in the spectrum.

As a result, the massless states are then given by
\begin{align}
\overline \a^i_{-1} |0\rangle_{\textrm{NS}_1,\textrm{NS}_2} \,&,\qquad (\mathbf{8_v},\mathbf{1},\mathbf{1}) \cr
\overline \lambda^A_{-1/2} \overline \lambda^B_{-1/2} |0\rangle_{\textrm{NS}_1,\textrm{NS}_2}\,&, \qquad  (\mathbf{1},\textbf{adj},\mathbf{1}) \ \textrm{or} \ (\mathbf{1},\mathbf{1},\textbf{adj})  \label{massless2}
\end{align}
where the bold letters are the representations of $\SO(8)\times \SO(16)_1\times \SO(16)_2$. Note that the GSO projection requires either $1\le A,B\le16$ or $17\le A,B\le32$ in \eqref{massless2}. The adjoint representation of $\SO(16)$ is of $16\times15/2=\textbf{120}$ dimensions.

In the $(\textrm{R}_1,\textrm{NS}_2)$ and $(\textrm{NS}_1,\textrm{R}_2)$ boundary conditions, the ground states are massless due to the zero normal ordering constant. The 16 $\overline \lambda_0^A$ zero modes can be expressed in terms of 8 raising and 8 lowering operators:
\[
\overline \lambda_0^{K\pm} = 2^{-1/2}(\overline \lambda_0^{2K-1} \pm i\overline \lambda_0^{2K})~ , \qquad  K = 1,\ldots, 8 \ \textrm{or} \ K = 9,\ldots, 16~,
\]
As a result, the $2^{8}=\textbf{256}$-dimensional spinor representation of $\mathrm{SO}(16)$ becomes massless. However, the GSO projection selects the positive chirality $\textbf{128}$ out of $\textbf{256}=\textbf{128}+\textbf{128}'$ in the Ramond sector. Therefore, the ground states in the $(\textrm{R}_1,\textrm{NS}_2)$ and $(\textrm{NS}_1,\textrm{R}_2)$ sectors are $(\mathbf{1},\textbf{128},\mathbf{1})$ and $(\mathbf{1},\mathbf{1},\textbf{128})$, respectively, under $\mathrm{SO}(8)\times \mathrm{SO}(16)_1\times \mathrm{SO}(16)_2$.

All in all, the left-moving massless states form the representations of $\SO(8)\times \SO(16)_1\times \SO(16)_2$
\[
\bf (8_v,1,1) + (1,120,1) + (1,1,120) + (1,128,1) + (1,1,128)
\]
This spectrum strongly suggests that gauge symmetry is enhanced $\SO(16)\to E_8$ because the dimension of the adjoint representation $E_8$ is  $\bf 120+128=248$. In fact, the $\mathrm{SO}(16)$ subgroup of $\mathrm{E}_8$ acts on the $\mathbf{248}$ adjoint representation as $\mathbf{120+128}$. Therefore, the massless spectrum of the Heterotic string theory includes the 496-dimensional adjoint representations of both $\mathrm{SO}(32)$ and $\mathrm{E}_8 \times \mathrm{E}_8$ gauge groups.

In summary, the fermionic construction of the Heterotic string theory reproduces the massless spectrum consisting of gravity, gauge bosons, and gauginos, which is consistent with the bosonic construction.

\subsubsection*{No supersymmetric D-branes in  Heterotic strings}
We have seen that D-branes are charged under R-R fields in Type II theories. However, there is no R-R field in Heterotic string theories because there is only world-sheet supersymmetry in the right-moving sector. In other words, although the R-R $(p+2)$-form field strength $G_{(p+2)}$ in Type II theories can be expressed as
\[
G_{(p+2)} \  \longleftrightarrow\ \bar\psi^L \G^{\mu_1\cdots \mu_{p+2}} \psi^R~,
\]
there is no left-moving fermionic field $\bar\psi^L$ present. Therefore, the R-R $(p+2)$-form field strength $G_{(p+2)}$ cannot be expressed in terms of fermionic fields in Heterotic string theories.
Hence, there is no D-brane in Heterotic string theories. Consequently, Heterotic string theories are the theories of closed strings\footnote{However, Polchinski discusses open Heterotic strings in \cite{Polchinski:2005bg}.}. Despite the absence of D-branes, Heterotic string theories do feature extended objects known as NS5-branes or Heterotic fivebranes in Heterotic string theories and they are magnetically charged under the $B$-field.

\section{Supergravity}\label{sec:supergravity}

String theory includes massless states as well as massive states.
However, in the energy regions much lower than $1/\ell_s$, we do not see the stringy massive states
because the mass $M^2 \sim \frac{1}{\alpha'} \sim \frac{1}{\ell_s^2}$ is assumed to be very heavy.
Therefore, in the low-energy region, we can describe the theory by an \textbf{effective theory},
which only contains the massless particles (lightest states).
The effective theory, of course, does not contain all the information of the original theory,
however, it does give us some information about the original theory.

For the bosonic string theory, the low-energy effective action for the massless fields is given in \eqref{NLSM-general}.
Below, we deal with its supersymmetric versions, type \textbf{IIA/IIB supergravity},
that are low-energy effective theories of IIA/IIB superstring theory. In \S\ref{sec:typeI-HO}, we will deal with the low-energy effective actions of Type I and Heterotic string theory. The theory of supergravity is very broad, and we refer to \cite{freedman2012supergravity} for more detail.

\subsection{Local supersymmetry}

Before going to Type IIA/IIB supergravity,
let us see the basic idea of supergravity.

Roughly speaking, a global supersymmetry algebra takes the form
\be
\{\e_2 Q, \overline \e_1 \overline Q\}=i \overline\e_1 \Gamma^\mu \e_2 P_\mu~.
\ee
(For instance, see \eqref{susy-algebra}.) The key idea in supergravity is that supersymmetry holds locally so that the spinor parameters $\epsilon$ are arbitrary functions of the spacetime coordinates $\epsilon \rightarrow \epsilon(x)$. The supersymmetry algebra will then involve local translation parameters $\bar{\epsilon}_{1} \gamma^{\mu} \epsilon_{2}$ which must be viewed as diffeomorphisms. Thus, local supersymmetry requires gravity. Note that we use $x^\mu$ for local coordinates of the spacetime instead of $X^\mu$, which has been adopted in the previous sections.

First of all, in order to define spinors in a curved space,
we need to introduce vielbeins $e_\mu^a(x^\rho)$. The metric can be the orthonormal one at one point by coordinate transformations, and  vielbeins transform a local coordinate $x^\mu$
into the orthonormal coordinate at the point:
\begin{align*}
 x^a = e^a_\mu x^\mu \ , \quad x^\mu = e^\mu_a x^a \ ,
\end{align*}
which is defined through spacetime metric $G_{MN}$ by
\begin{align*}
 G_{\mu\nu}(x^\rho) = \eta_{ab} e_\mu^a(x^\rho) e_\nu^b(x^\rho)  \ .
\end{align*}
Now we can use spinor representations and gamma matrices thanks to the vielbeins:
\begin{align*}
 \{\Gamma^a,\Gamma^b\} = 2\eta^{ab} \ .
\end{align*}

If we include gravity in a theory, the translational invariance is promoted to invariance under general coordinate transformations.
Similarly, global Lorentz symmetry is promoted to local Lorentz symmetry (this is done by the vielbeins):
\begin{align*}
 \Lambda^a_{\ b} e_a^\mu(x^\rho) e_\nu^b(x^\rho) \equiv \Lambda^\mu_{\ \nu}(x^\rho) \ .
\end{align*}
In the orthonormal basis, the covariant derivative is expressed by the spin connection
\bea 
D_\mu V^a=&\partial_\mu V^a+\omega_\mu{ }^a{ }_b V^b \cr 
D_\mu \psi=&\left(\partial_\mu+\frac{1}{4} \omega_\mu^{a b} \Gamma_{a b}\right) \psi
\eea

Gravity can be interpreted as the gauge field for general coordinate transformations.
Since the gauge transformation of a spin-1 gauge field can be generally written as
\begin{align}
 \delta A_\mu = D_\mu \lambda \ ,
\end{align}
the gauge transformation of the gravity field will take the form
\begin{align}
 \delta \omega_\mu{}^{a}{}_{b} = D_\mu \Theta{}^{a}{}_{b} \ ,
\end{align}
where $\Theta^a{}_{b} $ is a general coordinate transformation, and $D_\mu$ is a covariant derivative.

Similarly, we need to introduce a gauge field for local supersymmetry in supergravity, and the corresponding field is a \textbf{gravitino}. (Recall that gravitino appears as a massless field in superstring theory.)
Then, the supersymmetry transformations for the gravitino and vielbeins are given by
\begin{align}\label{local-susy}
 \delta \psi_\mu^\alpha = D_\mu \epsilon^\alpha \ ,\qquad \delta_\epsilon e^a_\mu = i\ol\epsilon \Gamma^a \psi_\mu~.
\end{align}
Then, the square of the supersymmetry transformation is as we want:
\bea
\left[\delta_{1}, \delta_{2}\right] e_{\mu}^{a}=&iD_\mu ( \bar{\epsilon}_{1} \Gamma^{a} \epsilon_{2})  \cr
\left[\delta_{1}, \delta_{2}\right] \psi_{\mu}=&i\bar{\epsilon}_{1} \Gamma^{\rho} \epsilon_{2}(D_{\rho} \psi_{\mu}-D_\mu \psi_\rho)~.
\eea

Supergravity theory always includes the two gauge fields, $e^a_\mu$ and $\psi^\alpha_\mu$,
and the action is given by
\begin{align}
 S = \frac{1}{2\kappa_D^2}\int d^D x\ e\ \left[ R -2i \psi_\mu \Gamma^{\m\n\rho} D_\nu \psi_\rho\right] \ ,
\end{align}
where $e = \det e_\mu^a$. One can check that the action is invariant under local supersymmetry transformation \eqref{local-susy} to leading order in the fermions. This action can be regarded as the supersymmetric version of the Einstein-Hilbert action.

\subsection{\texorpdfstring{$D=11$}{D=11} supergravity}

Supersymmetry puts a strong constraint on the spacetime dimension.
If we limit ourselves to consider fields up to spin-$2$,
then it is known that the highest dimension is $D=11$ \cite{Nahm:1977tg}.
Roughly speaking, this is because the degrees of freedom for fermions grow exponentially as $2^{[D/2]}$ whereas
those of bosons grow by the power law of $D$
(for instance, $\frac{(D-1)(D-2)}{2}-1$ for gravitons).
Therefore, to balance fermions and bosons with spin less than two, we cannot go arbitrarily higher.

The $D=11$ supergravity \cite{Cremmer:1978km} consists of three fields;
\begin{itemize}
  \item one is the graviton $G_{MN}$ (44 states)
\item three-form field $M_{(3)}$ (84 states)
  \item    gravitino $\psi_M$ (128 states)
\end{itemize}
We can see that the numbers of fermions and bosons are balanced.
Although the existence of fermions is crucial,
we write the bosonic part of the action of the $D=11$ supergravity:
\begin{align}
 2\kappa_{11}^2 S_{11} = \int d^{11}x \sqrt{ -G} \left[R -\frac{1}{2} K_{(4)}^{2}\right]
 -\frac{1}{6}\int d^{11}x\ M_{(3)} \wedge K_{(4)} \wedge K_{(4)} \ ,\label{11dsugra}
\end{align}
where $K_{(4)} = d M_{(3)}$ is the field strength,
and $K_{(4)}^{2} = K_{(4)} \wedge * K_{(4)}$. The fermionic part of the action follows from supersymmetry in principle.
Note that there is only one parameter $\kappa_{11}$ defined as
\be\label{11dcoupling}\frac{1}{2\kappa_{11}^2} = \frac{2\pi}{(2\pi \ell_p)^9}~,\ee
which is written in terms of Planck length
\[\ell_{\mathrm{P}}=\sqrt{\frac{\hbar G}{c^{3}}}~.\]

The critical dimension of superstring theory is $D=10$. Nonetheless, Type IIA $D=10$ supergravity can be obtained from the dimensional reduction of the $D=11$ supergravity on $S^1$, as we see below. 
In fact, about the $D=11$ supergravity, it was mentioned in \cite[\S13.1.1]{GSW} as follows:
\begin{quote}
The ten-dimensional supergravity theories are, of course, the low-energy limits of certain string theories. Further dimensional reduction can give a variety of supergravity theories in various dimensions less than ten. If string theory proves to be correct, it could be regarded as 'explaining' the existence of supergravity theories for $D \le 10$.
    Eleven-dimensional supergravity remains an enigma. It is hard to believe that its existence is just an accident, but it is difficult at the present time to state a compelling conjecture for what its role may be in the scheme of things.
\end{quote} 
The crucial understanding of this enigma was provided in \cite{Witten:1995ex} where the $D=11$ supergravity indeed gives an important clue to the strong coupling behavior of superstring theory.
%

\subsection{\texorpdfstring{$D=10$}{D=10} IIA supergravity}

Now we compactify the $D=11$ supergravity on $S^1$ of radius $R$. In the following, $C_{(p+1)}$ is the R-R $(p+1)$-form and $B_{(2)}$ is the $B$-field. Also, $G_{(p+2)}$ and $H_{(3)}$ are their field strengths. On the reduction,
the three-form field $M_{(3)}$ decomposes into the massless fields of Type IIA string theory in Table \ref{tab:masslessII}:
 \be M_{(3)} = C_{(3)} +B_{(2)} \wedge d\theta \ , \quad K_{(4)} = \wh G_{(4)} +H_{(3)} \wedge (d\theta +C_{(1)}) \ ,\ee
 where
 \be
\wh G_{(4)} = G_{(4)} -C_{(1)} \wedge H_{(3)}~.
\ee
Like \eqref{KK-metric}, the metric takes the form
\begin{align*}
 ds_{11}^2 = G_{MN} dx^M dx^N = G_{\mu\nu} dx^\mu dx^\nu +R^2 (d\theta+C_{(1)})^2 \ ,
\end{align*}
where the compactified $S^1$ direction is denoted as $\theta$.

Now we rewrite the action \eqref{11dsugra} in terms of the massless fields of Type IIA theory. (See the left of Table \ref{tab:masslessII}.) However, we want to obtain the action of the form \eqref{bosonic-NS} in the NS-NS sector. For this purpose, we consider that the radius of the circle depends on the spacetime coordinates, and it can be written in terms of the dilaton field as
\be\label{11d-radius}
R=\ell_p e^{\frac23\Phi}~.
\ee
Then, the action \eqref{11dsugra} is written as
\begin{align}\label{IIA-SUGRA}
 &S_\mathrm{A,NS} = \frac{1}{2\kappa_{10}^2} \int d^{10}x \sqrt{ -G} e^{-2\Phi} \left[
 R +4 \partial_\mu \Phi \partial^\mu \Phi -\frac{1}{2} H_{(3)}^{2} \right] \ ,  \cr
 &S_\mathrm{A,R} = -\frac{1}{4\kappa_{10}^2} \int d^{10}x \sqrt{ -G} \left[
G_{(2)}^{2} +\wh G_{(4)}^{2} \right] \ ,  \cr
 &S_\mathrm{A,CS} = -\frac{1}{4\kappa_{10}^2} \int B_{(2)} \wedge G_{(4)} \wedge G_{(4)} \ ,
\end{align}
where the coupling constants are related by
\be\label{kappa-11-10}
 \frac{2\pi R}{2\kappa_{11}^2} = \frac{e^{-2\Phi}}{2\kappa_{10}^2}~.
 \ee
Since it is the low-energy effective action of Type IIA string theory, the dimensional analysis of the coupling constant yields
\be
\frac{1}{2\kappa_{10}^2} = \frac{2\pi}{(2\pi \ell_s)^8}~.
\ee
Recalling that an expectation value of the dilaton is the string coupling $g_s=e^\Phi$, \eqref{11d-radius} gives
\be
 \left(\frac{R}{\ell_p}\right)^3 = g_s^2~. \ee
Also,
\eqref{kappa-11-10} provides the relation
\be
\frac{R}{\ell_p^3} = \frac{1}{\ell_s^2}\ee
Combining these two relations, we have
\be\label{Mcircle-radius} R = g_s \ell_s~.\ee

\subsection{\texorpdfstring{$D=10$}{D=10} IIB supergravity}\label{sec:IIB-SUGRA}

As seen in \S\ref{sec:Tdual}, Type IIA and IIB are T-dual to each other so that the action of Type IIB supergravity is consistent with the T-duality. The right of Table \ref{tab:masslessII} lists all the massless fields, and the supergravity action indeed takes a very similar form
\begin{align}\label{IIB-SUGRA}
 &S_\mathrm{B,NS} = \frac{1}{2\kappa_{10}^2} \int d^{10}x \sqrt{ -G}  e^{-2\Phi} \left[
 R +4 \partial_\mu \Phi \partial^\mu \Phi -\frac{1}{2} H_{(3)}^{2} \right] \ ,  \cr
 &S_\mathrm{B,R} = -\frac{1}{4\kappa_{10}^2} \int d^{10}x \sqrt{ -G} \left[
 G_{(1)}^{2}+ \wh G_{(3)}^{2} +\frac{1}{2} \wh G_{(5)}^{2} \right] \ ,  \cr
 &S_\mathrm{B,CS} = -\frac{1}{4\kappa_{10}^2} \int C_{(4)} \wedge H_{(3)} \wedge G_{(3)} \ ,
\end{align}
where
\bea
\wh G_{(3)} &= G_{(3)} -C_{(0)} H_{(3)}~,\cr
\wh G_{(5)} &=G_{(5)} -\frac{1}{2}C_{(2)} \wedge H_{(3)} +\frac{1}{2} B_{(2)} \wedge G_{(3)}~.
\eea
Note that the action for the NS sector is the same as that of Type IIA.

As in \eqref{SD-5form}, Type IIB supergravity has the self-dual 5-form $G_{(5)}$. In the notation here,
the self-dual condition is \[* \wh G_{(5)} = \wh G_{(5)}~,\]
which does not follow from the action.
In fact, if the field satisfies the self-dual condition, its kinetic action becomes trivial
\[
\int \wh G_{(5)}\wedge \ast \wh G_{(5)}=0~.
\]
Therefore, the self-duality condition needs to be imposed by hand.

As we will see below, Type IIB supergravity exhibits a hidden symmetry that manifests as the $S$-duality in Type IIB string theory. Moreover, Type IIB supergravity is subject to two types of $\mathbb{Z}_2$ actions. The first acts on the bosonic fields as follows:
\be\label{Z2-type1}
G_{(1,5)} \to -G_{(1,5)}, \quad H_{(3)} \to -H_{(3)}~.
\ee
The second $\mathbb{Z}_2$ action changes the sign of all the R-R field strengths:
\be\label{Z2-Heterotic}
G_{\textrm{odd}} \to -G_{\textrm{odd}}~.
\ee
Both $\mathbb{Z}_2$ actions identify two copies of the dilatino and gravitino in Type IIB theory (see Table \ref{tab:masslessII}). The first $\mathbb{Z}_2$ projection results in Type I supergravity, while the second leads to Heterotic supergravity.

\section{String dualities}\label{sec:dualities}

We have introduced superstring theories of five types: Type IIA, IIB, I, and Heterotic $\SO(32)$ and $E_8\times E_8$.
In fact, the seminal paper \cite{Witten:1995ex} reveals that these string theories are related by dualities, which led to the second string revolution. Indeed, we have already learned

\vspace{.3cm}
\noindent $\bullet$ Type IIA and IIB are T-dual to each other

\vspace{.3cm}
\noindent $\bullet$ Type I is the orientifold projection of Type IIB

\vspace{.3cm}
\noindent In this section, we will learn string dualities extensively studied in the second string revolution after \cite{Witten:1995ex}.

\vspace{.3cm}
\noindent $\bullet$  Type IIB has $\SL(2,\bZ)$ symmetry so that it is self-dual under S-duality

\vspace{.3cm}
\noindent $\bullet$  Type I to Heterotic $\SO(32)$ is S-dual

\vspace{.3cm}
\noindent $\bullet$  Heterotic $\SO(32)$ and $E_8\times E_8$ are T-dual to each other

\vspace{.3cm}
\noindent $\bullet$  The strong coupling regime of Type IIA is described by M-theory on $S^1$

\vspace{.3cm}
\noindent $\bullet$ The strong coupling regime of  Heterotic $E_8\times E_8$ is described by M-theory on $S^1/\bZ_2$

\vspace{.3cm}
\noindent $\bullet$  Heterotic string on $T^4$ is dual to Type IIA on K3

\begin{figure}[ht]\centering
\includegraphics[width=\textwidth]{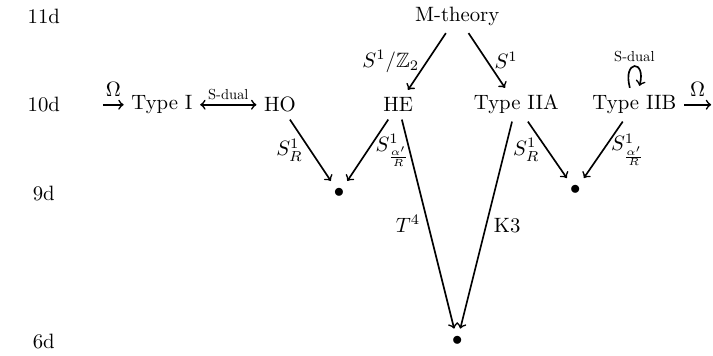}
\caption{Duality web of string theory}
\end{figure}

Even for these dualities, we can cover only key points in this section. More details can be found in  \cite{Polchinski,BBS}. Moreover, we just see the tip of the iceberg, and there are much more string dualities. Thus, we refer to good reviews  \cite{Aspinwall:1996mn,Forste:1996yd,Ooguri:1996ik,Polchinski:1996nb,Polchinski:1996na,Townsend:1996xj,Schwarz:1996bh,Sen:1996yy,Dijkgraaf:1997ip,Vafa:1997pm,Sen:1998kr,Johnson:2000ch} written during the second string revolution for this rich subject. All in all, these dualities tell us that quantum strings somehow see geometry from drastically different viewpoints. I hope you will get some feeling of it in this section.

\subsection{S-duality of Type IIB supergravity}\label{sec:Sdual}

The electromagnetic duality seen in \eqref{Maxwell} and \eqref{Maxwell2} is a basic example of duality. The generalization to Yang-Mills theory was proposed by Goddard-Nuyts-Olive \cite{Goddard:1976qe} and Montonen-Olive \cite{Montonen:1977sn}. The Montonen-Olive duality exchanges the Yang-Mills coupling by
\be
g_{\textrm{YM}}\leftrightarrow 1/g_{\textrm{YM}}~,
\ee
and it also exchanges a fundamental particle into a soliton. Thus, this duality is also called the \textbf{strong-weak duality} or \textbf{$S$-duality}.
Although standard techniques in quantum field theory cannot be applied to the strong coupling regime, the $S$-duality provides new insights into non-perturbative dynamics in the strong-coupling regime. One of the reasons why string theory sheds new light on physical theories is that the strong-weak dualities show up in string theory \cite{Sen:1994fa,Duff:1994zt}, which often have geometric origins.

In fact, Type IIB string theory enjoys the $S$-duality. To see it,
let us rewrite the action by rescaling the metric $G_{E,\mu\nu} = e^{-\Phi/2} G_{\mu\nu}$, and by introducing
\be\label{axiodilaton}\tau = C_{(0)} +ie^{-\Phi}~,\ee
and
\begin{align*}
 \mathbb M = \frac{1}{\Im \tau}
 \begin{pmatrix}
  |\tau|^2 & -\Re \tau \\ -\Re \tau & 1
 \end{pmatrix} \ , \qquad
 \mathbb F_{(3)} =
 \begin{pmatrix}
  H_{(3)} \\ G_{(3)}
 \end{pmatrix} \ ,
\end{align*}
Then, the action \eqref{IIB-SUGRA} is rewritten as
\begin{align}
 S_\mathrm{B} =& \frac{1}{2\kappa_{10}^2} \int d^{10}x \sqrt{ -G_E} \left[
 R_E - \frac{\partial_\mu \tau \partial^\mu \ol\tau}{2(\Im \tau)^2}
 -\frac{1}{2} \mathbb F_{(3)}^T \cdot \mathbb M \cdot \mathbb F_{(3)}-\frac{1}{4} \wt G_{(5)}^2
 \right]\cr
&  -\frac{1}{4\kappa_{10}^2} \int \epsilon_{ij} C_{(4)} \wedge \mathbb F_{(3)}^i \wedge  \mathbb F_{(3)}^j \ ,
\end{align}
where $\epsilon = \begin{pmatrix} 0 & 1 \\ -1 & 0 \end{pmatrix}$. Remarkably,
this action is invariant under the $\SL(2,\mathbb R)$ transformation:
\begin{align*}
 \tau' = \frac{a\tau +b}{c\tau +d} \ , \quad
 \mathbb M' = (\Lambda^{-1})^T \mathbb M \Lambda^{-1} \ , \quad
 \mathbb F_{(3)}' = \Lambda \mathbb F_{(3)} \ , \quad
 \Lambda =
 \begin{pmatrix}
  d & c \\ b & a
 \end{pmatrix}\in \SL(2,\bR)\ .
\end{align*}
On the other hand, the metric $G_E$ and $C_{(4)}$ are invariant under the transformation.

Let us consider its physical implications.
Since  $C_{(4)}$ is invariant, a D$3$-brane is invariant. On the other hand, since the  2-form fields $B_{(2)}$ and $C_{(2)}$ are transformed by $\SL(2,\bR)$, the extended objects coupled to these fields are transformed accordingly. Namely, F$1$ and D$1$-branes are electrically coupled to the 2-form fields while NS$5$ and D$5$-branes are magnetically coupled to them, respectively.
Thus, they are transformed under $\SL(2,\bR)$ as
\be  \begin{pmatrix}
  \mathrm{F1}' & \mathrm{D1}'
 \end{pmatrix} =
 \begin{pmatrix}
  \mathrm{F1} & \mathrm{D1}
 \end{pmatrix} \Lambda^{-1} ~,\qquad
\begin{pmatrix}
 \mathrm{NS5}' \\ \mathrm{D5}'
\end{pmatrix} = \Lambda
\begin{pmatrix}
 \mathrm{NS5} \\ \mathrm{D5}
\end{pmatrix} \ ,\qquad \textrm{as}\quad    \begin{pmatrix}
  B_{(2)}' \\ C_{(2)}'
  \end{pmatrix} = \Lambda
  \begin{pmatrix}
  B_{(2)} \\ C_{(2)}
 \end{pmatrix}\nonumber
\ee
Due to the Dirac quantization condition, the electric and magnetic charges of D-branes
must be integers so that the true symmetry in Type IIB string theory is indeed $\SL(2,\ZZ)$ \cite{Hull:1994ys}.

In particular, if we choose \[\Lambda = S =
\begin{pmatrix}
 0 & 1 \\ -1 & 0
\end{pmatrix}~,\]
the fields transform as
$\tau \leftrightarrow -1/\tau$, and the extended objects are exchanged as \[\mathrm{F1} \leftrightarrow \mathrm{D1}~, \qquad \mathrm{NS5} \leftrightarrow \mathrm{D5}~.\]
Assuming that $\langle C_0 \rangle = 0$, this results in the S-duality of the string coupling constant, $g_s \leftrightarrow 1/g_s$.
Other elements of $\SL(2,\ZZ)$ lead to infinitely many bound states of F$1$ and D$1$: $(p,q)$-string,
and those of NS$5$ and D$5$: $(p,q)$ 5-branes.

The combination \eqref{axiodilaton} of the fields behaves as a complex structure of a torus and the theory enjoys the modular transformations
\eqref{eq:ModularTra} and \eqref{eq:ModularTra2} of a torus. Hence, a $D=12$ dimensional theory, called \textbf{F-theory}, is proposed in \cite{Vafa:1996xn,Morrison:1996na,Morrison:1996pp} by considering a torus fibration over ten-dimensional spacetime of Type IIB theory, which can be interpreted as the geometric origin of the field combination \eqref{axiodilaton}.

\subsection{S--duality between Type I and Heterotic \texorpdfstring{$\SO(32)$}{SO(32)} }\label{sec:typeI-HO}

Now, let us explore the $S$-duality between Type I theory and Heterotic $\SO(32)$ theory \cite{Witten:1995ex,Dabholkar:1995ep,Hull:1995nu,Polchinski:1995df} from the perspective of their low-energy effective actions.

As seen in \eqref{massless-type1}, the massless fields in Type I string theory are obtained from Type IIB by the orientifold projection, which projects out $C_{(0)}$, $B_{(2)}$, and $C_{(4)}$. This orientifold projection reduces the amount of supersymmetry by half, resulting in a total of 16 supercharges in Type I. Consequently, the gravity part of the Type I supergravity action can be obtained by removing the part of $G_{(1)}$, $H_{(3)}$ and $G_{(5)}$ from \eqref{IIB-SUGRA}. Additionally, $\SO(32)$ gauge fields with appropriate dilaton dependence are included:
\begin{align}\label{TypeI}
S_{\textrm{I}}&=S_{\textrm{grav}}+S_{\textrm{YM}}\cr
S_{\textrm{grav}}&= \frac{1}{2\kappa_{10}^2} \int d^{10}x \sqrt{ -G}   \,\left[
e^{-2\Phi}( R +4 \partial_\mu \Phi \partial^\mu \Phi )-\frac{1}{2}| \wt G_{(3)}|^{2} \right] \cr
S_{\textrm{YM}}&=- \frac{1}{2g_{10}^2} \int d^{10}x \sqrt{ -G}\,  e^{-\Phi} \Tr  |F_{(2)}|^2
\end{align}
where $F_{(2)}$ is the $\SO(32)$ field strength and the trace is in the vector representation, $\Tr_{\rm adj}=30\,\Tr_{\square}$ (see \S\ref{sec:GS}). Here $G_{(3)}$ is the field strength of the R-R 2-form $C_{(2)}$
\be \label{gauge-inv-G3}
\wt G_{(3)}=dC_{(2)}-\frac{\kappa_{10}^2}{g_{10}^2}\omega_3
\ee
with  the Chern-Simons 3-form
\be \label{CS}
\omega_3=\Tr \Big(AdA+\frac{2}3 A^3\Big)-\Tr \Big(\omega d\omega+\frac{2}3 \omega^3\Big)~.
\ee 
The gauge coupling constant $g_{10}$ and the gravitational constant $\kappa_{10}$ are related by $\kappa_{10}^2/g_{10}^2 = \a'/4$, which is determined by anomaly cancellation. Under the gauge transformation $\delta A=d\lambda+[A,\lambda]$ and diffeomorphism $\delta \omega=d\Theta+[\omega,\Theta]$, the Chern-Simons term transforms as
\be \label{variation-CS}
\delta \omega_3=d \Tr(\lambda dA)-d \Tr(\Theta dA)
\ee 
up to a topological term.
Hence, the $C_{(2)}$ field is sensitive to the gauge transformation as
\be \label{variation-C2}
\d C_{(2)} = \frac{\kappa_{10}^2}{g_{10}^2} [d \Tr(\lambda dA)-d \Tr(\Theta dA)]~.
\ee

The difference between the massless fields in Type I and Heterotic strings involves the exchange between $C_{(2)}$ and $B_{(2)}$ (see Table \ref{tab:heterotic}). As mentioned earlier, $S$-duality in Type IIB exchanges $C_{(2)}$ and $B_{(2)}$. Therefore, it is natural to expect that Type I and Heterotic $\SO(32)$ (HO) are related by $S$-duality. However, in the absence of open strings or R-R fields, the dilaton dependence should be $e^{-2\Phi}$ throughout:
\begin{align}\label{Het}
S_{\textrm{Het}}&=S_{\textrm{grav}}+S_{\textrm{YM}}\cr
S_{\textrm{grav}}&= \frac{1}{2\kappa_{10}^2} \int d^{10}x \sqrt{ -G}   \,e^{-2\Phi}\left[
 R +4 \partial_\mu \Phi \partial^\mu \Phi -\frac{1}{2}| \wt H_{(3)}|^{2} \right] \cr
S_{\textrm{YM}}&=- \frac{1}{2g_{10}^2} \int d^{10}x \sqrt{ -G}\,  e^{-2\Phi} \Tr  |F_{(2)}|^2
\end{align}
where the 3-form $\wt H_3$ is the field strength of the $B$-field equipped with Chern-Simons form
\[
\wt H_{(3)}=dB_{(2)}-\frac{\kappa_{10}^2}{g_{10}^2}\omega_3~.
\]

Indeed the low-energy effective actions of Type I \eqref{TypeI} and Heterotic $\SO(32)$ \eqref{Het} are related by the following exchange
\begin{align}\label{TypeI-Het}
G_{\m\n}^{I} \leftrightarrow e^{-\Phi^{H}} G_{\m\n}^{H} ~,&\qquad  \Phi^I \leftrightarrow -\Phi^{H} \cr
\wt G_{(3)}^I \leftrightarrow \wt  H_{(3)}^{H}~ ,&\qquad  A^I \leftrightarrow A^{H} ~.
\end{align}
 Recalling that the vacuum expectation value of the dilaton is the string coupling $g_s=e^{\Phi}$, we see that the strong coupling limit of one theory is related to the weak coupling limit of the other theory and vice versa.

In Type I theory, D1-branes and
D5-branes are electrically and magnetically charged under $C_{(2)}$, respectively. In Heterotic $\SO(32)$ theory, fundamental strings and NS5-branes are electrically and magnetically charged under $B_{(2)}$, respectively. The S-duality
maps them as \cite{Polchinski:1995df}
\begin{table}[ht]\centering
\begin{tabular}{ccc}
Type I&$\leftrightarrow$& Heterotic SO(32)\\ \hline
D1-branes&$\leftrightarrow$&F-strings\\
D5-branes&$\leftrightarrow$&NS5-branes
\end{tabular}\end{table}

One can provide another evidence of this duality by looking at the massless spectrum. We have seen that Heterotic $\SO(32)$ has massless fields:
\begin{enumerate}\setlength{\parskip}{-0.1cm}
\item $\bf 8_v$ of SO(8): bosonic right-moving  $X^i(z)$
\item $\bf 8_c$ of SO(8):   fermionic right-moving $\psi^i(z)$
\item $\bf 32$ of SO(32):   left-moving Majorana-Weyl fermion $\bar\lambda^a(\bar z)$
\end{enumerate}
Correspondingly, one can see the massless BPS excitations from D1-strings stretched in the $x_1$-direction in Type I theory (exercise):
\begin{enumerate}\setlength{\parskip}{-0.1cm}
\item $\bf 8_v$ of SO(8): normal bosonic excitations of D1-D1 strings
\item $\bf 8_c$ of SO(8):  right-moving fermionic  excitations of D1-D1 strings
\item $\bf 32$ of SO(32):   left-moving fermionic  excitations of D1-D9 strings
\end{enumerate}

Further evidence of
this duality has been assembled by comparing tensions, $F_{(2)}^4$ interactions, and so on \cite[\S14.3]{Polchinski}.

\subsection{Heterotic T-duality} \label{sec:HeteroticT}

Let us consider the T-duality in Heterotic strings on a circle $S^1$ in \cite{Narain:1985jj,Narain:1986am,Ginsparg:1986bx}. In the bosonic construction, the bosonic left-moving sector is compactified on an even self-dual Euclidean lattice of 16-dimensions. There are only two such lattices: the weight lattice $\G_{\SO(32)}$ of $\SO(32)$ and the root lattice $\G_{E_8}\oplus \G_{E_8}$ of $E_8\times E_8$ as we have seen in \S\ref{sec:Heterotic-bosonic}.

One may also describe the compactification on a circle $S^1$ in terms of lattices. As we have seen, the left-moving and right-moving momenta of a compactified boson take values on the lattice $\Gamma^{1,1}$ of the Lorentzian signature. Hence, the circle compactification results in adding ($\oplus$) the lattice $\Gamma^{1,1}$ to the original lattice.

It is a useful mathematical fact that for Lorentzian
lattices, there is
only one even unimodular Lorentzian lattice for each rank. Therefore, the theorem implies
\[\G_{\SO(32)}\oplus \Gamma^{1,1} \cong\G^{1,17}
\cong  \G_{E_8}\oplus \G_{E_8}\oplus \Gamma^{1,1}~.\]
Together with the metric $G$ and the $B$-field,
they parameterize the
moduli space
\begin{equation}\label{eq:Mhettoroidal}
{\cal M}=\left. \frac{\mathrm{O}(1,17)}{\mathrm{O}(1)\times \mathrm{O}(17)}\right/ \mathrm{O}(1,17; \bZ) ,
\end{equation}
where $\mathrm{O}(1,17; \bZ)$ is the T-duality group. It is called \textbf{Narain moduli space}.
Different points in the moduli space correspond to physically distinct compactifications, e.g. the gauge groups can be different, although always of rank 18. At generic points, it is $\U(1)^{18}$ that corresponds to the fact that Wilson loops generically break the 10d gauge group to $\U(1)^{18}$.

However, there are special subspaces of the moduli space where it is enhanced.
This moduli space has exactly two asymptotic boundary points, one associated to the
decomposition $\G^{1,17}\cong \G_{E_8}\oplus \G_{E_8}\oplus\G^{1,1}$,
and the other to the decomposition $\G^{1,17}
\cong \G_{\SO(32)}\oplus\G^{1,1}$.
Therefore, these boundary points are associated to HE and HO string, or large and small radii.
T-duality will relate these boundary points.

In fact, starting from either Heterotic theory, there is a simple choice of
Wilson line which breaks the gauge group to
$\SO(16)\times \SO(16) \times \U(1)\times  \U(1)$.   If we leave this
group unbroken, then the only remaining parameter is the radius. An analysis of the massive states shows
that if we map $R\to 1/R$ while exchanging KK momenta
and winding modes, then the two Heterotic theories are exchanged \cite[\S11.6]{Polchinski}.

More generally, upon a compactification of Heterotic strings on a $d$-dimensional torus $T^d$, momenta take the values on an even self-dual Lorentzian lattice $\G^{d,d+16}$. Therefore, the Narain moduli space becomes
\be\label{Narain}
{\cal M}=\left. \frac{\mathrm{O}(d,d+16)}{\mathrm{O}(d)\times \mathrm{O}(d+16)}\right/ \mathrm{O}\left(d,d+16; \bZ\right)~.
\ee
which is  $d(d+16)$-dimensional.

\subsection{M-theory}

We would also like to understand the strong coupling region of Type IIA theory.
As we obtain Type IIA supergravity by compactification of the $D=11$ supergravity on $S^1$, the radius is proportional to the string coupling constant  \eqref{Mcircle-radius} in the unit of the string length.
This suggests that in the strong coupling region $g_s\gg1$ of Type IIA string theory, another spacetime direction
emerges (dimensional decompactification).
Although the strong coupling behavior may look rather bizarre, it turns out that a number of phenomena in string theory can be explained ``more naturally'' from the  $D=11$ strongly-coupled Type IIA ``string theory''.
Witten calls the $D=11$ ``string theory'' \textbf{M-theory} where M stands for Magic, Mystery, or Membrane, according to him. Thus, the $D=11$ supergravity is the low-energy effective theory of M-theory.

Since the $D=11$ supergravity is endowed with the 3-form fields $M_{(3)}$,
there must be corresponding objects coupled to the field electrically and magnetically.
They are called \textbf{M2-brane} and \textbf{M5-brane},
which are $(1+2)$- and $(1+5)$-dimensional objects, respectively.
Let us assume that they have the following tensions and charges
\begin{align*}
 T_{M2} = \mu_{M2} = \frac{2\pi}{(2\pi \ell_p)^3} \ , \qquad
 T_{M5} = \mu_{M5} = \frac{2\pi}{(2\pi \ell_p)^6} \ .
\end{align*}

The compactification of M-theory on a circle $S^1$ leads to
the IIA superstring theory. Since it is compactified on $S^1$, the momenta of a particle along $S^1$ are quantized as $n/R$ $(n\in \bZ)$, and it can be identified with D0-branes. An M2-brane on $S^1$ becomes a fundamental string while an M2-brane in $D=10$ reduces to a D2-brane. Similarly, an M5-brane on $S^1$ becomes a D4-brane whereas an M5-brane in $D=10$ reduces to an NS5-brane. A D6-brane arises geometrically as a Kaluza-Klein monopole, which we do not deal with in this lecture note. (See \cite[\S15.2]{johnson2002d} for instance.)
Let us see this by comparing the branes and their tensions in Table~\ref{table:tension}. One can convince oneself that the tensions of extended objects perfectly agree. Note that the tensions and charges of D-branes in Type IIA theory are related by the string coupling constant
\be
g_sT_{\textrm{D}p}^\textrm{eff}=\mu_p~.
\ee
In fact, charges of a D$p$ and D$(6-p)$-brane can be read off from Table~\ref{table:tension}, and it is easy to check that they are subject to the Dirac quantization condition \eqref{Dirac-quantization}. We will come back to this relation in \S\ref{sec:DBI}.

Note that M-theory is \emph{not} even defined in a sense that we do not know how to quantize
the M-branes. For instance, the world-volume theory on M5-branes is endowed with $\operatorname{OSp}(2,6|2)$ superconformal symmetry, called 6d $\cN=(2,0)$ superconformal field theory \cite{Witten:1995zh}. The field contents are as follows
\begin{itemize}\setlength{\parskip}{0.0cm}
 \item 2-form tensor field $B$ with self-dual field strength $dB=H=-* H$
\item spinors $\psi_{\alpha,a}$ with $\psi_{\alpha,a}=J_{\alpha \beta} J_{a b} \bar{\psi}^{\beta,b}$ ($\alpha,a=1,2,3,4$)
\item 5 scalars $\phi_{i}$ $(i=1,\ldots,5)$
\end{itemize}
The source for the tensor field is an M2-brane ending on the M5-brane, called self-dual string, $*dH=J$. However,
we do not know how to write down a Lagrangian of the theory for nonabelian cases because of the self-dual 3-form $H$:
\begin{equation}
  \frac{1}{g^2}\int d^6x~ H\wedge *H = - \frac{1}{g^2}\int d^6x ~H\wedge H=0~.
\end{equation}
It is believed that 6d $\cN=(2,0)$ SCFTs are classified by $A_{n}, D_{n}, E_{6,7,8}$ root system.

Nevertheless, the existence of such theories tells us a lot.
Especially, even though no effective description of M$5$-branes in flat space is known,
M$5$-branes wrapped on manifolds give rise to surprising dualities between
a $d$-dim topological theory and $(6-d)$-dim supersymmetric gauge theories.
A salient example is the AGT relation \cite{Alday:2009aq} where M$5$-branes are wrapped on a Riemann surface. Also, the highest dimension of SCFTs is $D=6$ \cite{Nahm:1977tg}, and string theory is indispensable for the study of 6d SCFTs. The reader is referred to \cite{Heckman:2018jxk} for this subject.

\begin{table}[htbp]
 \begin{center}
\begin{tabular}{c|cccccc}
 Dimension & $0$ & $1$ & $2$ & $4$ & $5$ & $6$ \\
 \hline
 M on $S^1$ & KK-mom. & M2/$S^1$ & M2 & M5/$S^1$ & M5 & KK-mono. \\
 & $\frac{1}{R}$ & $\frac{2\pi\cdot2\pi R}{(2\pi \ell_p)^3}$ & $\frac{2\pi}{(2\pi \ell_p)^3}$ & $\frac{2\pi\cdot2\pi R}{(2\pi \ell_p)^6}$ & $\frac{2\pi}{(2\pi \ell_p)^6}$ & $\frac{2\pi(2\pi R)^2}{(2\pi \ell_p)^9}$ \\
 \hline
 IIA & D$0$ & F$1$ & D$2$ & D$4$ & NS$5$ & D$6$ \\
 & $\frac{2\pi}{g_s (2\pi \ell_s)}$ & $\frac{2\pi}{(2\pi \ell_s)^2}$ & $\frac{2\pi}{g_s (2\pi \ell_s)^3}$ & $\frac{2\pi}{g_s (2\pi \ell_s)^5}$ & $\frac{2\pi}{g_s^2 (2\pi \ell_s)^6}$ & $\frac{2\pi}{g_s (2\pi \ell_s)^7}$ \\
\end{tabular}
\end{center}
\caption{Wrapped/unwrapped M-branes and the corresponding extended objects in Type IIA theory
with their effective tensions.}
\label{table:tension}
\end{table}

\subsection{Heterotic \texorpdfstring{$E_8\protect\times E_8$}{E8xE8} string from M-theory}

Now we shall consider the strong-coupling behavior of Heterotic $E_8\times E_8$ theory. Taking T-duality and S-duality, Heterotic $E_8\times E_8$ theory is dual to Type I theory. As seen in \S\ref{sec:TypeI'}, the T-dual to Type I theory is the Type I' theory, which is Type IIA theory on a line segment $S^1/\bZ_2$ where O8${}^-$-planes sit at the two ends, and $(16+16)$ D8-branes are distributed on  $S^1/\bZ_2$. In the strong coupling regime, the M-theory circle will emerge, and it is described as M-theory on $S^1\times S^1/\bZ_2$.

\begin{figure}[ht]\centering
\includegraphics[width=15cm]{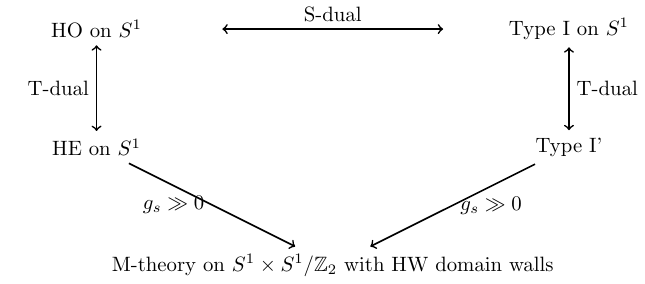}
\caption{Duality web for Heterotic M-theory}
\end{figure}

Interestingly enough, the relative position of O8${}^-$-planes and D8-branes in Type I' string theory may be adjusted. This freedom goes away in the M-theory limit; the D8-branes have to be stuck at the O8${}^-$-planes, and they become the domain walls of M-theory, which are called \textbf{Ho\v{r}ava-Witten domain walls} or \textbf{M9-branes} \cite{Horava:1995qa,Horava:1996ma}.

Its low-energy effective description is given by $D=11$ supergravity on $S^1/\bZ_2$ which gives rise to a gravitational anomaly \cite{AlvarezGaume:1983ig}. In order to cancel this
anomaly, non-Abelian gauge fields have to be present at the boundaries in order to employ a Green-Schwarz mechanism \cite{Green:1984sg,Green:1984qs}. This mechanism that bulk anomaly cancels with boundary anomaly is called \textbf{anomaly inflow}. Indeed the low-energy effective theory at the Ho\v{r}ava-Witten domain wall is described by 10d $\cN=1$ SYM with $E_8$ gauge group that cancels anomaly.

As in Type IIA, the distance between the two boundaries is related to Heterotic coupling $R = g_{\rm het}^{\frac{2}{3}}\ell_p$.
Hence, the line segment $S^1/\bZ_2$ shrinks at the weak coupling regime, leading to Heterotic $E_8\times E_8$ string theory.
The reason to pick $E_8 \times E_8$ is that the anomalies must be canceled on both boundaries, and there is no way to distribute $\SO(32)$ between two boundaries (it is a simple group with no factors). Using the previous terminology Heterotic $E_8\times E_8$
string theory can be viewed as M-theory compactified
on $S^1/\bZ_2$. This setup is called \textbf{Ho\v{r}ava-Witten M-theory}  or \textbf{Heterotic M-theory}  \cite{Horava:1995qa,Horava:1996ma}.

\begin{figure}[ht]\centering
\includegraphics[width=13cm]{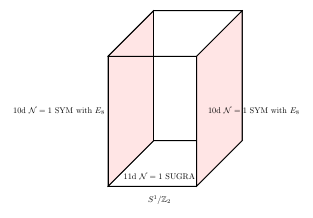}
\caption{Low-energy effective description of Heterotic M-theory. Ho\v{r}ava-Witten domain walls at the two boundaries give rise to $\cN=1$ SYM with $E_8$ gauge group and they cancel bulk anomaly.}
\end{figure}

\subsection{Duality between Heterotic on \texorpdfstring{$T^4$}{T4} and Type IIA on K3}
Let us now see one more non-trivial duality. Although we have studied only toroidal compactifications, we have seen a rich web of dualities. In string theory, a theory is consistent if we compactify it on a Calabi-Yau manifold. Since there are wide varieties of Calabi-Yau manifolds, string dualities involving them are much richer. It has still been an active research area both in physics and mathematics. Here, we deal with the next simplest Calabi-Yau manifold, called a \textbf{K3 surface}.

\subsubsection*{K3 surface}

A K3 surface is a resolution of $T^4/\bZ_2$.  We write a 4-torus as
\[
T^4=\bR^4/\bZ^4=\{\bfx=(x_1,x_2,x_3,x_4)\in\bR^4| x_i\sim x_i+1\}
\]
and  the $\bZ_2$ action is a reflection $x_i\rightarrow -x_i$.
Note that this action has $2^4=16$ fixed points given by choice of midpoints or the origin in any of the four $x_i$.
Thus, the resulting space $T^4/\bZ_2$ is singular at any of these 16 fixed points. The neighborhood of a singular point is indeed a cone of $\bR P^3$.
To make it smooth,  let us consider the set of vectors of length $\le 1$ in the tangent bundle of $TS^2$
\[
V=\{ (v_1,v_2)\in S^2\times T_{v_1}S^2| ~|v_2|\le 1\}~.
\]
Then the boundary of $V$  is $\partial V=\bR P^3$ so that you can replace the neighborhood of each singular point by $V$. Since $V$ is a smooth manifold, the resulting space is smooth, and it is a K3 surface. This smoothing procedure is called \textbf{resolution} or \textbf{blow-up}.

Although the construction of a K3 surface is rather simple, its geometry is surprisingly fertile. First of all, it is a Calabi-Yau manifold, namely a Ricci-flat K\"ahler manifold.
In real four dimensions, there are only two topologically inequivalent compact closed Calabi-Yau manifolds, $T^4$ and K3.  Moreover, it is a hyper-K\"ahler manifold. (Let us not go into detail about hyper-K\"ahler manifolds.)

Let us now briefly look at the topological property of K3 surfaces. The resolution of the 16 singular points provides 16 elements of $H^2(K3,\bZ)$ in addition to $6={}_4C_{2}$ tori in $T^4$. Therefore, we have $H^2(K3;\bZ)\cong \bZ^{22}$. Moreover, the Hodge diamond as a complex manifold turns out to be
\[
\begin{array}{ccccc}
            &              &  h^{0,0} &            &            \\
            & h^{1,0}\! &              &\! h^{0,1}\!&            \\
  h^{2,0}\! & \!             & h^{1,1} & \!            &\! h^{0,2}\\
            & h^{2,1} &              &h^{1,2}&             \\
             &              &h^{2,2}&              &
\end{array} =
\begin{array}{cccccc}
\ &\ &\ 1\ &\ &\\
\ &\ 0\ &\ & \ 0\ &\\
\ 1\ &\ &\ 20\ &\ &\ 1\\
\ &\ 0\ &\ &\ 0\ &\\
\ &\ &\ 1\ &\ &
\end{array}\  .
\]
Since it is a real 4-dimensional manifold, one can consider the intersection matrix of rank 22
\[
Q(\a_i,\a_j) =\a_i \cap  \a_j \qquad \a_i\in H_2(K3;\bZ)
\]
In fact, in a certain nice basis,  the intersection  matrix can be written as follows
\[
Q(\a_i,\a_j) \sim 2(-E_8)\oplus 3\begin{pmatrix} 0&1\\1&0\end{pmatrix}
\]
where $-E_8$ denotes the $8\times8$ matrix given by minus the Cartan
matrix of the Lie algebra $E_8$. Hence, we may decompose
\[
  H^2(K3,\bR) = H^+\oplus H^-,
\]
where $H^\pm$ represents the cohomology of the space of
(anti-)self-dual 2-forms. We then see that
\[
  \dim H^+=3,\quad\dim H^-=19~.
\]

The moduli
space of non-trivial metric deformations on a K3
is 58-dimensional and given by the coset space \cite{Aspinwall:1996mn}
\[
{\cal M}_{\textrm{K3}}\ =\  \bR^+\ \times\
\left.\frac{\mathrm{O}(3,19)}{\mathrm{O}(3)\times \mathrm{O}(19)}\right/ \mathrm{O}(3,19,\bZ)\ ,
\]
where the second factor is the Teichm\"uller space
for Ricci-flat metrics of
volume one on a K3 surface and
the first factor is associated with the
size of the K3.

This is not the end of the story if we consider string propagation
on $K3$. For each element of $H_2(K3,\bZ)$, we can turn
on the $B$-field. Because of $H_2(K3,\bZ)=\bZ^{22}$, we have 22 additional real parameters and that makes the total
dimension of moduli space $58+22=80$.  It turns out that this moduli
space is isomorphic to
\begin{equation}
{\cal M}_{\textrm{K3}}^{\textrm{stringy}} =
\left. \frac{\mathrm{O}(4,20)}{\mathrm{O}(4)\times \mathrm{O}(20)}\right/
\mathrm{O}(4,20,\bZ)\ .
\end{equation}
Substituting $d=4$ into \eqref{Narain}, one can see that this is exactly the same as the Narain moduli space for Heterotic string on $T^4$!

\subsubsection*{Heterotic on $T^4$/Type IIA on K3}
The duality between Heterotic on $T^4$ and Type IIA on K3 can be seen by comparing the effective actions in $D= 6$.

On the Heterotic side, when generic Wilson lines are considered, the gauge symmetry $E_8 \times E_8$ or $\mathrm{SO}(32)$ is broken to $\mathrm{U}(1)^{16}$. By including the Kaluza-Klein (KK) gauge bosons resulting from the compactification on $T^4$, the gauge group is enhanced to $\mathrm{U}(1)^{24}$. The effective action for the six-dimensional supergravity in Heterotic string theory is given by
\[
S_{\textrm{Het}}= \frac{1}{2\kappa_{6}^2} \int d^{6}x \sqrt{ -G}   \,e^{-2\Phi}\left[
 R +4 \partial_\mu \Phi \partial^\mu \Phi -\frac{1}{2}| \wt H_{(3)}|^{2}- \frac{\kappa_{6}^2}{2g_{6}^2} \sum_{I=1}^{24} |F_{(2)}^I|^2 \right] ~.
\]
where $\kappa_{6}$ represents the six-dimensional gravitational coupling, $G$ is the determinant of the six-dimensional metric, $\Phi$ denotes the dilaton field, $\widetilde{H}_{(3)}$ is the field strength associated with the antisymmetric tensor field $B_{(2)}$, and $F_{(2)}^I$ corresponds to the field strength of the $\mathrm{U}(1)^{24}$ gauge fields.

On the Type IIA side, the compactification of the superstring theory on K3 leads to the breaking of half of the supersymmetries, resulting in 16 preserved supercharges, which is the same as in the Heterotic theory.  The compactification also gives rise to $\mathrm{U}(1)^{24}$ gauge fields, all originating from the R-R sector.  One comes from the one-form field $C_{(1)}$ with indices in the non-compact six-dimensional spacetime, while the other gauge fields arise from the three-form field $C_{(3)}$, which is Hodge-dual to a massless vector in six dimensions. Since $H_2(K3,\mathbb{Z})=\mathbb{Z}^{22}$, the three-form field $C_{(3)}$ on the two-cycles of K3 provides 22 vectors, resulting in a total of $1+1+22=24$ $\mathrm{U}(1)$ gauge fields. The effective action for Type IIA superstring theory compactified on K3 is given by
\[
S_{\textrm{IIA}}= \frac{1}{2\kappa_{6}^2} \int d^{6}x \sqrt{ -G}   \left[e^{-2\Phi}\Big(
 R +4 \partial_\mu \Phi \partial^\mu \Phi -\frac{1}{2}| \wt H_{(3)}|^{2} \Big)- \frac{\kappa_{6}^2}{2g_{6}^2} \sum_{I=1}^{24} |F_{(2)}^I|^2 \right] ~.
\]

 It is straightforward to see that these two actions are equivalent through a straightforward field redefinition,
 \begin{align}\nonumber
\Phi^{\textrm{H}}=-\Phi^{\textrm{IIA}}~,&\qquad G^{\textrm{H}}=e^{-2\Phi^{\textrm{IIA}}}G^{\textrm{IIA}}\cr
A^{\textrm{H}}=A^{\textrm{IIA}}~,&\qquad \wt H^{\textrm{H}}=e^{-2\Phi^{\textrm{IIA}}}\ast \wt H^{\textrm{IIA}}~.
\end{align}

\subsubsection*{More dualities}

Indeed, the examples presented above provide a glimpse into the rich landscape of string dualities. However, these are just a few representative instances of the many dualities that have been discovered in string theory. By considering different compactifications of M-theory on tori with orbifold actions, one can uncover numerous other dualities. Similar arguments and techniques as discussed earlier have been employed to establish these dualities. Here are some conjectured dualities in this context, supported by references \cite{Dasgupta:1995zm,Witten:1995em,Sen:1996zq}:
\begin{eqnarray}
\hbox{M-theory on} && \nonumber \\
\textrm{K3} \qquad  &\leftrightarrow& \qquad\hbox{Heterotic/Type I on $T^3$} \nonumber
\\
T^5/\bZ_2 \qquad
&\leftrightarrow&\qquad \hbox{IIB on K3} \nonumber \\
T^8/\bZ_2 \qquad &\leftrightarrow&\qquad \hbox{Type I/Heterotic on $T^7$}
\nonumber \\
T^9/\bZ_2 \qquad &\leftrightarrow&\qquad \hbox{Type IIB on
$T^8/\bZ_2$}\nonumber
\end{eqnarray}
In each case, the action of the $\mathbb{Z}_2$ orbifold is to reverse the sign of all the coordinates of $T^D$. These proposed dualities have been extensively examined and verified by taking the Type IIA limit of M-theory.

\section{Green-Schwarz anomaly cancellation}\label{sec:GS}
In \S\ref{sec:TypeI}, we observed from the perspective of string amplitudes that unoriented superstrings can be well-defined only for SO(32) gauge groups. In this section, we consider the celebrated \textbf{Green-Schwarz anomaly cancellation} for Type I supergravity from the viewpoint of anomaly polynomials.

Quantum anomalies arise when the partition function is not invariant under gauge transformations in the presence of background gauge fields. More concretely, this is expressed as:
\be
Z_{\text{QFT}}[A^g] = e^{i\alpha(A,\epsilon)} Z_{\text{QFT}}[A]~,
\ee
where $A^g$ and $A$ are background gauge fields of the symmetry group $G$, related by a gauge transformation $\epsilon$. An anomaly is identified when $\alpha(A, \epsilon)$ cannot be eliminated by local counterterms, indicating a global topological (cohomological) issue.

Anomalies for global symmetries, known as 't Hooft anomalies, are a fundamental aspect of quantum field theory (QFT). According to \cite{tHooft:1979rat,Freed:2023snr}, an 't Hooft anomaly is a feature, not a bug. It also represents an obstruction to gauging the symmetry $G$. Therefore, the gauge anomaly must vanish for a well-defined QFT. This section focuses on the cancellation of gauge anomalies in Type I and heterotic string theory.

The anomaly for a QFT on a manifold $M_{2n}$ is given by the integral of a local density. Mathematically, this is expressed as:
\be
\alpha(A, \epsilon) = \delta_\epsilon W[A] = 2\pi \int_{M_{2n}} \omega_{2n},
\ee
where $\omega_{2n}$ is the local density contributing to the anomaly. In this section, we introduce the powerful framework of \textbf{anomaly polynomials} to derive $\omega_{2n}$. The anomaly polynomial is a characteristic class of a degree 2 higher than the spacetime dimension of the theory under consideration. By performing a simple algebraic operation called the \textbf{descent equation} on it, the anomaly $\omega_{2n}$ of the theory can be obtained. For details, we refer to \cite[\S13]{GSW}, \cite{Bilal:2008qx,Alvarez-Gaume:2022aak}.

Our objective is to study chiral anomalies in string theory. A necessary preliminary is an understanding of chiral gauge theories.

\subsection{Anomalies and the Wess-Zumino condition}

Let us focus on an even-dimensional space-time $D=2n$. The Clifford algebra 
\begin{equation}
\{\Gamma_\mu,\Gamma_\nu\} = 2\eta_{\mu\nu}~,
\end{equation}
has a spinor representation of dimension $\mathbf{2}^n$, which corresponds to the Dirac spinor. Defining the chirality operator 
\begin{equation} 
\frac{1\pm\Gamma_{2n+1}}{2}~, \qquad \Gamma_{2n+1} = i^{n-1}\Gamma_0\Gamma_1\cdots \Gamma_{2n-1}~,
\end{equation}
we can decompose it into a sum of irreducible representations:
\begin{equation}
\mathbf{2}^n = \mathbf{2}^{n-1}_+ \oplus \mathbf{2}^{n-1}_-~.
\end{equation}
Thus, if $\psi$ is a $\mathbf{2}^n$ component Dirac spinor, it can be decomposed into a sum of chiral (Weyl) spinors $\psi_\pm$ where
\begin{equation}
\psi_\pm = \frac{1\pm\Gamma_{2n+1}}{2}\psi~.
\end{equation}
When chiral fermions propagate in the background of gauge and gravitational fields, it is essential to consider the potential for gauge and gravitational anomalies.

Now we consider a chiral theory in which the gauge field $A$ couples only to the Weyl fermion $\psi_+$ where the effective action $W[A]$ is given by
$$
\mathrm{e}^{-W[A]}=\int D \psi D \bar{\psi} \exp \left(-\int d x \bar{\psi}_{+} i \slashed{D}\psi_{+}\right)~.
$$
The covariant derivative $D_\mu=\partial_\mu +A_\mu$ depends on the representation of the chiral fermion $\psi_+$. In this section, unless it is specified, we assume the adjoint representation. 
Let us consider an infinitesimal gauge transformation
\be 
\delta A_\mu^a =D_\mu \theta^a
\ee 
where $T_a$ is the basis of the gauge Lie algebra $\frakg$, and $\theta=\theta^a T_a$ is an infinitesimal gauge transformation parameter. Under this transformation, we have
\bea\label{deltaW}
\delta W =&\int  d^{2n}x \delta A^a_\mu(x) \frac{\delta W}{\delta  A^a_\mu(x) }\cr
=&\int  d^{2n}x  D_\mu \theta^a \frac{\delta W}{\delta  A^a_\mu(x) }\cr
=&- \int  d^{2n}x \theta^a\langle D_\mu j^\mu_a \rangle\cr
\eea
where 
\be
j^\mu{ }_a= i \bar{\psi}_+ \gamma^\mu T_a \psi_+
\ee 
If $\langle D_\mu J^\mu_a\rangle \neq 0$, then the effective action $W[A]$ is not gauge invariant, and the theory is anomalous. Our aim is to find an expression for and examine the conditions under which the quantum theory maintains the gauge symmetries of the classical theory defined by the action $S$.

On the other hand, the effective action changes as
\bea
\delta W =&\int  d^{2n}x \delta A^a_\mu(x) \frac{\delta W}{\delta  A^a_\mu(x) }\cr
=&\int  d^{2n}x (\partial_\mu \theta^a +f^{abc}A_\mu^b
\theta^c) \frac{\delta W}{\delta A^a_\mu(x)}\cr
=&\int  d^{2n}x \theta^a \Bigl( -\partial_\mu\frac{\delta W}{\delta A^a_\mu(x)}-f^{abc}A_\mu^b(x)\frac{\delta W}{\delta A^c_\mu(x)}\Bigr)
\eea

Comparing this with \eqref{deltaW}, we have the equality
\be
-\delta W =\int d^{2 n} x \theta^{a}(x) \frac{\delta}{\delta \theta^a}(x) W=\int d^{2 n} x  \theta^{a}(x) \langle D_{\mu} J^{\mu}_{a}(x) \rangle
\ee
where we define
$$
\frac{\delta}{\delta \theta^a} =\partial_\mu \frac{\delta}{\delta  A^a_\mu(x)} + f_{ab}{}^cA_\mu^b
(x)\frac{\delta }{\delta A^c_\mu(x)}
$$
This gives
\begin{equation}
\langle D_{\mu} J^{\mu}_{a}(x) \rangle =\frac{\delta}{\delta \theta^a}(x) W:=G_a(x)~.
\end{equation}
It is easy to see that
\be \label{WZ}
\left[\frac{\delta}{\delta \theta^a}(x), \frac{\delta}{\delta \theta^b}(x^{\prime})\right]=f_{abc} \frac{\delta}{\delta \theta^c}(x) \delta_{2 n}(x-x^{\prime})~.
\ee
It follows that
\begin{equation}
\frac{\delta}{\delta \theta^a}(x) G_{b}(x^{\prime})-\frac{\delta}{\delta \theta^b}(x^{\prime}) G_{a}(x)=f_{abc} G_{c}(x) \delta_{2 n}(x-x^{\prime})
\end{equation}
This is known as the \textbf{Wess-Zumino consistency condition}, and consistent anomalies obey these equations.

\subsection*{Rewriting Using BRST Transformation}
The Wess-Zumino consistency condition can be more concisely written by using the Grassmann odd ghost field $c = c^a T^a$. This will lead to the BRST transformation for a small gauge transformation.

The BRST transformation $\delta$ for a gauge field $A$ is defined using the ghost field $c = c^a T^a$ as follows:
\bea 
\delta A^a =& c_x^b \frac{\delta}{\delta \theta_x^b} A^a \cr 
\delta c^a =& -\frac{1}{2} f^{abc} c^b c^c
\eea
Let us check that the BRST transformation is nilpotent $\delta^2 A=0$. Applying the BRST transformation twice, we obtain
\be
\delta^2 A^a = \delta c_x^b \frac{\delta}{\delta \theta_x^b} A^a - c_x^b \frac{\delta}{\delta \theta_x^b} \delta A^a
\ee
For the first term, we have
\be
\delta c_x^b \frac{\delta}{\delta \theta_x^b} A^a = -\frac{1}{2} f^{bcd} c_x^c c_x^d \frac{\delta}{\delta \theta^b} A^a = -\frac{1}{2} f^{dbc} c_x^b c_x^c \frac{\delta}{\delta \theta_x^d} A^a
\ee
For the second term, we have
\be
c_x^b \frac{\delta}{\delta \theta_x^b} \delta A^a = c_x^b \frac{\delta}{\delta \theta_x^b} c_y^c \frac{\delta}{\delta \theta_y^c} A^a = \frac{1}{2} c_x^b c_x^c \left[ \frac{\delta}{\delta \theta_x^b}, \frac{\delta}{\delta \theta_x^c} \right] A^a~.
\ee
Therefore, due to the Wess-Zumino integrability condition, it becomes zero
\begin{equation}
\delta^2 A^a=\frac{1}{2} c_x^b c_x^c\left(-f^{d b c} \frac{\delta}{\delta \theta_x^d}+\left[\frac{\delta}{\delta \theta_x^b}, \frac{\delta}{\delta \theta_x^c}\right]\right) A^a =0~.
\end{equation}

The above calculations can be similarly applied to any functional $ F[A] $. By formally replacing $ A $ with $ F[A] $ in the above calculations:
\be
\delta^2 F[A] = \frac{1}{2} c_x^b c_x^c \left( -f^{dbc} \frac{\delta}{\delta \theta_x^d} + \left[ \frac{\delta}{\delta \theta_x^b}, \frac{\delta}{\delta \theta_x^c} \right] \right) F[A]=0~.
\ee
Thus, the Wess-Zumino consistency condition \eqref{WZ} is equivalent to the nilpotency $\delta^2 = 0$. Introducing the BRST transformation,  the Wess-Zumino consistency condition becomes surprisingly simple.

Consequently, the anomaly is obtained by the BRST transformation of the effective action, and this will lead to a cohomological problem for the anomaly as follows:
\be\label{anomaly-inte}
\delta W[A] =2\pi \int_{M_{2n}} \omega_{2n}(A, c)~.
\ee
Note that due to the nilpotency, we have
\be
\delta^2 W[A] =2\pi  \int_{M_{2n}} \delta \omega_{2n}(A, c) = 0~.
\ee

\subsection{Anomaly polynomials and descent equations}
If the spacetime $M_{2n}$ is closed (i.e., without boundary) and oriented, it can be bounded by a $(2n+1)$-dimensional oriented manifold ${M}_{2n+1}$. By extending the fields (including $c$) defined on $M_{2n}$ to ${M}_{2n+1}$, the Stokes theorem tells us
$$
\delta_c W=2\pi \int_{M_{2n+1}} d \omega_{2n}(c).
$$
The Wess-Zumino consistency condition now states
$$
\left[\delta_{c_1}, \delta_{c_2}\right] W=2\pi \int_{M_{2n+1}}\left[\delta_{c_1} d \omega_{2n}(c_2)-\delta_{c_2} d \omega_{2n}(c_1)\right]=2\pi \int_{M_{2n+1}} d \omega_{2n}\left(\left[c_1, c_2\right]\right).
$$
Since the above equation holds for any arbitrary extension of the fields into the interior, we have
$$
\delta_{c_1} d \omega_{2n}(c_2)-\delta_{c_2} d \omega_{2n}(c_1)=d \omega_{2n}\left(\left[c_1, c_2\right]\right).
$$
This implies the existence of a $2n+1$-form $\omega_{2n+1}^{\textrm{CS}}$ (independent of $c$) defined over $M_{2n+1}$ that satisfies
$$
\delta_c \omega_{2n+1}^{\textrm{CS}}=d \omega_{2n}(c).
$$
This $\omega_{2n+1}^{\textrm{CS}}$ is commonly referred to as a Chern-Simons form. Taking an exterior derivative, we define a $(2n+2)$-form
\be
I_{2n+2}=d \omega_{2n+1}^{\textrm{CS}}~,
\ee 
which is closed, $dI_{2n+2}=0$. Moreover, the form $I_{2n+2}$ is gauge invariant
$$
\delta I_{2n+2}=d \delta \omega_{2n+1}^{\textrm{CS}}=d^2 \omega_{2n}(c)=0
$$
since $d$ and $\delta$ commute.

The natural candidate for $I_{2n+2}$ is a characteristic class, called \textbf{anomaly polynomial}, constructed from the curvatures $F=dA+A^2$ for the gauge field $A$. More precisely, the anomaly polynomial is a specific polynomial expression (see \eqref{anomaly-poly}) built from the curvatures whose specific form depends on the space-time dimension, the global symmetries, and the field content of the theory.

In summary, starting from the anomaly polynomial $I_{2n+2}$, the Wess-Zumino consistency condition ensures that we can obtain $\omega_{2n}$ by the \textbf{descent equation}
\begin{align}\label{descent}
I_{2n+2} =& d  \omega_{2n+1}^{\textrm{CS}} \cr 
\delta \omega_{2n+1}^{\textrm{CS}} =& d\omega_{2n}
\end{align}
The anomaly under the variation of the gauge field is obtained by integrating $\omega_{2n}$ over the space-time $M_{2n}$ as in \eqref{anomaly-inte}. The descent equation can be illustrated by the following diagram:
\be 
\begin{tikzcd}
    & I_{2n+2}\arrow[d, "\delta"]&  \arrow[ l, "d"'] \omega_{2n+1}^{\mathrm{CS}} \arrow[d, "\delta"]&& \\
  &0 & \arrow[l, "d"']  d\omega_{2n} \arrow[d, "\delta"] & \arrow[l, "d"']  \omega_{2n} \arrow[d, "\delta"] &\\
  &  &0 & {} & {}
\end{tikzcd}
\ee

As a toy example, we consider the Chern character 
\begin{equation}
I_{2n+2}=\operatorname{ch}_{n+1}(F)=\frac{i^{n+1}}{(2 \pi)^{n+1} (n+1)!} \Tr F^{n+1}
\end{equation}
as an anomaly polynomial. 
It is easy to find the explicit forms for $n=1$:
\be 
\begin{tikzcd}
\operatorname{ch}_2(F) = -\frac{1}{8 \pi^2} \Tr F^2 &  \arrow[ l, "d"'] \omega_3^{\mathrm{CS}} = -\frac{1}{8 \pi^2} \Tr\left(AdA+\frac23 A^3\right)\arrow[d, "\delta"] \\
& \omega_{2}(c)= -\frac{1}{8 \pi^2}  \Tr (c\cdot dA)
\end{tikzcd}
\ee
In this case, $\omega_3^{\mathrm{CS}}$ is literally the Chern-Simons 3-form so that the variation under the gauge transformation is given by \eqref{variation-CS}.

More generally, we have the following expressions
$$
\begin{aligned}
\Tr F^n & = \int_0^1 d t \frac{d}{d t}\left[\Tr F_t^n\right] \\
& = \int_0^1 d t n \cdot \Tr\left[\left(D_t A\right) F_t^{n-1}\right] \\
& = \int_0^1 d t n \cdot \Tr\left[D_t\left(A F_t^{n-1}\right)\right] \\
& = \int_0^1 d t n \cdot \Tr\left[d\left(A F_t^{n-1}\right) + t A \left(A F_t^{n-1}\right) + \left(A F_t^{n-1}\right) t A\right] \\
& = \int_0^1 d t n \cdot d\left[\Tr A F_t^{n-1}\right]
\end{aligned}
$$
Here, let $ A_t = t A $ and the field strength be $ F_t = t F + \left(t^2 - t\right) A^2 $. Also, the covariant derivative can be written as $ D_t C = d C + A_t C + C A_t $, and the Bianchi identity is $ D_t F_t = 0 $.
$$
\frac{d}{d t} F_t = d A + 2 t A^2 = D_t A
$$
By noting this, the above expressions can be confirmed. More generally,
\be \label{omegaCS}
\omega_{2 n+1}^{\mathrm{CS}} = \frac{i^{n+2}}{(2 \pi)^{n+2}(n+1)!} \int_0^1 d t \Tr\left[A F_t^{n+1}\right]
\ee 
is given.

To transition from the first step to the second step of the descent equation \eqref{descent}, one should apply the BRST transformation $\delta$ to the Chern-Simons form $\omega_{2n+1}^{\mathrm{CS}}$. In the case of a $U(N)$ gauge theory, by applying the BRST transformation to \eqref{omegaCS}, we obtain
$$
\omega_{2 n} = -\frac{i^{n+2}}{(2 \pi)^{n+2}(n+1)!} \int_0^1 d t (1-t) \Tr \sum_{r=0}^{n-1}\left[c \cdot d\left(F_t^r A F_t^{n-r-1}\right)\right].
$$

So far, we have treated the case in which the chiral fermion is coupled to the Yang-Mills gauge field, but if the manifold $M_{2n}$ has a curvature, the covariant derivative involves the spin connection
$$
D_{\mu} \psi_{+}=\left(\partial_{\mu}+\frac14 \omega_{\mu}^{ab}\Gamma_{ab}+A_{\mu}\right) \psi_{+}
$$
Then, the anomaly polynomial involves characteristic classes with the spacetime curvature so that it generally can be written as 
\bea \label{anomaly-poly}
I_{2n+2}(R, F) =&
P\left(\Tr R^2, \ldots, \Tr R^{n+1} ; \Tr F, \ldots, \Tr F^{n+1}\right) \cr 
=&\sum \beta_{k;a_1\cdots a_m}\, \Tr R^{2k}\,\Tr F^{a_1}\cdots\Tr F^{a_m} ,\qquad 2k+a_1+\cdots+a_m=n+1,
\eea
where the sum runs over products of traces of total degree $n+1$ (for example $\Tr F^2\Tr F^4$ appears in $D=10$) and $\beta$ are some constants. 
Note that the spacetime holonomy of $M_{2n}$ is $\SO(2n)$ in general, so the curvature form $R$ takes 
\be \label{Chern-roots}
\frac{R}{2 \pi}=\left(\begin{array}{ccccc}
0 & x_1 & 0 & 0 & \cdots \\
-x_1 & 0 & 0 & 0 & \cdots \\
0 & 0 & 0 & x_2 & \cdots \\
0 & 0 & -x_2 & 0 & \cdots \\
\vdots & \vdots & \vdots & \vdots & \ddots
\end{array}\right)
\ee
where $x_j$ are Chern roots. These generators are $2n \times 2n$ antisymmetric matrices, which implies that $\Tr R^{2k+1} = 0$.

\subsubsection{Anomaly polynomials}
The evaluation of anomaly polynomials involves the use of the index theorem. Instead of elaborating on the details here, we refer the reader to \cite{Harvey:2005it,Bilal:2008qx} for the derivation. To facilitate the expression of anomaly polynomials, we briefly introduce the definition of characteristic classes.

Given the Chern roots \eqref{Chern-roots}, the Pontryagin classes are defined by
\begin{equation}
p(R) = \prod_{i=1}^n \left(1 + x_i^2\right) \equiv 1 + p_1(R) + p_2(R) + \ldots + p_n(R),
\end{equation}
where the $\ell$-th Pontryagin class ($4\ell$-form) is defined as the homogeneous polynomial 
\begin{equation}
p_{\ell}(R) = \sum_{i_1 < \ldots < i_{\ell}}^n x_{i_1}^2 \ldots x_{i_{\ell}}^2.
\end{equation}
For the first few cases, 
it is straightforward to express $ p_{\ell}(R) $ in terms of the curvature two-form
\begin{equation}
\begin{aligned}
& p_1 = -\frac{1}{8 \pi^2} \Tr R^2, \\
& p_2 = \frac{1}{128 \pi^4} \left[\left(\Tr R^2\right)^2 - 2 \Tr R^4\right], \\
& p_3 = -\frac{1}{3072 \pi^6} \left[\left(\Tr R^2\right)^3 - 6 \left(\Tr R^2\right)\left(\Tr R^4\right) + 8 \Tr R^6\right],
\end{aligned}
\end{equation}

Moreover, let us introduce another important characteristic class, the $ \widehat{A} $-roof genus:
\begin{equation}
\begin{aligned}
\widehat{A}(R) & \equiv \prod_{a=1}^n \frac{x_a / 2}{\sinh \left(x_a / 2\right)} \\
& = 1 - \frac{1}{24} p_1 + \frac{1}{5760} \left(7 p_1^2 - 4 p_2\right) - \frac{1}{967680} \left(31 p_1^3 - 44 p_1 p_2 + 16 p_3\right) + \ldots
\end{aligned}
\end{equation}
where we have expressed it in terms of the Pontryagin classes in the second line.

Now, let us provide the contribution of each chiral fermion to the anomaly polynomial using these characteristic classes.  For a complex Weyl gravitino of a given chirality, the anomaly polynomial is given by:
\bea\label{spin3/2}
I^{\frac32}(R) &= \widehat{A}(R) \left(\Tr e^{{R}/{2\pi}} - 1\right)\\
&= 2n - 1 + \frac{25 - 2n}{24} p_1 + \frac{1}{5760}\left[(233 + 14n) p_1^2 - 4 (239 + 2n) p_2\right] \\
&\quad + \frac{1}{967680}\left[(535 - 62n) p_1^3 - 4 (559 - 22n) p_1 p_2 + 16 (505 - 2n) p_3\right] + \ldots
\eea

For a complex Weyl spinor of a given chirality coupled to a gauge field, the anomaly polynomial is given by:
\begin{equation}\label{spin1/2}
\begin{aligned}
I^{\frac12}(R, F) &= \widehat{A}(R) \textrm{ch}(F) 
\end{aligned}
\end{equation}
In these expressions, $R$ represents the curvature 2-form in the fundamental representation of $\SO(2n)$. In $D = 2n$ dimensions, we extract the rank $2n+2$ terms from the above expressions. The trace $\Tr_\cR e^{\frac{iF}{2\pi}}$ should be computed in the representation to which the fermions belong.

In $D = 4k + 2$ dimensions, in addition to chiral fermions, self-dual (or anti-self-dual) antisymmetric tensor fields also contribute to pure gravitational anomalies. These fields are of the form
\be
G_{(2k+1)} = dC_{(2k)} = \pm  \ast G_{(2k+1)},
\ee
where the Hodge star operator picks up a minus sign under parity. The (anti)self-dual condition is not parity invariant, and these fields must be considered when analyzing anomalies. The contribution of a self-dual tensor field to the anomaly is given in terms of the Hirzebruch polynomial $L(R)$ as:
\begin{equation}\label{SD}
\begin{aligned}
I^{\mathrm{sd}}(R) &= -\frac{1}{8} L(R) \\
&= -\frac{1}{8} \prod_{a=1}^n \frac{x_a}{\tanh x_a} \\
&= -\frac{1}{8} - \frac{1}{24} p_1 + \frac{1}{360}\left(p_1^2 - 7 p_2\right) - \frac{1}{7560}\left(2 p_1^3 - 13 p_1 p_2 + 62 p_3\right) + \ldots
\end{aligned}
\end{equation}

\subsubsection{Anomaly cancellation of IIB supergravity}
As an illustrative example of anomaly cancellation, let us examine $D=10$ $\mathcal{N}=2$ supergravity. As we recall from \S\ref{sec:supergravity}, there are two $\mathcal{N}=2$ supergravity models in $D=10$, corresponding to the low-energy limits of type IIA and type IIB superstring theories. The type IIA theory is non-chiral in terms of the spacetime fermions and thus free from anomalies. (See Table \ref{tab:masslessII}.) In contrast, the type IIB theory is chiral in terms of the spacetime fermions, and this chiral theory lacks gauge interactions but can exhibit a gravitational anomaly. The fields that contribute to gravitational anomaly include: 
\begin{itemize}\setlength{\itemsep}{0.1pt}
    \item two left-handed spin-$\frac{3}{2}$ gravitini $\psi_m^-$
    \item two right-handed spin-$\frac{1}{2}$ dilatinos $\lambda^+$
    \item a self-dual four-form field $C_{(4)}^+$
\end{itemize}
All chiral fermions also satisfy the Majorana condition. In ten dimensions, the relevant anomaly polynomial is a 12-form. 
For the gravitino, we have:
\be\label{spin32-10}
I_{12}^{\frac32}(R) = \frac{1}{107520} \left( 25 p_1^3 - 180 p_1 p_2 + 880 p_3 \right),
\ee
For the spin-$\frac{1}{2}$ fields:
\be\label{spin12-10}
I_{12}^{\frac12}(R) = -\frac{1}{967680} \left( 31 p_1^3 - 44 p_1 p_2 + 16 p_3 \right),
\ee
For the self-dual four-form field:
\be\label{sd-10}
I_{12}^{\mathrm{sd}}(R) = -\frac{1}{7560} \left( 2 p_1^3 - 13 p_1 p_2 + 62 p_3 \right).
\ee

As a result, the total anomaly polynomial $I_{12}$ is given by:
\be
I_{12} = 2 \times \frac{1}{2} I_{12}^{\frac32}(R) -2 \times \frac{1}{2} I_{12}^{\frac12}(R)  + I_{12}^{\mathrm{sd}}(R)
\ee
The factors of 2 account for the two copies of gravitino and dilatino, while the $\frac{1}{2}$ terms reflect the Majorana-Weyl condition, halving the degrees of freedom compared to a complex Weyl fermion. The signs correspond to their chiralities. Using the above expressions, we can verify the vanishing of the anomaly
\be
I_{12}^{\frac32}(R) -I_{12}^{\frac12}(R) + I_{12}^{\mathrm{sd}}(R) = 0~.
\ee
Thus, all gravitational anomalies cancel in Type IIB supergravity. Since this theory does not include gauge fields, gauge or mixed anomalies need not be considered.

\subsection{Green-Schwarz mechanism}
Finally, let us consider Type I supergravity. As shown in \eqref{massless-type1}, the orientifold projection of the Type IIB theory removes the self-dual field. Consequently, in Type I theory, only the left-handed gravitino and the right-handed dilatino contribute to the gravitational anomaly. Therefore, this theory is not free from gravitational anomalies \cite{AlvarezGaume:1983ig}. Specifically, using \eqref{spin12-10} and \eqref{spin32-10}, the total anomaly polynomial is found to be nonzero:
\be
\begin{aligned}
I_{12} & = \frac{1}{2} I_{12}^{\frac32}(R) - \frac{1}{2} I_{12}^{\frac12}(R) \\
   & = \frac{1}{15120}\left(2 p_1^3 - 13 p_1 p_2 + 62 p_3\right) \neq 0,
\end{aligned}
\ee
where the $\frac{1}{2}$ factors arise from the Majorana condition. Nonetheless, the massless spectrum of Type I superstrings includes, besides an $\mathcal{N}=1$ supergravity multiplet, an $\mathcal{N}=1$ super-Yang-Mills multiplet \eqref{TypeI}. This multiplet includes its gaugino, a Majorana-Weyl left-handed fermion, which transforms in the adjoint representation of the gauge group $G$.

To verify anomaly cancellation in Type I string theory, we must add the gaugino's contribution 
\be
I_{12}^{\frac12}(R,F) = \operatorname{ch}_6 - \frac{1}{24} p_1 \operatorname{ch}_4 + \frac{1}{5760} \left(7 p_1^2 - 4 p_2\right) \operatorname{ch}_2  - \frac{\dim G}{967680} \left(31 p_1^3 - 44 p_1 p_2 + 16 p_3\right).
\ee
with all Chern characters evaluated in the adjoint representation. 
By incorporating these contributions, the anomaly polynomial for Type I superstrings amounts to
\be\label{type1-anomaly}
\begin{aligned}
2 I_{12} & =\operatorname{ch}_6-\frac{1}{24} p_1 \operatorname{ch}_4+\frac{1}{5760}\left(7 p_1^2-4 p_2\right) \mathrm{ch}_2 \\
& +\frac{256-31 \dim G}{967680} p_1^3+\frac{11 \dim G-416}{241920} p_1 p_2+\frac{496-\dim G}{60480} p_3,
\end{aligned}
\ee
At first glance, it might still seem impossible to construct an anomaly-free $\cN=1$ supergravity plus Yang-Mills theory in $ D=10 $. However, this conclusion is incorrect! The Green-Schwarz mechanism is the way out.

The astute reader immediately notices that the $p_3$ term drops when either $G=\SO(32)$ or $G=E_8\times E_8$ since $\dim G=496$. (See \S\ref{sec:Heterotic}.) Moreover, for SO($n$) gauge group, the characteristic classes in the adjoint representation and fundamental (vector) representation are related by
\begin{equation}
\Tr _{\mathbf{adj}} F^6=(n-32) \Tr _{\square} F^6+15\left(\Tr _{\square} F^2\right)\left(\Tr _{\square} F^4\right)
\end{equation}
so that the  $\Tr _{\square} F^6$ term drops for SO(32). Even for $E_8$, we have
\begin{equation}
\Tr_{\mathbf{adj}} {F}^6=\frac{1}{7200}\left(\Tr_{\mathbf{adj}} {F}^2\right)^3~.
\end{equation}
Consequently, for $G=\SO(32)$ or $G=E_8\times E_8$, the  $\operatorname{ch}_6$ term in \eqref{type1-anomaly} splits into
\begin{equation}
\operatorname{ch}_6=\frac{1}{720}\left(\operatorname{ch}_2 \operatorname{ch}_4-\frac{1}{1800} \operatorname{ch}_2^3\right)~.
\end{equation}
Therefore, for these gauge groups, the anomaly polynomial \eqref{type1-anomaly} can be factorized into an exterior product of a 4-form and 8-form
\be 
I_{12}=X_{(4)}\wedge Y_{(8)}~,
\ee 
where 
\be 
X_{(4)}=-\frac{1}{30} \mathrm{ch}_2+p_1~,\qquad  Y_{(8)}=-\frac{1}{48}\left(\mathrm{ch}_4-\frac{1}{1800} \mathrm{ch}_2^2-\frac{1}{60} p_1 \mathrm{ch}_2+\frac{3}{8} p_1^2-\frac{1}{2} p_2\right) .
\ee 
This factorization is the key point of the anomaly cancellation as we will see below.

Since both $X_{(4)}$ and $Y_{(8)}$ are closed forms, they can be written locally as
\be
X_{(4)}=d \Omega_{(3)}~,  \qquad Y_{(8)}=d \Omega_{(7)}~,  \qquad  I_{12}=d \omega_{11}^{\textrm{CS}}
\ee
so that
\bea
I_{12}=&d \Omega_{(3)}\wedge Y_{(8)}=X_{(4)}\wedge d \Omega_{(7)}\cr
=&d\left(c~\Omega_{(3)} \wedge Y_{(8)}+(1-c)~ X_{(4)} \wedge \Omega_{(7)}\right)
\eea
where $c$ is any real constant.
$$\omega_{11}^{\textrm{CS}}=c~\Omega_{(3)} \wedge Y_{(8)}+(1-c)~ X_{(4)} \wedge \Omega_{(7)}$$
Now, the descent equation for each Chern-Simons term yields
\be\label{descent-each}
\delta\Omega_{(3)} =d \mu_{(2)} ~, \qquad \delta\Omega_{(7)} =d \mu_{(6)} ~.
\ee 
The anomaly can be obtained from the gauge variation  of $\omega_{11}^{\textrm{CS}}$
$$\delta \omega_{11}^{\textrm{CS}}=d \omega_{10}=c~ d \mu_{(2)} \wedge Y_{(8)}+(1-c) X_{(4)} \wedge d \mu_{(6)}
$$
Consequently, the gauge variation of the effective action is given by
\be\label{GS-anomaly}
\delta W  =\int \omega_{10}  =\int\left[c~  \mu_{(2)} \wedge  Y_{(8)}+(1-c) X_{(4)} \wedge \mu_{(6)}\right].
\ee

To the effective action, we can add a local counterterm defined by
\be \label{GS-counter-term}
W_{\textrm{ct}}= -(1-c)\int\Omega_{(3)}  \wedge \Omega_{(7)}+ \frac{4}{\alpha^\prime}\int C_{(2)}\wedge Y_{(8)}
\ee
where $C_{(2)}$ is the R-R 2-form in the $\cN=1$ $D=10$ supergravity. Now, let us examine the response of $\widetilde{W}=W+ W_{\textrm{ct}}$ with respect to gauge transformation:
\bea 
\delta \widetilde{W}=&\delta W-(1-c) \int\left(d \mu_{(2)} \wedge \Omega_{(7)}+\Omega_{(3)} \wedge d \mu_{(6)}\right) +\frac{4}{\alpha^\prime}\int{\delta C_{(2)}\wedge Y_{(8)}} \cr
=&\int\left(\mu_{(2)} \wedge Y_{(8)}+\frac{4}{\alpha^\prime}\, \delta C_{(2)}  \wedge Y_{(8)}\right).
\eea
If we choose 
\be 
\delta C_{(2)}=-\frac{\alpha^\prime}{4}\mu_{(2)}~,
\ee
we obtain $\delta \widetilde{W}=0$. Thus, the theory defined by $\widetilde{W}$ is anomaly-free. This method of achieving anomaly cancellation is called the \textbf{Green-Schwarz mechanism}. Recalling that $\delta \Omega_{(3)}=d\mu_{(2)}$, we have the gauge-invariant combination
\bea
\wt G_{(3)}=dC_{(2)} +\Omega_{(3)} =& dC_{(2)} -
\frac{\alpha^{\prime}}{4}\left[\frac{1}{30}\Tr_{\textbf{adj}}\left(A d A+\frac{2}{3} A^3\right)-\operatorname{Tr}\left(\omega d \omega+\frac{2}{3} \omega^3\right)\right]~.
\eea
In fact, the gauge-invariant combination \eqref{gauge-inv-G3} that appears in the action of Type I supergravity can be obtained by using the relation of the Chern-Simons terms in the adjoint and fundamental representation of SO(32):
\be 
\Tr_{\textbf{adj}}\left(A d A+\frac{2}{3} A^3\right)=30\Tr_{\square}\left(A d A+\frac{2}{3} A^3\right).
\ee
The counterterm \eqref{GS-counter-term} added is often called the Green-Schwarz counterterm, which is a higher derivative correction to the low-energy effective action.

\begin{figure}[ht]
\centering
\includegraphics[width=0.85\textwidth]{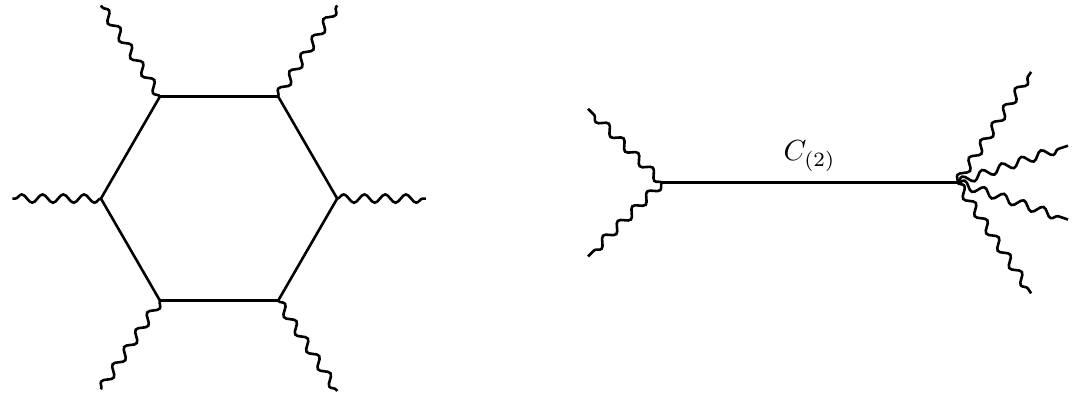}
\caption{Green-Schwarz anomaly cancellation. The left diagram gives rise to the gravitational anomaly, and the right diagram is from the counterterm to cancel the anomaly.}
\label{fig:hexagon}
\end{figure}

In fact, the anomaly arises from the hexagon one-loop diagram in $D=10$. This is canceled by the tree diagram where the $C_{(2)}$ field propagates in the intermediate state (see Figure \ref{fig:hexagon}). The kinetic term of $\widetilde{G}_{(3)}$ in the supergravity action \eqref{TypeI} gives the left vertex with two gravitons and/or gauge bosons, while the Green-Schwarz term \eqref{GS-counter-term} gives rise to the right vertex with four gravitons and/or gauge bosons.

The Green-Schwarz anomaly cancellation predicts string theory with an $E_8 \times E_8$ gauge group. This leads to the formulation of Heterotic string theory \cite{Gross:1984dd,Gross:1985fr,Gross:1985rr}. In fact, as we saw in \S\ref{sec:typeI-HO}, the low-energy effective action of the Heterotic string theory is given in \eqref{Het}, which is related to Type I supergravity by \eqref{TypeI-Het}.

\section{D-brane dynamics}\label{sec:D-brane}
So far, D-branes are treated as static rigid objects that are associated to boundary conditions of an open string.
Actually, it has dynamics like a fundamental string, and this section introduces the dynamics of D-branes.
After we derive the action for the world-volume theory of D$p$-branes in the first half, we discuss the dynamics of D-branes. Branes control non-perturbative dynamics in string theory, which reveals the richness and depth of string theory.
Therefore, the study of D-branes is remarkably broad and we cover a tiny part of it. The reader is referred to \cite{johnson2006d} for more detail.

\subsection{D-brane action}\label{sec:DBI}

When we quantize a world-sheet in \S\ref{sec:bosonic}, we use the string sigma action \eqref{string-sigma} instead of the Nambu-Goto action \eqref{NG-action} to avoid the complexity of the square root.
There, the dimension of a world-sheet to be two is crucial for quantization, and we cannot simply follow the same procedure for the fundamental string to quantize a D$p$-brane, in general.
Thus, we will first learn the effective action of a D-brane in this section.

Let us summarize the properties the D-brane action should be endowed with:
\vspace{-4pt}
\begin{enumerate}
 \setlength{\itemsep}{0pt}
 \item it contains scalars as the spacetime coordinates of the world-volume of a D-brane
 \item it includes gauge fields living on a D-brane, which arises from an open string massless spectrum
 \item it involves the $B$-field because open strings can end on a D-brane
 \item it couples to the R-R-field $C_{(p+1)}$ via \eqref{RR-coupling}
 \item it can possess supersymmetry (though we focus only on the bosonic part in this section)
\end{enumerate}
\vspace{-4pt}

The first point will be addressed by the Nambu-Goto action \eqref{NG-action} for a D-brane, namely the volume of the D-brane. We write the world-volume coordinates of D$p$-branes by $\sigma^a (a=0,1,\ldots,p)$, and $X^\mu(\sigma^a)$ maps from the world-sheet to the spacetime. Then, the Nambu-Goto action \eqref{NG-action} for a D-brane can be written as
\begin{align*}
 S_\textrm{D$p$} &= -T_\textrm{D$p$}^\mathrm{eff} \int d^{p+1}\sigma
 \sqrt{-\det \left( G_{\mu\nu} \frac{\partial X^\mu}{\partial \sigma^a} \frac{\partial X^\nu}{\partial \sigma^b} \right)} \
\end{align*}
where $T_\textrm{D$p$}^\mathrm{eff}$ is an effective D$p$-brane tension which we will discuss shortly. Starting from the Nambu-Goto action with the R-R coupling \eqref{RR-coupling}, we will generalize the action by using the T-duality and gauge invariance. In this way, we will incorporate the second and third points.

\begin{figure}[htb]
\centerline{\includegraphics[width=7cm]{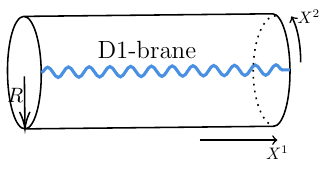}}
\caption{Vibrating D$1$-brane.}
\label{fig:vibD1}
\end{figure}

Let us consider the simple setup as in Figure \ref{fig:D1-D2-T-dual} where D1 and D2-brane are related by the T-duality. Here we are interested in a part $\RR_t \times \RR \times S^1$ of the spacetime where the D-brane is located, and we assume that its metric is flat  $ds^2 = \eta_{\mu\nu} dx^\mu dx^\nu$ (i.e. $G_{\mu\nu} = \eta_{\mu\nu}$). As explained in \S\ref{sec:Tdual-open}, the position $X_2$ ($\sim X_2 +2\pi R$) of the D$1$-brane, and the gauge field  $A_2$ ($\sim A_2 +1/R$) on the D$2$-brane are related under the T-duality by
\begin{align*}
 X_2 = 2\pi \alpha' A_2 \ .
\end{align*}
Now let us consider a situation in which the D$1$-brane has dynamics, namely it vibrates as $X^2 = X^2(X^1)$ (see Figure \ref{fig:vibD1}).
Then, the vibration of the D1-brane is translated as non-trivial field strength $F_{12} = \partial_1 A_2(X^1) \neq 0$ on the D2-brane. Parametrizing the world-volume coordinates as $X^0 = \sigma^0$ , $X^1 = \sigma^1$, the Nambu-Goto action of the D$1$-brane becomes
\begin{align*}
 S_\textrm{D1} &= -T_\textrm{D1}^\mathrm{eff} \int d^2\sigma \sqrt{-\det \left( G_{\mu\nu} \frac{\partial X^\mu}{\partial \sigma^a}
 \frac{\partial X^\nu}{\partial \sigma^b} \right)} \cr
 &= -T_\textrm{D1}^\mathrm{eff} \int d\sigma^0 d\sigma^1 \sqrt{1 +\left(\frac{\partial X^2}{\partial \sigma^1} \right)^2} \ .
\end{align*}
In order for the D-brane action to be invariant under the T-duality, we incorporate the field strength of the gauge field in such a way that
\begin{align*}
 S_\textrm{D2} &= -T_\textrm{D2}^\mathrm{eff} \int d^3\sigma \sqrt{-\det \left( G_{\mu\nu} \frac{\partial X^\mu}{\partial \sigma^a}
 \frac{\partial X^\nu}{\partial \sigma^b} +2\pi \alpha' F_{ab} \right)}  \cr
 &= -T_\textrm{D2}^\mathrm{eff} \cdot 2\pi \wt R\cdot \int d\sigma^0 d\sigma^1 \sqrt{1 +\left(2\pi\alpha'F_{12} \right)^2} \ .
\end{align*}

\subsubsection*{D-brane tension}

The dimension analysis tells us that a D-brane tension should be proportional to
\begin{align*}
 T_{\mathrm Dp} \sim \frac{\mathrm{mass}}{p\textrm{-dim vol}} \quad\Rightarrow\quad
 T_{\mathrm Dp} \sim \frac{1}{\ell_s^{p+1}} \ .
\end{align*}
From the argument above in order for the two actions to coincide
we need $T_\textrm{D1}^\mathrm{eff} = 2\pi \wt R T_\textrm{D2}^\mathrm{eff}$.
On the other hand,
$T_{\textrm{D}p}$ should be independent of the spacetime geometry including the radius $R$.
Note that D-brane effective theory is supposed to reproduce open string amplitude,
whose leading contribution is the disk amplitude $\sim e^{-\langle \Phi \rangle}$.
Thus, we reach the following form
\begin{align}\label{DBI-pre}
 S_{\textrm{D}p} &= -T_{\textrm{D}p} \int d^{p+1}\sigma\, e^{-\Phi(X)} \sqrt{-\det \left( G_{\mu\nu}
 \partial_a X^\mu \partial_b X^\nu +2\pi \alpha' F_{ab} \right)} \ .
\end{align}
Then, the ratio of the tensions of D1 and D2-brane is
\begin{align*}
 2\pi \wt R= \frac{T_{\textrm{D}1}^\mathrm{eff}}{T_{\textrm{D}2}^\mathrm{eff}}
 = \frac{T_{\textrm{D}1} e^{-\Phi}}{T_{\textrm{D}2} e^{-\wt\Phi}}
 =  \frac{T_{\textrm{D}1}}{T_{\textrm{D}2}} \cdot \frac{\wt R}{\ell_s}
 \quad\Rightarrow\quad T_{\textrm{D}1} = 2\pi \ell_s \cdot T_{\textrm{D}2} \ .
\end{align*}
Note that the dilaton field transforms under the T-duality as $e^{-\wt\Phi} = e^{-\Phi} \frac{\ell_s}{\wt R}$ (exercise).
For a D-brane in superstring theory, the correct normalization is
\be\label{Dbrane-tension} T_{\textrm{D}p} = \frac{2\pi}{(2\pi \ell_s)^{p+1}} ~.\ee

\subsubsection*{The $B$-field}

In \eqref{NLSM-general}, we see the action of the bosonic closed string. For an open string, we need to add a boundary term in order for the action to be invariant under gauge transformations:
\begin{align}\label{NLSM-open}
 S_{\textrm{open}} =& \frac{1}{4\pi \alpha'} \int_\Sigma d^2\sigma \left( \sqrt h h^{ab} \partial_a X^\mu \partial_b X^\nu G_{\mu\nu}(X)
 +i \varepsilon^{ab} \partial_a X^\mu \partial_b X^\nu B_{\mu\nu}(X)
 +\alpha' \sqrt h R^{(2)} \Phi(X)
 \right) \cr
 &\qquad +i\int_{\partial\Sigma} d\sigma^0 \partial_0 X^\mu A_\mu  \cr
 &= \cdots +\frac{1}{2\pi\alpha'} \int_{\Sigma} B_{(2)} +\int_{\partial\Sigma} A_{(1)} \ .
\end{align}
Here the gauge transformation  \eqref{B-gauge} of the $B$-field $\delta_B B_{(2)} = d \Lambda_{(1)}$
is compensated by that of the gauge field $A_{(1)}$, which is
\begin{align*}
 \delta_B A_{(1)} = -\frac{\Lambda_{(1)}}{2\pi\alpha'} \ .
\end{align*}

Therefore, \eqref{DBI-pre} can be generalized by incorporating the $B$-field as
\be
S_{\textrm{D}p} = -T_{\textrm{D}p} \int d^{p+1}\sigma\, e^{-\Phi(X)} \sqrt{-\det \left( G_{ab}
+2\pi \alpha' F_{ab} +B_{ab} \right)}~.
\ee
This is called the \textbf{Dirac-Born-Infeld (DBI) action}.

\subsubsection*{Generalization of R-R coupling and the DBI action}

In addition, a D$p$-brane couples to the R-R field $C_{(p+1)}$ via the action \eqref{RR-coupling}, which can be written in terms of local coordinates as
\begin{align*}
 S_{\textrm{R-R D}p}= \mu_{p} \cdot \int_{\textrm{D}p}  C_{(p+1)}   =  \mu_{p} \cdot \int d^{p+1}\sigma\,
 C_{\mu_1\cdots\mu_{p+1}}(X) \frac{\partial X^{\mu_1}}{\partial \sigma^1 } \cdots
 \frac{\partial X^{\mu_{p+1}}}{\partial \sigma^{p+1} }\ .
\end{align*}
As above, let us consider the T-duality for the vibrating D$1$-brane in the R-R coupling.
Recalling that R-R fields are transformed under the T-duality by \eqref{RR-Tduality}, the T-duality connects the following two expressions
\begin{align*}
 &S_{\textrm{R-R D}1} =
 \mu_{1} \cdot \int d\sigma^0 d\sigma^1
 \left( C_{01} +C_{02} \frac{\partial X^2}{\partial \sigma^1}\right) \ , \\
 &S_{\textrm{R-R D}2} =\mu_{2} \cdot \int d\sigma^0 d\sigma^1 d\sigma^2
 \left( \wt C_{012} +\wt C_{0} \cdot 2\pi\alpha' F_{12} \right) \ ,
\end{align*}
where $C_{01} \lra \wt C_{012}$, $C_{02} \lra \wt C_{0}$, and $X^2 \lra 2\pi\alpha' A_2$.
This can be understood as follows. The vibrating D$1$-brane consists of a straight D$1$-brane along $X^1$
and local vibration along $X^2$.
After T-duality along $X^2$, the vibration part gives
non-trivial flux $F_{12}\neq 0$ or equivalently gives D$0$-branes. (See left Figure \ref{fig:Myers}.)
Therefore, the generalization of the R-R coupling \eqref{RR-Tduality} is
\begin{align}\label{RR-general}
 S_{\textrm{D}p} =\mu_{p} \cdot \int C_\mathrm{RR} \wedge \exp(2\pi\alpha' F_{(2)}+B_{(2)}) \ ,
\end{align}
where $C_{RR} = \sum_{n} C_{(n)}$.
Note that the $B$-field appears with $F_{(2)}$ due to the gauge invariance.

Finally, the fully general form of D$p$-brane action is given as follows:
\begin{align}
 S_{\textrm{D}p} &= -T_{\textrm{D}p} \int d^{p+1}\sigma\, e^{-\Phi(X)} \sqrt{-\det \left( G_{ab}
 +2\pi \alpha' F_{ab} +B_{ab} \right)} \cr
 &\qquad +\mu_{p} \cdot \int C_{RR} \wedge \exp(2\pi\alpha' F_{(2)}+B_{(2)}) \ ,
\end{align}
where $G_{ab} = G_{\mu\nu} \partial_a X^\mu \partial_b X^\nu$ and
$B_{ab} = B_{\mu\nu} \partial_a X^\mu \partial_b X^\nu$.

Though we have focused on the bosonic part so far, there is a fermionic part
so that they form spacetime supersymmetry.
Here, we only write down the leading fluctuation:
\begin{align*}
 -i\int d^{p+1}\sigma\, \Tr \left( \ol\psi \Gamma^a D_a \psi \right) \ .
\end{align*}
For the full nonlinear supersymmetric form, one should consult with, for example, \cite{Tseytlin:1999dj}.

So far, we consider the world-volume effective action of a single D-brane, and non-Abelian generalization for multiple D-branes remains an open problem although there is a proposal \cite{Tseytlin:1997csa,Myers:1999ps}.

\subsubsection*{D-brane tensions and charges}

As seen in \eqref{Dirac-quantization}, D-brane charges must satisfy the Dirac quantization condition. Similarly, the D-brane tension \eqref{Dbrane-tension} is also governed by the Dirac quantization condition. Thus, we can express $\mu_{p}$ as $\mu_{p} = \frac{2\pi}{(2\pi \ell_s)^{p+1}} = T_{\textrm{D}p}$. The tension between D$p$-branes induces an attractive gravitational force (mediated by gravitons and dilatons), whereas the R-R charge induces a repulsive force. Therefore, the equality of the charge and the tension is essential for the stable existence of multiple D$p$-branes.

\begin{figure}[ht]\centering
  \includegraphics[width=8cm]{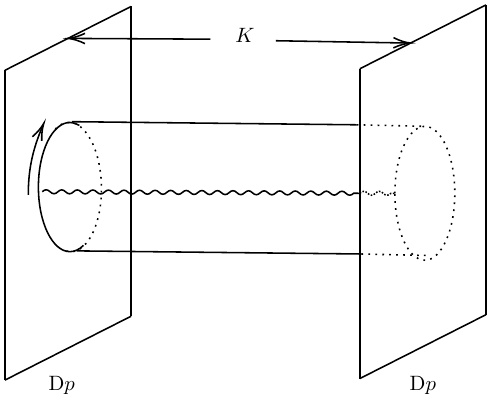}
  \caption{One-loop open string annulus diagram between two D$p$-branes.}
  \label{fig:1loop}
\end{figure}

Let us evaluate the gravitational force between two D$p$-branes by following the seminal paper \cite{Polchinski:1995mt}. As illustrated in Figure \ref{fig:openClosedAmp}, the closed string amplitude between the D$p$-branes can be interpreted as the one-loop amplitude of an open string connecting the D$p$-branes. The contribution of a particle with mass $m$ to the free energy at one-loop can be evaluated as follows:
\begin{equation}
\begin{aligned}
F &=-V_{p+1}\frac12\int \frac{d^{p+1} k}{(2 \pi)^{p+1}} \log \left(k^{2}+m^{2}\right) \\
&=-V_{p+1}\int \frac{d t}{2t} \int \frac{d^{p+1} k}{(2 \pi)^{p+1}} e^{-2\pi \alpha^{\prime}\left(k^{2}+m^{2}\right) t} \\
&=-iV_{p+1}\int \frac{d t}{2t}(8\pi^2\ell_{s}^{2} t)^{-\frac{p+1}{2}} e^{-2\pi \alpha^{\prime} m^{2} t}
\end{aligned}
\end{equation}
Since the mass of an open string mode between the branes at distance $K$ is given by
\begin{equation}
\alpha^{\prime} m^{2}=L_0+\frac{K^{2}}{\ell_{s}^{2}}~,
\end{equation}
the free energy takes the form
\begin{equation}
F=-iV_{p+1}\int \frac{d t}{2t}(8\pi^2\ell_{s}^{2} t)^{-\frac{p+1}{2}} q^{\frac{K^{2}}{\ell_{s}^{2}}} \Tr (q^{L_0})~.
\end{equation}
where
\begin{equation}
\Tr (q^{L_0}) =q^{-\frac12} \prod_{m=1}^{\infty} \frac{(1-q^{m-1 / 2})^8}{\left(1-q^{m}\right)^{8}}=\frac{\eta(i t/2)^{8}}{\eta(i t )^{16}}~.
\end{equation}
Note that the denominator is the bosonic and the numerator is the fermionic contribution. As we learn in \eqref{cylinder2}, the leading contribution from a closed string can be read off in the limit $t\to0$ when the cylinder becomes a thin tube. Therefore, we make the modular transformation \eqref{S-T} and we take the leading order
\bea\label{openclosed}
F &=-iV_{p+1}\frac{1}{(2\pi\ell_{s})^{p+1}} \int \frac{d t}{2t} (2t)^{\frac{7-p}{2}} e^{-2\pi \frac{K^2}{\ell_{s}^{2}} t} \\
&=-iV_{p+1}\frac{1}{(2\pi\ell_{s})^{p+1}}\left(\frac{\ell_{s}^{2}}{\pi K^2}\right)^{\frac{7-p}{2}} \int \frac{d x}{x} x^{\frac{7-p}{2}} e^{-x}\cr
&=-iV_{p+1}\frac{1}{(2\pi\ell_{s})^{p+1}}\left(\frac{\ell_{s}^{2}}{\pi K^2}\right)^{\frac{7-p}{2}} \Gamma\left(\frac{7-p}{2}\right)\cr
&=\textrm{const}\cdot  \ell_s^{6-2p}G_{9-p}(K^2)
\eea
where $G_d(K)$ is the massless scalar Green's function in $d$ dimensions, the inverse of $-\nabla^2$. In the $D=10$ supergravity \S\ref{sec:supergravity}, the Newton constant is given by $2\kappa_{10}^2=(2\pi \ell_s)^8/2\pi$ so that the Newton's potential is given by
\begin{equation}\label{Newton}
V_{\mathrm{G}}=-\textrm{const}\cdot\kappa_{10}^2 T_{\mathrm{D} p}^{2} G_{9-p}(r)~.
\end{equation}
Therefore, we can conclude that $T_{\mathrm{D} p} \propto \ell_s^{-(p+1)}$, which justifies \eqref{Dbrane-tension}.

For the $B$-field, the normalization of the quantization condition is different from \eqref{Dirac-quantization}
\begin{align*}
 T_\mathrm{F1} \cdot T_\mathrm{NS5} \cdot 2\kappa_{10}^2 g_s^2 \in 2\pi \ZZ \ .
\end{align*}
Since $T_\mathrm{F1} = \frac{2\pi}{(2\pi \ell_s)^2}$, we have
\begin{align*}
 T_\mathrm{NS5} = \frac{2\pi}{T_\mathrm{F1} \cdot 2\kappa_{10}^2 g_s^2}
 = \frac{2\pi}{(2\pi \ell_s)^6 g_s^2} \ .
\end{align*}
See also Table \ref{table:tension}.

\subsection{Branes and strings ending on branes}

Now we consider D-brane systems in Type IIA theory. Let us recall that Maxwell equations lead to the charge conservation law:
\begin{align*}
 \begin{array}{l}
  d *\!F_{(2)} = J_e \\
  d F_{(2)} = J_m
 \end{array} \qquad\Rightarrow\qquad
 \begin{array}{l}
  d J_e = 0 \\
  d J_m = 0
 \end{array} \ .
\end{align*}
Because of the charge conservation, a world-line of a charged particle cannot have an endpoint and therefore it must be either a closed path or an infinitely long line.
If we apply this logic to branes, we may find the same result for branes.
However, the charge conservation in supergravity is quite non-trivial due to the non-linearity of the equations of motion. In this subsection, we will study the physics of D-branes from the equations of motion in (massive) Type IIA supergravity.

\subsubsection*{Massive IIA supergravity}

When Type IIA supergravity action is introduced in \eqref{IIA-SUGRA}, we do not include the R-R field corresponding to D$8$-brane. As explained in \S\ref{sec:intro-Dbrane}, D8-branes are non-dynamical, and their R-R field is a $9$-form so that its field strength is $10$-form $G_{(10)}$. Consequently, $d *\! G_{(10)} = 0$ leads to $*G_{(10)} = G_{(0)} \equiv m$, which clearly shows that D8-brane is non-dynamical. However, it actually has a constant but non-trivial contribution to the action, called \textbf{massive IIA supergravity}. Since it is a constant, it contributes as a mass term, called \textbf{Romans mass} to the action \cite{Romans:1985tz}.

Let us write the massive IIA supergravity action (we omit the wedge product $\wedge$ and the Hodge star $*$ here):
\begin{align*}
 &S_\mathrm{A,NS} = \frac{1}{2\kappa_{10}^2} \int d^{10}x \sqrt{ -G} e^{-2\Phi} \left[
 R +4 \partial_\mu \Phi \partial^\mu \Phi -\frac{1}{2} H_{(3)}^{2} \right] \ ,  \\
 &S_\mathrm{A,R} = \frac{1}{2\kappa_{10}^2} \int d^{10}x \sqrt{ -G} \left[
 -\frac{1}{2} m^{2} -\frac{1}{2} G_{(2)}^{2} -\frac{1}{2} G_{(4)}^{2} \right] \ ,  \\
 &S_\mathrm{A,CS} = \frac{1}{2\kappa_{10}^2} \int \left[
 -\frac{1}{2} B_{(2)} G_{(4)} G_{(4)}
 +\frac{1}{2} B_{(2)}^2 G_{(2)} G_{(4)}
 -\frac{1}{6} B_{(2)}^3 G_{(2)}^2 \right. \cr
 &\hspace{100pt} \left.
 -\frac{m}{6} B_{(2)}^3 G_{(4)}
 +\frac{m}{8} B_{(2)}^4 G_{(2)}
 -\frac{m^2}{40} B_{(2)}^5
 \right] \ ,
\end{align*}
where the field strengths of the R-R fields receive some modifications (here we omit the tilde notation for simplicity)
\begin{align}
 \begin{array}{l}
  H_{(3)} = dB_{(2)} \ , \\
  G_{(2)} = dC_{(1)} + m B_{(2)} \ , \\
  G_{(4)} =dC_{(3)} +dC_{(1)} B_{(2)} +\frac{1}{2} m B_{(2)}^2 \ .
 \end{array} \label{eq:fsGf}
\end{align}

From the action, we obtain the following equations of motion for the fields $C_{(1)}$ and $C_{(3)}$
\begin{align}\label{EOM}
 -d*\!G_{(2)} &= H_{(3)} *\!G_{(4)} \ , \cr
 d*\!G_{(4)} &= H_{(3)} G_{(4)} \ .
\end{align}
Given the relationship between the field strengths and their corresponding gauge fields \eqref{eq:fsGf}, we derive the following Bianchi identities
\begin{align}\label{Bianchi}
 d G_{(2)} &= m H_{(3)} \ , \cr
 d G_{(4)} &= H_{(3)} G_{(2)} \ .
\end{align}
If we relabel $m$ as $G_{(0)}$ and define the dual field strengths:
\begin{align*}
 G_{(10)} = * G_{(0)} \ , \qquad
 G_{(8)} = -*\! G_{(2)} \ , \qquad
 G_{(6)} = * G_{(4)} \ ,
\end{align*}
then the equations of motion \eqref{EOM} and the Bianchi identities \eqref{Bianchi} can be unified into
\begin{align*}
 d G_\mathrm{even} =  H_{(3)} G_\mathrm{even} \ ,
\end{align*}
where
\begin{align}
   C_\mathrm{odd} =& C_{(1)} + C_{(3)} + C_{(5)} + C_{(7)} + C_{(9)} \ ,\cr
 G_\mathrm{even} =& G_{(0)} + G_{(2)} + G_{(4)} + G_{(6)} + G_{(8)} + G_{(10)} \ .
\end{align}
This equation can be solved by setting
\begin{align*}
 G_\mathrm{even} = e^{B_{(2)}} \left( m + dC_\mathrm{odd} \right) \ .
\end{align*}

The equation of motion for the $B$-field is given by
\begin{align*}
 d \left(e^{-2\Phi}*\!  H_{(3)} \right) = m *\!G_{(2)} + *G_{(4)}G_{(2)} - \frac{1}{2} G_{(4)}^2 \ .
\end{align*}
Defining the dual field strength $H_{(7)} = e^{-2\Phi}*\!  H_{(3)}$, the equation of motion becomes
\begin{align}
 dH_{(7)} = \frac{1}{2} \left[ (* G_\mathrm{even}) G_\mathrm{even} \right]_{(8)} \ .
\end{align}
The Bianchi identity is straightforward: $d  H_{(3)} = 0$.
Note that these field strengths are invariant under the gauge transformations of the $B$-field and the R-R fields
\begin{align*}
 &\delta_B B_{(2)} = d \lambda_{(1)} \ , \qquad \delta_B C_\mathrm{odd} = -\lambda_{(1)} \left(m + dC_\mathrm{odd}\right) \ , \\
 &\delta_C B_{(2)} = 0 \ , \qquad \delta_C C_\mathrm{odd} = d \lambda_\mathrm{even} \ ,
\end{align*}
where we introduced a formal sum of gauge parameters
\begin{align*}
 \lambda_\mathrm{even} = \lambda_{(0)} + \lambda_{(2)} + \lambda_{(4)} + \lambda_{(6)} + \lambda_{(8)} \ .
\end{align*}

Now, let us introduce brane currents $J_{(8)}^\mathrm{F1}$, $J_{(4)}^\mathrm{NS5}$, and
\begin{align}
 J_\mathrm{odd} = J_{(1)}^\mathrm{D8} + J_{(3)}^\mathrm{D6} + J_{(5)}^\mathrm{D4} + J_{(7)}^\mathrm{D2} + J_{(9)}^\mathrm{D0} \ .
\end{align}
We can then include these in the equations of motion
\begin{align*}
 d H_{(3)} &= J_{(4)}^\mathrm{NS5} \ , \\
 d H_{(7)} &= J_{(8)}^\mathrm{F1} + \frac{1}{2} \left[(* G_\mathrm{even}) G_\mathrm{even} \right]_{(8)} \ , \\
 d G_\mathrm{even} &= J_\mathrm{odd} + H_{(3)} G_\mathrm{even} \ .
\end{align*}
From these, we can derive the following ``conservation'' laws
\begin{align*}
 d J_{(4)}^\mathrm{NS5} &= 0 \ , \\
 d J_{(8)}^\mathrm{F1} &= -\left[ J_\mathrm{odd} (* G_\mathrm{even}) \right]_{(9)} \ , \\
 d J_\mathrm{odd} &= -J_{(4)}^\mathrm{NS5} G_\mathrm{even} - J_\mathrm{odd} H_{(3)} \ .
\end{align*}
These equations imply several important facts:
\vspace{-4pt}
\begin{itemize}
 \setlength{\itemsep}{0pt}
 \item NS$5$-branes cannot have boundaries.
 \item F$1$ strings can end on any D-branes.
 \item D$p$-branes can end on NS$5$-branes for $p \leq 6$ (D$8$ cannot).
 \item D$p$-branes can end on D$(p+2)$-branes.
\end{itemize}

A similar analysis can be performed in Type IIB supergravity, or we can apply T-duality appropriately. The results are summarized in Table \ref{table:branes-branes}.

\begin{table}[htbp]
 \begin{center}
\begin{tabular}{c|c}
 Brane & Branes end on \\\hline
 F1 & nothing \\
 NS5-brane & D0, D2, D4, D6 \\
 D0-brane & F1 \\
 D2-brane & F1, D0 \\
 D4-brane & F1, D2 \\
 D6-brane & F1, D4 \\
 D8-brane & F1, D6
\end{tabular}
\hspace{3cm}
\begin{tabular}{c|c}
 Brane & Branes end on \\\hline
 F1 & nothing \\
 NS5-brane & D1, D3, D5 \\
 D1-brane & F1 \\
 D3-brane & F1, D1 \\
 D5-brane & F1, D3 \\
 D7-brane & F1, D5 \\
\end{tabular}
\caption{Which branes can end on a brane in Type IIA (left) and IIB (right).}\label{table:branes-branes}
\end{center}
\end{table}


\subsubsection*{D0-D2 bound states and Myers effect}

The general R-R coupling \eqref{RR-general} includes all possible R-R fields $ C_\mathrm{RR}$ in the theory. This implies that D$p$-branes can include lower-dimensional D$(p-2n)$-branes ($n \in \ZZ_{\ge0}$), forming a bound state.

\begin{figure}[htb]
\centerline{\includegraphics[width=\textwidth]{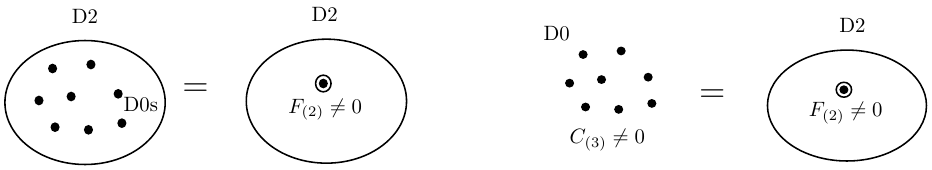}}
\caption{Myers effect: transition from D$0$-branes to D$0$-D$2$ .}
\label{fig:Myers}
\end{figure}

Let us consider a concrete example of a D$2$-brane supported on $\Sigma$ where the R-R coupling at $B_{(2)}=0$ is
\begin{align*}
 S \sim \frac{1}{2\pi} \int_{\RR_t \times \Sigma} \left( C_{(3)} + F_{(2)} C_{(1)}\right) \ ,
\end{align*}
Note that the $F_{(2)}$ flux needs to be quantized
\begin{align*}
 \frac{1}{2\pi}\int_{\Sigma} F_{(2)} =  n \in \ZZ \ .
\end{align*}
If $C_{(3)}$ is zero, the D2-brane tries to shrink to a point due to its tension.
On the other hand, $C_{(1)}$ part remains finite because the flux $F_{(2)}$ is quantized as
\begin{align*}
 S \sim \frac{1}{2\pi} \int_{\RR_t \times \Sigma} \left( F_{(2)} C_{(1)}\right) \to n \int_{\RR_t} C_{(1)} \ .
\end{align*}
This is equivalent to $n$ D$0$-branes.
Namely, when the D$2$-brane with $n$ flux shrinks to a point, $n$ D$0$-branes remain. This can be understood as a bound state of D$2$- and D$0$-branes (see Figure \ref{fig:Myers}).

Let us consider the opposite process of the previous argument.
When there are $n$ D$0$-branes in the non-trivial $C_{(3)}$ background flux, they can become a D$2$-brane with the $F_{(2)}$ flux.
Due to the non-trivial $C_{(3)}$ flux, being a D$2$-brane is energetically more preferable than being the D$0$-branes (see Figure \ref{fig:Myers}).
This is similar to the polarization phenomenon in electromagnetism,
and in this case, it is called \textbf{Myers effect} \cite{Myers:1999ps}.

%
%

\subsection{Brane creations/annihilations}\label{sec:HW}

\begin{figure}[htb]
\centerline{\includegraphics[width=8cm]{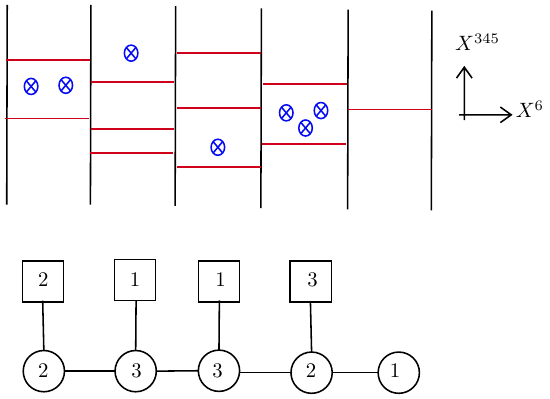}}
\caption{Type IIB brane system where NS5/D5/D3-branes are colored by black/blue/red,  and the corresponding 3d $\cN=4$ quiver theory where circles/squares represent unitary gauge/flavor groups.}
\label{fig:quiver}
\end{figure}

In fact, the last two Type IIB D-brane configurations in Table \ref{tab:Tdual} preserve a quarter of supersymmetries, and they give rise to 3d $\cN=4$ theories in $X^{012}$. Let us first consider the third configuration of Table \ref{tab:Tdual}. As in Figure \ref{fig:quiver}, $N$ D3-branes suspended by NS5-branes give rise to $\U(N)$ gauge group (or vector multiplet), and F1-strings between D3- and D5-branes give rise to matters (hypermultiplet) in the fundamental representation. Also, bifundamental matter (hypermultiplet) arises from F1-strings between two adjacent D3-branes. The resulting theory is often represented by a quiver diagram.
3d $\cN=4$ supersymmetry is endowed with $\SU(2)_C\times \SU(2)_H$ R-symmetry where $\SU(2)_C$ acts on the $X^{345}$ direction and  $\SU(2)_H$ acts on the $X^{789}$ direction in the Type IIB setup. Also, 3d $\cN=4$ supersymmetric theories have moduli spaces of vacua, called Coulomb and Higgs branch. Both are hyper-K\"ahler manifolds.
In the brane realization, the Coulomb branch corresponds to the motion of D3-branes along $X^{345}$ and the Higgs branch corresponds to the motion of D3-branes along $X^{789}$. Therefore, a large class of 3d $\cN=4$ theories can be studied by using Type IIB brane configurations in the third configuration of Table \ref{tab:Tdual}.

\begin{figure}[htb]
\centerline{\includegraphics[width=8cm]{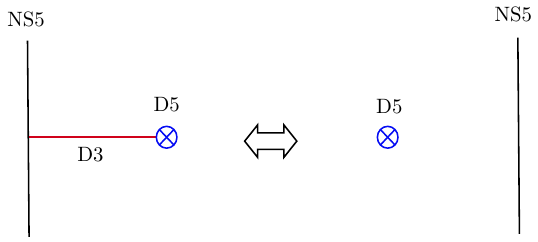}}
\caption{The crossing of NS5 and D5-brane leads to D3-brane creation/annihilation.}
\label{fig:HW}
\end{figure}

What makes this system very intriguing is a brane creation/annihilation process called \textbf{Hanany-Witten transition} \cite{Hanany:1996ie}. As a D5-brane crosses an NS5-brane, a D$3$-brane is created or annihilated. See Figure \ref{fig:HW}. In the Hanany-Witten transition, the s-rule constrains that at most one D3-brane can be suspended between an NS5 and a D5-brane.


\begin{figure}[htb]
\centerline{\includegraphics[width=\textwidth]{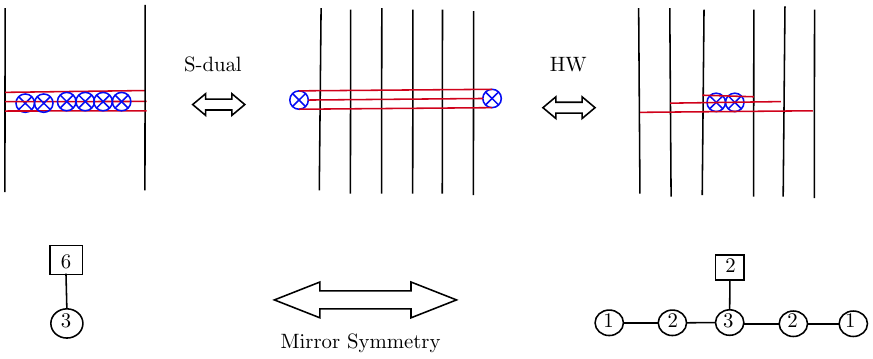}}
\caption{The S-duality maps from the left to middle brane configuration. Moving the two D5-branes into the middle involves the Hanany-Witten transition, and D3-branes are annihilated. The 3d $\cN=4$ theories from the left and the right brane configuration are dual to each other.}
\label{fig:MS}
\end{figure}

As in Table \ref{tab:Tdual}, Type IIB theory enjoys the S-duality which exchanges NS5 and D5-branes.  The combination of the S-duality and the Hanany-Witten transition predicts non-trivial duality in 3d $\cN=4$ theories, called the 3d $\cN=4$ \textbf{mirror symmetry} \cite{Intriligator:1996ex}. One example of mirror symmetry is illustrated in Figure \ref{fig:MS}. Since the S-duality exchanges NS5 and D5-branes, so are the roles of the Coulomb and Higgs branches (namely the roles of $X^{345}$ and $X^{789}$). Therefore, 3d $\cN=4$ supersymmetric theories can be studied from many perspectives such as branes, dualities, geometry of moduli spaces,  and quantum field theories. They also admit mathematically very deep interpretations in representation theory, the geometry of hyper-K\"ahler manifolds, and equivalences of categories.

The study of supersymmetric theories by D-branes is very fruitful and successful. Brane dynamics provides profound insights into supersymmetric theories. Since the subject is very rich and broad, the reader is referred to, for instance,  \cite{Hanany:1996ie,Douglas:1996sw,Witten:1997sc,Aharony:1997bh,Taylor:1997dy,Giveon:1998sr,Johnson:2000ch} as a starting point.

\section{Black holes in string theory}\label{sec:BH}

In the previous section, we explored the remarkable properties of D-branes, including their dynamical nature and their ability to form bound states. These features make D-branes an ideal tool for studying black holes.

When a large number of D-branes is present, their collective mass can give rise to a black hole. In particular, wrapping a cycle in a compact manifold can generate a black hole configuration. The presence of open strings attaching to the D-branes leads to a significant degeneracy, which provides a statistical interpretation of the thermodynamic entropy associated with the black hole. This remarkable result, first derived by Strominger-Vafa \cite{Strominger:1996sh}, provides a microscopic explanation for the Bekenstein-Hawking entropy of supersymmetric black holes.

The investigation of black holes within the framework of string theory using D-branes has played a crucial role in the development of the AdS/CFT correspondence \cite{Maldacena:1997re}. This groundbreaking duality, first proposed by Maldacena, establishes a deep connection between the quantum theory of gravity in Anti-de Sitter (AdS) spacetime and conformal field theories (CFTs) defined on the boundary of the AdS space. Maldacena's Ph.D. thesis \cite{Maldacena:1996ky} is an excellent reference for further exploration of this correspondence.

Before delving further into these aspects, it is important to review the basics of black holes in the framework of general relativity and the laws of black hole thermodynamics established in the early 1970s \cite{Bekenstein:1972tm,Bekenstein:1973ur,Bardeen:1973gs,Hawking:1974sw}. A comprehensive understanding of these concepts can be found in an excellent lecture note by Townsend \cite{Townsend:1997ku}, and for a historical account of black hole entropy, I recommend the work by Weinstein \cite{Weinstein:2021dop}.

\subsection{Black holes}

To begin with, we consider the Einstein-Maxwell action
\begin{equation}\label{EMaction}
\frac{1}{16\pi} \int  d^4x  \sqrt{g} \Bigl(\frac1G R-  F_{\mu\nu}F^{\mu\nu}\Bigr)~,
\end{equation}
where $G$ is Newton's constant. In this subsection, we shall review black hole solutions to the action \eqref{EMaction} and see that they are characterized by mass $M$, charge $Q$ and angular momentum $J$.

\subsubsection*{Schwarzschild metric}

If there are no electromagnetic fields $F=0$ in the action \eqref{EMaction}, the equation of motion is $R_{\mu\nu}-{\frac12}g_{\mu\nu}R = 0$, which has a spherically symmetric, static solution
\[
ds^2 \equiv g_{\mu\nu} dx^\mu dx^\nu =  - \Bigl(1 - \frac{2GM}{r}\Bigr) dt^2
+ \Bigl(1 - \frac{2GM}{r}\Bigr)^{-1} dr^2 + r^2 d \Omega^2,
\]
where $t$ is the time, $r$ is the radial coordinate, and $d\Omega$
is the canonical metric of a 2-sphere.
This metric describes the spacetime outside a gravitationally collapsed
non-rotating star with zero electric charge, called \textbf{Schwarzschild metric}.
It is well-known that the \textbf{event horizon} appears at
\[
g^{rr} = 0 \,,
\]
and  the sphere $r = 2GM$ is indeed the {event horizon} of the
Schwarzschild black hole with mass $M$.

It turns out that much of the interesting physics having to
do with the quantum properties of black holes comes from the
region near the event horizon.
To examine the region \emph{near $r=2GM$}, we analytically continue to the Euclidean metric $t= -it_E$, and we set
\[
r-2GM=\frac{x^2}{8GM}~.
\]
Then, the metric near the event horizon $r=2GM$ becomes
\[
ds^2_{\textrm{E}} \approx  (\kappa x)^2dt_E^2+dx^2
+\frac{1}{4\kappa^2}d\Omega^2~,
\]
where $\kappa=\frac1{4GM}$ is called the \textbf{surface gravity} because it is indeed the acceleration of a static
particle near the horizon as measured at spatial infinity. Note that the surface gravity is defined by using Killing vector at the horizon, precisely speaking \cite{Townsend:1997ku}.
The first part of the metric is just $\bR^2$ with polar coordinates if we make the
{periodic identification}
\[
t_E \sim t_E +\frac{2\pi}{\kappa}~.
\]
Using the relation between Euclidean periodicity and temperature,
we can deduce \textbf{Hawking temperature} of the Schwarzschild black hole
\begin{equation}\label{hawktemp}
k_B T_H = \frac{\hbar \kappa}{2 \pi c}=\frac{\hbar c^3}{8\pi GM}~.
\end{equation}
Here we restore the Boltzmann constant $k_B$, and the speed of light $c$.
This is a very heuristic way to introduce the Hawking temperature which was not originally found in this way.

\subsubsection*{Reissner-Nordstr\"om  black hole}

The most general static, spherically symmetric, charged  solution
of the Einstein-Maxwell theory (\ref{EMaction}) is
\begin{equation}\label{rn}
ds^2 = -\left(1 - \frac{2GM}{r} + \frac{G Q^2}{r^2}\right) dt^2 +
\left(1 - \frac{2GM}{r} + \frac{G Q^2}{r^2}\right)^{-1} dr^2 + r^2 d
\Omega^2,
\end{equation}
with the electromagnetic field strength
\[
F_{tr} = \frac{Q}{r^2}~.
\]
This solution is called the \textbf{
Reissner-Nordstr\"om (RN) black hole} with mass $M$ and charge $Q$.
{}From the metric \eqref{rn} we see that there are two event horizons for this solution where
$g^{rr} =0$ at
\[
r_\pm = GM \pm \sqrt{(GM)^2 - GQ^2}~.
\]
Thus, $r_+$ defines the outer horizon of the black hole and $r_-$
defines the inner horizon of the black hole. The area of the black
hole is  $4\pi r_+^2$. It turns out that the Hawking temperature of the RN black hole is
\[
T_H = \frac{\sqrt{(GM)^2-GQ^2}}{2\pi
\left(GM+\sqrt{(GM)^2-GQ^2}\right)^2} \,.
\]

For a physically sensible definition of temperature, the mass must satisfy the bound $GM^2 \geq Q^2$, and  the two horizons
coincide $r_+ = r_- = GM$ when  this bound is saturated.
In this case, the temperature of the black hole is zero and it is called an \textbf{extremal black hole}.

\subsubsection*{Kerr-Newman black hole}

If we relax the static condition, black holes can have angular momentum. Hence, general stationary solutions, called \textbf{Kerr-Newman black holes}, to the action \eqref{EMaction}
are described with three parameters. In \textbf{Boyer-Lindquist
coordinates}, the KN metric is
\bea\nonumber
ds^2 & =  -\frac{ \left(\Delta -a^2\sin^2\theta\right)}{\Sigma}dt^2 - 2 a
\sin^2\theta \frac{ \left(r^2+a^2-\Delta\right)}{\Sigma}dt\,d\phi \\
 &  +\left( \frac{ \left(r^2+a^2\right)^2-\Delta a^2\sin^2\theta}{\Sigma}
\right)\sin^2\theta d\phi^2 +\frac{\Sigma}{\Delta}dr^2+\Sigma d\theta^2
\eea
where
\bea\nonumber
\Sigma & =  r^2+a^2\cos^2\theta \cr
\Delta & =  r^2-2Mr+a^2+e^2~. \eea
The three parameters are $M$, $a$, and $e$.  It can be shown that
\[
a=\frac{J}{M}
\]
where $J$ is the total angular momentum, while
\[
e = \sqrt{ Q^2+P^2}
\]
where $Q$ and $P$ are the electric and magnetic (monopole) charges,
respectively.  The Maxwell 1-form of the KN solution is
\[
A_\mu dx^\mu= \frac{ Qr\left(dt-a\sin^2\theta d\phi\right)-P\cos\theta
\left[a dt-\left(r^2+a^2\right)d\phi\right] }{\Sigma} ~.
\]

\subsection{Black hole thermodynamics and Bekenstein-Hawking entropy}

Classically, a stationary black hole is characterized by its mass $M$, angular momentum $J$, and charge $Q$. This property is known as the ``black hole no hair theorem''. However, Bekenstein raised an intriguing question in \cite{Bekenstein:1972tm}: if we consider a black hole as a purely geometric object, what happens to the entropy of the universe when we throw an object with entropy (such as a cup of tea) into the black hole? Naively, the total entropy outside the black hole would seem to decrease, which contradicts the second law of thermodynamics that states the total entropy of a closed system never decreases. To reconcile this contradiction, it suggests that black holes must possess their own entropy. The challenge then becomes how to quantify this entropy.

The key insight came from the area theorem of black holes \cite{Hawking:1971tu,Hawking:1971vc}, which states that the total area of the black hole horizons never decreases in any physical process. For example, when two Schwarzschild black holes with masses $M_1$ and $M_2$ merge to form a larger black hole with mass $M = M_1 + M_2$, the total area of the resulting black hole is greater than or equal to the sum of the areas of the individual black holes, as the area is proportional to the square of the mass, namely, $(M_1 + M_2)^2 \geq M_1^2 + M_2^2$. Conversely, the reverse process of a larger black hole splitting into two smaller ones is not allowed by this theorem. Motivated by the area theorem, Bekenstein proposed in \cite{Bekenstein:1972tm} that black holes possess entropy proportional to their horizon area.

Shortly thereafter, Bardeen, Carter, and Hawking established the analogy between the laws of black hole mechanics and the laws of thermodynamics in \cite{Bardeen:1973gs}. They found precise correspondences between the laws of black hole dynamics and the three laws of thermodynamics:
\begin{enumerate}
\item[{(0)}]
Zeroth Law: In thermodynamics, the zeroth law states that the temperature $T$ of a thermal equilibrium object is constant throughout the body. Correspondingly, for a stationary black hole, its surface gravity $\kappa=1/4GM$ is constant over the event horizon.

\item[{(1)}]
First Law: The first law of thermodynamics states that energy is conserved, and the variation of energy is given by
\be dE = TdS + \mu dQ + \Omega dJ \ee
where $E$ is the energy, $Q$ is the charge with chemical potential $\mu$ and $J$ is the angular momentum with chemical potential
$\Omega$ in the system.
Correspondingly, for a black hole, the variation of its mass is given by
\be dM = \frac{\kappa}{8\pi G} dA + \mu dQ + \Omega dJ \ee
where $A$ is the area of the horizon,  and $\kappa$ is the surface gravity, $\mu$ is the chemical potential conjugate to $Q$, and $\Omega$ is the angular velocity conjugate to $J$.

\item[{(2)}]
Second Law: The second law of thermodynamics states that the total entropy $S$ never decreases, $\delta S \geq 0$.  Correspondingly, for a black hole, the area theorem states that the total area of a black hole in any process never decreases, $\delta A \geq 0$.
\end{enumerate}

\begin{table}[ht]
\centering
\begin{tabular}{c|c}
\hline
\textbf{Laws of thermodynamics} & \textbf{Laws of black hole mechanics}\\
 \hline
 Temperature is constant & Surface gravity is constant \\
 throughout a body at equilibrium. &  on the event horizon.\\
 $T$=constant. & $\kappa$ =constant.\\
 \hline
Energy is conserved.& Energy is conserved. \\
$dE = T dS + \mu dQ + \Omega dJ. $& $dM = \frac{\kappa}{8\pi G} dA + \mu dQ + \Omega dJ . $\\
 \hline
 Entropy never decreases.  & Area never decreases.\\
 $\delta S \geq 0$. & $ \delta A \geq 0 $. \\
 \hline
\end{tabular}
\caption{\small{Laws of black hole thermodynamics}}
\label{blackholelaws}
\end{table}

This result can be understood as one of the highlights of general relativity. Classically, a black hole is not only a geometric but also a thermodynamic object.
If a black hole has energy $E$ and entropy $S$, then it must also have temperature $T$ given by
\[
\frac{1}{T} = \frac{\partial S}{\partial E}.
\]
For example, for a Schwarzschild black hole, the area and the entropy are proportional to $M^2$. Hence, we can derive
\[
\frac{1}{T} = \frac{\partial S}{\partial M} \sim \frac{\partial M^2
}{
\partial M} \sim M.
\]
Therefore, black hole temperature is inversely proportional to mass $M$. The smaller a black hole is, the hotter it is!
Moreover, if the black hole has temperature, it must thermally radiate like any hot body. The understanding of the thermal properties of black holes requires treatment beyond classical general relativity.

Hawking has applied techniques of quantum field theories on a curved background to the near-horizon region of a black hole and showed that a black hole indeed radiates \cite{Hawking:1974sw}.
This can be intuitively understood as follows: in a quantum theory, particle-antiparticle creations constantly occur in the vacuum. Around the horizon, after pairs are created, antiparticles fall into a black hole due to the gravitational attraction whereas particles escape to infinity.  Although we do not deal with Hawking's calculation unfortunately (see \cite{Townsend:1997ku}), it indeed justifies this picture. Moreover, it revealed that the spectrum emitted by the black hole is precisely subject to the thermal radiation with temperature \eqref{hawktemp}. Indeed, a black hole is not black at quantum level.
Hence, we can treat a black hole as a thermal object, and the analogy of the laws in Table \ref{blackholelaws} can be understood as the natural consequence of the laws of thermodynamics. As a result,
the formula for the Hawking temperature \eqref{hawktemp} and the first law of thermodynamics
\[
c^2 dM = T_HdS = \frac{\kappa c^2}{8\pi G} dA,
\]
lead to the precise relation between entropy and the area of the black hole:
\be \label{BH-entropy}
 S = \frac{k_B c^3 A}{4G\hbar} \, .
\ee
This is a universal result for any black hole, and this remarkable relation between the thermodynamic properties
of a black hole and its geometric properties is called the celebrated \textbf{Bekenstein-Hawking entropy formula}.
This formula involves all four fundamental constants of nature; $(G,c,k_B,\hbar)$. Also, this is the first place where the Newton constant $G$ meets with the Planck constant $\hbar$.
Thus, this formula shows a deep connection between black hole geometry, thermodynamics and quantum mechanics.

For ordinary objects, Boltzmann has given the statistical interpretation of the thermodynamic entropy of a system.
We fix the macroscopic parameters (e.g. total electric charge, energy etc.) and count the number $\O$ of quantum states, known as microstates, each of which has the same values for the macroscopic parameters, and the entropy is expressed as
\[
S = k_B \log \O~.
\] 
Since the Bekenstein-Hawking entropy \eqref{BH-entropy} behaves as the ordinary thermodynamic entropy in every aspect, it is therefore natural to ask
whether the black hole entropy admits a statistical interpretation in the same way.

Furthermore, one of the most profound implications of Hawking's work is the notion that black holes are associated with information loss. In quantum mechanics, information is associated with pure states. If we throw a pure quantum state, such as an s-wave, into a black hole, it eventually emerges as a thermal (mixed) state. As a result, the evolution of a pure quantum state into a mixed state violates the law of quantum mechanics known as unitarity. This conundrum is referred to as the \textbf{information paradox} \cite{Hawking:1976ra}. The paradox arises due to Hawking's semi-classical analysis, where the background is fixed and particles are quantized.  In fact, the information paradox stems from the absence of such a microscopic description of gravity.

In order to investigate the microscopic description of black hole entropy, we need quantum theory of gravity. This is precisely what string theorists have attempted to do and have been partially successful.

\subsection{Black holes in string theory}
In string theory on a $d$-dimensional compact manifold, branes can be wrapped in a cycle of the compact manifold and it looks like a point-like object in $(10-d)$-dimensional spacetime. In the regime where supergravity approximation is valid, configurations of this kind give rise to black hole solutions of the corresponding low-energy supergravity theory. Moreover, if a brane configuration preserves supersymmetry, then the corresponding solution will be an extremal supersymmetric black hole. Extremal black holes are interesting because they are stable against Hawking radiation and nevertheless have a large entropy.
On the other hand, configurations without supersymmetry yield non-extremal black holes.

In general, the regime of the parameter space in which supergravity is valid is different from the regime in which the microstate counting can be performed. Thus, even if we know that a given brane configuration becomes a black hole when we go from weak to strong coupling, it is generally difficult to extract microscopic information of the black hole from the brane configuration.

For supersymmetric black holes, however, one can count the number of states at weak coupling and extrapolate the result to the black hole phase due to the BPS property. We will see that in this way, one derives the Bekenstein-Hawking entropy formula (including the precise numerical coefficient) for a 5d supersymmetric black hole \cite{Strominger:1996sh}. (For more detail, I refer to \cite{David:2002wn,Dabholkar:2012zz}.)

\subsubsection*{D1-D5-P brane system}

\begin{figure}[ht]\centering
\includegraphics[width=13cm]{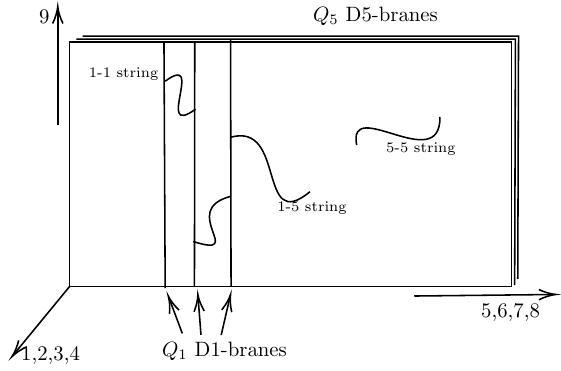}
\caption{The D1-D5-P brane system.}
\end{figure}

Here we follow the example treated in \cite{Callan:1996dv}, called the D1-D5-P brane system on $\bR^{1,4}\times T^5$.
Let us consider Type IIB compactified on a five-torus $T^5=T^4\times S^1$, which spans the $(x_5 \cdots x_9)$ coordinates, with $Q_1$ D1-branes and $Q_5$ D5-branes in the following configuration.
\begin{table}[ht]\centering\begin{tabular}{c|cccccccccc}
 & 0 & 1 & 2 & 3 & 4 & 5 & 6 & 7 & 8 & 9\tabularnewline
\hline
$Q_1$ D1 & $\times$ &  &  &  & &  &  &  &  & $\times$\tabularnewline
$Q_5$  D5& $\times$ &  &  &  &  & $\times$ & $\times$ & $\times$ & $\times$ & $\times$\tabularnewline
$Q_P$ mom &  &  &  &  & &  &  &  &  & $\rightsquigarrow$\tabularnewline
\end{tabular}\end{table}
We consider that the volume of  $T^4$ is $(2\pi)^4V$ and the radius of $S^1$ is $R$.
Here we also assume that there is an excitation by open strings carrying momenta $Q_P/R$ in the $x_9$-direction.
This system preserves 4 real supercharges since each constituent breaks a half of supersymmetry.

\subsubsection*{Black hole in 5d supergravity}

If there are large enough D-brane charges $(Q_1,Q_5,Q_P)$ and the five-torus is sufficiently small, the configuration produces a 5d black hole. We would like to compute the Bekenstein-Hawking entropy \eqref{BH-entropy} of the black hole by evaluating the area of the event horizon.
When 5d supergravity analysis is valid, this brane system gives rise to a $1/8$-BPS black hole configuration. Ignoring the R-R field and the $B$-field configuration, the 5d Einstein frame metric of this solution then becomes
\bea\nonumber
ds_5^2 = &\!
-\lambda(r)^{-\frac23} dt^2 + \lambda(r)^{\frac13}
\left[ dr^2 + r^2 d\Omega_3^2 \right] \,,
\eea
where the harmonic functions are
\[
\lambda(r)=H_1(r)H_5(r) H_P(r)  = \Bigl(1 + {\frac{r_1^2}{r^2}} \Bigr)\Bigl( 1 + {\frac{r_5^2}{r^2}}\Bigr)\Bigl(  1+{\frac{r_P^2}{r^2}}\Big )\,,
\]
with
\[
r_1^2 = {\frac{g_s Q_1\ell_s^6}{V}} \,, \qquad
r_5^2 = g_s Q_5 \ell_s^2 \, \qquad
r_P^2 = {\frac{g_s^2 Q_P \ell_s^8}{R^2 V}} \,.
\]
Let us briefly evaluate the validity of the supergravity analysis. In order for the $\alpha^\prime$ corrections to geometry to be small, the radius parameters have to be large
with respect to the string unit, $r_{1,5,P}\!\gg\!\ell_s$. Since we assume $V^{1/4}$, $R$ are an order of the string length, this implies
\be \label{sugra-limit}
g_s Q_1\gg1~, \qquad g_s Q_5\gg1~, \qquad g_s^2 Q_P\gg1~.
\ee
To suppress string loop corrections, we need $g_s$ to be small (but finite) so that the D-brane charges must be sufficiently large for supergravity analysis.

It turns out that the surface gravity and therefore the Hawking temperature of this black hole is zero, $T_{{H}}=0$, as expected.  The metric shows that the event horizon is located at $r=0$ and
the Bekenstein-Hawking entropy \eqref{BH-entropy} is
\bea\label{sugra-result}
S_{\textrm{macro}} &= {\displaystyle{
 {\frac{A}{4 G_5}}  = {\frac{1}{4G_5}}2
\pi^2\left[r^2\lambda(r)^{\frac13}\right]^{\frac32} }}
\ \ {\rm{at\ }} r=0 \,\cr
&= {\displaystyle{
 {2\frac{\pi^2}{4\left[\pi g_s^2\ell_s^8/(4VR)\right]}}
  r_1 r_5 r_P
= {\frac{2\pi{}VR}{g_s^2\ell_s^8}}
\left(
{\frac{g_s{}Q_1\ell_s^6}{V}}\,g_s{}Q_5\ell_s^2\,
{\frac{g_s^2{}Q_P\ell_s^8}{R^2V}} \right)^{\frac12} }} \cr
 &=  2\pi\sqrt{Q_1 Q_5 Q_P} \,,
\eea
where we use  $G_5=\frac{G_{10}}{(2\pi)^5VR}$ and $16\pi G_{10}=(2\pi)^7g_s^2\ell_s^8$.
Notice that it is also independent of $R$ and of $V$ while the ADM mass depends on $R$, $V$ explicitly.
\[
M = {\frac{Q_P}{R}} + {\frac{Q_1R}{g_s\ell_s^2}}
+ {\frac{Q_5RV}{g_s\ell_s^6}}\,.
\]

\subsubsection*{Counting microstates}

The next step is to identify the degeneracy of open string states of the D1-D5-P system, which can be analyzed at the limit opposite to \eqref{sugra-limit}, i.e.
\be\label{CFT-regime}
g_s Q_1\ll1~, \qquad g_s Q_5\ll1~, \qquad g_s^2 Q_P\ll1~.
\ee
Further simplification can be made by taking the limit that the
volume of $T^4$ is small as compared to the radius of the circle $S^1$,
\[
V^{\frac14}\ll R \,.
\]
In this limit,  the theory on the D-branes is an effective $2d$ theory in the $(x_0,x_9)$-directions. Moreover,  the smeared D1-branes plus D5-branes have a
symmetry group
$SO(1,1)\!\times\!\SO(4)_\parallel\!\times\!\SO(4)_\perp$ where $\SO(4)_\parallel\!\times\!\SO(4)_\perp$ becomes $R$-symmetry of the $2d$ theory which we call $\cN=(4,4)$ 2d CFT. In the supersymmetric configuration, the left-movers are in their ground states so that we count excited right-movers.

Because the D1-branes are instantons in the D5-brane theory, the
low-energy theory of interest is actually a $\sigma$-model on the
moduli space of instantons
\[{\cal{M}}=\textrm{Sym}^{Q_1Q_5}(T^4)=(T^{4})^{Q_1Q_5}/S_{Q_1Q_5}~.\]
The central charge of this 2d CFT is
\be \label{D1D5Pc} c=n_{\rm
bose}{+}{\frac12}n_{\rm fermi}=6Q_1Q_5~.\ee
Roughly, this central charge $c$ can be thought of as coming from having $Q_1Q_5$ 1-5
strings that can move in the 4 directions of $T^4$. Although this orbifold theory has many twisted sectors, the special point of the moduli space corresponds to a single string winding $Q_1Q_5$ times. It turns out that counting the excitations of this \textbf{long string} is only relevant in the limit of large D-brane charges.  For this long string, the level-matching condition is
\[N - \overline N =  {Q_P} W  ~, \quad W=Q_1Q_5~, \quad \to \quad N= Q_PQ_1Q_5\]
where the left-movers are in the ground states $\overline N=0$.

If $N_m^i$ and $n^i_m$ denote occupation numbers of the four compact ($T^4$) bosonic and fermionic oscillators, respectively, then evaluation of $N$ gives
\be\label{occupation}
Q_PW=\sum_{i=1}^4\sum_{m=1}^\infty m(N_m^i+n^i_m)
\ee
The degeneracy $\O(Q_1, Q_5, Q_P)$ is then given by the number of choices for $N_m^i$ and $n^i_m$ subject to \eqref{occupation}.

The partition function of this system is the partition function for
4 bosons and an equal number of fermions
\[
Z = \left[\prod_{m=1}^\infty {\frac{1+q^{m}}{1-q^{m}}}
\right]^{4} \equiv \sum \Omega(Q_1, Q_5, Q_P) q^{N} \,,
\]
where $\Omega(Q_1,Q_5,Q_P)$ is the degeneracy of states at the level $N= {Q_PQ_1Q_5}$. The Cardy formula \cite{Cardy:1986ie} can be applied in the regime in which the KK momentum is much larger than the
central charge, $Q_P\gg Q_1Q_5$, \eqref{D1D5Pc} while assuming \eqref{CFT-regime}:
\[
\Omega(Q_1,Q_5,Q_P)\!\sim\! \exp\left(2\pi\sqrt{Q_1Q_5Q_P} \right)
= \exp\left(2\pi\sqrt{\frac{c}{6}\,Q_P}\right)
\]
Therefore the microscopic D-brane statistical entropy is
\[
S_{\rm{micro}} = \log\left(\Omega(Q_1,Q_5,Q_P)\right) = 2\pi\sqrt{Q_1Q_5Q_P}
\,.
\]
This agrees exactly with the black hole result \eqref{sugra-result}!

\begin{figure}[ht]\centering
\includegraphics[width=16cm]{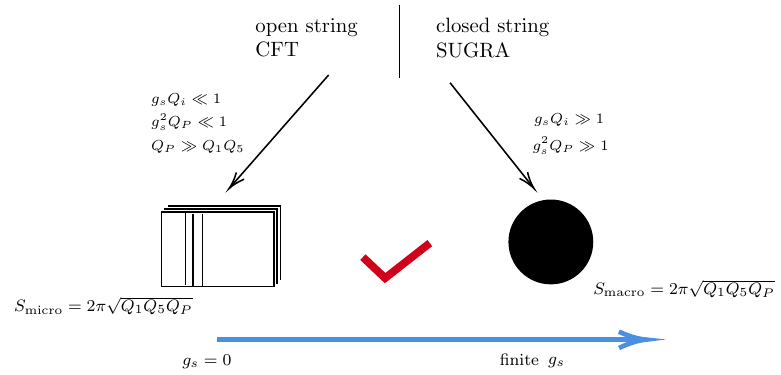}\caption{The BPS property protects the state counting as the coupling is varied.}
\end{figure}

\subsubsection*{More results}

By coupling the low energy degrees of freedom in the D1-D5-p system to supergravity modes (therefore perturbing the extremal condition), one can also compute the rate of Hawking radiation from a black hole that agrees precisely with the Hawking calculation. Thus this provides a microscopic explanation of Hawking radiation. (See \cite[\S8]{David:2002wn}.)

In fact, vigorous research in the last decade has shown that one can show the exact match between macroscopic and microscopic calculations of black hole entropy even at finite D-brane charges.

Moreover, a generalization of Bekenstein-Hawking entropy has been proposed in \cite{Ryu:2006bv} that connects quantum theory of gravity and quantum information theory.  A recent study has clearly suggested that quantum entanglement must have something to do with quantum physics of spacetime.

\section{Introduction to AdS/CFT correspondence}\label{sec:AdSCFT}

The study of black holes in string theory by using D-branes has led to the celebrated AdS/CFT correspondence \cite{Maldacena:1997re}. The AdS/CFT correspondence is the equivalence between a string theory or M-theory on an anti-de Sitter background and a conformal field theory. It has shed new light on quantum gravity as well as strongly coupled quantum field theories. Although it was proposed in the framework of string theory, it has already been studied beyond string theory, influencing other physical theories. It has attracted a large number of researchers, and it is connected to many branches of physics. For the basics of the AdS/CFT correspondence, the most famous review is \cite{Aharony:1999ti}, and there are many other reviews and books, \textit{e.g.} \cite{DHoker:2002nbb,Maldacena:2003nj,Ramallo:2013bua,nuastase2015introduction}.

We first study conformal symmetry and anti-de Sitter geometry, then derive the D3-brane example of the correspondence. We develop the field/operator dictionary into a prescription for correlation functions and apply it to thermal states and external probes. Throughout, we distinguish the conjectured exact duality from the limits in which classical bulk calculations are reliable.

\subsection{Conformal field theory}\label{sec:conformal}
A conformal transformation preserves the metric up to a local scale factor. In flat space, an infinitesimal transformation $x'^\mu=x^\mu+\epsilon^\mu(x)$ obeys the conformal Killing equation
\begin{equation}\label{eq:def1}
\partial_\mu\epsilon_\nu+\partial_\nu\epsilon_\mu
=\frac{2}{d}\eta_{\mu\nu}\partial\cdot\epsilon.
\end{equation}
For $d>2$, differentiating and permuting indices gives
\[
\partial_\mu\partial_\nu\epsilon_\rho
=\eta_{\mu\rho}\partial_\nu\sigma
+\eta_{\nu\rho}\partial_\mu\sigma
-\eta_{\mu\nu}\partial_\rho\sigma,
\qquad \sigma=\frac1d\partial\cdot\epsilon.
\]
Compatibility of the third derivatives implies $\partial_\mu\partial_\nu\sigma=0$. Thus $\epsilon^\mu$ is at most quadratic in $x$ and has the form
\[
\epsilon^\mu=a^\mu+\omega^\mu{}_{\nu}x^\nu+\alpha x^\mu
+2(b\cdot x)x^\mu-b^\mu x^2,
\qquad \omega_{\mu\nu}=-\omega_{\nu\mu}.
\]
The parameters describe translations, rotations or Lorentz transformations, dilations and special conformal transformations. In $d=2$, the compatibility argument is different, allowing the local holomorphic transformations studied in \S\ref{sec:2dcft}.

\begin{table}[htbp]\centering
\begin{tabular}{ll}
\toprule
Transformation & Finite coordinate map \\
\midrule
Translation & $x'^\mu=x^\mu+a^\mu$ \\
Rotation or Lorentz transformation & $x'^\mu=\Lambda^\mu{}_{\nu}x^\nu$ \\
Dilation & $x'^\mu=a x^\mu$, $a>0$ \\
Special conformal transformation & $x'^\mu=(x^\mu-b^\mu x^2)/(1-2b\cdot x+b^2x^2)$ \\
\bottomrule
\end{tabular}
\caption{Finite conformal transformations in flat space. Special conformal maps are defined away from their coordinate singularities.}\label{table:conformal-group}
\end{table}
A special conformal transformation is an inversion, followed by a translation and a second inversion. Inversion itself is discrete and is not a generator of the connected Lie algebra. Counting the continuous parameters gives $(d+1)(d+2)/2$ generators. Their Lie algebras are
\begin{equation}\label{conformal-group}
\operatorname{conf}(\mathbb R^{1,d-1})\cong\mathfrak{so}(2,d),\qquad
\operatorname{conf}(\mathbb R^d)\cong\mathfrak{so}(1,d+1).
\end{equation}
These are statements about Lie algebras; global groups can differ by covers and discrete quotients. The same algebras will appear as AdS isometries.

\subsubsection*{Primary operators and radial quantisation}
To discuss local operators and unitarity, Wick rotate to Euclidean space and use radial quantisation. Spheres centred on the origin are constant-time slices, and logarithmic radius is time. We choose the radial-quantisation generators with no factors of $i$ in
\begin{equation}\label{eq:commD}
[D,P_\mu]=P_\mu,\qquad [D,K_\mu]=-K_\mu.
\end{equation}
Here $D$ is the dilation Hamiltonian, $P_\mu$ generates descendants and $K_\mu$ lowers dimension. This convention is for operators on the radially quantised Hilbert space, rather than differential operators acting on coordinates. In a unitary theory $D^\dagger=D$ and $P_\mu^\dagger=K_\mu$.

A local primary at the origin creates a state $|\mathcal O\rangle$ satisfying
\[
D|\mathcal O\rangle=\Delta|\mathcal O\rangle,\qquad
K_\mu|\mathcal O\rangle=0.
\]
It also transforms in a representation of the rotation group. Acting with $n$ translation generators gives descendants of dimension $\Delta+n$. Although the spin representation is finite-dimensional, the resulting conformal representation is generally infinite-dimensional. A lower bound on dimensions permits a lowest-weight state; finiteness of the spin representation alone does not establish this.

For a scalar primary, the finite dilation law is
$\mathcal O'(a x)=a^{-\Delta}\mathcal O(x)$. Translation and rotation symmetry make a scalar two-point function depend only on $|x-y|$. Dilations fix its power, and special conformal invariance requires equal dimensions for a nonzero two-point function of primaries. In a basis diagonalising their two-point functions,
\[
\langle\mathcal O_i(x)\mathcal O_j(y)\rangle
=\frac{C_i\delta_{ij}}{|x-y|^{2\Delta_i}}.
\]
The constants $C_i$ depend on operator normalisation. Computing these constants from a bulk action will provide our first quantitative application of the dictionary.

Positivity of descendant norms gives the unitarity bounds \cite{Qualls:2015qjb}. For nonidentity scalar primaries in $d\geq3$, $\Delta\geq(d-2)/2$; for spinor primaries, $\Delta\geq(d-1)/2$; and for symmetric traceless tensor primaries of integer spin $s\geq1$, $\Delta\geq d+s-2$. The identity has $\Delta=0$. A conserved current and stress tensor saturate the spin-one and spin-two bounds, respectively, with dimensions $d-1$ and $d$. These dimensions will be reflected in the gauge invariances of their dual bulk fields.

\subsection{Anti-de Sitter space}

An anti-de Sitter (AdS) space is a maximally symmetric manifold with constant negative scalar curvature.
It is a solution of Einstein's equations for an empty universe with negative cosmological constant.
The easiest way to understand it is as follows.

A Lorentzian AdS$_{d+1}$ space can be illustrated by the hyperboloid in $(2,d)$ Minkowski space:
\begin{align}
 X_0^2 +X_{d+1}^2 -\sum_{i=1}^{d} X_i^2 &= R^2 \ .
 \label{embedding}
\end{align}
The metric can be naturally induced from the Minkowski space
\begin{align}\nonumber
 ds^2 &= -dX_0^2 -dX_{d+1}^2 +\sum_{i=1}^{d} dX_i^2 \ .
\end{align}
By construction, it has $\SO(2,d)$ isometry, which is the first connection to
the conformal group in $d$-dim.

\subsubsection*{Global coordinate}

A simple solution to \eqref{embedding} is given as follows.
\begin{align}\nonumber
 &X_0^2 +X_{d+1}^2 = R^2 \cosh^2 \rho \ , \cr
 &\sum_{i=1}^{d} X_i^2 = R^2 \sinh^2 \rho \ .
\end{align}
Or, more concretely,
\begin{align}\nonumber
 X_{0} &= R \cosh \rho\ \cos \tau \ , \qquad
 X_{d+1} = R \cosh \rho\ \sin \tau \ ,  \nonumber \cr
 X_i &= R \sinh \rho\ \Omega_{i} \quad (i=1,\ldots,d,
 \text{ and } \sum_i \Omega_i^2 = 1).
\end{align}
These are $S^{1}$ and $S^{d-1}$ with radii $R\cosh\rho$ and $R\sinh\rho$, respectively.
The metric is
\begin{align}\nonumber
 ds^2 &= R^2 \left( -\cosh^2 \rho \ d\tau^2 +d\rho^2 +\sinh^2 \rho \ d\Omega_{(d-1)}^2 \right) \ .
\end{align}
Note that $\tau$ is a periodic variable and if we take $0 \le \tau <2\pi$,
the coordinate wraps the hyperboloid precisely once.
This is why this coordinate is called the \textbf{global coordinate}.
The manifest sub-isometries are $\SO(2)$ and $\SO(d)$ of $\SO(2,d)$.
To obtain a causal spacetime, we simply unwrap the circle $S^1$,
namely, we take the region $-\infty < \tau < \infty$ with no identification,
which is called the \textbf{universal cover} of the hyperboloid.

In the literature, another global coordinate is also used,
which can be derived by redefinitions
$r \equiv R \sinh \rho$ and $dt \equiv R d\tau$:
\begin{align}\nonumber
 ds^2 &= - f(r) dt^2 +\frac{1}{f(r)} dr^2 + r^2 d\Omega_{(d-1)}^2 \ , \qquad
 f(r) = 1+\frac{r^2}{R^2} \ .
\end{align}

\subsubsection*{Poincar\'e coordinates}

There is yet another coordinate system, called the \textbf{Poincar\'e coordinates}.
As opposed to the global coordinate, this coordinate covers only half of the hyperboloid.
The patch used below is the region $X_0-X_d>0$. Its null boundary is a coordinate horizon, not a curvature singularity.

We define the coordinate as follows.
\begin{align}\nonumber
 X_{0} &= \frac{1}{2u} \left( 1+u^2 \left( R^2 +x_i^2-t^2 \right) \right) \ , \cr
 X_{i} &= R u x_i \qquad (i=1,\ldots,d-1) \ , \cr
 X_{d} &= \frac{1}{2u} \left( 1-u^2 \left( R^2 -x_i^2+t^2 \right) \right) \ , \cr
 X_{d+1} &= R u t \ ,
\end{align}
where $u > 0$.
As it is stated, the coordinate covers half of the hyperboloid; in the region, $X_0 > X_{d}$.
The metric is
\begin{align}\label{ads-poincare}
 ds^2 &= R^2 \left( \frac{du^2}{u^2} +u^2 (-dt^2 +dx_i^2) \right)
 = R^2 \left( \frac{du^2}{u^2} +u^2 dx_\mu^2 \right) \ .
\end{align}
The coordinates $(u,t,x_i)$ are called the \textbf{Poincar\'e coordinates}.
This metric has manifest $ISO(1,d-1)$ and $\SO(1,1)$ sub-isometries of $\SO(2,d)$;
the former is the Poincar\'e transformation and the latter corresponds to the dilatation
\begin{align}\nonumber
 (u,t,x_i) \to (\lambda^{-1}u,\lambda t,\lambda x_i) \ .
\end{align}

If we further define $z=1/u$ ($z>0$), then,
\begin{align}\label{ads-poincare2}
 ds^2 &= \frac{R^2}{z^2} \left( dz^2 +dx_\mu^2 \right) \ .
\end{align}
This is called \textbf{the upper (Poincar\'e) half-plane model}.
The hypersurface given by $z=0$ is called the \textbf{(asymptotic) boundary} of the AdS space,
which corresponds to $u\to\infty$. A conformal rescaling by $z^2/R^2$ gives the boundary metric $dx_\mu^2$; only its conformal class is intrinsic. In global coordinates the conformal boundary is instead the cylinder $\mathbb R\times S^{d-1}$ on the universal cover.

The isometries act on the boundary as conformal transformations. In the Poincar\'e metric \eqref{ads-poincare2}, the dilation $\xi=x^\mu\partial_\mu+z\partial_z$ is a Killing vector, since $z$ and $dx^\mu$ scale together. A direct calculation of $\mathcal L_\xi g=0$ shows that, for every constant $b^\mu$, so is
\begin{equation}\label{ads-sct}
\xi_{(b)}=\big(2(b\cdot x)x^\mu-b^\mu(x^2+z^2)\big)\partial_\mu+2(b\cdot x)\,z\,\partial_z .
\end{equation}
At $z=0$ it reduces to $\epsilon^\mu=2(b\cdot x)x^\mu-b^\mu x^2$, the special conformal transformation of Table~\ref{table:conformal-group}, which solves \eqref{eq:def1}. Together with translations and Lorentz transformations these give $(d+1)(d+2)/2$ Killing vectors, the dimension of $\mathfrak{so}(2,d)$ in \eqref{conformal-group}. This is the geometric content of the statement that the conformal group of the boundary is the isometry group of the bulk.

\subsection{Introduction to \texorpdfstring{AdS${}_5$/CFT${}_4$}{AdS5/CFT4} correspondence}

Now let us study the most well-studied example of AdS/CFT correspondence, which relates the interacting sector of 4d $\U(N)$ $\cN=4$ super-Yang-Mills (SYM) and Type IIB string theory on AdS${}_5\times S^5$, which arises from the large number of D3-branes. Type IIB string theory with D3-branes contains two kinds of perturbative excitations, closed strings and open strings.
 If we consider the system at low energies, energies lower than the string scale $1/\ell_s$, then only the massless string states
can be excited. The closed string massless states give a gravity supermultiplet in $D=10$ in Type IIB supergravity as in \S\ref{sec:IIB-SUGRA}. The open string massless
states give an $\cN=4$ vector multiplet in $D=4$, and their low-energy effective theory is $\cN=4$ $\U(N)$
SYM. Therefore, the duality can also be understood as \textbf{open/closed duality}.

\subsubsection*{$\cN=4$ super-Yang-Mills theory}

The low-energy theory on $N$ D3-branes is four-dimensional $\mathcal N=4$ $\U(N)$ SYM. To fix all numerical factors in the dictionary, use Hermitian generators with $\operatorname{Tr}(T^aT^b)=\delta^{ab}$, as for the Chan--Paton factors in \S\ref{sec:intro-Dbrane}, $D_M=\partial_M-i[A_M,\,\cdot\,]$ and $F_{MN}=\partial_MA_N-\partial_NA_M-i[A_M,A_N]$. The bosonic part of the action obtained by dimensional reduction is
\begin{align}\label{YMaction}
S_{\rm bos}=-\frac{1}{2g_{YM}^2}\int d^4x\,\operatorname{Tr}\left[
\frac12F_{\mu\nu}F^{\mu\nu}+D_\mu X_mD^\mu X_m
-\frac12[X_m,X_n][X_m,X_n]\right].
\end{align}
Here the six $X_m$ descend from the internal gauge components. Reduction of a ten-dimensional Majorana--Weyl gaugino gives four four-dimensional Weyl fermions, whose kinetic and Yukawa terms complete the supersymmetric action. The scalar potential is nonnegative for Hermitian $X_m$ and vanishes when they commute. Their eigenvalues then describe separated D3-branes. We work at the conformal vacuum where these expectation values vanish.

In Euclidean signature the theta term is
\[
S_{E,\theta}=-i\theta k,\qquad
k=\frac{1}{8\pi^2}\int\operatorname{Tr}(F\wedge F)\in\mathbb Z
\]
for an $\SU(N)$ instanton with the chosen orientation. Comparing the D3-brane gauge action and axion coupling with the bulk parameters fixes
\begin{equation}\label{ads-couplings}
g_{YM}^2=2\pi g_s,\qquad
\lambda=2g_{YM}^2N,\qquad
\tau_{YM}=\frac{\theta}{2\pi}+\frac{2\pi i}{g_{YM}^2}
=C_0+\frac{i}{g_s}.
\end{equation}
With $\operatorname{Tr}(T^aT^b)=\delta^{ab}$ the canonically normalised coupling of a single D3-brane is $g_{YM}^2=2\pi g_s$ (\S\ref{sec:DBI}); the factor $2$ in the definition of $\lambda$ is chosen so that $\lambda=4\pi g_sN$ is the standard 't~Hooft coupling, with $R^4=\lambda\alpha'^2$. All quantitative results in this chapter use \eqref{ads-couplings}. The free centre-of-mass $\U(1)$ multiplet decouples. The usual bulk correspondence concerns the interacting $\SU(N)$ sector, with the global form of the gauge group and allowed line operators specified when needed.

It is easy to see that 4d $\cN=4$ SYM is classically \textbf{conformal invariant} because the mass dimensions of the fields are
\be
[A_\mu ]=[X^i] =1~, \qquad
[\lambda _a] = \frac{3 }{ 2} ~,
\ee
so that the coupling constant is dimensionless: $[g]=[\theta]=0$. However, one has to be careful at the quantum level because quantum correction generally breaks the conformal invariance. To be conformal invariant at the quantum level, the beta function of the coupling constant has to vanish $\b=0$. It turns out that 4d $\cN=4$ SYM is the case, and hence, it is quantum mechanically conformal.
The $\cN=4$ supersymmetry combined with conformal symmetry forms the superconformal group $\mathrm{PSU}(2,2|4)$ which consists of the following generators

\begin{itemize}
\item \textbf{Conformal Symmetry} is $\mathfrak{so}(2,4) \cong \mathfrak{su}(2,2)$ at the Lie-algebra level in $d=4$, as we have seen in \S\ref{sec:conformal}. The generators consist of translations $P^\mu$, Lorentz transformations $L_{\mu\nu}$, dilatations $D$ and special conformal transformations $K^\mu$ with Lie algebra \eqref{conformal-group};

\item \textbf{R-symmetry} is $\mathrm{Spin}(6)_R \cong \SU(4)_R$ which is manifest from the 10d viewpoint, and R-symmetry rotates the 6 scalar $X^m$ ($m=1,\ldots ,6$);

\item \textbf{Poincar\'e supersymmetries} are generated  by the supercharges $Q^I_\alpha,\bar Q _{\dot \alpha }^I$,
$I=1,\ldots,4$ that  transform under
the {\bf 4} of $\SU(4)_R$. They can be understood as a ``square root'' of $P_\mu$. Type IIB string theory has 32 supercharges, and D3-branes break half of the supersymmetries. Consequently, the 16 preserved supercharges are indeed $Q^I_\alpha,\bar Q _{\dot \alpha }^I$, which form
$\cN=4$ Poincar\'e supersymmetry;

\item \textbf{Conformal supersymmetries}: are generated by the fermionic generators
$S_{\alpha }^I$ and  $\bar S ^I _{\dot \alpha}$ that are superconformal partners of $Q^I_\alpha,\bar Q _{\dot \alpha }^I$. They can be understood as a ``square root'' of $K_\mu$.
\end{itemize}

The 32 supercharges $Q,\overline Q,S,\overline S$ obey the following anticommutators:
\bea\label{susy-algebra}
&\{Q^{\alpha}_A, {\overline Q}^{\dot \alpha B} \} = P^{\alpha {\dot \alpha}} \delta_A^B, \cr
&\{ S_{\alpha }^A , {\overline S}_{\dot \alpha B}\} = K_{\alpha {\dot \alpha}} \delta^A_B,  \cr
& \{S_{\alpha }^A,Q^{\beta }_B \} = \delta^{A}_{B} M^{~\beta}_{\alpha} +\delta^{\beta}_{\alpha} R^{A}_{B}+ \delta^{A}_{B}\delta^{\beta}_{\alpha} \frac{D}{2}  , \cr
  &\{\overline S_{\dot\alpha A},\overline Q^{\dot\beta B} \} = \delta_{A}^{B} \overline M^{\dot\beta}_{~\dot\alpha} -\delta^{\dot\beta}_{\dot\alpha} R_{A}^{B} + \delta_{A}^{B}\delta^{\dot\beta}_{\dot\alpha} \frac{D}{2} ~.
\eea

The $\cN=4$ SYM enjoys \textbf{S-duality}  \cite{Goddard:1976qe,Montonen:1977sn} that is the $\SL(2,\bZ)$ action on the complexified coupling constant $\tau \equiv \frac{\theta}{2 \pi} + \frac{ 2 \pi i}{g_{YM}^2}$
\be\label{SL2-SYM}
\tau \to \frac{a \tau + b}{c \tau + d}
\qquad \begin{pmatrix}a&b\\ c&d\end{pmatrix} \in \SL(2,\bZ)~.
\ee
The $\cN=4$ SYM is the world-volume theory on a stack of D3-branes, and D3-branes are invariant under the $\SL(2,\bZ)$ symmetry of Type IIB theory as seen in \S\ref{sec:Sdual}. In fact, this is the origin of the  $\SL(2,\bZ)$ symmetry \eqref{SL2-SYM}.

There is another way to realize the $\cN=4$ SYM from string theory. Indeed, the M5-branes wrapped on a torus with complex structure $\tau$ give rise to the  $\cN=4$ SYM with the complexified coupling constant $\tau$. Here, $\tau$ manifestly admits a geometric origin as the complex structure of a torus.
Note that when $\theta =0$, the S-duality transformation amounts to
$g_{YM}\to 2\pi/g_{YM}$, thereby exchanging strong and weak coupling.\footnote{Precisely speaking, the electromagnetic duality of the $\cN=4$ SYM depends on a choice of gauge groups, and the duality group is usually a congruence subgroup of $\SL(2,\bZ)$. For more detail, we refer to \cite{Aharony:2013hda}.}

\subsubsection*{Near-horizon geometry of D3-branes}

To gain a deeper understanding of geometry, let us now study the closed string side of the system. A system of $N$ coincident D3-branes is a classical solution of the low-energy string effective action, characterized by the metric and R-R field.

The metric of the D3-brane solution is given by:
\bea
\label{D3metric}
ds^2 = H(y)^{-\frac{1}{2}} \eta_{\mu\nu} dx^\mu dx^\nu + H(y)^{\frac{1}{2}} (dy^2 + y^2 d\Omega_5^2),
\eea
where $y$ is the coordinate transverse to the brane and $d\Omega_5^2$ represents the metric on a five-sphere. The function $H(y)$ is given by:
\bea\label{acca}
H(y) = 1 + \frac{R^4}{y^4}, \quad \textrm{where} \quad R^4 = 4\pi g_s N (\alpha')^2,
\eea
with $g_s$ being the string coupling constant and $\alpha'$ representing the Regge slope.

\begin{figure}[htp]
\centering
\includegraphics[width=10cm]{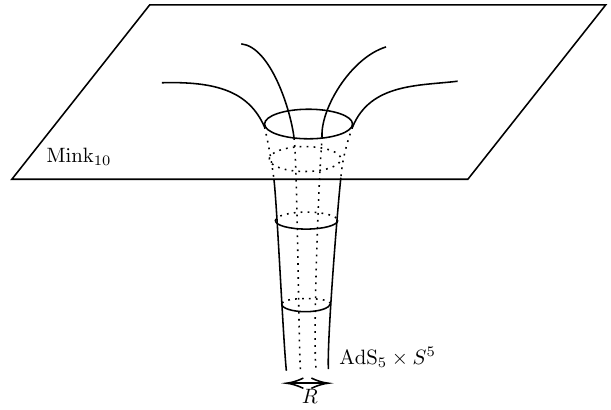}
\caption{Minkowski region and near-horizon region}
\label{fig:throat}
\end{figure}

In the region where $y \gg R$, the D3-brane solution smoothly transitions to flat spacetime $\mathbb{R}^{1,9}$, which is the Minkowski space.

On the other hand, when $y < R$, the geometry undergoes a significant transformation and is often referred to as the ``throat''. Initially, it may seem that the geometry becomes singular as $y$ becomes much smaller than $R$. However, this is not the case. By examining the near-horizon region described by the following limit
\be
y \rightarrow 0 \hspace{2cm} \alpha ' \rightarrow 0 \hspace{2cm} u \equiv  y/R^2
\label{lim}
\ee
where $\alpha'$ is taken to zero while $u$ remains fixed. In this limit, we can safely neglect the factor $1$ in the function $H(y)$ appearing in \eqref{acca}. Consequently, the metric simplifies to:
\bea\label{nmet}
ds^2 = R^2 \biggl [ u^2 \eta _{\mu\nu} dx^\mu dx^\nu  + \frac{du^2}{u^2}  + d\Omega _5 ^2  \biggr ]
\eea
As we have previously observed in \eqref{ads-poincare}, the first part of the metric corresponds to AdS$_5$, while the second part corresponds to $S^5$. Remarkably, in the vicinity of the brane where $y\sim 0$ or equivalently $u\sim 0$, the geometry exhibits remarkable regularity and symmetry, manifesting as AdS$_5\times S^5$ with the same radii
\[
R_{\textrm{AdS}_5}^{2} = R_{S^5}^{2}= \alpha' \sqrt{4 \pi N g_{s}} ~.
\]
where $N$ represents the number of coincident D3-branes and $g_s$ denotes the string coupling constant.

In this limit described by \eqref{lim}, the dynamics in the asymptotically flat region of the D3-brane geometry decouples from the theory. The interacting sector is described by the near-horizon geometry; the asymptotically flat sector remains as a decoupled free theory. Moreover, the interactions between the bulk and the brane dynamics become negligibly small. This limiting regime is aptly referred to as the ``decoupling limit''.
By studying the decoupling limit, we can focus solely on the AdS region of the D3-brane and unravel its remarkable connections to conformal field theory (CFT) through the AdS/CFT correspondence.

\subsubsection*{The AdS/CFT correspondence}

As mentioned, the world volume theory of $N$ coincident D3-branes is 4d ${\cal{N}} =4$
SYM with $\U(N)$ gauge group. On the other hand, the classical solution in \eqref{nmet} is a good
approximation when the radii of AdS$_5$ and $S^5$ are very big:
\be \label{same-radii}
\frac{R^{2}}{\alpha^{\prime}} \gg 1 \ \Longrightarrow \  2N g_{Y M}^{2} \equiv \lambda \gg 1
\ee
The fact that those two descriptions are simultaneously consistent for
large values of the coupling constant $\lambda$ led Maldacena
to formulate the conjecture that the strongly interacting
${\cal{N}}=4$ SYM with gauge group $\SU(N)$ at large $N$ is equivalent
to Type IIB supergravity compactified on AdS$_5 \times S^5$. However, supergravity is not a consistent quantum theory and it is just a low-energy effective theory of string theory. Hence, the natural way to extend the equivalence at any value of $\lambda$ is therefore  that ${\cal{N}}=4$ SYM is equivalent to Type IIB string theory on
AdS$_5 \times S^5$~\cite{Maldacena:1997re}. Namely, the following two theories are dual to each other:
\begin{itemize}
\item $\cN=4$ super-Yang-Mills theory in 4-dimensions with gauge group
$\SU(N)$, after removing the free centre-of-mass multiplet.
\item Type IIB superstring theory on  AdS$_5\times S^5$
with the same radius $R$ as in \eqref{same-radii}, with $N$ units of quantised self-dual five-form flux through $S^5$.
\end{itemize}
The coupling constants are related by \eqref{ads-couplings}. The precise formulation of this duality will follow in the next subsection.
In these two theories, we can immediately find the symmetry correspondence in the first rows of Table~\ref{table:dictionary}. The remaining rows record the quantitative entries of the dictionary, which are derived in the following subsections.

\begin{table}[ht]
\begin{center}
\begin{tabular}{c|c}
4d $\cN=4$ SYM & Type IIB on AdS$_{5}\times S^5$ \\
\hline
32 supercharges & 32 supercharges \\
$\SO(2,4)$ conformal group& $\SO(2,4)$ isometry of AdS$_{5}$ \\
$\SU(4)_R$ symmetry & $\SO(6)$ isometry of $S^5$\\
$\SL(2,\bZ)$ symmetry of coupling constants & $\SL(2,\bZ)$ symmetry of axio-dilaton\\
\hline
$\sim N^2$ adjoint degrees of freedom & $R^3/G_5=2N^2/\pi$ \eqref{ads-newton}\\
$\lambda=2g_{YM}^2N$ & $R^4/\alpha'^2$ \eqref{ads-expansions}\\
Scalar operator of dimension $\Delta$ & Scalar field with $\Delta(\Delta-d)=m^2R^2$ \eqref{dimenmass}\\
Conserved current, $\Delta=d-1$ & Massless gauge field\\
Stress tensor, $\Delta=d$ & Metric fluctuation\\
Thermal state at temperature $T$ & Planar black brane \eqref{ads-black-brane}\\
Wilson loop & Fundamental-string world-sheet
\end{tabular}
\end{center}
\caption{Dictionary for the AdS${}_5$/CFT${}_4$ correspondence}
\label{table:dictionary}
\end{table}

This conjecture is the most general statement, which is valid at any values of coupling and rank $N$. However, it is still difficult to quantize string theory at any value of $g_s$ on a general manifold, including asymptotic AdS space. Hence, it is still an open problem to prove this general statement of the conjecture. Nevertheless, taking various limits of the conjecture, we can show a variety of non-trivial evidence for the conjecture, providing new physical insight.

\vspace{.5cm}

\begin{figure}[ht]\centering
\includegraphics[width=8cm]{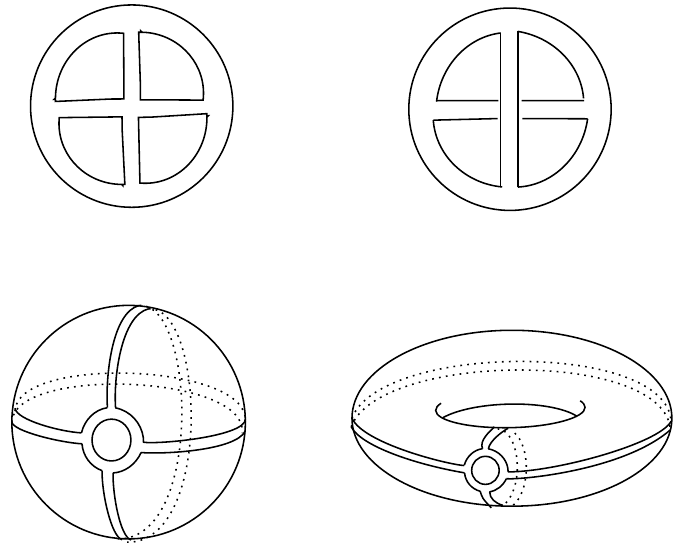}
\caption{Planar and non-planar diagram}\label{fig:planar-non}
\end{figure}

\noindent\textbf{The 't~Hooft Limit}: The 't~Hooft limit \cite{tHooft:1973alw} is the limit in which we keep the \textbf{'t~Hooft coupling} $\lambda \equiv 2g_{YM}^2 N = 4\pi g_s N$ fixed and letting $N\to \infty$, $g_s\to 0$. As in Figure \ref{fig:planar-non}, the planar
diagrams become dominant in this limit on the Yang-Mills side. On the AdS side, since the string coupling can be re-expressed in terms of the 't~Hooft coupling as $g_s = \lambda /(4\pi N)$, the 't~Hooft limit corresponds to the regime where weak coupling string perturbation theory is valid.
At small $\lambda$ the AdS radius is smaller than the string length, so the bulk is highly curved and the gauge theory is the useful description; the string side becomes a classical geometry only at large $\lambda$. Put differently, this limit of the AdS/CFT correspondence can be understood as the incarnation of the idea of 't~Hooft \cite{tHooft:1973alw}.

\vspace{.5cm}

\noindent\textbf{Supergravity limit}: While we take $N\to\infty$, in the regime that
the 't Hooft coupling is large $\lambda=4\pi g_s N \gg 1$, supergravity description becomes reliable.
On the gauge theory side, the theory is strongly-coupled so that perturbation techniques cannot be applied.  The two theories are conjectured to be the same, but when one side is weakly coupled, the other is strongly coupled.
This is a salient feature of \textbf{duality}. Thus, using the AdS/CFT correspondence, supergravity calculations give access to strongly coupled correlation functions and thermal observables. Undeformed $\mathcal N=4$ SYM on Minkowski space in its conformal vacuum has neither a confinement scale nor a mass gap; holographic models of confinement require a different background or deformation.

\subsection{The parameters and the radial energy scale}
\label{sec:ads-regimes}
Combining \eqref{acca} and \eqref{ads-couplings} makes the two expansions transparent:
\begin{equation}\label{ads-expansions}
\frac{R^4}{\alpha'^2}=\lambda,\qquad
\frac{\alpha'}{R^2}=\lambda^{-1/2},\qquad
g_s=\frac{\lambda}{4\pi N}.
\end{equation}
Large $\lambda$ suppresses curvature corrections in string units, while small $g_s$ suppresses string loops. A useful sufficient regime for classical Type IIB supergravity is $1\ll\lambda\ll N$. The condition $\lambda\ll N$ is not necessary for the classical gravity approximation, since the bulk loop-counting parameter $G_5/R^3\sim1/N^2$ is small at any $\lambda$; it only guarantees $g_s\ll1$, so that no S-duality transformation is needed to describe the string theory. In the planar limit at fixed $\lambda$, handles are suppressed but curvature corrections need not be small. A typical massive string excitation has $m\sim1/\ell_s$, hence $mR\sim\lambda^{1/4}$; such states become heavy relative to the AdS scale only at strong 't Hooft coupling. Kaluza--Klein masses are instead of order $1/R$ and cannot all be discarded when computing general CFT observables.

The rank also controls the gravitational action. At constant dilaton, choose the metric normalisation in which
\[
16\pi G_{10}=(2\pi)^7g_s^2\alpha'^4,\qquad
\operatorname{Vol}(S^5)=\pi^3R^5.
\]
Reduction of the Einstein term gives
\begin{equation}\label{ads-newton}
G_5=\frac{G_{10}}{\pi^3R^5}
=\frac{\pi R^3}{2N^2},\qquad
\frac{R^3}{G_5}=\frac{2N^2}{\pi}.
\end{equation}
Thus the classical bulk action scales as $N^2$, in agreement with the number of adjoint degrees of freedom. The exact gauge-theory count is $N^2-1$ for $\SU(N)$; a classical calculation captures its leading large-$N$ term.

A sharper check of \eqref{ads-newton} is the Weyl anomaly. For a conformal field theory dual to Einstein gravity in AdS$_5$, the two anomaly coefficients are equal, $a=c=\pi R^3/(8G_5)$ \cite{Skenderis:2002wp}, in the normalisation in which a real scalar contributes $c=1/120$. Inserting \eqref{ads-newton} gives $c=N^2/4$. For free fields $c=(N_s+6N_D+12N_V)/120$, with $N_s$ real scalars, $N_D$ Dirac fermions and $N_V$ gauge fields. $\mathcal N=4$ $\SU(N)$ SYM has $N_s=6(N^2-1)$, $N_D=2(N^2-1)$ and $N_V=N^2-1$, hence $c=(N^2-1)/4$. The agreement at leading order in $N$ is expected to hold at every $\lambda$, because $a$ and $c$ are protected by supersymmetry. This contrasts with the entropy density below, which is not protected and differs from its free value by a factor of $3/4$ at infinite coupling.

The near-horizon limit has a useful energy interpretation. A fundamental string stretched a transverse distance $y$ has energy $y/(2\pi\alpha')$. Keeping $U=y/\alpha'$ fixed while $\alpha'\to0$ therefore retains finite brane excitation energies. At fixed $g_sN$, this agrees with holding $u=y/R^2=U/\sqrt\lambda$ fixed in \eqref{lim}. It is not a limit at fixed $y$.

In Poincar\'e coordinates, a stationary bulk excitation of local proper energy $E_{\rm proper}$ has boundary-coordinate energy
\begin{equation}\label{ads-redshift}
E=\sqrt{-g_{tt}}\,E_{\rm proper}=\frac{R}{z}E_{\rm proper}.
\end{equation}
The dilation isometry is $(x^\mu,z)\mapsto(a x^\mu,a z)$. Boundary distances and radial depth scale together: small $z$ probes ultraviolet scales, whereas larger $z$ probes infrared scales. This is the \textbf{UV/IR relation}. It does not mean that each point in the bulk is a single boundary energy eigenstate; a field profile generally samples a range of scales.

\subsection{GKPW relation and boundary conditions}
\label{sec:ads-gkpw}
The prescription of Gubser, Klebanov, Polyakov and Witten \cite{Gubser:1998bc,Witten:1998qj} promotes the symmetry dictionary to an equality of generating functionals. For a real scalar operator and a Euclidean source $\phi_0$, choose the convention
\begin{equation}\label{GKPW}
Z_{\rm string}[\phi_0]
=\left\langle\exp\left(\int d^dx\,\phi_0\mathcal O\right)\right\rangle_{\rm CFT},
\qquad W[\phi_0]=\log Z_{\rm string}[\phi_0].
\end{equation}
The boundary condition specifies the leading asymptotic coefficient of the bulk field, not usually its unrescaled value at $z=0$. The string path integral is over all bulk fields compatible with the sources. In a regime with a dominant classical saddle,
\[
W[\phi_0]\simeq-S_{\rm ren}[\phi_{\rm cl}[\phi_0]],
\qquad
\langle\mathcal O(x_1)\cdots\mathcal O(x_n)\rangle_{\rm conn}
=\left.\frac{\delta^n W}{\delta\phi_0(x_1)\cdots\delta\phi_0(x_n)}\right|_{\phi_0=0}.
\]
Here $S_{\rm ren}$ is the renormalised on-shell action. When several saddles compete, their contributions must be compared before making this approximation. The choice of state and interior regularity condition is part of the prescription; in Lorentzian signature it also determines the appropriate causal correlator; for instance, retarded correlators at finite temperature require ingoing conditions at the horizon.

\subsubsection*{Deriving the scalar mass--dimension relation}
For a scalar of mass $m$ in Euclidean AdS$_{d+1}$, take
\begin{equation}\label{ads-scalar-action}
S_E=\frac{\mathcal K}{2}\int d^{d+1}x\sqrt g\,
\left(g^{MN}\partial_M\phi\partial_N\phi+m^2\phi^2\right),
\qquad \mathcal K>0.
\end{equation}
The constant $\mathcal K$ fixes the operator normalisation. Using $\sqrt g=(R/z)^{d+1}$, its equation of motion becomes
\begin{equation}\label{ads-scalar-wave}
\left[z^2\partial_z^2-(d-1)z\partial_z+z^2\partial_\mu\partial_\mu-m^2R^2\right]\phi=0.
\end{equation}
Near the boundary the radial terms determine the leading powers. Substituting $\phi\sim z^a$ gives $a(a-d)=m^2R^2$, with roots
\begin{equation}\label{dimenmass}
\Delta_\pm=\frac d2\pm\nu,\qquad
\nu=\sqrt{\frac{d^2}{4}+m^2R^2},\qquad
\Delta\equiv\Delta_+.
\end{equation}
For nondegenerate roots the expansion has the form
\begin{equation}\label{ads-source-response}
\phi(z,x)=z^{d-\Delta}\bigl(\phi_0(x)+\cdots\bigr)
+z^\Delta\bigl(A(x)+\cdots\bigr).
\end{equation}
The omitted terms include derivatives of the source and, for special masses, logarithms. In standard quantisation $\phi_0$ is the source and $A$ determines the response. The coupling $\int d^dx\,\phi_0\mathcal O$ is dimensionless, so $[\phi_0]=d-\Delta$ and $[\mathcal O]=\Delta$. The formula is for scalars; fields with spin obey different mass--dimension relations.

Negative $m^2$ is not automatically an instability in AdS. Reality of the exponents gives the Breitenlohner--Freedman bound
\begin{equation}\label{ads-bf}
m^2R^2\geq-\frac{d^2}{4}.
\end{equation}
For a free scalar with the standard AdS boundary conditions this is also the stability bound. Boundary conditions matter: when $0<\nu<1$, both falloffs are normalisable in the appropriate renormalised norm, and an \textbf{alternative quantisation} with operator dimension $\Delta_-$ is possible \cite{Klebanov:1999tb}. Source and response exchange roles, and the generating functional is related by a Legendre transform. At $\nu=0$ the roots coincide and the second solution is $z^{d/2}\log z$; the endpoints of the alternative window require separate treatment. Thus the labels ``non-normalisable'' and ``normalisable'' for the two branches are useful in the generic standard case but are not universal.

\subsubsection*{Other fields and their operators}
The boundary metric sources the stress tensor, and a boundary value of a massless bulk gauge field sources a conserved current. Conservation is enforced by the corresponding bulk gauge invariance. A bulk gauge redundancy therefore encodes a global symmetry of the boundary theory under the usual boundary conditions; it is not an additional gauge redundancy acting on boundary observables.

The dimensions of these operators follow by the same asymptotic analysis. Take $S=\frac1{4g_5^2}\int d^{d+1}x\sqrt g\,F_{MN}F^{MN}$ in Poincar\'e AdS$_{d+1}$ with $d>2$, in the gauge $A_z=0$. Since $\sqrt g\,g^{\mu\rho}g^{\nu\sigma}=R^{d-3}z^{3-d}\eta^{\mu\rho}\eta^{\nu\sigma}$, the $\mu$-components give $\partial_z(z^{3-d}\partial_zA_\mu)+z^{3-d}\partial^\nu F_{\nu\mu}=0$, whose near-boundary solution is
\[
A_\mu=a_\mu(x)+z^{d-2}j_\mu(x)+\cdots .
\]
The coupling $\int d^dx\,a_\mu J^\mu$ is dimensionless and $[a_\mu]=1$, so the dual current has $\Delta=d-1$. The $z$-component of Maxwell's equations, $\partial^\mu\partial_zA_\mu=0$, then gives $\partial^\mu j_\mu=0$: current conservation is the boundary shadow of the constraint associated with bulk gauge invariance, and $a_\mu\to a_\mu+\partial_\mu\lambda$ is consistent with it. Likewise the source for the stress tensor is the boundary metric $g_{(0)}^{\mu\nu}$, which is dimensionless, so $\Delta=d$. Both are special cases of the formula $(\Delta-p)(\Delta+p-d)=m^2R^2$ for a $p$-form, at $m=0$ with $p=1$ for the gauge field.

In the large-$N$ description, elementary bulk single-particle fields correspond to appropriate single-trace operators, while multiparticle states are represented by products of operators. This distinction prevents interpreting the dictionary as a bijection between elementary bulk fields and every local CFT operator. A consistent truncation to a few bulk fields can compute a restricted class of observables; it does not remove the rest of the dual theory.

\subsection{A scalar two-point function from the bulk}
\label{sec:ads-two-point}
We now implement the prescription on $\mathbb R^d$ for $\Delta>d/2$ in standard quantisation. The solution regular in the interior is obtained with a bulk-to-boundary kernel,
\begin{align}\label{ads-kernel}
\phi(z,x)&=\int d^dy\,\mathcal B_\Delta(z,x;y)\phi_0(y),\nonumber\\
\mathcal B_\Delta(z,x;y)&=C_\Delta
\left(\frac{z}{z^2+|x-y|^2}\right)^\Delta,
\qquad
C_\Delta=\frac{\Gamma(\Delta)}{\pi^{d/2}\Gamma(\Delta-d/2)}.
\end{align}
Substitution verifies \eqref{ads-scalar-wave}. Its normalisation follows by setting $y=x+zv$:
\[
\int d^dy\,\mathcal B_\Delta(z,x;y)
=C_\Delta z^{d-\Delta}\int d^dv\,(1+v^2)^{-\Delta}
=z^{d-\Delta}.
\]
The last integral is $\pi^{d/2}\Gamma(\Delta-d/2)/\Gamma(\Delta)$. Consequently $z^{\Delta-d}\mathcal B_\Delta$ approaches a delta function, which gives the required boundary source for smooth test functions. At separated points, the coefficient of $z^\Delta$ in the solution is
\[
A(x)=C_\Delta\int d^dy\,\frac{\phi_0(y)}{|x-y|^{2\Delta}},
\]
understood as a regulated distribution at $x=y$.

Integration by parts in \eqref{ads-scalar-action} leaves the boundary term
\begin{equation}\label{ads-regulated-action}
S_{E,\epsilon}=-\frac{\mathcal K R^{d-1}}2
\int_{z=\epsilon}d^dx\,\epsilon^{1-d}\phi\partial_z\phi.
\end{equation}
The sign comes from the outward normal $n^z=-z/R$ for the retained region $z\geq\epsilon$. Substitution of the asymptotic expansion produces divergences local in $\phi_0$. These must be subtracted before taking $\epsilon\to0$; simply keeping a finite-looking term gives the wrong normalisation.

For example, the leading counterterm is
\[
S_{\rm ct}^{(0)}=\frac{\mathcal K(d-\Delta)}{2R}
\int_{z=\epsilon}d^dx\sqrt\gamma\,\phi^2,
\]
where $\gamma$ is the induced metric. It cancels the leading power divergence. Additional derivative counterterms, and logarithmic ones when required, cancel the remaining local divergences \cite{Skenderis:2002wp}. The nonlocal finite part is
\[
S_{\rm ren}^{\rm nonlocal}
=-\frac{\mathcal K R^{d-1}}2(2\Delta-d)\int d^dx\,\phi_0(x)A(x).
\]
With the sign convention in \eqref{GKPW}, differentiating $W=-S_{\rm ren}$ yields
\begin{equation}\label{ads-two-point}
\langle\mathcal O(x)\mathcal O(y)\rangle
=\mathcal K R^{d-1}(2\Delta-d)
\frac{\Gamma(\Delta)}{\pi^{d/2}\Gamma(\Delta-d/2)}
\frac{1}{|x-y|^{2\Delta}},\qquad x\ne y.
\end{equation}
Its power is fixed by conformal invariance; the bulk calculation supplies the normalisation once $\mathcal K$ is chosen. Finite local counterterms change contact terms, not this separated-point answer. The derivation assumes a free bulk scalar at quadratic order, vanishing background source and a fixed AdS geometry. Bulk interactions generate higher connected correlators through Witten diagrams, with bulk vertices integrated over AdS.

\subsection{Kaluza--Klein modes and protected operators}
Consider a minimally coupled ten-dimensional massless scalar on AdS$_5\times S^5$, such as the linearised dilaton about the constant-dilaton solution. Expand it in scalar spherical harmonics:
\begin{equation}\label{spherical-harmonics}
\phi(x,z,\Omega)=\sum_l\phi_l(x,z)Y_l(\Omega),\qquad
\nabla_{S^5}^2Y_l=-\frac{l(l+4)}{R^2}Y_l,\quad l=0,1,\ldots.
\end{equation}
The ten-dimensional wave equation gives $m_l^2R^2=l(l+4)$, and hence
\begin{equation}\label{Delta-mass-4d}
\Delta_l=2+\sqrt{4+l(l+4)}=l+4.
\end{equation}
For $l=0$, the dilaton couples to the supersymmetric Lagrangian operator, which includes $\operatorname{Tr}F_{\mu\nu}F^{\mu\nu}$ together with scalar and fermion terms. Being an R-symmetry singlet does not forbid such scalar or fermion contributions. The axion couples to the corresponding topological density.

Higher harmonics transform as symmetric traceless tensors of $\SO(6)$. Their duals contain schematic terms $\operatorname{Tr}(X_{(i_1}\cdots X_{i_l)}F^2)$ with traces subtracted, but require supersymmetric completion and, where appropriate, operator mixing. Their protected dimensions cannot be established by naive engineering dimensions alone. This tower should not be confused with the chiral primaries $\operatorname{Tr}(X_{(i_1}\cdots X_{i_k)})$, whose dimensions are $k$ and whose bulk scalars arise from mixtures of metric and five-form fluctuations. The two towers are nevertheless related by supersymmetry. The chiral primary $\operatorname{Tr}X^{(i_1}\cdots X^{i_k)}$ is half-BPS with $\Delta=k$ in the $[0,k,0]$ of $\SU(4)_R$. Acting with four supercharges raises the dimension by two and lowers the $\SO(6)$ spin by two units, giving $\Delta=k+2$ in $[0,k-2,0]$. For $k=l+2$ this is exactly the quantum numbers $\Delta=l+4$ of the $l$-th dilaton harmonic. For $\SU(N)$ the series starts at $k=2$, because $\operatorname{Tr}X$ vanishes; the $l=0$ mode is the Lagrangian density, the fourth descendant of the stress-tensor multiplet. Matching masses, representations and supermultiplets gives a stronger test than matching a list of symmetries.

\subsection{Thermal states and black-hole entropy}
\label{sec:ads-thermal}
A thermal boundary state is described by a Euclidean time circle of period $\beta=1/T$. On an infinite flat spatial boundary, the relevant classical solution is the planar AdS$_5$ black brane,
\begin{equation}\label{ads-black-brane}
ds_5^2=\frac{r^2}{R^2}\left[-f(r)dt^2+d\vec x^{\,2}\right]
+\frac{R^2}{r^2f(r)}dr^2,
\qquad f(r)=1-\frac{r_h^4}{r^4}.
\end{equation}
It is accompanied by the $S^5$ and five-form flux of the vacuum solution. The Hawking temperature is fixed by smoothness of the Euclidean metric. Setting $r=r_h+\delta r$ gives $f\simeq4\delta r/r_h$. With $\delta r=r_h\rho^2/R^2$, the radial-time part becomes
\[
ds_{E,2}^2\simeq d\rho^2+\rho^2\left(\frac{2r_h}{R^2}dt_E\right)^2.
\]
Avoiding a conical singularity requires
\[
\beta=\frac{\pi R^2}{r_h},\qquad T=\frac{r_h}{\pi R^2}.
\]
The horizon area per boundary coordinate volume $V_3$ is $A_h/V_3=(r_h/R)^3$. Using \eqref{ads-newton} and the Bekenstein--Hawking formula gives
\begin{equation}\label{ads-entropy}
s\equiv\frac{S}{V_3}=\frac{r_h^3}{4G_5R^3}
=\frac{\pi^2}{2}N^2T^3.
\end{equation}
This is a strong-coupling, leading-large-$N$ result. At zero coupling, the eight bosonic and eight fermionic physical degrees of freedom per adjoint generator instead give
\[
s_{\rm free}=\frac{2\pi^2}{45}\left(8+\frac78\,8\right)(N^2-1)T^3
=\frac{2\pi^2}{3}(N^2-1)T^3.
\]
Thus $s/s_{\rm free}\to3/4$ at large $N$ and infinite $\lambda$. The two values refer to different couplings; their difference is not a contradiction. String corrections change the coefficient away from infinite coupling \cite{Gubser:1998nz}.

\subsubsection*{A boundary sphere and competing saddles}
On $\mathbb R\times S^3$ with sphere radius $R$, thermal AdS competes with a global Schwarzschild--AdS$_5$ black hole. Its metric is $ds^2=-F(r)dt^2+dr^2/F(r)+r^2d\Omega_3^2$, where
\begin{equation}\label{ads-global-thermal}
F(r)=1+\frac{r^2}{R^2}-\frac{\mu}{r^2},\quad
\mu=r_h^2\left(1+\frac{r_h^2}{R^2}\right),\quad
T=\frac{2r_h^2+R^2}{2\pi R^2r_h}.
\end{equation}
Relative to thermal AdS with the same boundary geometry, the energy and entropy are $E=3\pi\mu/(8G_5)$ and $S=\pi^2r_h^3/(2G_5)$. Their combination is
\[
\mathcal F_{\rm BH}-\mathcal F_{\rm thermal\ AdS}
=E-TS=\frac{\pi r_h^2}{8G_5}\left(1-\frac{r_h^2}{R^2}\right).
\]
The sign changes at $r_h=R$. This Hawking--Page transition is a change in the dominant saddle of the canonical partition function and has a large-$N$ gauge-theory interpretation \cite{Witten:1998zw}. The finite boundary radius supplies a scale. It must not be confused with a confinement scale in the zero-temperature conformal theory on infinite Minkowski space.

\subsection{Wilson loops and the force between external probes}
\label{sec:ads-wilson}
A further test uses a heavy external quark and antiquark separated by $L$. In $\mathcal N=4$ SYM the standard string construction computes a Wilson operator with a specified coupling to the six adjoint scalars, often called the Maldacena--Wilson loop \cite{Maldacena:1998im}. Fixing the endpoint position on $S^5$ fixes this scalar coupling. For a rectangular contour of Euclidean time extent $\mathcal T\gg L$, its expectation value behaves as $\exp[-\mathcal T V(L)]$ after subtracting the rest masses of the probes.

At large $N$ and large $\lambda$, the leading contribution is the area of a fundamental-string world-sheet ending on the contour. Use $t_E=\tau$, $x=\sigma$, and $z=z(x)$ in Euclidean Poincar\'e AdS. The Nambu--Goto action per unit time is
\[
\frac{S_{\rm NG}}{\mathcal T}
=\frac{R^2}{2\pi\alpha'}\int_{-L/2}^{L/2}dx\,
\frac{\sqrt{1+(dz/dx)^2}}{z^2}.
\]
The absence of explicit $x$ dependence gives a first integral. At the midpoint, let the maximum depth be $z_*$, with $z'(0)=0$. Then
\begin{equation}\label{ads-string-profile}
\frac{1}{z^2\sqrt{1+z'^2}}=\frac{1}{z_*^2},\qquad
\frac L2=z_*\int_0^1\frac{v^2\,dv}{\sqrt{1-v^4}}
=z_*\frac{\sqrt\pi\,\Gamma(3/4)}{\Gamma(1/4)}.
\end{equation}
The unrenormalised area diverges at the endpoints. Each straight string extending from the cutoff $z=\epsilon$ to the Poincar\'e horizon has energy $R^2/(2\pi\alpha'\epsilon)$. Subtracting two such strings gives
\begin{align}
V(L)&=\frac{R^2}{\pi\alpha'z_*}
\left[\int_0^1\frac{dv}{v^2}\left(\frac{1}{\sqrt{1-v^4}}-1\right)-1\right]\nonumber\\
&=-\frac{4\pi^2}{\Gamma(1/4)^4}\frac{\sqrt\lambda}{L}.
\label{ads-potential}
\end{align}
The subtraction removes the infinite external masses while retaining their interaction. The potential has the $1/L$ dependence demanded by conformal symmetry, but its coefficient scales as $\sqrt\lambda$ rather than the weak-coupling $\lambda$. A confining flux tube would instead give a potential growing linearly with $L$. Thus a fundamental string in the bulk need not imply confinement of boundary probes: the shape of the geometry is essential.

A second, simpler example is the circular loop of radius $a$. The surface $z^2+r^2=a^2$ in Euclidean Poincar\'e AdS$_5$ solves the Nambu--Goto equations, and with the cutoff $z\geq\epsilon$ its area is $2\pi R^2(a/\epsilon-1)$. Dividing by $2\pi\alpha'$ and subtracting the term proportional to the perimeter $2\pi a$, which is the infinite mass of the probe, leaves $S=-R^2/\alpha'=-\sqrt\lambda$, hence
\begin{equation}\label{ads-circular-wilson}
\langle W_{\circ}\rangle\simeq e^{\sqrt\lambda}.
\end{equation}
The exact planar result of the gauge theory, $\langle W_\circ\rangle=\frac2{\sqrt\lambda}I_1(\sqrt\lambda)$ \cite{Erickson:2000af,Drukker:2000rr}, expands as $1+\lambda/8+\cdots$ at weak coupling and behaves as $\sqrt{2/\pi}\,\lambda^{-3/4}e^{\sqrt\lambda}$ at strong coupling. The classical string thus reproduces the exponent; the power-law prefactor requires the one-loop fluctuation determinant around the classical surface.

The examples in this chapter link world-sheet dynamics, brane geometry and field-theory observables. They also show the limits of each calculation: the exact duality is a conjecture, whereas the explicit formulas above use classical bulk fields or a classical string world-sheet in regimes with controlled curvature and loop corrections. Further applications, including holographic renormalisation-group flows and entanglement entropy, can be approached using these same boundary conditions and variational methods.

\bibliography{references}
\bibliographystyle{hyperamsalpha}

\bigskip

\noindent
Satoshi Nawata, {\sf Department of Physics and Center for Field Theory and Particle Physics, Fudan University,
Songhu Road 2005, 200438 Shanghai, China}, snawata@gmail.com

\

\noindent
Runkai Tao, {\sf Department of Physics and Astronomy, Rutgers University,
126 Frelinghuysen Road, Piscataway NJ 08854, USA}, runkaitao@gmail.com

\

\noindent
Daisuke Yokoyama, {\sf Department of Physics, Meiji University, Kanagawa 214-8571, Japan},  ddyokoyama@meiji.ac.jp

\end{document}